\documentclass[11pt,a4paper]{article}
\usepackage[utf8]{inputenc}
\usepackage[T1]{fontenc}
\usepackage[normalem]{ulem} 
\usepackage{amsmath}
\usepackage{amsfonts}
\usepackage{amssymb}
\usepackage{amsthm}
\usepackage{mathtools}
\usepackage{bbm}
\usepackage{setspace}
\usepackage{graphicx}
\usepackage{upquote}
\usepackage{hyperref}
\usepackage{tikz}
\usetikzlibrary{arrows}
\hypersetup{
        citecolor=black,
	colorlinks=true,
	linkcolor=black,
	filecolor=black,
	urlcolor=black,
}
\usepackage{cleveref}
\usepackage[authoryear]{natbib}
\usepackage{soul}
\usepackage{float}
\usepackage{verbatim}
\usepackage{longtable,booktabs}
\usepackage[font=footnotesize,labelfont=bf]{caption}
\usepackage{subcaption} 
\usepackage{makecell}
\usepackage{xcolor}
\usepackage[margin=0.75in]{geometry}
\usepackage{xr}
\usepackage{caption}
\usepackage{subcaption}
\usepackage{color}
\renewcommand{\baselinestretch}{1}

\author{Rui Feng\footnote{Feng is PhD student, Department of Statistics, University of Warwick, Email: rui.feng.1@warwick.ac.uk} 
 \and  Chenlei Leng\footnote{Leng is Professor, Department of Statistics, University of Warwick, Email: c.leng@warwick.ac.uk. Corresponding author.}}
\title{Regression Analysis of Reciprocity in Directed Networks}
\date{}

\providecommand{\keywords}[1]
{
	\small	
	\textbf{\textit{Key words:}} #1
}
\DeclareMathOperator*{\argmin}{arg\,min}

\newtheorem{theorem}{Theorem}
\newtheorem{Prop}{Proposition}
\newtheorem{Lem}{Lemma}

\theoremstyle{definition}

\newtheorem{Assum}{Assumption}

\newtheorem*{Prop*}{Proposition}
\newtheorem*{theorem*}{Theorem}

\tikzset{
    vertex/.style={circle,draw,minimum size=1.5em, inner sep=0pt}
}

\begin{document}
\maketitle
	
\begin{abstract}
\begin{singlespace}
Reciprocity--the tendency of individuals to form mutual ties--is a fundamental structural feature of many directed networks. Despite its ubiquity, reciprocity remains insufficiently integrated into statistical network models, particularly in relation to covariate information. In this paper, we introduce the \textit{R\textsuperscript{2}-Model}, a novel and flexible framework that explicitly models reciprocity while incorporating covariate effects. Built upon a generalized $p_1$ model, our framework accommodates both network sparsity and node heterogeneity, offering the most comprehensive parametrization of reciprocity to date--capturing not only its baseline level but also how it systematically varies with observed covariates. 
To address the challenges posed by high dimensionality and nuisance parameters, we develop a conditional likelihood estimator that isolates and consistently estimates the reciprocity effects. We establish its theoretical guarantees, including consistency, asymptotic normality, and minimax optimality under broad sparsity regimes. Extensive simulations and real-world applications demonstrate the \textit{R\textsuperscript{2}-Model}'s flexibility, interpretability, and strong finite-sample performance, highlighting its practical utility for uncovering covariate-driven patterns of reciprocity in directed networks.

\end{singlespace}
\end{abstract}

\noindent
\keywords{Directed network;  Reciprocity;  Conditional likelihood estimator; Sparse networks; Asymptotic normality; Minimax optimality.}

\section{Introduction}\label{Sec: Introduction}
Many types of relational data are inherently directional, capturing interactions such as follower-followee relationships on social media, import-export flows in international trade, and citation links in scientific literature. A salient feature of such networks is the frequent occurrence of mutual connections: when node $i$ sends a link to node $j$, node $j$ is more likely to reciprocate. This propensity for links to form in both directions is widely referred to as \textit{reciprocity} in network science.

From a statistical perspective, reciprocity reflects a departure from within-dyad independence, as the formation of a link from $j$ to $i$ is no longer independent of the existence of a link from $i$ to $j$. Moreover, whether a link is reciprocated is often influenced by observable characteristics of the interacting pair. This motivates a central statistical question:
\begin{center}
\emph{How can we develop a regression framework that explains reciprocity in terms of observed covariates?}
\end{center}

Let $V_{ij}$ denote a vector of covariates associated with the dyad $(i, j)$. Our goal is to formulate a model in which the reciprocity between nodes $i$ and $j$, denoted by $\rho_{ij}$, depends systematically on these covariates through a regression structure:
\begin{equation}
\rho_{ij} = \rho_n + V_{ij}^\top \gamma, \label{eq:regression}
\end{equation}
where $\rho_n$ is a baseline parameter representing the average tendency toward mutual connection in a network of size $n$, and $\gamma$ is a vector of coefficients quantifying how covariates modulate reciprocity. The subscript $n$ highlights that the baseline reciprocity $\rho_n$ may vary with network size to reflect the asymptotic regime of interest, whereas $\gamma$ is assumed to remain stable across network sizes.

Despite the familiar regression-like form of $\rho_{ij}$ in \eqref{eq:regression}, incorporating covariates into a principled framework for modeling reciprocity presents substantial challenges. First, the quantity $\rho_{ij}$ is only meaningful when embedded within a coherent model for network formation. It is therefore essential to ground the regression formulation in a well-defined statistical model--one that reflects plausible assumptions about the unobserved social processes driving link formation. Such a foundation enables principled inference while supporting the flexible inclusion of covariates at both the node and dyadic levels.

Second, real-world networks exhibit structural features that must be respected in any realistic model. A key characteristic is \emph{sparsity}: the number of observed links typically grows sub-quadratically with the number of possible node pairs. Our notation $\rho_n$, with its explicit dependence on the network size $n$, reflects this scaling behavior and allows for reciprocity to weaken as networks grow. Another crucial feature is \emph{heterogeneity}: nodes can differ substantially in their propensity to send or receive links. This heterogeneity is often captured by node-specific parameters governing outgoing and incoming tendencies. Models that fail to account for sparsity or heterogeneity are inherently limited and may miss essential aspects of network structure. To ensure general applicability, we allow for unrestricted node-level heterogeneity by assigning each node its own sender and receiver parameters.

\noindent
\textbf{The model.}  
We observe a single snapshot of a directed network, represented by a graph $G_n = (V, E)$, where $V = \{1, \ldots, n\}$ is the set of nodes and $E \subseteq V \times V$ is the set of directed edges. Each ordered pair $(i, j) \in E$ indicates a directed link from node $i$ to node $j$. We restrict attention to simple graphs without self-loops, so $(j, j) \notin E$ for all $j \in V$. 
Let $A = (A_{ij})_{1 \leq i, j \leq n}$ denote the adjacency matrix of the observed network, where $A_{ij} \in \{0,1\}$ indicates the presence of a directed edge from node $i$ to node $j$. We assume that dyads $\{(A_{ij}, A_{ji})\}_{1 \leq i < j \leq n}$ are mutually independent conditional on observed covariates and unobserved node heterogeneity. For each unordered node pair $(i, j)$, let $V_{ij} \in \mathbb{R}^q$ denote a vector of dyad-specific covariates. 
 
We model the joint distribution of the dyad $(A_{ij}, A_{ji})$ as a multinomial over the four possible directed outcomes:  
\begin{align}\label{def: model}
\begin{aligned}
p_{ij}(0,0) &\propto 1, 
&\quad p_{ij}(1,0) &\propto \exp(\mu_n + \alpha_i + \beta_j), \\
p_{ij}(0,1) &\propto \exp(\mu_n + \alpha_j + \beta_i), 
&\quad p_{ij}(1,1) &\propto \exp\!\big(2\mu_n + \alpha_i + \alpha_j + \beta_i + \beta_j + \rho_n + V_{ij}^\top \gamma\big).
\end{aligned}
\end{align}
where $p_{ij}(v_1, v_2) = \mathbb{P}(A_{ij} = v_1, A_{ji} = v_2 \mid V_{ij})$ denotes the conditional probability of observing configuration $(v_1, v_2) \in \{0,1\}^2$ for dyad $(i,j)$.  
Each parameter in the model plays a specific structural role, as explained below. The global density parameter $\mu_n \in \mathbb{R}$ controls the baseline propensity for link formation across the network, while the global reciprocity parameter $\rho_n \in \mathbb{R}$ captures the additional log-odds associated with the formation of mutual links. The vector of regression coefficients $\gamma \in \mathbb{R}^q$ links the dyadic covariates $V_{ij}$ to the strength of reciprocity, allowing observed characteristics of node pairs to modulate their tendency to reciprocate. Finally, the node-specific parameters $\alpha = (\alpha_1, \ldots, \alpha_n)^\top$ and $\beta = (\beta_1, \ldots, \beta_n)^\top$ account for individual heterogeneity, representing each node's propensity to initiate and to receive links, respectively. Together, these parameters yield a flexible yet interpretable representation of both global and local structural features in directed networks.

We refer to this model as the \textit{R\textsuperscript{2}-Model} (\emph{Reciprocal Regression for Directed Networks}). It offers a flexible and principled framework for modeling mutual link formation, seamlessly integrating covariate effects while explicitly accounting for both sparsity and heterogeneity in directed networks. By capturing how reciprocity varies with observed characteristics, the \textit{R\textsuperscript{2}-Model} extends classical formulations and provides a richer, more interpretable representation of network structure.

Importantly, the dyadic independence assumption in \eqref{def: model} holds only conditionally on the edge covariates $\{V_{ij}\}_{i \neq j}$ and the latent node-specific parameters $\alpha$ and $\beta$. To illustrate, consider that node $i$ may inherently have a high propensity to send links (a large $\alpha_i$). Even if two edges $(i,j)$ and $(i,k)$ are conditionally independent given $\alpha_i$, marginalizing over $\alpha_i$ induces dependence: observing a link from $i$ to $j$ increases the likelihood of a link from $i$ to $k$, because both depend on the same unobserved trait $\alpha_i$.

When $\gamma = 0$ or in the absence of edge covariates $V_{ij}$, the \textit{R\textsuperscript{2}-Model} reduces to the classical $p_1$ model introduced by \citet{holland1981exponential}, a foundational yet complex model for reciprocity in directed networks. Despite its significance, rigorous statistical theory for the $p_1$ model has long been elusive, with major progress primarily limited to the case without reciprocity effects ($\rho_n = 0$) \citep{yan2016asymptotics, yan2019statistical}. In our prior work \citep{feng2025modelling}, we mitigated this challenge by parametrizing the heterogeneity terms $\alpha_i$ and $\beta_j$ via node covariates as $X_i^\top \gamma_1$ and $Y_j^\top \gamma_2$, thereby reducing dimensionality. By contrast, the current model \eqref{def: model} retains unrestricted heterogeneity, involving at least $2n$ parameters, which precludes traditional fixed-dimensional asymptotic analysis.

Our framework also relates to the $p_2$ model of \citet{van2004p2}, which treats $\alpha$ and $\beta$ as random effects. In contrast, we treat these node-specific parameters as fixed but unknown, enabling a distinct asymptotic regime that supports inference under network sparsity and covariate-driven reciprocity.

For the model specified in \eqref{def: model}, two primary challenges arise in parameter estimation. The first challenge concerns the high and growing dimensionality of the parameter space. Specifically, the total number of parameters is  
\(
d_{\text{total}} = 2n + 2 + q,
\)  
where $2n$ corresponds to the latent node-specific heterogeneity parameters $\alpha$ and $\beta$, $2$ accounts for the global parameters $\mu_n$ and $\rho_n$, and $q$ is the dimension of the covariate vector $V_{ij}$. We work within an asymptotic regime where only a single network is observed, but the number of nodes $n$ tends to infinity. Consequently, $d_{\text{total}}$ diverges with $n$, which poses fundamental difficulties for classical inference techniques developed under fixed-dimensional assumptions.

The second challenge arises from the fundamentally different roles these parameters play in network formation. The global parameters $\mu_n$, $\rho_n$, and $\gamma$ influence all dyads simultaneously, while the node-specific parameters $\alpha_i$ and $\beta_i$ affect only edges incident to node $i$. This intrinsic heterogeneity naturally leads to distinct convergence rates for global versus local parameters. Moreover, under sparse network regimes, the asymptotic behavior of even the global parameters can vary substantially; see \citet{feng2025modelling} for a comprehensive discussion of these effects.

\noindent
\textbf{Contributions}. Our primary contribution is the introduction of the \textit{R\textsuperscript{2}-Model}, a novel and flexible framework for regression analysis of reciprocity in directed networks, supported by a rigorous theoretical foundation. While the classical $p_1$ model captures baseline tendencies for edge formation and mutual connections, it does not allow reciprocity to vary systematically with observable covariates. In contrast, the proposed \textit{R\textsuperscript{2}-Model} generalizes the $p_1$ model by allowing the reciprocity parameter to depend on node- or dyad-specific covariates, yielding a richer and more interpretable representation of reciprocal behavior. This extension bridges the gap between structural network models and covariate-driven regression approaches, providing a principled way to study how individual attributes and contextual factors jointly shape patterns of reciprocity.

For parameter estimation, we develop a conditional likelihood approach to estimate $\rho_n$ and $\gamma$, while accommodating sparse networks. The key idea is to profile out the two infinite-dimensional nuisance parameters, $\alpha$ and $\beta$, by conditioning on their sufficient statistics--namely, the node-specific out-degree and in-degree counts. This conditioning removes the nuisance effects of $\alpha$ and $\beta$ from the likelihood, resulting in a reduced likelihood that depends solely on the global parameters $\rho_n$ and $\gamma$. Consequently, the dimensionality of the estimation problem is greatly reduced, while preserving the essential information needed to capture reciprocity and covariate effects.

We conduct a thorough theoretical analysis in a high-dimensional asymptotic regime, where the number of nodes $n$ grows and the parameter dimension diverges accordingly. Under mild regularity and sparsity conditions, we establish the consistency and asymptotic normality of the conditional likelihood estimator for $(\rho_n, \gamma^\top)$. In addition, we derive minimax lower bounds for estimating these parameters and show that our estimator achieves these bounds, thereby proving its optimality. This framework advances likelihood-based inference for network models by systematically addressing covariate-driven reciprocity and tackling the challenges posed by high-dimensional nuisance parameters and network sparsity.

An important technical challenge arises because the conditioning argument, while reducing dimensionality, induces additional dependence in the likelihood function. This dependence is intrinsically linked to the presence and arrangement of network motifs and requires careful, model-specific analysis. To address this, we identify the leading term governing the asymptotic behavior of the Fisher score function via a projection-based approach, enabling the application of U-statistics theory. 

Moreover, the parameters $\rho_n$ and $\gamma$ exhibit distinct convergence rates under different sparsity regimes, necessitating appropriate scaling factors when deriving their asymptotic distributions. Our proof strategy departs from conventional M-estimation techniques and instead leverages novel tools built around the crucial notion of \emph{effective sample size} for statistical inference \citep{feng2025modelling}. Since multiple conditional likelihood constructions are possible, we further justify our specific formulation by establishing its minimax optimality. Beyond the specific model studied here, our theoretical results provide a transferable framework for analyzing conditional likelihood estimators in a broad class of network models.

Finally, we complement the theoretical analysis with extensive simulations across various network sizes and sparsity levels, demonstrating the finite-sample accuracy and robustness of our estimator. We further illustrate the practical effectiveness of the \textit{R\textsuperscript{2}-Model} through applications to two real-world directed network datasets, highlighting its ability to uncover reciprocal relationships driven by covariates.

\noindent
\textbf{Literature review}.  
Relational data are ubiquitous, fueled by the proliferation of interactions between entities in modern scientific and societal datasets. Research on network analysis is rapidly evolving and inherently interdisciplinary, situated at the crossroads of statistics \citep{kolaczyk2009statistical}, economics \citep{jackson2008social}, physics \citep{newman2018networks}, social sciences \citep{wasserman1994social}, and applied mathematics \citep{dorogovtsev2003evolution}. 

For asymmetrical relational data, reciprocity--the tendency of entities to form mutual connections--is widely observed in many real-world directed networks \citep{newman2002email, garlaschelli2004patterns, jiang2015reciprocity, ji2016coauthorship}. Studying reciprocity not only helps model pairwise interactions but also sheds light on higher-order network structures such as triadic effects \citep{kolaczyk2015question, lou2013learning}. Degree heterogeneity, reflecting variation in entities' propensity to form connections, is ubiquitous in real networks \citep{barabasi1999emergence, fienberg2012brief}. Moreover, empirical networks tend to be sparse \citep[cf.][]{newman2018networks}.

Despite its importance, the modeling of reciprocity remains limited. The seminal $p_1$ model of \cite{holland1981exponential} is one of the earliest and most influential approaches for capturing reciprocity, yet its parameter estimation and theoretical understanding remain incomplete. The existence and uniqueness of the MLE have been addressed in \cite{rinaldo2010existence} and \cite{petrovic2010algebraic}, while rigorous statistical inference is available only under strong assumptions, either without degree heterogeneity \citep{kolaczyk2015question, feng2025modelling} or without reciprocity \citep{yan2016asymptotics, yan2019statistical}. Thus, the rigorous estimation of reciprocity remains an open challenge, which our theory addresses decisively.

Our estimation strategy is based on conditional likelihood methods to eliminate the nuisance parameters $\alpha$ and $\beta$. This conditioning principle has a long history, originating at least with \cite{bartlett1937properties}, and further developed by \cite{barndorff1983formula}, \cite{cox1987parameter}, and \cite{barndorff1993note}. In network contexts, conditioning approaches have been used by \cite{graham2017econometric}, \cite{charbonneau2017multiple}, and \cite{jochmans2018semiparametric}, but none explicitly incorporated reciprocity effects in directed networks. While related to \cite{graham2017econometric}, our model is significantly more complex, and crucially, we establish the minimax optimality of our estimator--an advancement not known in prior work. Existing likelihood-based methods often require relatively dense networks \citep{yan2013central, yan2016asymptotics} or sparsity assumptions on nuisance parameters \citep{chen2021analysis, stein2024sparse}, whereas our approach accommodates sparsity without such assumptions. By filtering out nuisance parameters, our method avoids the incidental parameter problem \citep{fernandez2016individual, dzemski2019empirical} and the need for bias correction.

Addressing the challenge of diverging parameters, prior work such as \cite{yan2013central, yan2016asymptotics, yan2019statistical} analyzed maximum likelihood estimators via a detailed study of the Fisher information matrix and its inverse. Penalized likelihood methods have also been developed for sparse networks \citep{chen2021analysis, stein2024sparse, shao20212}, though some assume sparsity in heterogeneity parameters \citep{chen2021analysis, stein2024sparse}, and others do not incorporate covariates \citep{shao20212}. Notably, these works mostly focus on undirected networks or directed models without reciprocity, relying heavily on Fisher matrix approximations. Incorporating reciprocity parameters significantly complicates the analysis and, to the best of our knowledge, remains analytically intractable. The only exception is the study by \cite{yan2015simulation} for the $p_1$ model, which offers numerical justification but no theoretical guarantees.

\noindent
\textbf{Organization of the paper}. The remainder of the paper is organized as follows. In Section \ref{Sec: Main results}, we introduce our conditional likelihood estimator, establish its asymptotic properties, and compare it to alternative estimators, culminating in a formal proof of its minimax optimality. Section \ref{Sec: Numerical} presents simulation studies and two real data applications. We conclude with remarks and discussions in Section \ref{Sec: Discussion}. All the proofs, technical details, and additional simulation results can be found in the supplementary material.

\section{$R^2$-Model: Estimation, Asymptotics, and Optimality}\label{Sec: Main results}

In this section, we first develop a conditional likelihood approach to estimate the reciprocity parameters $\rho_n$ and $\gamma$. The main idea is to condition on a set of graphs sharing key characteristics with the observed graph, thereby eliminating nuisance parameters $\mu_n$, $\alpha$, and $\beta$. This set is combinatorially large and infeasible to enumerate exhaustively. Instead, we focus on specific tetrads -- subgraphs or network motifs consisting of four nodes. We choose tetrads because they are the smallest subgraphs that simultaneously eliminate irrelevant parameters and preserve the reciprocity parameter after conditioning.

We then establish the consistency and asymptotic normality of our estimator. Crucially, we show that our estimator is minimax optimal in the sense that it attains the optimal convergence rate among all estimators based on conditioning. This result provides a framework for assessing the optimality of both our proposed estimator and the estimator introduced by \cite{graham2017econometric} as well, offering new insights into the efficiency of conditional likelihood estimation.

\noindent
\textbf{Notations}. 
Throughout the paper, a subscript $n$ on a parameter (e.g., $\mu_n$, $\rho_n$) indicates its dependence on the network size $n$, while parameters without this subscript are independent of $n$. True parameter values are denoted by a subscript $0$, such as $\gamma_{0}$ for the true value of $\gamma$, and $\rho_{n0}$ for the true value of $\rho_n$. We use $C$ to denote a generic constant that may change across contexts. 
For non-negative sequences $a_n$ and $b_n$, we write $a_n = O(b_n)$ if there exist constants $C > 0$ and $N$ such that $a_n \leq C b_n$ for all $n \geq N$. We write $a_n \sim b_n$ if both $a_n = O(b_n)$ and $b_n = O(a_n)$. For vectors, $\mathbf{0}_q$ and $\mathbf{1}_q$ denote the $d$-dimensional all-zeros and all-ones vectors, respectively. For a matrix $M$, $M_{i,j}$ denotes the $(i,j)$-th entry, and $M_{r_1:r_2, c_1:c_2}$ denotes the submatrix formed by rows $r_1$ to $r_2$ and columns $c_1$ to $c_2$. For matrices $A$ and $B$ of the same dimension, the Hadamard product $A \odot B$ is defined elementwise as $(A \odot B)_{i,j} = A_{i,j} B_{i,j}$. For brevity, we write $\vartheta_n = (\rho_n, \gamma^\top)^\top$ and $\vartheta = (\rho, \gamma^\top)^\top$.

\subsection{Estimation: Conditional likelihood} \label{sec: Conditional likelihood estimator}

Let $d_i = \sum_{j \neq i} A_{ij}$ and $b_i = \sum_{j \neq i} A_{ji}$ denote the out-degree and in-degree of node $i$, respectively. Define the out-degree sequence of the network $G_n$ as $d = (d_1, \ldots, d_n)^\top$ and the in-degree sequence as $b = (b_1, \ldots, b_n)^\top$. Let $t = \sum_{i} d_i = \sum_{i} b_i$ 
be the total number of edges in the network, and $m = \sum_{i<j} A_{ij} A_{ji}$ 
be the total number of mutual (reciprocal) connections. 
To simplify notation, define the bivariate indicator functions
\[
\phi_{ij}(0,0) = (1 - A_{ij})(1 - A_{ji}), \quad \phi_{ij}(1,0) = A_{ij}(1 - A_{ji}), \quad \phi_{ij}(0,1) = A_{ji}(1 - A_{ij}), \quad \phi_{ij}(1,1) = A_{ij} A_{ji}.
\]
Thus, $\phi_{ij}(v_1,v_2)$ equals 1 if and only if $(A_{ij}, A_{ji}) = (v_1, v_2)$ and 0 otherwise.

Given the covariates $\{V_{ij}\}$, the probability of observing the network $G_n$ can be expressed as
\begin{align*}
&\prod_{i<j} p_{ij}(0,0)^{\phi_{ij}(0,0)} p_{ij}(1,0)^{\phi_{ij}(1,0)} p_{ij}(0,1)^{\phi_{ij}(0,1)} p_{ij}(1,1)^{\phi_{ij}(1,1)} \\
=& \, K(\mu_n, \rho_n, \alpha, \beta, \gamma) \cdot \exp \left( t \mu_n + \alpha^\top d + \beta^\top b \right) \cdot \exp \left( m \rho_n + s^\top \gamma \right),
\end{align*}
where $K(\mu_n, \rho_n, \alpha, \beta, \gamma)$ is a normalizing constant independent of $G_n$, and
\[
s = \sum_{i<j} \phi_{ij}(1,1) V_{ij}.
\]
Here, the triplet $\{t, d, b\}$ forms the sufficient statistics for the parameters $\{\mu_n, \alpha, \beta\}$; in other words, the bi-degree sequence $(b, d)$ is sufficient for these three non-reciprocal parameters due to the definition of $t$.

While maximum likelihood estimation might appear natural for the model in \eqref{def: model}, statistical inference remains a challenging open problem except in the special case without reciprocity parameters (i.e., $\rho_n = \gamma = 0$) \citep{yan2016asymptotics}. The primary difficulty arises from the complexity of inverting the Fisher information matrix, which complicates direct estimation. Our estimation approach leverages the sufficiency of the bi-degree sequence: graphs that share the same bi-degree sequence but differ in their mutual connection patterns provide the necessary variation to identify the reciprocity parameters $\rho_n$ and $\gamma$.

Let
\[
\mathcal{S}^A = \{ B : B \in \mathcal{A}, \, b(B) = b(A), \, d(B) = d(A) \},
\]
where $\mathcal{A}$ is the set of all $2^{n \times (n-1)}$ possible adjacency matrices, and, with some abuse of notation, $b(\cdot)$ and $d(\cdot)$ denote the mappings from a graph to its in-degree and out-degree sequences, respectively.

The conditional probability of the observed graph $A$ given its degree sequences and covariates is then
\[
\Pr(\mathbf{A} = A \mid \{V_{ij}\}, \mathcal{S}^A) = \frac{\exp\left( m \rho_n + \sum_{i<j} A_{ij} A_{ji} V_{ij}^\top \gamma \right)}{\sum_{B \in \mathcal{S}^A} \exp\left( m_B \rho_n + \sum_{i<j} B_{ij} B_{ji} V_{ij}^\top \gamma \right)},
\]
where $m_B$ is the number of mutual connections in graph $B$. 
Hence, provided that $\rho_n$ and $\gamma$ do not cancel out in this conditional probability, we obtain a likelihood function for the reciprocity parameters based on conditioning.

\begin{figure}[htbp]
\centering
\begin{tikzpicture}
\node[vertex] (6) at (2,0) {$i$};
\node[vertex] (7) at (4,0) {$j$};
\node[vertex] (8) at (2,-2) {$l$};
\node[vertex] (9) at (4,-2) {$k$};

\node[vertex] (1) at (6,0) {$i$};
\node[vertex] (2) at (8,0) {$j$};
\node[vertex] (3) at (6,-2) {$l$};
\node[vertex] (4) at (8,-2) {$k$};

\node[vertex] (10) at (10,0) {$i$};
\node[vertex] (11) at (12,0) {$j$};
\node[vertex] (12) at (10,-2) {$l$};
\node[vertex] (13) at (12,-2) {$k$};

\node (14) at (3,-2.6) {\small (I)};
\node (15) at (7,-2.6) {\small (II)};
\node (16) at (11,-2.6) {\small (III)};

\draw[->, bend left=10, thick,line width=0.25mm] (1) to (2);
\draw[->,  bend left=10, thick,line width=0.25mm] (2) to (1);
\draw[->, bend left=10, thick,line width=0.25mm] (4) to (3);
\draw[->,  bend left=10, thick,line width=0.25mm] (3) to (4);

\draw[->, bend left=10, dashed, thick,line width=0.25mm] (1) to (4);
\draw[->, bend left=10, dashed, thick,line width=0.25mm] (4) to (1);
\draw[->, bend left=10, dashed, thick,line width=0.25mm] (2) to (3);
\draw[->, bend left=10, dashed, thick,line width=0.25mm] (3) to (2);

\draw[->, thick,line width=0.25mm] (7) to (9);
\draw[->, thick,line width=0.25mm] (6) to (7);
\draw[->, thick,line width=0.25mm] (8) to (6);
\draw[->, thick,line width=0.25mm] (9) to (8);
\draw[->, bend left=10, dashed, thick,line width=0.25mm] (6) to (9);
\draw[->, bend left=10, dashed, thick,line width=0.25mm] (9) to (6);
\draw[->, bend left=10, dashed, thick,line width=0.25mm] (7) to (8);
\draw[->, bend left=10, dashed, thick,line width=0.25mm] (8) to (7);

\draw[->, bend left=10, thick,line width=0.25mm] (10) to (12);
\draw[->,  bend left=10, thick,line width=0.25mm] (12) to (10);
\draw[->, bend left=10, thick,line width=0.25mm] (11) to (13);
\draw[->,  bend left=10, thick,line width=0.25mm] (13) to (11);
\draw[->, bend left=10, dashed, thick,line width=0.25mm] (10) to (13);
\draw[->, bend left=10, dashed, thick,line width=0.25mm] (13) to (10);
\draw[->, bend left=10, dashed, thick,line width=0.25mm] (11) to (12);
\draw[->, bend left=10, dashed, thick,line width=0.25mm] (12) to (11);

\end{tikzpicture}
\caption{Diagrams illustrating the values of $S_{ijkl}$: $S_{ijkl} = 0$ (left), $S_{ijkl} = 1$ (middle), and $S_{ijkl} = -1$ (right). Dashed lines indicate no restrictions on the presence of edges.}
\label{Sijkl}
\end{figure}
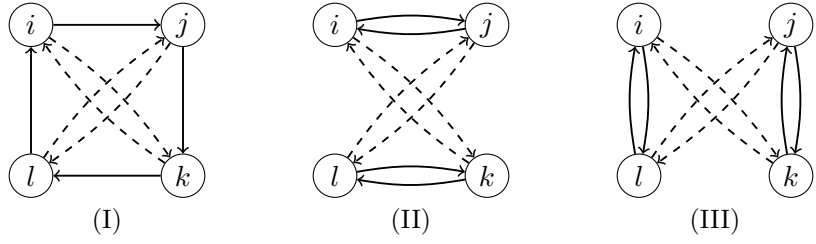

However, although this idea is intuitive, enumerating the set $\mathcal{S}^A$ is a combinatorial problem and can be numerically prohibitive, as similarly noted by \cite{graham2017econometric}. To address this, we restrict our focus to a computationally tractable subset of $\mathcal{S}^A$. The smallest subgraphs that preserve the degree sequences of $A$ while allowing variation in the total number of mutual links and the term involving $\gamma$ are tetrads -- subgraphs consisting of four nodes. Specifically, we consider the tetrads depicted in Figure \ref{Sijkl} and define the indicator random variable $S_{ijkl}$ as follows:
\[
S_{ijkl} =  
\begin{cases} 
0 & \text{if } \phi_{ij}(1,0)\phi_{jk}(1,0)\phi_{kl}(1,0)\phi_{li}(1,0) = 1, \\[6pt]
1 & \text{if } \phi_{ij}(1,1)\phi_{jk}(0,0)\phi_{kl}(1,1)\phi_{li}(0,0) = 1, \\[6pt]
-1 & \text{if } \phi_{ij}(0,0)\phi_{jk}(1,1)\phi_{kl}(0,0)\phi_{li}(1,1) = 1, \\[6pt]
\infty & \text{otherwise}.
\end{cases}
\]

Each subgraph in Figure \ref{Sijkl} can be obtained by rewiring the edges of the other two. Apparently, such rewiring does not change the out-degree and in-degree sequences. That is, for all three configurations in Figure \ref{Sijkl}, the out-degree and in-degree sequences restricted to the tetrad $(i,j,k,l)$ are all $(1,1,1,1)$ (ignoring the unspecified dashed lines). Therefore, by comparing the frequency of these three configurations in $G_n$, we can isolate the non-reciprocal parameters and identify only $\rho_n$ and $\gamma$. We present this argument rigorously in Lemma \ref{sufficiency}. 

Let $\mathbf{V}_{ijkl} = \{V_{ij}, V_{jk}, V_{kl}, V_{li}, V_{ik}, V_{jl}\}$, define 
\[
f_{ijkl}(\rho_n, \gamma) = \exp\big(2\rho_n + (V_{ij}^\top + V_{kl}^\top) \gamma \big), \quad
g_{ijkl}(\rho_n, \gamma) = \exp\big(2\rho_n + (V_{il}^\top + V_{jk}^\top) \gamma \big),
\]
and 
\[
r_{ijkl}(\rho_n, \gamma) = f_{ijkl}(\rho_n, \gamma) + g_{ijkl}(\rho_n, \gamma).
\]
We now state Lemma \ref{sufficiency}.

\begin{Lem}[Sufficiency]\label{sufficiency}
We have
\begin{align*}
\mathbb{P}\big(S_{ijkl} = 0 \mid \mathbf{V}_{ijkl}, S_{ijkl} \in \{0,1,-1\}\big) &= \frac{1}{1 + r_{ijkl}(\rho_n, \gamma)}, \\
\mathbb{P}\big(S_{ijkl} = 1 \mid \mathbf{V}_{ijkl}, S_{ijkl} \in \{0,1,-1\}\big) &= \frac{f_{ijkl}(\rho_n, \gamma)}{1 + r_{ijkl}(\rho_n, \gamma)}, \\
\mathbb{P}\big(S_{ijkl} = -1 \mid \mathbf{V}_{ijkl}, S_{ijkl} \in \{0,1,-1\}\big) &= \frac{g_{ijkl}(\rho_n, \gamma)}{1 + r_{ijkl}(\rho_n, \gamma)}.
\end{align*}
\end{Lem}

This lemma states that the conditional likelihood of observing one realization of $S_{ijkl}$, given the three possible configurations in Figure \ref{Sijkl}, is a function of $\rho_n$ and $\gamma$ and is independent of the non-reciprocal parameters. Our estimation approach builds on a similar idea from \cite{graham2017econometric} for the $\beta$-model with covariates but extends it in several key directions. 
First, our model accommodates directed networks, which are inherently more complex than undirected ones. Second, the reciprocity parameters $\rho_n$ and $\gamma$ we consider may exhibit distinct optimal convergence rates in certain regimes, requiring a more refined and carefully designed conditioning strategy. Further discussion on this point is provided in Section \ref{sec:rate}. We now formally introduce our conditional likelihood estimator.

We have emphasized that our model aims to capture sparse networks. To this end, we allow $\mu_n$ and $\rho_n$ to vary with $n$, by reparametrizing them as 
\[
\mu_n = -a \log n + \mu, \quad \rho_n = b \log n + \rho,
\]
where $\mu$ and $\rho$ are fixed parameters belonging to a closed interval $[-C, C]$ for some constant $C < \infty$ independent of $n$. 
This $-\log n$ transformation introduces a polynomial order scaling in $n$, which is commonly employed in the literature. We remark that this polynomial order scaling is mainly for ease of presentation, highlighting the order of the scaling factors in Theorem \ref{AN}, and is not essential for the statistical inference procedure described in Theorem \ref{Inference}. We emphasize that under this reparametrization, the newly introduced parameters are not identifiable; for example, $a$ and $\mu$ are confounded, as are $b$ and $\rho$. Nonetheless, introducing these scalings clarifies the theoretical analysis by allowing an \textit{exact} characterization of the effective sample sizes for statistical inference of $\mu_n$ and $\rho_n$. In practice, when these parameters are unknown, we have developed a feasible inference procedure as detailed in Theorem \ref{AN}.

Let $\vartheta = (\rho, \gamma^\top)^\top$ and $W_{ij}= (1, V_{ij}^\top)^\top$. The negative conditional log-likelihood function for the tetrad $(i,j,k,l)$ is seen as
\begin{align*}
\ell_{ijkl}(\vartheta) = \mathbb{I}\left(S_{ijkl} \in \{0, \pm 1\}\right) \log \big(1 + r_{ijkl}(\vartheta)\big) - \left[\mathbb{I}(S_{ijkl} = 1)(W_{ij}^\top + W_{kl}^\top) + \mathbb{I}(S_{ijkl} = -1)(W_{jk}^\top + W_{il}^\top)\right] \vartheta.
\end{align*}
To impose symmetry, we define
\[
g_{ijkl}(\vartheta) = \frac{1}{4!} \sum_{\pi \in \Pi_4} \ell_{\pi(i,j,k,l)}(\vartheta),
\]
where $\Pi_4$ denotes the group of all $4! = 24$ permutations of a 4-tuple.
The overall negative conditional log-likelihood function is
\[
\mathcal{L}_{n}(\vartheta) = \binom{n}{4}^{-1} \sum_{i < j < k < l} g_{ijkl}(\vartheta).
\]

Our estimator $\hat{\vartheta}$ is defined as the minimizer of $\mathcal{L}_{n}(\vartheta)$ over the parameter space
\[
\Theta = [-C, C] \times \Theta_1,
\]
where $\Theta_1$ is the compact parameter space for $\gamma$ in $\mathbb{R}^q$. {Before presenting the asymptotic properties of $\hat{\vartheta}$, we first note that, unlike the standard log-likelihood function, the components of $\mathcal{L}_{n}(\vartheta)$ are not mutually independent. In particular, $g_{i_1 j_1 k_1 l_1}$ and $g_{i_2 j_2 k_2 l_2}$ are correlated whenever they share two or more indices. This dependence arises as a consequence of the conditioning argument and must be handled with care in the subsequent analysis.
 Further discussions are provided in Section \ref{sec: Asymptotic results}.}

\subsection{Asymptotics}\label{sec: Asymptotic results}

We impose the following assumptions to establish the asymptotic properties of our estimator.

\begin{Assum}\label{covass}
The covariates $\{V_{ij}\}_{i<j}$ are centered, i.i.d., and uniformly bounded. Furthermore, the true parameter $\vartheta_0$ lies in the interior of the compact parameter space $\Theta_1$. 
\end{Assum}

\begin{Assum}[Moderate degree heterogeneity]\label{nodeass}
The node-specific heterogeneity parameters satisfy the uniform boundedness condition
\[
\max_{i} \{ |\alpha_i|, |\beta_i| \} \leq C,
\]
for some constant $C < \infty$ independent of $n$.
\end{Assum}

\begin{Assum}[Sparse network regime]\label{sparseass}
The sparsity parameters $a$ and $b$ satisfy 
\[
0 < a < 1, \quad 0 < 2a - b < 2.
\]
\end{Assum}

Assumption \ref{covass} is a standard regularity condition in nonlinear network models and ensures the covariate effects are well-behaved and identifiable within the parameter space. While the i.i.d. assumption simplifies the theoretical analysis, in many practical applications the covariates $V_{ij}$ may exhibit local dependence structures, for example, correlation between $V_{ij}$ and $V_{ik}$ for $j \neq k$. While our theory can be extended to accommodate such dependence, the resulting proofs would be technically demanding. For clarity, we retain Assumption \ref{covass} throughout the paper.

Assumption \ref{nodeass} restricts the degree heterogeneity parameters to a bounded regime, which significantly simplifies the technical arguments by avoiding extreme sparsity or hubs with exceptionally high connectivity. This assumption strikes a balance between realism and tractability, as many real-world networks exhibit moderate degree heterogeneity.

Assumption \ref{sparseass} characterizes the network sparsity regime under which the number of tetrads that are informative for identifying the reciprocity parameters grows sufficiently fast with $n$. More concretely, the expected number of tetrads with configuration $S_{ijkl}=0$ is of order $n^{4-4a}$, and those with $S_{ijkl}=\pm 1$ are of order $n^{4-2(2a - b)}$. The inequalities in Assumption \ref{sparseass} ensure that both types of tetrads appear frequently enough to provide asymptotic information. This condition excludes overly sparse networks where the data would be insufficient to identify the reciprocity parameters reliably.

Together, these assumptions establish the framework for our asymptotic analysis, allowing us to leverage concentration results and large sample theory to study the properties of the conditional likelihood estimator. In what follows, we present Lemma \ref{Lem_score}, which characterizes the score function associated with the conditional likelihood for a single tetrad.

\begin{Lem}[Bounds for $\nabla_{\vartheta} \ell_{ijkl}(\vartheta)$]\label{Lem_score} Under Assumptions \ref{covass} and \ref{nodeass}, for the gradient of the conditional log-likelihood contribution from the tetrad $(i,j,k,l)$, the following bounds hold:
\begin{align*}
&\left| \nabla_{\rho}  \ell_{ijkl}(\vartheta) \right| \leq C_1 \, n^{2 \min\{b, 0\}} \, \mathbb{I}\left(S_{ijkl} = 0\right) + C_2 \, n^{-2 \max\{b, 0\}} \, \mathbb{I}\left(S_{ijkl} = \pm 1\right), \\
&\left\| \nabla_{\gamma} \ell_{ijkl}(\vartheta) \right\|_2 \leq C_3 \, n^{2 \min\{b, 0\}} \, \mathbb{I}\left(S_{ijkl} = 0\right) + C_4 \, \mathbb{I}\left(S_{ijkl} = \pm 1\right),
\end{align*}
where $C_1, C_2, C_3,$ and $C_4$ are positive constants independent of $n$.
\end{Lem}

The proof of Lemma \ref{Lem_score} is straightforward and is provided in the supplementary material. Note that for any tetrad $(i,j,k,l)$, the probabilities satisfy
\[
\mathbb{P}\{S_{ijkl} = 0\} \asymp n^{-4a}, \quad \text{and} \quad \mathbb{P}\{S_{ijkl} = \pm 1\} \asymp n^{-4a + 2b}.
\]
By Lemma \ref{Lem_score}, when $b > 0$, it follows that
\[
\nabla_{\rho} \mathcal{L}_{n}(\vartheta) = O(n^{-4a}) \quad \text{and} \quad \nabla_{\gamma_m} \mathcal{L}_{n}(\vartheta) = O(n^{-4a + 2b}) \quad \text{for each } m = 1, \ldots, q.
\]
When $b < 0$, both components satisfy
\[
\nabla_{\rho} \mathcal{L}_{n}(\vartheta) = O(n^{-4a + 2b}), \quad \text{and} \quad \nabla_{\gamma_m} \mathcal{L}_{n}(\vartheta) = O(n^{-4a + 2b}).
\]

To guarantee that the scaled score vector of $\mathcal{L}_n(\vartheta)$ converges to a non-random limit, we impose an appropriate normalization as detailed in Assumption \ref{Hessian}. Similarly, the Hessian matrix, whose blocks may differ in magnitude, is assumed to converge to a non-random positive definite matrix after suitable scaling. A thorough analysis of the Hessian structure under various sparsity regimes-- characterized by the parameters $a$ and $b$ -- is provided in the proof of Theorem \ref{AN}. The positivity of the smallest eigenvalue of $H_0(\vartheta_0)$ is a standard regularity condition in the theory of M-estimation.

\begin{Assum}\label{Hessian}
Define the scaling vector
\[
\tau_n= 
\begin{cases}
\bigl(n^{4a}, \, n^{4a - 2b} \mathbf{1}_q^\top\bigr)^\top, & \text{if } b > 0, \\
n^{4a - 2b} \mathbf{1}_{q+1}, & \text{if } b \leq 0,
\end{cases}
\]
where $\mathbf{1}_k$ denotes the $k$-dimensional vector of ones. Suppose that
\[
\lim_{n \to \infty} \mathbb{E}\big(\tau_n \odot \nabla_{\vartheta} \mathcal{L}_n(\vartheta)\big) = \Psi(\vartheta),
\]
and
\[
\lim_{n \to \infty} \big(\tau_n \mathbf{1}_{q+1}^\top\big) \odot \mathbb{E}\big(\nabla_{\vartheta \vartheta} \mathcal{L}_n(\vartheta)\big) = H_0(\vartheta),
\]
for any $\vartheta \in \Theta$. Furthermore, we assume that the smallest eigenvalue of $H_0(\vartheta_0)$ is strictly positive.
\end{Assum}
The scaling vector $\tau_n$ reflects the differing effective sample sizes associated with the parameters $\rho$ and $\gamma$ under various sparsity regimes characterized by $(a,b)$. Intuitively, the order of the score components corresponds to the expected number of informative tetrads that contribute to the estimation of each parameter. For instance, when $b > 0$, the components related to $\rho$ primarily depend on tetrads with $S_{ijkl} = 0$, which occur with probability on the order of $n^{-4a}$, leading to a natural scaling by $n^{4a}$. Conversely, the components related to $\gamma$ are driven by tetrads with $S_{ijkl} = \pm 1$, whose probability is $n^{-4a + 2b}$, requiring a different scaling by $n^{4a - 2b}$. When $b \leq 0$, both parameters are effectively estimated from the same sparse set of tetrads, leading to a unified scaling $n^{4a - 2b}$. 

This adaptive scaling is critical to obtain non-degenerate limiting distributions of the score function and Hessian matrix, thereby enabling valid statistical inference. The positive definiteness of the limiting Hessian $H_0(\vartheta_0)$ ensures local identifiability and the asymptotic normality of the estimator. Such differential scaling is a distinguishing feature of our approach, highlighting the interplay between sparsity, reciprocity parameters, and network heterogeneity, which is not captured in simpler models for undirected or dense networks.

We now establish the consistency of the conditional likelihood estimator \(\hat{\vartheta}\) in the following theorem. The proof is provided in the supplementary material.

\begin{theorem}[Consistency of \(\hat{\vartheta}\)]\label{consistency}
Suppose Assumptions \ref{covass}, \ref{nodeass}, \ref{sparseass}, and \ref{Hessian} hold. Then the estimator \(\hat{\vartheta}\) defined as the minimizer of \(\mathcal{L}_n(\vartheta)\) over \(\Theta\) is consistent; that is,
\[
\hat{\vartheta} \xrightarrow{\mathbb{P}} \vartheta_0 \quad \text{as} \quad n \to \infty.
\]
\end{theorem}

Next, we turn to establishing the asymptotic normality of the estimator \(\hat{\vartheta}\), which requires a more delicate and nuanced analysis. Two primary challenges arise in this endeavor.  {First, the likelihood contributions \(g_{i_1 j_1 k_1 l_1}(\vartheta)\) and \(g_{i_2 j_2 k_2 l_2}(\vartheta)\) are not independent when their corresponding tetrads \(\{i_1,j_1,k_1,l_1\}\) and \(\{i_2,j_2,k_2,l_2\}\) share two or more common nodes. This intrinsic dependence imparts a U-statistic-like structure to the score vector \(\nabla_{\vartheta} \mathcal{L}_n(\vartheta)\), necessitating the use of a Hoeffding-type decomposition to disentangle the dependence and accurately characterize the variance. See Proposition \ref{variance} in the supplementary material for more details.} Second, the convergence rates of \(\hat{\rho}\) and \(\hat{\gamma}\) depend critically on the sparsity regime of the network and may differ substantially under certain parameter scalings, further complicating the asymptotic analysis. To rigorously handle these issues, we impose a strengthened sparsity condition (see Assumption \ref{sparseass2}), which ensures sufficient sample size and regularity for asymptotic normality to hold.

\begin{Assum}[Strengthened sparsity condition]\label{sparseass2}
The sparsity parameter $a$ further satisfies \(0 < a < 2/3\).
\end{Assum}

Assumption \ref{sparseass2} ensures that the H\'ajek-type projection of the score function \(\nabla_{\vartheta} \mathcal{L}_n(\vartheta)\), which is a summation over all dyads, dominates the variance decomposition. Specifically, the covariance matrix corresponding to the projection onto dyad \((i,j)\) is given by
\[
\Gamma_{ij}(\vartheta_0) = \binom{n-2}{2}^{-2} \sum_{(k_1,l_1), (k_2,l_2) \in \mathcal{H}_{ij}} \mathbb{E}\left[\psi_{ijk_1l_1}(\vartheta_0)\, \psi^\top_{ijk_2l_2}(\vartheta_0)\right],
\]
where \(\psi_{ijkl}(\vartheta_0) = \nabla_{\vartheta} g_{ijkl}(\vartheta_0)\) and 
\[
\mathcal{H}_{ij} = \{(k,l): k < l, \{k,l\} \cap \{i,j\} = \emptyset \}.
\]
To guarantee a well-defined asymptotic variance, Assumption \ref{score} requires that the appropriately scaled sum of these covariance matrices is positive definite.

\begin{Assum}\label{score}
Define
\[
\eta_n = 
\begin{cases}
\bigl(n^{\frac{6a + \min\{a,b\}}{2}}, \; n^{\frac{6a - 3b}{2}} \mathbf{1}_q^\top\bigr)^\top, & b > 0, \\
n^{\frac{6a - 3b}{2}} \mathbf{1}_{q+1}, & b \leq 0.
\end{cases}
\]
Suppose that
\[
\lim_{n \to \infty} 36 \binom{n}{2}^{-1} \left( \eta_n \eta_n^\top \right) \odot \bigg( \sum_{i < j} \Gamma_{ij}(\vartheta_0) \bigg) = \Omega(\vartheta_0),
\]
where \(\Omega(\vartheta_0)\) is strictly positive definite.
\end{Assum}

Now, we state the asymptotic normality of \(\hat{\vartheta}\) in Theorem \ref{AN}.

\begin{theorem}[Asymptotic normality of \(\hat{\vartheta}\)] \label{AN}
Under Assumptions \ref{covass}, \ref{nodeass}, \ref{Hessian}, \ref{sparseass2}, and \ref{score}, we have
\[
\left(
\sqrt{n^{2 - 2a + \min\{a, b\}}} (\hat{\rho} - \rho_0), \quad
\sqrt{n^{2 - 2a + b}} (\hat{\gamma} - \gamma_0)^\top
\right)
\overset{d}{\longrightarrow} N(0, \Sigma_0),
\]
where 
\[
\Sigma_0 = H(\vartheta_0)^{-1} \, \Omega(\vartheta_0) \, H(\vartheta_0)^{-1}.
\]
When \(b \leq a\), \(H = H_0\). When \(b > a\), 
\[
H = \begin{pmatrix}
H_{0,1,1} & \mathbf{0}_q^\top \\
\mathbf{0}_q & H_{0,2:(q+1), 2:(q+1)}
\end{pmatrix}.
\]
\end{theorem}

Theorem \ref{AN} reveals that the convergence rate of \(\hat{\gamma}\) can exceed that of \(\hat{\rho}\) when \(b > a\). Furthermore, the asymptotic covariance matrix depends on the sparsity parameters, reflecting different information levels across parameters. Intuitively, as illustrated in Figure \ref{Sijkl}, both comparisons between configurations (II) and (I), and configurations (III) and (I), provide information for identifying \(\rho\) and \(\gamma\). However, the comparison between configurations (II) and (III) informs only \(\gamma\). Consequently, more data effectively inform the estimation of \(\gamma\) than \(\rho\), especially when the reciprocity effect is strong (i.e., \(b > a\)), causing configurations (II) and (III) to appear more frequently.

Despite these theoretical insights, direct statistical inference is complicated by the unknown sparsity parameters \(a\) and \(b\). We address this challenge through the following approach.

Recall that \(\vartheta_n = (\rho_n, \gamma^\top)^\top\), and denote by \(\ell_{ijkl}(\vartheta_n)\) the negative conditional log-likelihood for the tetrad \((i,j,k,l)\), defined with respect to \(\vartheta_n\). The associated functions \(g_{ijkl}(\vartheta_n)\), \(\psi_{ijkl}(\vartheta_n)\), and the full conditional likelihood \(\mathcal{L}_n(\vartheta_n)\) are defined analogously.

Define the empirical covariance matrix of the projected score:
\[
\Omega_n(\vartheta_n) = 36 \binom{n}{2}^{-1} \binom{n-2}{2}^{-2} \sum_{i<j} \sum_{(k_1,l_1), (k_2,l_2) \in \mathcal{H}_{ij}} \psi_{ijk_1l_1}(\vartheta_n) \psi_{ijk_2l_2}^\top(\vartheta_n),
\]
and let \(H_n(\vartheta_n)\) denote the Hessian matrix of \(\mathcal{L}_n(\vartheta_n)\).

Our estimator \(\hat{\vartheta}_n\) is defined as
\begin{align} \label{CLL_inference}
\hat{\vartheta}_n := \argmin_{\vartheta_n \in \mathbb{R}^{q+1}} \mathcal{L}_n(\vartheta_n).
\end{align}
The following theorem establishes that valid statistical inference on \(\vartheta_n\) is possible without requiring prior knowledge of the sparsity parameters \(a\) and \(b\).

\begin{theorem}[Inference for \(\vartheta_n\)]\label{Inference}
Under Assumptions \ref{covass}, \ref{nodeass}, \ref{Hessian}, \ref{sparseass2}, and \ref{score}, we have
\begin{align*}
n \cdot \frac{\hat{\rho}_n - \rho_{n0}}{\sqrt{\hat{\Sigma}_{1,1}}} &\overset{d}{\longrightarrow} \mathcal{N}(0, 1), \\
n \cdot \frac{\hat{\gamma}_{k} - \gamma_{k0}}{\sqrt{\hat{\Sigma}_{1+k,1+k}}} &\overset{d}{\longrightarrow} \mathcal{N}(0, 1), \quad \text{for } k = 1, 2, \dots, q,
\end{align*}
where \(\hat{\Sigma} = H_n(\hat{\vartheta}_n)^{-1} \Omega_n(\hat{\vartheta}_n) H_n(\hat{\vartheta}_n)^{-1}\).
\end{theorem}

Without the existence of the local reciprocity effect \(V^\top_{ij}\gamma\), model~\eqref{def: model} reduces to the classical \(p_1\) model \citep{holland1981exponential}. In this simplified setting, our estimator \(\hat{\rho}_n\) defined in~\eqref{CLL_inference} admits a closed-form solution, stated below.

\begin{Prop}[$\hat{\rho}_n$ in the \(p_1\) model]\label{p_one}
In the \(p_1\) model, the estimator in~\eqref{CLL_inference} is given by
\[
\hat{\rho}_n = \frac{1}{2}\log\left(\frac{\sum_{(i,j,k,l)}\mathbb{I}(S_{ijkl} = \pm 1)}{\sum_{(i,j,k,l)} \mathbb{I}(S_{ijkl} = 0)}\right).
\]
where for the summation, $(i,j,k,l)$ and $(j,i,k,l)$ are treated as two distinct tuples despite containing the same set of elements. Other permutations are similar. 
\end{Prop}

Proposition~\ref{p_one} also links our estimator to the method-of-moments estimator proposed by \citet{chang2024edge}, where the log odds ratio is proposed to identify the degree parameters for undirected networks. Moreover, it aligns with the cycle-based statistics used in \citet{jin2024optimal}, where the estimation of dyadic interaction parameters is based on counts of oriented 4-cycles. In both cases, the estimation reduces to evaluating ratios of specific configuration counts, highlighting the interpretability and computational simplicity of the pointer estimator under our framework in the absence of covariate effects. {However, we are currently not aware of any further simplification for the asymptotic variance of $\hat{\rho}_n$, leaving it for future research.}

\subsection{Minimax optimality}\label{sec:rate}

\begin{figure}[htbp]
\centering
\begin{tikzpicture}

\node[vertex] (1) at (8,0) {$i$};
\node[vertex] (2) at (10,0) {$j$};
\node[vertex] (3) at (8,-2) {$l$};
\node[vertex] (4) at (10,-2) {$k$};

\node[vertex] (6) at (2,0) {$i$};
\node[vertex] (7) at (4,0) {$j$};
\node[vertex] (8) at (2,-2) {$l$};
\node[vertex] (9) at (4,-2) {$k$};

\draw[->, bend left=10, thick,line width=0.25mm] (1) to (2);
\draw[->,  bend left=10, thick,line width=0.25mm] (2) to (1);
\draw[->, thick,line width=0.25mm] (3) to (4);

\draw[->, bend left=10, dashed, thick,line width=0.25mm] (1) to (4);
\draw[->, bend left=10, dashed, thick,line width=0.25mm] (4) to (1);
\draw[->, bend left=10, dashed, thick,line width=0.25mm] (2) to (3);
\draw[->, bend left=10, dashed, thick,line width=0.25mm] (3) to (2);

\draw[->, thick,line width=0.25mm] (7) to (9);
\draw[->, thick,line width=0.25mm] (6) to (7);
\draw[->, thick,line width=0.25mm] (8) to (6);
\draw[->, bend left=10, dashed, thick,line width=0.25mm] (6) to (9);
\draw[->, bend left=10, dashed, thick,line width=0.25mm] (9) to (6);
\draw[->, bend left=10, dashed, thick,line width=0.25mm] (7) to (8);
\draw[->, bend left=10, dashed, thick,line width=0.25mm] (8) to (7);
\end{tikzpicture}
\caption{Illustration of configurations corresponding to $\tilde{S}_{ijkl} = 0$ (left) and $\tilde{S}_{ijkl} = 1$ (right). Solid arrows represent required edge directions, while dashed lines indicate unrestricted edges.}
\label{tildeS}
\end{figure}
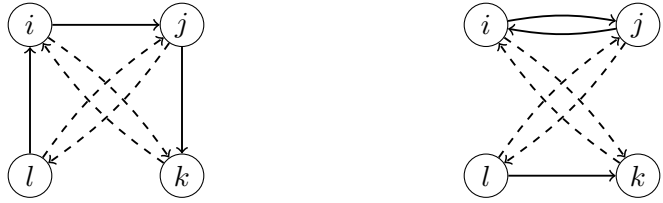

Theorem~\ref{AN} establishes the convergence rate and asymptotic normality of \(\hat{\vartheta}\). However, it remains to justify the specific construction of the tetrad configurations in Figure~\ref{Sijkl}, and to understand how our proposed estimator compares with alternative conditional likelihood approaches. To this end, consider an alternative configuration based on a new statistic \(\check{S}_{ijkl}\), defined as
\[
\check{S}_{ijkl} =  \begin{cases} 
0 & \text{if } \phi_{ij}(1,0)\phi_{jk}(1,0)\phi_{kl}(0,0)\phi_{li}(1,0)=1, \\
1 & \text{if } \phi_{ij}(1,1)\phi_{jk}(0,0)\phi_{kl}(0,1)\phi_{li}(0,0)=1, \\
\infty & \text{otherwise}.
\end{cases}
\]
See Figure~\ref{tildeS} for an illustration of these configurations. As in Lemma~\ref{sufficiency}, conditional on the event \(\check{S}_{ijkl} \in \{0,1\}\), we have
\begin{align*}
\mathbb{P}\big(\check{S}_{ijkl}=0 \mid \textbf{V}_{ijkl}, \check{S}_{ijkl}\in \{0,1\}\big) &= \frac{1}{1+\exp(\rho_n + V^\top_{ij}\gamma)}, \\
\mathbb{P}\big(\check{S}_{ijkl}=1 \mid \textbf{V}_{ijkl}, \check{S}_{ijkl}\in \{0,1\}\big) &= \frac{\exp(\rho_n + V^\top_{ij}\gamma)}{1+\exp(\rho_n + V^\top_{ij}\gamma)},
\end{align*}
both of which depend on \(\rho_n\) and \(\gamma\). Consequently, we may define the corresponding log-likelihood contributions \(\check{\ell}_{ijkl}(\vartheta)\), score terms \(\check{g}_{ijkl}(\vartheta)\), the total likelihood \(\check{\mathcal{L}}_{n}(\vartheta)\), and the associated estimator \(\check{\vartheta}\), in full analogy to the construction of \(\hat{\vartheta}\).

Following similar arguments as in Theorems~\ref{consistency} and~\ref{AN}, we obtain the asymptotic normality:
\[
\sqrt{n^{2 - 2a + \min\{a, b\}}}(\check{\vartheta} - \vartheta_0) \overset{d}{\longrightarrow} N(0, \Sigma_1),
\]
for some positive definite matrix \(\Sigma_1\). Crucially, this rate reflects a key difference in the performance of \(\hat{\vartheta}\) and \(\check{\vartheta}\). When the reciprocity effect is strong (\(b > a\)), the convergence rate of \(\hat{\gamma}\) exceeds that of \(\check{\gamma}\), due to the additional identification provided by comparing configurations (II) and (III) in Figure~\ref{Sijkl}, which \(\check{S}_{ijkl}\) fails to exploit. On the other hand, the convergence rates for \(\hat{\rho}\) and \(\check{\rho}\) are always the same. 

We conclude that $\hat{\vartheta}$ achieves a strictly faster convergence rate for estimating $\gamma$ when $b > a$. Crucially, we establish that this estimator is minimax rate-optimal among all possible estimations.

We next formally show that $\hat{\vartheta}$ achieves the minimax lower bound in terms of mean squared error. Let $\theta_n = (\mu_n, \rho_n, \gamma^\top, \alpha^\top, \beta^\top)^\top$, and define the parameter class
\[
\Theta_n = \left\{ \theta_n \colon \mu_n \in I_1, \rho_n \in I_2, \max\left\{ \Vert \gamma \Vert_2, \max\nolimits_{i} \{ \alpha_i, \beta_i \} \right\} \leq C_3 \right\},
\]
where $I_1 = [-a \log n - C_1, -a \log n + C_1]$ and $I_2 = [b \log n - C_2, b \log n + C_2]$ for some positive constants $C_1, C_2, C_3$. In what follows, we treat the covariates as fixed and assume they are uniformly bounded. Let $\mathcal{P} = \{ (p_{ij}(0,0), p_{ij}(1,0), p_{ij}(0,1), p_{ij}(1,1))_{i<j} \colon \theta_n \in \Theta_n \}$ denote the family of distributions governing the observed network $(A_{ij}, A_{ji})_{i<j}$. The following theorem establishes a minimax lower bound for estimating $\rho_n$ and $\gamma$.

\begin{theorem}[Rate optimality]\label{Rate}
For any estimators $\tilde{\rho}_n$ and $\tilde{\gamma}$, we have
\begin{align*}
& \sup _{P \in \mathcal{P}}\mathbb{E}_{P}(\tilde{\rho}_n - \rho_{n0})^2  \gtrsim n^{-2 + 2a - \min\{a, b\}}, \\
& \sup _{P \in \mathcal{P}}\mathbb{E}_{P}\Vert \tilde{\gamma} - \gamma_0 \Vert_{2}^2  \gtrsim n^{-2 + 2a - b},
\end{align*}
where $\mathbb{E}_P$ denotes expectation under distribution $P$.
\end{theorem}
For our estimator defined in (\ref{CLL_inference}), we have $\hat{\vartheta}_n - \vartheta_{n0} = \hat{\vartheta} - \vartheta_0$. By Theorem \ref{AN}, we obtain $(\hat{\rho}_n - \rho_{n0})^2 = O_P(n^{-2 + 2a - \min\{a, b\}})$ and $\Vert \hat{\gamma} - \gamma_0 \Vert_2^2 = O_P(n^{-2 + 2a - b})$. Comparing these with the minimax lower bounds established in Theorem \ref{Rate}, we conclude that our estimator $\hat{\vartheta}_n$ attains the optimal rates of convergence and is thus minimax rate-optimal.

Theorem \ref{Rate} shows that no estimator of $\rho_n$ and $\gamma$ can achieve mean squared error smaller than order $n^{-2 + 2a - \min\{a, b\}}$ and $n^{-2 + 2a - b}$, respectively, over the parameter class $\Theta_n$. Combined with the matching upper bounds from Theorem \ref{AN}, this implies that $\hat{\vartheta}_n$ is statistically efficient in the minimax sense. In particular, when the reciprocity effect dominates ($b > a$), our choice of tetrad configuration leads to a strictly faster convergence rate for $\hat{\gamma}$ than other conditional likelihood approaches, highlighting a key advantage of our estimator. 

Moreover, the framework developed here can be applied more broadly to study the rate optimality of other estimators constructed via conditional likelihood. For instance, it applies to the tetrad estimator, denoted $\hat{\beta}_{\text{TL}}$ in \cite{graham2017econometric}, which serves as an estimator of the homophily parameter associated with the covariates in their $\beta$-model with covariates. 
Following Corollary~1 in \cite{graham2017econometric}, the author remarked that the tetrad estimator $\hat{\beta}_{\text{TL}}$ converges at the usual parametric rate under dense graphs, whereas its convergence rate is considerably slower in sparse networks. Building on an argument similar to that used in Theorem~\ref{Rate}, we can further show that $\hat{\beta}_{\text{TL}}$, when all unobserved agent-level attributes are treated as fixed, is also minimax rate-optimal even in sparse networks. This result provides additional evidence for the appeal of the conditional likelihood approach across various network models.


\section{Numerical Study}\label{Sec: Numerical}

\subsection{Simulation}\label{simulation}

We investigate the finite-sample performance of our estimator defined in (\ref{CLL_inference}) through extensive simulation studies. Specifically, we consider networks with node sizes \(n = 50, 100, 150,\) and \(200\) to assess how estimation accuracy scales with network size. The sparsity indices are set to \(a = 0.2\) and \(b = 0.1\), and the true underlying parameter vector is fixed at \((\mu_0, \rho_0) = (0.5, 0.5)\) with \(\gamma_0 = (\gamma_{10}, \gamma_{20})^\top = (1, 1)^\top\). Additional simulations exploring a broader range of sparsity regimes and parameter settings are reported in the supplementary material to demonstrate the robustness of our methodology.

For heterogeneity parameters, each \(\alpha_i\) and \(\beta_i\) is independently drawn from a uniform distribution \(\text{Unif}[0,1]\). Covariates \(V_{ij} = (V_{ij,1}, V_{ij,2})^\top\) are generated independently with \(V_{ij,1} \sim \text{Unif}[-1,1]\) and \(V_{ij,2} \sim N(0,1)\), reflecting a mixture of bounded and unbounded continuous predictors. For each network size, we simulate 1000 independent datasets to provide reliable empirical performance metrics.

\begin{figure}[htbp]
\centering
\subfloat[Absolute error for $\hat{\rho}_n$]{
\includegraphics[width=.3\textwidth]{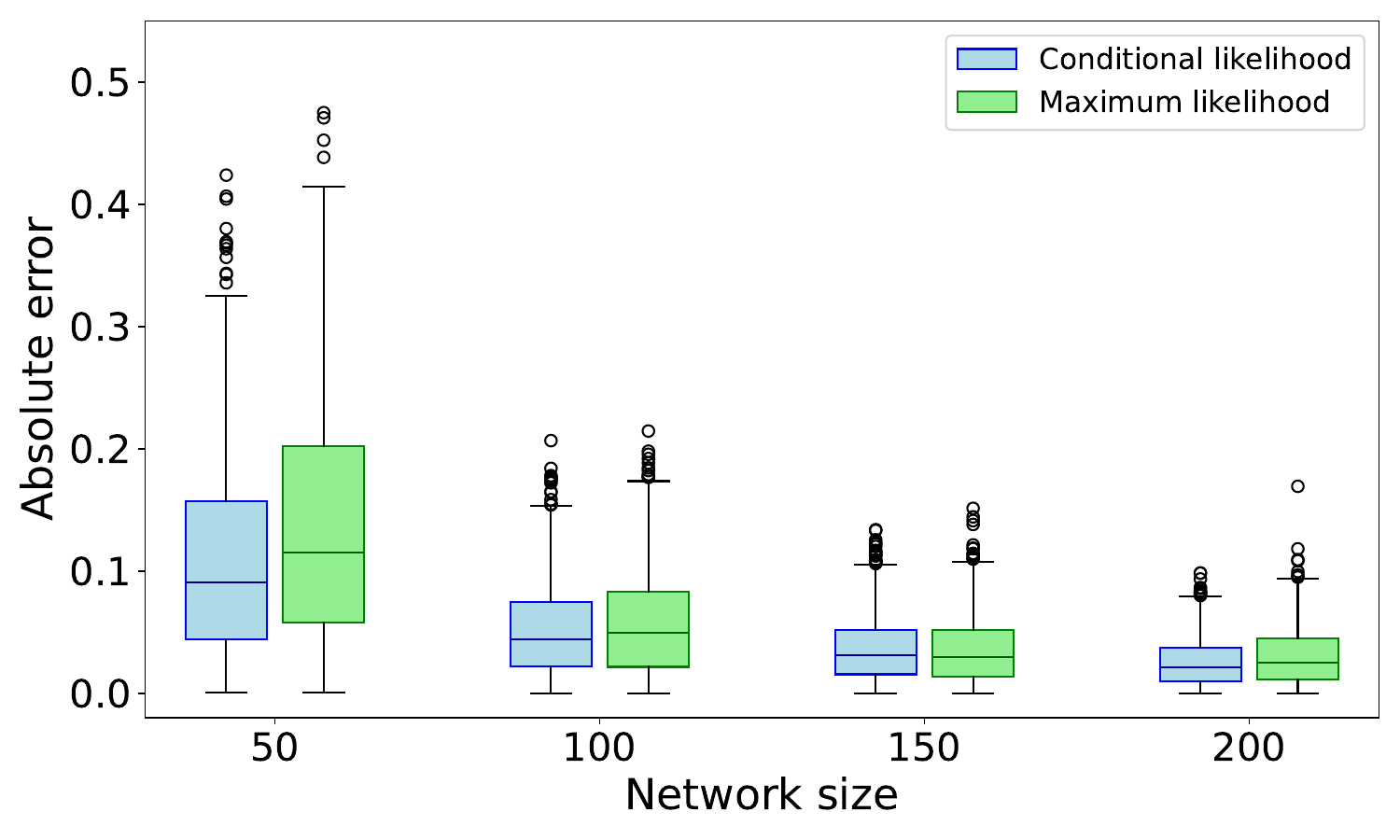}
}
\hfill
\subfloat[Absolute error for  $\hat{\gamma}_1$]{
\includegraphics[width=.3\textwidth]{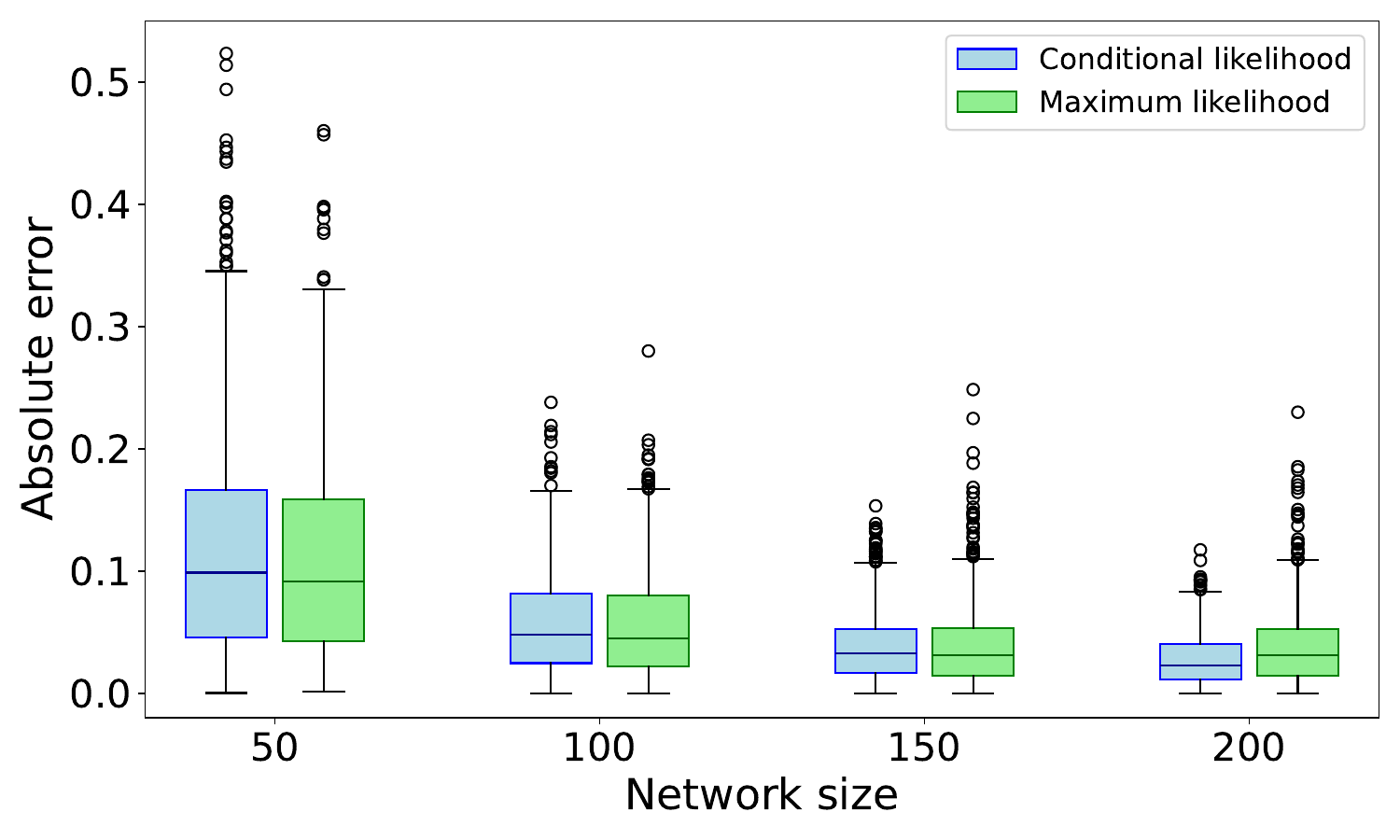}
}
\hfill
\subfloat[Absolute error for  $\hat{\gamma}_2$]{
\includegraphics[width=.3\textwidth]{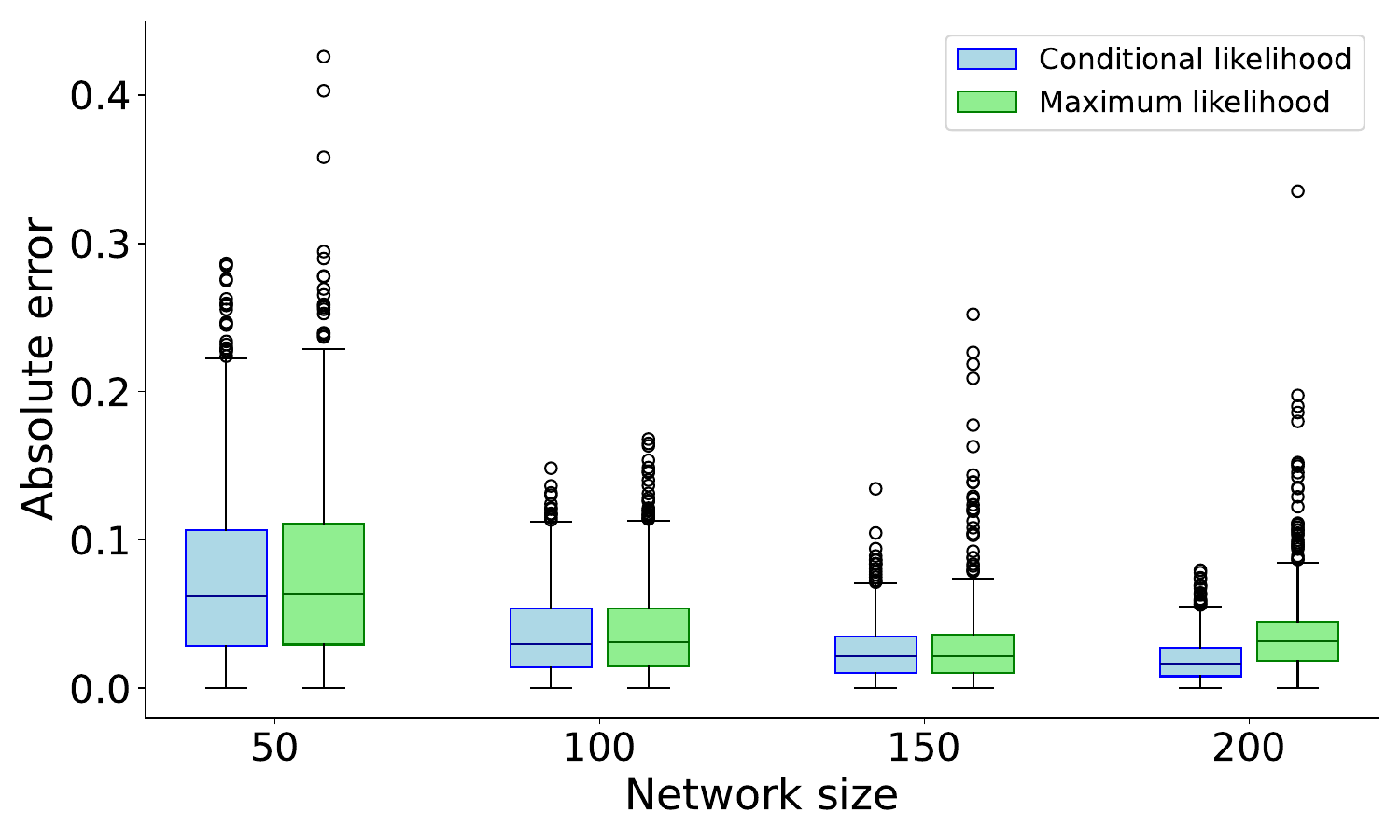}
}
\caption{MAE of our estimator and the MLE for the true parameters \(\rho_n\) and \(\gamma_k\) (\(k=1,2\)) across different network sizes. The decreasing trend illustrates improved estimation accuracy as the number of nodes \(n\) increases.}
\label{error_plot_1}
\end{figure}

To quantify estimation accuracy, we compute the mean absolute error (MAE) of \(\hat{\rho}_n\) and each component \(\hat{\gamma}_k\) \((k=1,2)\). The results, depicted in Figure \ref{error_plot_1}, exhibit a clear decreasing trend in estimation error as \(n\) grows, consistent with the theoretical consistency established in Theorem \ref{consistency}. {Additionally, we compute the MLE in~(\ref{def: model}), whose asymptotic properties remain largely unexplored \citep{yan2015simulation}, and compare it with our proposed method. As shown in Figure~\ref{error_plot_1}, our estimator achieves empirically comparable error rates to the MLE, thereby demonstrating its efficiency to some extent.} 

\begin{figure}[htbp]
\centering
\subfloat[Normal QQ plot for $\hat{\rho}_n$]{
\includegraphics[width=.28\textwidth,]{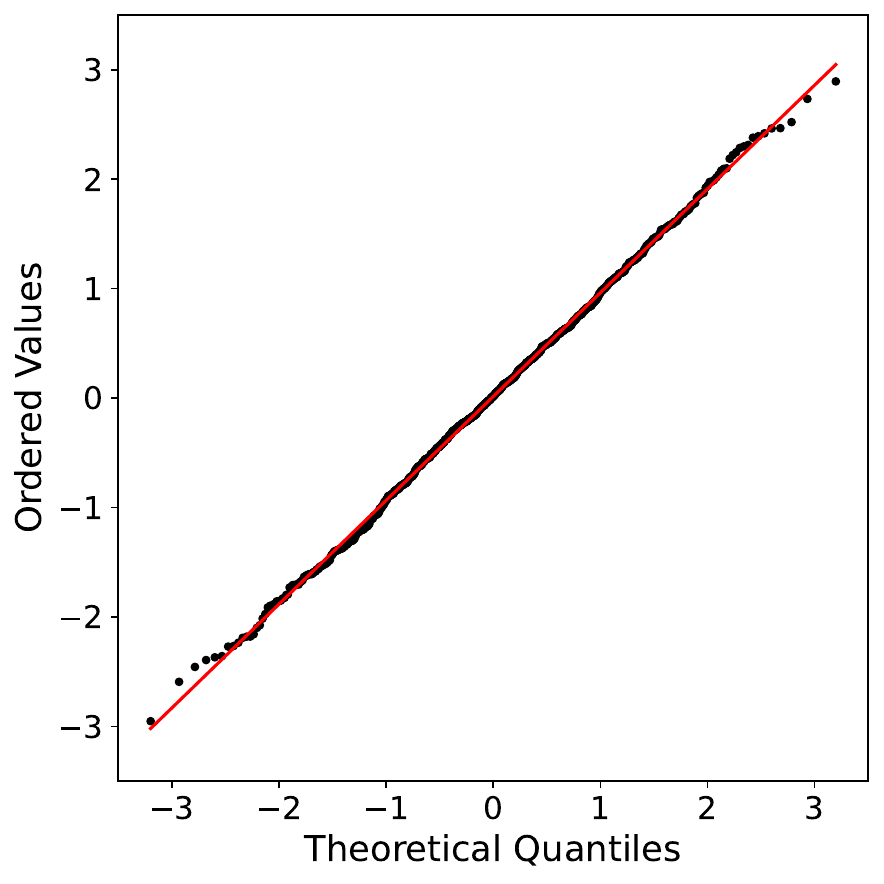}
\label{fig:subfigure1}
}
\hfill
\subfloat[Normal QQ plot for $\hat{\gamma}_1$]{
\includegraphics[width=.28\textwidth]{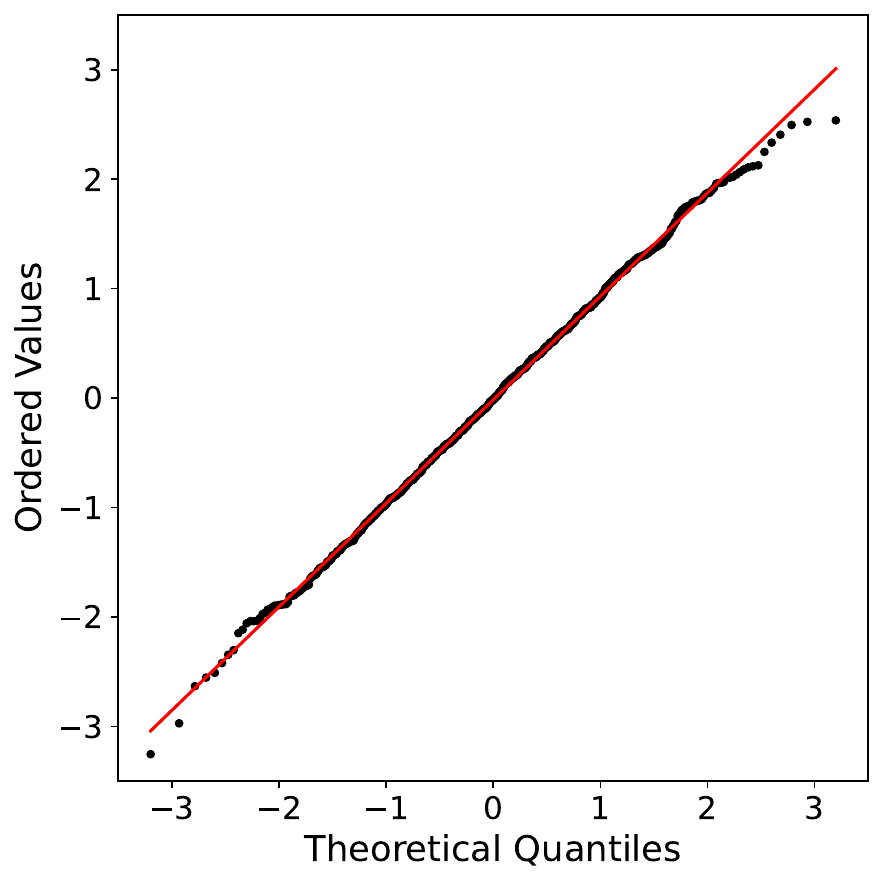}
\label{fig:subfigure2}
}
\hfill
\subfloat [Normal QQ plot for $\hat{\gamma}_2$]{
\includegraphics[width=.28\textwidth]{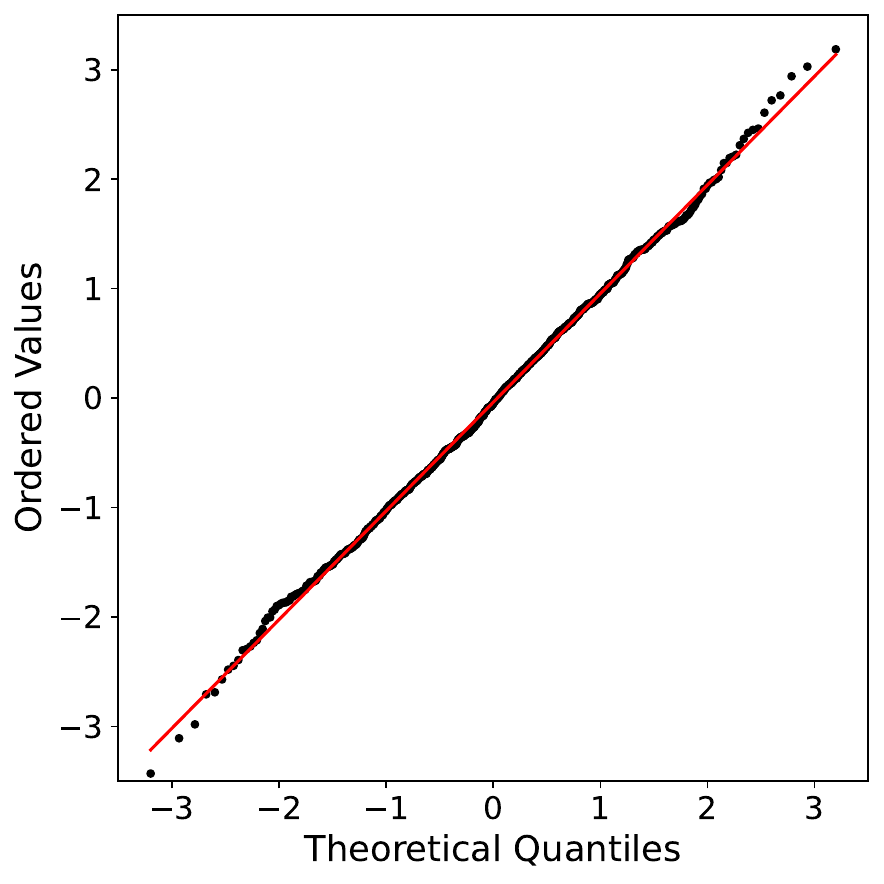}
\label{fig:subfigure3}
}
\caption{Quantile-quantile (QQ) plots of the standardized estimators for \(\rho_n\) and \(\gamma_k\) (\(k=1,2\)) when \(n=200\). The alignment of empirical quantiles with the theoretical normal quantiles supports the asymptotic normality of the estimators.}
\label{qq}
\end{figure}

Beyond point estimation accuracy, we assess the validity of the asymptotic normality result stated in Theorem \ref{AN} via quantile-quantile (QQ) plots. Due to space constraints, we present the QQ-plots for \(n=200\) in Figure \ref{qq}. These plots compare the empirical quantiles of the scaled estimation errors with the corresponding quantiles of the standard normal distribution. The close alignment between the empirical points and the reference line \(y = x\) across all parameters provides strong evidence that the sampling distribution of the estimator closely approximates the normal distribution in moderate to large samples.

Moreover, we evaluate the practical utility of our theoretical inference framework by constructing \(95\%\) confidence intervals for \(\rho_n\) and \(\gamma_k\) using the variance estimates from Theorem \ref{Inference}. Table \ref{Table: coverage} reports the empirical coverage probabilities and median lengths of these confidence intervals. The coverage probabilities are uniformly close to the nominal \(95\%\) level across all network sizes, empirically validating the asymptotic normality and variance estimation procedures. Moreover, the median interval lengths decrease as \(n\) increases, reflecting the improved precision of our estimator with growing sample size.

\begin{table}[htbp]
    \centering
    \begin{tabular}[t]{lrrrrrrrrrrrr}
        \toprule
             & \multicolumn{2}{c}{{$n = 50$}} && \multicolumn{2}{c}{{$n = 100$}} && \multicolumn{2}{c}{{$n = 150$}} && \multicolumn{2}{c}{{$n = 200$}}\\
        \midrule
        & Coverage & Width && Coverage & Width && Coverage & Width && Coverage & Width \\
        \midrule
        \addlinespace[0.3em]
        $\hat{\rho}_n$  & $98.4\%$ & 0.601 && $97.5\%$ & 0.275 && $95.8\%$ & 0.179&& $95.9\%$ & 0.132 \\
        $\hat{\gamma}_1$ & $98.0\%$ & 0.700 && $98.5\%$ & 0.306 && $95.9\%$ & 0.195 && $96.5\%$ & 0.144 \\
        $\hat{\gamma}_2$  & $97.2\%$ & 0.461 && $96.8\%$ & 0.201&& $95.4\%$ & 0.128 && $96.1\%$ & 0.094 \\
        \bottomrule
    \end{tabular}
    \caption{Empirical coverage probabilities at the nominal 95\% level and median lengths of the confidence intervals for the estimator \(\hat{\vartheta}_n\) across different network sizes. }

    \label{Table: coverage}
\end{table}

\begin{figure}[htbp]
\centering
\subfloat[Absolute error for $\hat{\rho}_n$]{
\includegraphics[width=.3\textwidth]{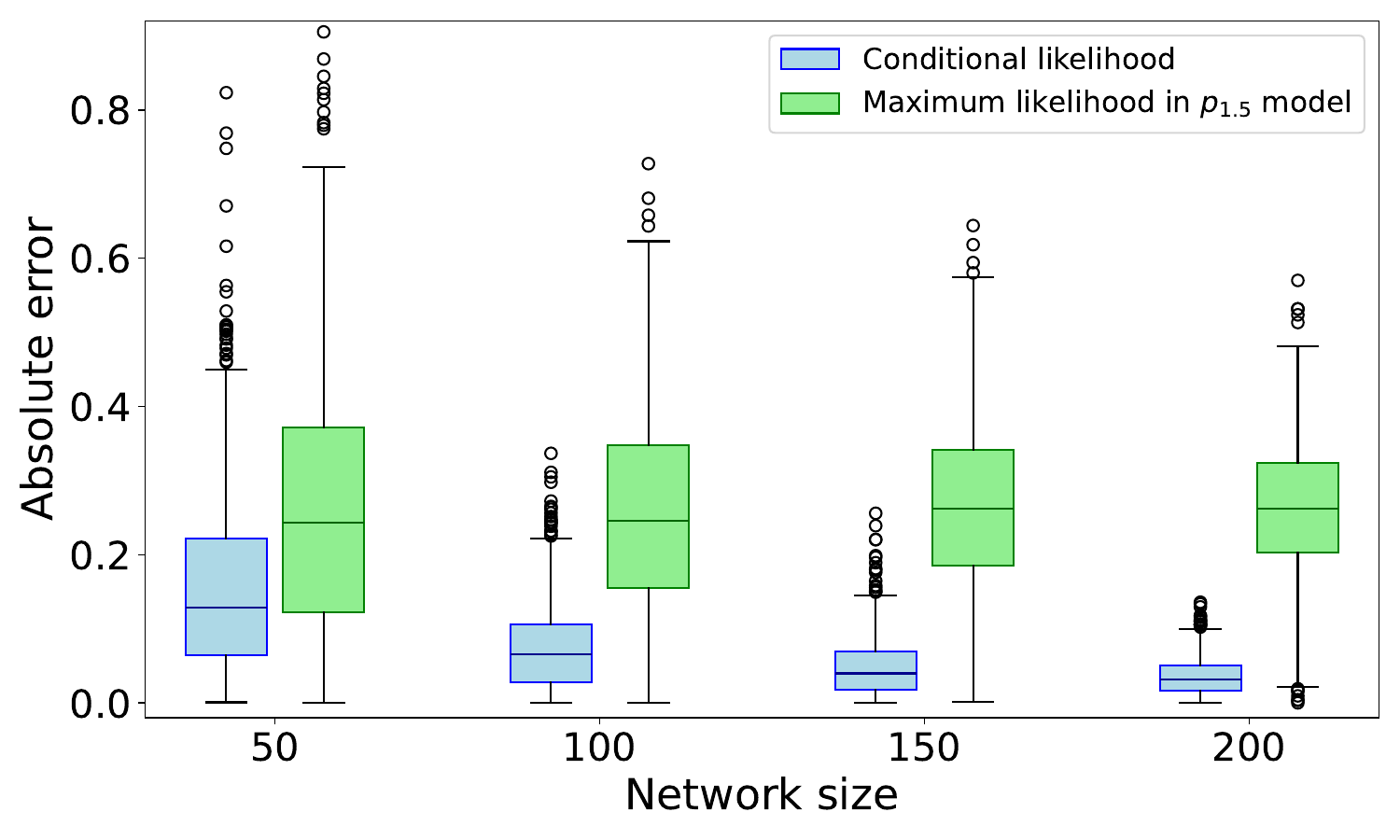}
}
\hfill
\subfloat[Absolute error for  $\hat{\gamma}_1$]{
\includegraphics[width=.3\textwidth]{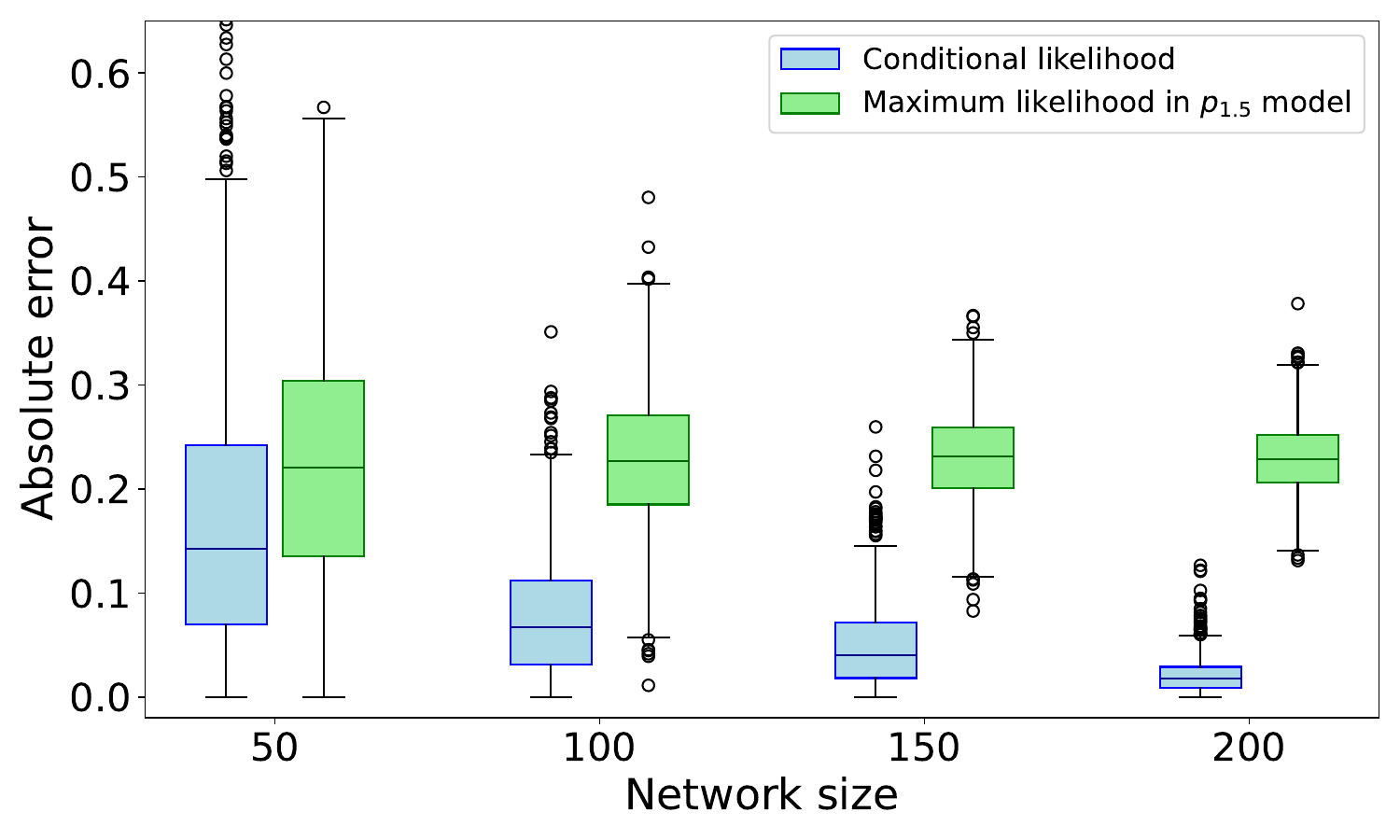}
}
\hfill
\subfloat[Absolute error for  $\hat{\gamma}_2$]{
\includegraphics[width=.3\textwidth]{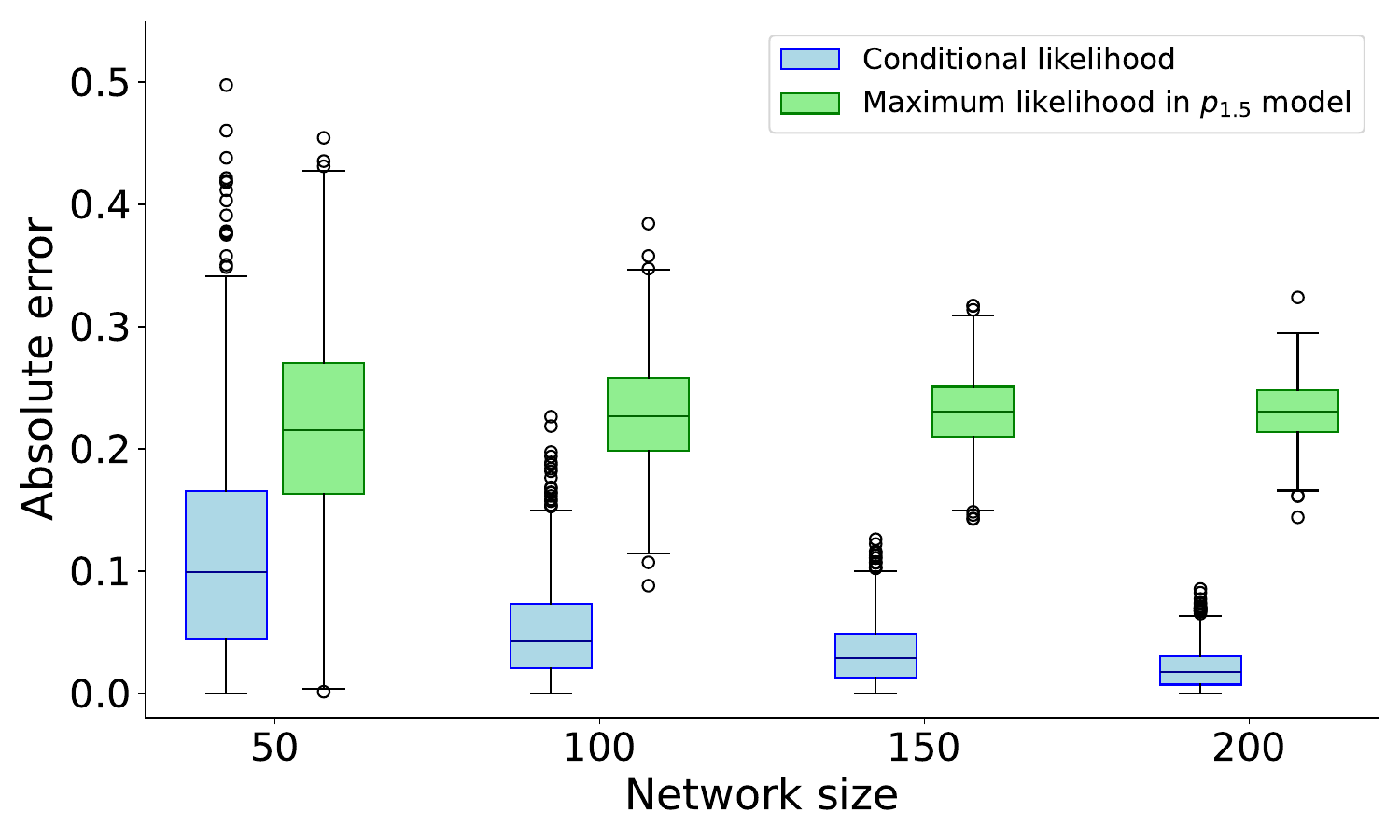}
}
\caption{MAE of our estimator and MLE in the $p_{1.5}$ model across different network sizes. With unobserved degree heterogeneity, fitting the data using the $p_{1.5}$ model will be biased.}
\label{error_p1.5_1}
\end{figure}

Finally, we investigate a setting in which the degree heterogeneity parameters are modeled as \(\alpha_i = X_i \delta_1 + u_i\) and \(\beta_j = Y_j \delta_2 + v_j\). Here, the covariates \(X_i\) and \(Y_j\) are independently drawn from a uniform distribution \(\mathrm{Unif}[-1,1]\) to capture observed heterogeneity, while the latent heterogeneity components \(u_i\) and \(v_j\) are generated from a standard normal distribution. We fix \(\delta_1 = \delta_2 = 1\) and keep all other parameters unchanged. To evaluate our estimator under this setting, we apply the maximum likelihood estimator (MLE) based on the $p_{1.5}$ model \citep{feng2025modelling} and compare its performance with that of our proposed method. Note that the $p_{1.5}$ model is mis-specified in this scenario. The results, displayed in Figure~\ref{error_p1.5_1}, demonstrate that the $p_{1.5}$ model produces biased estimates of the reciprocity parameter in the presence of unobserved heterogeneity, even as the network size increases. In contrast, our proposed estimator consistently yields accurate estimates, as evidenced by the decreasing error rates.

Overall, these simulation results demonstrate not only the consistency and efficiency of our estimator in finite samples but also the reliability of the proposed inference procedures, providing strong support for its application in practical network analysis scenarios.

\subsection{Data analysis}

\textbf{Analysis of Lazega‘s Lawyer Dataset.}  
We apply our proposed model to the well-studied Lazega's lawyer dataset \citep{lawyernetwork}, which has served as a benchmark in numerous network analyses \citep{hunter2006inference, handcock2010modeling, ma2020universal}. This dataset comprises relational data among 71 lawyers (36 partners and 35 associates) at a New England law firm. We focus on the advice-seeking network, represented as a directed graph where an edge from lawyer \(i\) to lawyer \(j\) indicates that \(i\) has sought basic professional advice from \(j\). The network exhibits a density of 0.18, with out-degree ranging from 0 to 30 and in-degree from 0 to 37. Table \ref{tab: degree distribution} summarizes the in-degree and out-degree distributions, highlighting pronounced bi-degree heterogeneity within the network. The rows `Difference' and `Absolute difference' quantify, respectively, the signed and absolute differences between each node's in-degree and out-degree. Notably, the network contains 175 reciprocated dyads, corresponding to an average of roughly five mutual ties per node, suggesting a strong underlying reciprocity effect.

\begin{table}[htbp]
	\centering
	\begin{tabular}{lrrrrr}
		\toprule
		 & Minimum & $25\%$ Quantile & Mean & $75\%$ Quantile & Maximum\\
                 \midrule
		In-degrees & $0    $ & $8.5$ &$12.6$ & $19.5$ & $37$  \\
		Out-degrees & $0   $ & $7.0$ & $12.6$ & $17.0$ & $30$ \\
		Difference & $-3   $ & $-1.0$ & $0$ & $1$ & $7$ \\
		Absolute difference & $0 $  & $0.5   $ & $1.2$ & $2$ & $7$ \\		
				\bottomrule
	\end{tabular}
\caption{Degree distributions in Lazega's advice network.}
	\label{tab: degree distribution}
\end{table}

Besides network structure, several node-level covariates are available: professional status (partner vs.\ associate), gender (male vs.\ female), office location (Boston, Hartford, or Providence), years of tenure, age, practice area (litigation vs.\ corporate), and law school attended (Harvard, Yale, UConn, or other). We begin by fitting a null model without covariates, obtaining an estimated reciprocity parameter \(\hat{\rho}_n = 3.00\) with a 95\% confidence interval of \((2.12, 3.89)\), confirming a statistically significant and substantial reciprocity effect in this professional advice network.

To motivate covariate inclusion, we conduct preliminary exploratory analyses. For example, considering the contingency table formed by dyad types \((1,0)\) and \((1,1)\) and an indicator of whether the lawyers share the same professional status, a chi-squared test yields a p-value of \(2.62 \times 10^{-9}\), strongly justifying the inclusion of this covariate in our model. See Table \ref{tab: contingency_status} for the contingency table. Conversely, for the covariate `same office', the corresponding chi-squared test p-value is \(6.02 \times 10^{-2}\), suggesting a weaker, though possibly relevant, effect.

\begin{table}[htbp]
	\centering
	\begin{tabular}{lrr}
		\toprule
		 & Different status & Same status \\
                 \midrule
		(1,0) & $237$ & $305$   \\
		(1,1) & $32$ & $143$ \\	
				\bottomrule
	\end{tabular}
	\caption{Contingency table of dyad configurations by professional status. The chi-squared test yields a highly significant p-value of $2.62 \times 10^{-9}$, indicating a strong association between dyad types and shared status.}
	\label{tab: contingency_status}
\end{table}

Subsequently, we incorporate indicator variables reflecting whether lawyers share the same status, gender, office location, practice area, or law school into the model. The estimated coefficients and their 95\% confidence intervals are reported in Table \ref{tab: lawyer results}. Our findings largely conform to substantive expectations: lawyers are more likely to seek advice from colleagues who share their office, professional status, or practice area, consistent with homophily and organizational proximity effects. While the point estimates for sharing the same gender and law school are positive, their confidence intervals include zero, indicating that these effects are not statistically significant at conventional levels. Thus, no definitive conclusions can be drawn about the roles of gender or alma mater in shaping advice-seeking ties within this network.

\begin{table}[htbp]
	\centering
	\begin{tabular}{lrr}
		\toprule
		Covariate & Estimate & Confidence Interval\\
                 \midrule
                 Intercept $\hat{\rho}_n$ & $-3.10$ & $(-4.91, -1.31)$\\
		Same status & $2.40   $ & $(1.53 ,  3.27)$ \\
		Same gender & $0.55    $ & $( -0.44 ,  1.54)$\\
		Same office & $3.72     $ & $( 2.52 ,  4.91)$\\
		Same practice & $2.59   $ & $(1.71,  3.47)$\\
		Same law school & $0.72$ & $( -0.01,  1.46)$\\
		\bottomrule
	\end{tabular}
	\caption{Estimation results with corresponding 95\% confidence intervals for Lazega's lawyer advice network model.}
\label{tab: lawyer results} 
\end{table}

Overall, our results illustrate the practical utility of the proposed model in capturing complex dependency structures in directed social networks, particularly with respect to reciprocity and covariate effects. The ability to jointly model these factors provides richer insights into network formation mechanisms compared to simpler models that ignore reciprocity or covariate information.

\noindent \textbf{Analysis of the trade network.} We now turn our attention to the trade partnerships network data collected by \cite{Silva:Tenreyro:2006}, and previously analyzed by \cite{jochmans2018semiparametric} and \cite{stein2024sparse}. This dataset comprises bilateral export flows among 136 countries in the year 1990. Several dyadic covariates capturing relational closeness between country pairs are available, including indicators for a shared language, a common border, the presence of a preferential trade agreement, and historical colonial ties. Additionally, the dataset contains the logarithm of bilateral distance, defined as the geographical distance between the capitals of the two countries.

For our analysis, we define a directed edge from country \(i\) to country \(j\) if the export volume from \(i\) to \(j\) constitutes at least \(1\%\) of country \(i\)'s total exports. This criterion identifies country \(j\) as a significant export partner for country \(i\). The resulting directed network consists of 136 nodes and 1,880 edges, corresponding to a density of approximately \(10.2\%\). Notably, the network contains 234 reciprocated (mutual) trade links, indicating a substantial level of bilateral trade relationships.

\begin{figure}[htbp]
	\centering
	\includegraphics[width=.30\textwidth]{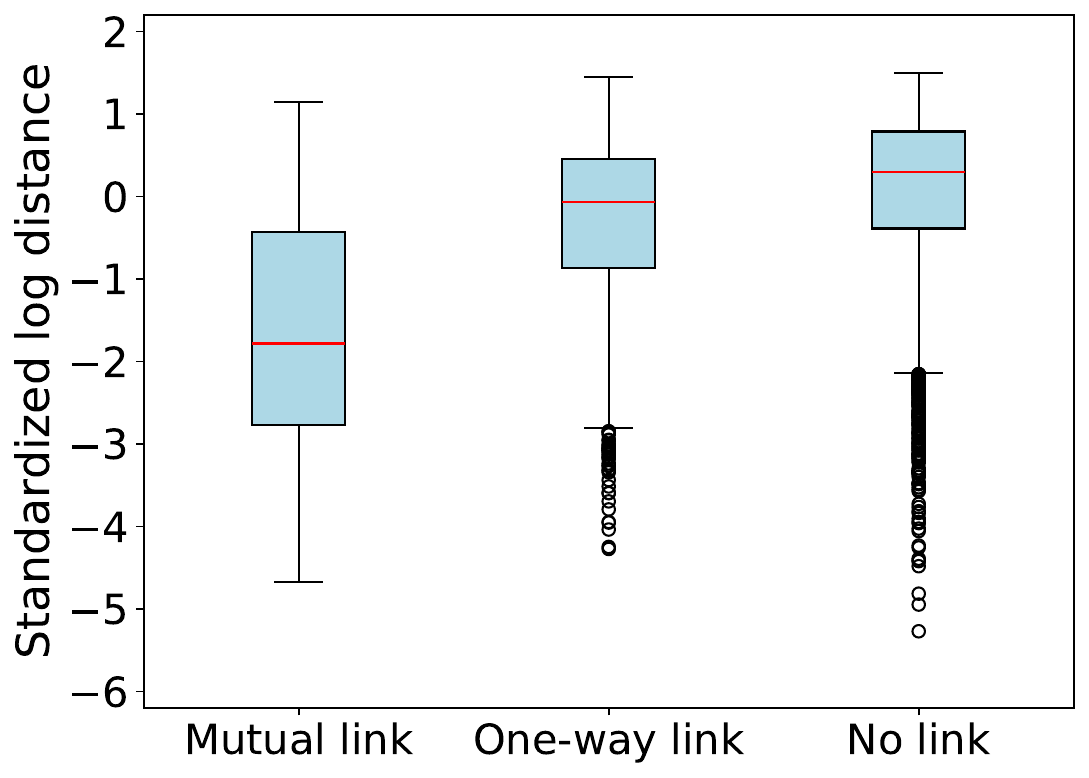}
	\caption{Boxplot of the log distance covariate grouped by dyad configurations. Dyads with mutual trade links exhibit lower average log distances, highlighting the role of geographic proximity in reciprocal trade relationships.}
	\label{fig: Mutual links}
\end{figure}

We first fit our model without covariates. The estimated reciprocity parameter is \(\hat{\rho}_n = 3.38\), with a 95\% confidence interval of \((2.46, 4.30)\), providing strong evidence of positive reciprocity in this trade network. To investigate covariate effects, we explore the continuous covariate (log distance) via boxplots stratified by dyad types, shown in Figure \ref{fig: Mutual links}. The figure reveals that dyads with mutual trade links tend to have lower average log distances, consistent with the intuition that geographic proximity fosters reciprocal trade. For binary covariates, we perform chi-squared tests similar to those in the analysis of Lazega's network. For example, the common language indicator yields a highly significant p-value of \(1.16 \times 10^{-5}\), supporting its inclusion in the model.

Finally, we fit the full model incorporating the log distance and four binary covariates: common border, common language, colonial ties, and preferential trade agreement. The estimation results are presented in Table \ref{tab:trade}. Our findings confirm that preferential trade agreements strongly facilitate bilateral trade partnerships, while greater geographical distance significantly reduces the likelihood of trade links. Although the point estimates for common border, common language, and colonial ties are positive, their confidence intervals include zero, preventing firm conclusions regarding their effects on network formation.

\begin{table}[htbp]
	\centering
	\begin{tabular}{lrr}
		\toprule
		Covariate & Estimate & 95\% Confidence Interval \\
		\midrule
		Intercept \(\hat{\rho}_n\) & 1.62 & (0.53, 2.71) \\
		Log distance & -1.61 & (-2.21, -1.01) \\
		Common border & 1.07 & (-1.12, 3.26) \\
		Common language & 0.13 & (-1.39, 1.66) \\
		Colonial ties & 0.97 & (-0.60, 2.53) \\
		Preferential trade agreement & 2.51 & (0.73, 4.28) \\
		\bottomrule
	\end{tabular}
	\caption{Parameter estimates and 95\% confidence intervals for the world trade network model.}
	\label{tab:trade}
\end{table}

The results obtained from the trade network analysis provide valuable insights into the structural and relational factors shaping international trade partnerships. The strong positive reciprocity effect, evidenced by a substantially positive \(\hat{\rho}_n\), suggests that trade relationships tend to be mutual rather than one-sided, reflecting the interdependent nature of global commerce. The significant negative coefficient for log distance aligns with well-established economic theories emphasizing the frictional costs of geographical separation on trade volume \citep{anderson2011gravity}. Meanwhile, the pronounced positive effect of preferential trade agreements underscores the importance of formal institutional arrangements in facilitating cross-border trade by reducing barriers and fostering cooperation.

\section{Discussion}\label{Sec: Discussion}

In this paper, we introduced a new statistical model for directed networks that explicitly captures the effect of reciprocity while accommodating bi-degree heterogeneity. Our estimator, based on a novel conditional likelihood approach, achieves the minimax optimal rate, and we established its consistency and asymptotic normality under the condition that the network is moderately sparse.

Beyond the core model, we discussed two possible generalizations of our methodology. First, although our current estimator targets the reciprocity parameter $\rho_n$ and the covariate effects $\gamma$, it is also possible to infer the node-specific heterogeneity parameters $\alpha_i$ and $\beta_j$ by leveraging differences across suitable triad configurations. This provides a potential alternative to the tetrad-based estimation framework. Second, our conditioning argument extends naturally to incorporate homophily effects through dyadic covariates \citep{yan2019statistical}. In particular, consider the following model for dyadic outcomes:
\[
\begin{aligned}
&p_{ij}(0,0) \propto 1, \quad
p_{ij}(1,0) \propto \exp\left(\mu_n + \alpha_i + \beta_j + Z_{ij}^\top \delta\right),\\
&p_{ij}(0,1) \propto \exp\left(\mu_n + \alpha_j + \beta_i + Z_{ji}^\top \delta\right), \quad
p_{ij}(1,1) \propto \exp\left(2\mu_n + \rho_n + V_{ij}^\top \gamma + \alpha_i + \alpha_j + \beta_i + \beta_j + (Z_{ij}^\top + Z_{ji}^\top)\delta\right),
\end{aligned}
\]
where $p_{ij}(v_1, v_2) = \mathbb{P}(A_{ij} = v_1, A_{ji} = v_2 \mid V_{ij}, Z_{ij}, Z_{ji})$ and $Z_{ij}$ is a covariate capturing edgewise homophily -- the tendency for nodes with similar attributes to form connections. Under suitable regularity conditions, a sufficiency result analogous to Lemma~\ref{sufficiency} can be established for this extended model, ensuring identifiability of the parameters $\rho_n$, $\gamma$, and $\delta$. These observations suggest that our approach can be generalized to a broader class of directed network models with more complex dependencies.

We also acknowledge two limitations of the current methodology. First, the computational cost is high due to the need to enumerate all valid tetrads and check their configurations. The resulting $O(n^4)$ time complexity limits practical applicability to medium-sized networks. {Developing more efficient algorithms or sampling-based approximations \citep{chen2019randomized, shao2025u} is an important avenue for future research.} Second, while we proved the rate-optimality of our estimator, we did not fully characterize the efficiency of its asymptotic variance. This characterization is essential for constructing tight and accurate confidence intervals. Addressing this gap remains an open problem.

\renewcommand{\baselinestretch}{1}
\normalsize
\bibliographystyle{apalike}
\bibliography{reference}

\begin{thebibliography}{}

\bibitem[Anderson, 2011]{anderson2011gravity}
Anderson, J.~E. (2011).
\newblock The gravity model.
\newblock {\em Annual Review of Economics}, 3(1):133--160.

\bibitem[Barab{\'a}si and Albert, 1999]{barabasi1999emergence}
Barab{\'a}si, A.-L. and Albert, R. (1999).
\newblock Emergence of scaling in random networks.
\newblock {\em Science}, 286(5439):509--512.

\bibitem[Barndorff-Nielsen, 1983]{barndorff1983formula}
Barndorff-Nielsen, O. (1983).
\newblock On a formula for the distribution of the maximum likelihood
  estimator.
\newblock {\em Biometrika}, 70(2):343--365.

\bibitem[Barndorff-Nielsen and McCullagh, 1993]{barndorff1993note}
Barndorff-Nielsen, O.~E. and McCullagh, P. (1993).
\newblock A note on the relation between modified profile likelihood and the
  cox-reid adjusted profile likelihood.
\newblock {\em Biometrika}, 80(2):321--328.

\bibitem[Bartlett, 1937]{bartlett1937properties}
Bartlett, M.~S. (1937).
\newblock Properties of sufficiency and statistical tests.
\newblock {\em Proceedings of the Royal Society of London. Series
  A-Mathematical and Physical Sciences}, 160(901):268--282.

\bibitem[Chang et~al., 2024]{chang2024edge}
Chang, J., Hu, Q., Kolaczyk, E.~D., Yao, Q., and Yi, F. (2024).
\newblock Edge differentially private estimation in the $\beta$-model via
  jittering and method of moments.
\newblock {\em The Annals of Statistics}, 52(2):708--728.

\bibitem[Charbonneau, 2017]{charbonneau2017multiple}
Charbonneau, K.~B. (2017).
\newblock Multiple fixed effects in binary response panel data models.
\newblock {\em The Econometrics Journal}, 20(3):S1--S13.

\bibitem[Chen et~al., 2021]{chen2021analysis}
Chen, M., Kato, K., and Leng, C. (2021).
\newblock Analysis of networks via the sparse $\beta$-model.
\newblock {\em Journal of the Royal Statistical Society Series B: Statistical
  Methodology}, 83(5):887--910.

\bibitem[Chen and Kato, 2019]{chen2019randomized}
Chen, X. and Kato, K. (2019).
\newblock Randomized incomplete $u$-statistics in high dimensions.
\newblock {\em The Annals of Statistics}, 47(6):3127--3156.

\bibitem[Cox and Reid, 1987]{cox1987parameter}
Cox, D.~R. and Reid, N. (1987).
\newblock Parameter orthogonality and approximate conditional inference.
\newblock {\em Journal of the Royal Statistical Society: Series B
  (Methodological)}, 49(1):1--18.

\bibitem[Dorogovtsev and Mendes, 2003]{dorogovtsev2003evolution}
Dorogovtsev, S.~N. and Mendes, J. F.~F. (2003).
\newblock {\em Evolution of Networks: From Biological Nets to the Internet and
  WWW}.
\newblock Oxford University Press.

\bibitem[Dzemski, 2019]{dzemski2019empirical}
Dzemski, A. (2019).
\newblock An empirical model of dyadic link formation in a network with
  unobserved heterogeneity.
\newblock {\em Review of Economics and Statistics}, 101(5):763--776.

\bibitem[Feng and Leng, 2025]{feng2025modelling}
Feng, R. and Leng, C. (2025).
\newblock Modelling directed networks with reciprocity.
\newblock {\em Biometrika}, 112(2):asaf035.

\bibitem[Fern{\'a}ndez-Val and Weidner, 2016]{fernandez2016individual}
Fern{\'a}ndez-Val, I. and Weidner, M. (2016).
\newblock Individual and time effects in nonlinear panel models with large n,
  t.
\newblock {\em Journal of Econometrics}, 192(1):291--312.

\bibitem[Fienberg, 2012]{fienberg2012brief}
Fienberg, S.~E. (2012).
\newblock A brief history of statistical models for network analysis and open
  challenges.
\newblock {\em Journal of Computational and Graphical Statistics},
  21(4):825--839.

\bibitem[Garlaschelli and Loffredo, 2004]{garlaschelli2004patterns}
Garlaschelli, D. and Loffredo, M.~I. (2004).
\newblock Patterns of link reciprocity in directed networks.
\newblock {\em Physical Review Letters}, 93(26):268701.

\bibitem[Graham, 2017]{graham2017econometric}
Graham, B.~S. (2017).
\newblock An econometric model of network formation with degree heterogeneity.
\newblock {\em Econometrica}, 85(4):1033--1063.

\bibitem[Handcock and Gile, 2010]{handcock2010modeling}
Handcock, M.~S. and Gile, K.~J. (2010).
\newblock Modeling social networks from sampled data.
\newblock {\em The Annals of Applied Statistics}, 4(1):5--25.

\bibitem[Holland and Leinhardt, 1981]{holland1981exponential}
Holland, P.~W. and Leinhardt, S. (1981).
\newblock An exponential family of probability distributions for directed
  graphs.
\newblock {\em Journal of the American Statistical Association},
  76(373):33--50.

\bibitem[Horn and Johnson, 2012]{horn2012matrix}
Horn, R.~A. and Johnson, C.~R. (2012).
\newblock {\em Matrix analysis}.
\newblock Cambridge university press.

\bibitem[Hunter and Handcock, 2006]{hunter2006inference}
Hunter, D.~R. and Handcock, M.~S. (2006).
\newblock Inference in curved exponential family models for networks.
\newblock {\em Journal of Computational and Graphical Statistics},
  15(3):565--583.

\bibitem[Jackson, 2008]{jackson2008social}
Jackson, M.~O. (2008).
\newblock {\em Social and Economic Networks}.
\newblock Princeton University Press.

\bibitem[Ji and Jin, 2016]{ji2016coauthorship}
Ji, P. and Jin, J. (2016).
\newblock Coauthorship and citation networks for statisticians.
\newblock {\em The Annals of Applied Statistics}, 10(4):1779--1812.

\bibitem[Jiang et~al., 2015]{jiang2015reciprocity}
Jiang, B., Zhang, Z.-L., and Towsley, D. (2015).
\newblock Reciprocity in social networks with capacity constraints.
\newblock In {\em Proceedings of the 21th ACM SIGKDD International Conference
  on Knowledge Discovery and Data Mining}, pages 457--466.

\bibitem[Jin et~al., 2025]{jin2024optimal}
Jin, J., Ke, Z.~T., Luo, S., and Ma, Y. (2025).
\newblock Optimal network pairwise comparison.
\newblock {\em Journal of the American Statistical Association},
  120(550):1048--1062.

\bibitem[Jochmans, 2018]{jochmans2018semiparametric}
Jochmans, K. (2018).
\newblock Semiparametric analysis of network formation.
\newblock {\em Journal of Business $\&$ Economic Statistics}, 36(4):705--713.

\bibitem[Kolaczyk, 2009]{kolaczyk2009statistical}
Kolaczyk, E.~D. (2009).
\newblock {\em Statistical Analysis of Network Data: Methods and Models}.
\newblock Springer.

\bibitem[Krivitsky and Kolaczyk, 2015]{kolaczyk2015question}
Krivitsky, P.~N. and Kolaczyk, E.~D. (2015).
\newblock On the question of effective sample size in network modeling: an
  asymptotic inquiry.
\newblock {\em Statistical Science}, 30(2):184--198.

\bibitem[Lazega, 2001]{lawyernetwork}
Lazega, E. (2001).
\newblock {\em The Collegial Phenomenon: The Social Mechanisms of Cooperation
  Among Peers in a Corporate Law Partnership}.
\newblock Oxford University Press.

\bibitem[Lou et~al., 2013]{lou2013learning}
Lou, T., Tang, J., Hopcroft, J., Fang, Z., and Ding, X. (2013).
\newblock Learning to predict reciprocity and triadic closure in social
  networks.
\newblock {\em ACM Transactions on Knowledge Discovery from Data (TKDD)},
  7(2):1--25.

\bibitem[Ma et~al., 2020]{ma2020universal}
Ma, Z., Ma, Z., and Yuan, H. (2020).
\newblock Universal latent space model fitting for large networks with edge
  covariates.
\newblock {\em Journal of Machine Learning Research}, 21(4):1--67.

\bibitem[Newman, 2018]{newman2018networks}
Newman, M. (2018).
\newblock {\em Networks}.
\newblock Oxford university press.

\bibitem[Newman et~al., 2002]{newman2002email}
Newman, M.~E., Forrest, S., and Balthrop, J. (2002).
\newblock Email networks and the spread of computer viruses.
\newblock {\em Physical Review E}, 66(3):035101.

\bibitem[Petrovic et~al., 2010]{petrovic2010algebraic}
Petrovic, S., Rinaldo, A., and Fienberg, S.~E. (2010).
\newblock Algebraic statistics for a directed random graph model with
  reciprocation.
\newblock {\em Contemporary Mathematics}, 516:261–283.

\bibitem[Rinaldo et~al., 2010]{rinaldo2010existence}
Rinaldo, A., Petrovi{\'c}, S., and Fienberg, S. (2010).
\newblock On the existence of the $\text{MLE}$ for a directed random graph
  network model with reciprocation.
\newblock {\em arXiv}, 2010.

\bibitem[Shao et~al., 2025]{shao2025u}
Shao, M., Xia, D., and Zhang, Y. (2025).
\newblock U-statistic reduction: Higher-order accurate risk control and
  statistical-computational trade-off.
\newblock {\em Journal of the American Statistical Association}.

\bibitem[Shao et~al., 2021]{shao20212}
Shao, M., Zhang, Y., Wang, Q., Zhang, Y., Luo, J., and Yan, T. (2021).
\newblock L-2 regularized maximum likelihood for $\beta$-model in large and
  sparse networks.
\newblock {\em arXiv preprint arXiv:2110.11856}.

\bibitem[Silva and Tenreyro, 2006]{Silva:Tenreyro:2006}
Silva, J. M. C.~S. and Tenreyro, S. (2006).
\newblock The log of gravity.
\newblock {\em The Review of Economics and Statistics}, 88(4):641--658.

\bibitem[Stein et~al., 2025]{stein2024sparse}
Stein, S., Feng, R., and Leng, C. (2025).
\newblock A sparse beta regression model for network analysis.
\newblock {\em Journal of the American Statistical Association},
  120(550):1281--1293.

\bibitem[Tsybakov, 2009]{tsybakov2009nonparametric}
Tsybakov, A.~B. (2009).
\newblock {\em Introduction to Nonparametric Estimation}.
\newblock Springer Series in Statistics. Springer.

\bibitem[Van~der Vaart, 2000]{van2000asymptotic}
Van~der Vaart, A.~W. (2000).
\newblock {\em Asymptotic statistics}, volume~3.
\newblock Cambridge university press.

\bibitem[Van~Duijn et~al., 2004]{van2004p2}
Van~Duijn, M.~A., Snijders, T.~A., and Zijlstra, B.~J. (2004).
\newblock $p_2$: a random effects model with covariates for directed graphs.
\newblock {\em Statistica Neerlandica}, 58(2):234--254.

\bibitem[Wainwright, 2019]{wainwright2019high}
Wainwright, M.~J. (2019).
\newblock {\em High-dimensional statistics: A non-asymptotic viewpoint},
  volume~48.
\newblock Cambridge university press.

\bibitem[Wasserman and Faust, 1994]{wasserman1994social}
Wasserman, S. and Faust, K. (1994).
\newblock {\em Social Network Analysis: Methods and Applications}.
\newblock Cambridge University Press.

\bibitem[Yan et~al., 2019]{yan2019statistical}
Yan, T., Jiang, B., Fienberg, S.~E., and Leng, C. (2019).
\newblock Statistical inference in a directed network model with covariates.
\newblock {\em Journal of the American Statistical Association},
  114(526):857--868.

\bibitem[Yan and Leng, 2015]{yan2015simulation}
Yan, T. and Leng, C. (2015).
\newblock A simulation study of the $p_{1}$ model for directed random graphs.
\newblock {\em Statistics and Its Interface}, 8(3):255--266.

\bibitem[Yan et~al., 2016]{yan2016asymptotics}
Yan, T., Leng, C., and Zhu, J. (2016).
\newblock Asymptotics in directed exponential random graph models with an
  increasing bi-degree sequence.
\newblock {\em The Annals of Statistics}, 44(1):31--57.

\bibitem[Yan and Xu, 2013]{yan2013central}
Yan, T. and Xu, J. (2013).
\newblock A central limit theorem in the $\beta$-model for undirected random
  graphs with a diverging number of vertices.
\newblock {\em Biometrika}, 100(2):519--524.

\end{thebibliography}

\newpage
\begin{center}
{\large\bf SUPPLEMENTARY MATERIAL}
\end{center}

This supplementary material provides complete technical proofs and additional simulation results to support the findings presented in the main paper.

\section*{Proofs}

\begin{proof}[Proof of Lemma \ref{sufficiency}]
Since all dyads $(A_{ij}, A_{ji})$ are assumed to be independent, and by the construction of $S_{ijkl} = 0$, a direct calculation yields:
\begin{align*}
\mathbb{P}\left(S_{ijkl} = 0 \mid \textbf{V}_{ijkl}\right) 
&= \mathbb{E}\left[\phi_{ij}(1,0)\phi_{jk}(1,0)\phi_{kl}(1,0)\phi_{li}(1,0) \mid \textbf{V}_{ijkl}\right] \\
&= p_{ij}(1,0) \cdot p_{jk}(1,0) \cdot p_{kl}(1,0) \cdot p_{li}(1,0) \\
&\propto \exp(4\mu_n + (\alpha_i + \alpha_j + \alpha_k + \alpha_l)+(\beta_i + \beta_j + \beta_k + \beta_l)).  \qquad \qquad \qquad
\end{align*}
Similarly,
\begin{align*}
\mathbb{P}\left(S_{ijkl} = 1 \mid \textbf{V}_{ijkl}\right) 
&= \mathbb{E}\left[\phi_{ij}(1,1)\phi_{jk}(0,0)\phi_{kl}(1,1)\phi_{li}(0,0) \mid \textbf{V}_{ijkl}\right] \\
&= p_{ij}(1,1) \cdot p_{jk}(0,0) \cdot p_{kl}(1,1) \cdot p_{li}(0,0) \\
&\propto f_{ijkl}(\rho_n, \gamma) \cdot \exp(4\mu_n + (\alpha_i + \alpha_j + \alpha_k + \alpha_l)+(\beta_i + \beta_j + \beta_k + \beta_l)), \quad
\end{align*}
and
\begin{align*}
\mathbb{P}\left(S_{ijkl} = -1 \mid \textbf{V}_{ijkl}\right) 
&= \mathbb{E}\left[\phi_{ij}(0,0)\phi_{jk}(1,1)\phi_{kl}(0,0)\phi_{li}(1,1) \mid \textbf{V}_{ijkl}\right] \\
&= p_{ij}(0,0) \cdot p_{jk}(1,1) \cdot p_{kl}(0,0) \cdot p_{li}(1,1) \\
&\propto g_{ijkl}(\rho_n, \gamma) \cdot \exp(4\mu_n + (\alpha_i + \alpha_j + \alpha_k + \alpha_l)+(\beta_i + \beta_j + \beta_k + \beta_l)).
\end{align*}
The result follows immediately.
\end{proof}

\begin{proof}[Proof of Lemma \ref{Lem_score}]
We begin by computing the gradient of $\ell_{ijkl}(\vartheta)$. A direct calculation gives:
\begin{align*}
\nabla_{\rho} \ell_{ijkl}(\vartheta) 
&= 2\mathbb{I}(S_{ijkl}=0) - \frac{2\mathbb{I}(S_{ijkl} = 0, \pm 1)}{1 + r_{ijkl}(\vartheta)}, \\
\nabla_{\gamma} \ell_{ijkl}(\vartheta) 
&= \frac{\mathbb{I}(S_{ijkl} = 0, \pm 1) \nabla_{\gamma} r_{ijkl}(\vartheta)}{1 + r_{ijkl}(\vartheta)} 
- (V_{ij} + V_{kl}) \mathbb{I}(S_{ijkl} = 1) 
- (V_{il} + V_{jk}) \mathbb{I}(S_{ijkl} = -1),
\end{align*}
where
\[
\nabla_{\gamma} r_{ijkl}(\vartheta) 
= (V_{ij} + V_{kl}) \exp\big(2b \log n + 2\rho + (V_{ij}^{\top} + V_{kl}^{\top}) \gamma\big) 
+ (V_{il} + V_{jk}) \exp\big(2b \log n + 2\rho + (V_{il}^{\top} + V_{jk}^{\top}) \gamma\big).
\]

We now analyze each case $S_{ijkl} = 0$, $1$, and $-1$ separately.

\medskip
\noindent \textbf{Case 1: $S_{ijkl} = 0$.} Then
\[
\left|\nabla_{\rho} \ell_{ijkl}(\vartheta)\right| 
= \frac{2 r_{ijkl}(\vartheta)}{1 + r_{ijkl}(\vartheta)} = 
O(n^{2b}) \text{ when } b \leq 0, \quad O(1) \text{ when } b > 0
\]
by Assumptions \ref{covass} and \ref{nodeass}. Similarly,
\[
\left\| \nabla_{\gamma} \ell_{ijkl}(\vartheta) \right\|_2 
= \left\| \frac{\nabla_{\gamma} r_{ijkl}(\vartheta)}{1 + r_{ijkl}(\vartheta)} \right\|_2 
= O(n^{2b}) \text{ when } b \leq 0, \quad O(1) \text{ when } b > 0.
\]

\medskip
\noindent \textbf{Case 2: $S_{ijkl} = 1$.} Then under Assumptions \ref{covass} and \ref{nodeass},
\[
\left|\nabla_{\rho} \ell_{ijkl}(\vartheta)\right| 
= \frac{2}{1 + r_{ijkl}(\vartheta)} = 
O(1) \text{ when } b \leq 0, \quad O(n^{-2b}) \text{ when } b > 0.
\]
Moreover,
\begin{align*}
\left\| \nabla_{\gamma} \ell_{ijkl}(\vartheta) \right\|_2 
&= \frac{\big\Vert (V_{il} + V_{jk} - (V_{ij} + V_{kl}))\exp(2b\log n+ 2\rho + (V^{\top}_{il}+V^{\top}_{jk})\gamma) - (V_{ij} + V_{kl})\big\Vert_2}{1+r_{ijkl}(\vartheta)},
\end{align*}
which remains bounded uniformly in $n$ for any $b$.

\medskip
\noindent \textbf{Case 3: $S_{ijkl} = -1$.} By Assumptions \ref{covass} and \ref{nodeass}, we have,
\[
\left|\nabla_{\rho} \ell_{ijkl}(\vartheta)\right| 
= \frac{2}{1 + r_{ijkl}(\vartheta)} = 
O(1) \text{ when } b \leq 0, \quad O(n^{-2b}) \text{ when } b > 0.
\]
Similarly,
\begin{align*}
\left\| \nabla_{\gamma} \ell_{ijkl}(\vartheta) \right\|_2 
&= \frac{\big\Vert (V_{ij} + V_{kl} - (V_{il} + V_{jk}))\exp(2b\log n+ 2\rho + (V^{\top}_{ij}+V^{\top}_{kl})\gamma) - (V_{il} + V_{jk})\big\Vert_2}{1+r_{ijkl}(\vartheta)},
\end{align*}
which is again uniformly bounded for any $b$.

\end{proof}

\begin{Prop}\label{variance} Under Assumptions \ref{covass}, \ref{nodeass}, \ref{sparseass}, we have
\begin{align*}
\operatorname{Var}\big(\tau_n\odot \tau_n \odot \Psi_{n}(\vartheta) \big)=o(1)
\end{align*}
\end{Prop}
 
 \begin{proof}
Recall that 
\[
\Psi_{n}(\vartheta) = \binom{n}{4}^{-1} \sum_{i<j<k<l} \psi_{ijkl}(\vartheta).
\]
We analyze $\operatorname{Cov}(\psi_{ijkl}(\vartheta), \psi_{pqtr}(\vartheta))$ according to the cardinality of $\{i, j, k, l\} \cap \{p, q, t, r\}$.

\textbf{Case 1:} When $|\{i, j, k, l\} \cap \{p, q, t, r\}| = 0$ or $1$, $\psi_{ijkl}(\vartheta)$ and $\psi_{pqtr}(\vartheta)$ are independent. Thus,
\[
\operatorname{Cov}(\psi_{ijkl}(\vartheta), \psi_{pqtr}(\vartheta)) = 0.
\]

\textbf{Case 2:} When $|\{i, j, k, l\} \cap \{p, q, t, r\}| = 2$, due to permutation invariance of $\psi_{ijkl}(\vartheta)$ with respect to $(i,j,k,l)$, without loss of generality, we consider:
\[
\operatorname{Cov}(\psi_{ijk_1l_1}(\vartheta), \psi_{ijk_2l_2}(\vartheta)),
\]
where $(k_1,l_1), (k_2,l_2) \in \mathcal{H}_{ij} := \{(k,l): k<l, \{k, l\} \cap \{i, j\} = \emptyset\}$ and $\{k_1, l_1\} \cap \{k_2, l_2\} = \emptyset$. To this end, we consider
\[
\operatorname{Cov}( \nabla_{\vartheta} \ell_{\pi(i,j,k_1,l_1)}(\vartheta), \nabla_{\vartheta} \ell_{\pi'(i,j,k_2,l_2)}(\vartheta))
\]
for any permutations $\pi, \pi' \in \Pi_4$. Due to symmetry, it suffices to consider $\pi = \text{id}$ and $\pi'$ arbitrary, yielding 24 cases.

\textbf{(Case 2.1):} $\pi'(i,j,k_2,l_2) = (i,j,k_2,l_2)$. See Figure \ref{Case 2.1} for an illustration. Using the law of total expectation:
\begin{align*}
&\mathbb{E}\left( \nabla_{\rho} \ell_{ijk_1l_1}(\vartheta) \nabla_{\rho} \ell_{ijk_2l_2}(\vartheta) \right) \\
=~ & \mathbb{E}\left( \nabla_{\rho} \ell_{ijk_1l_1}(\vartheta) \nabla_{\rho} \ell_{ijk_2l_2}(\vartheta) \,\middle|\, S_{ijk_1l_1} = 1, S_{ijk_2l_2} = 1 \right) \mathbb{P}(S_{ijk_1l_1} = 1, S_{ijk_2l_2} = 1) \\
+~ & \mathbb{E}\left( \nabla_{\rho} \ell_{ijk_1l_1}(\vartheta) \nabla_{\rho} \ell_{ijk_2l_2}(\vartheta) \,\middle|\, S_{ijk_1l_1} = -1, S_{ijk_2l_2} = -1 \right) \mathbb{P}(S_{ijk_1l_1} = -1, S_{ijk_2l_2} = -1) \\
+~ & \mathbb{E}\left( \nabla_{\rho} \ell_{ijk_1l_1}(\vartheta) \nabla_{\rho} \ell_{ijk_2l_2}(\vartheta) \,\middle|\, S_{ijk_1l_1} = 0, S_{ijk_2l_2} = 0 \right) \mathbb{P}(S_{ijk_1l_1} = 0, S_{ijk_2l_2} = 0),
\end{align*}
where we used the fact that $\mathbb{P}(S_{ijk_1l_1} \neq S_{ijk_2l_2}) = 0$.
\begin{figure}[htbp]
\centering
\begin{tikzpicture}
\node[vertex] (6) at (2,0) {$i$};
\node[vertex] (7) at (4,0) {$j$};
\node[vertex] (8) at (2,-2) {$l_2$};
\node[vertex] (9) at (4,-2) {$k_2$};
\node[vertex] (16) at (2,0.7) {$i$};
\node[vertex] (17) at (4,0.7) {$j$};
\node[vertex] (18) at (2,2.7) {$l_1$};
\node[vertex] (19) at (4,2.7) {$k_1$};

\node[vertex] (1) at (6,0) {$i$};
\node[vertex] (2) at (8,0) {$j$};
\node[vertex] (3) at (6,-2) {$l_2$};
\node[vertex] (4) at (8,-2) {$k_2$};
\node[vertex] (11) at (6,0.7) {$i$};
\node[vertex] (12) at (8,0.7) {$j$};
\node[vertex] (13) at (6,2.7) {$l_1$};
\node[vertex] (14) at (8,2.7) {$k_1$};

\node[vertex] (21) at (10,0) {$i$};
\node[vertex] (22) at (12,0) {$j$};
\node[vertex] (23) at (10,-2) {$l_2$};
\node[vertex] (24) at (12,-2) {$k_2$};
\node[vertex] (31) at (10,0.7) {$i$};
\node[vertex] (32) at (12,0.7) {$j$};
\node[vertex] (33) at (10,2.7) {$l_1$};
\node[vertex] (34) at (12,2.7) {$k_1$};

\draw[->, bend left=10, thick,line width=0.25mm] (1) to (2);
\draw[->,  bend left=10, thick,line width=0.25mm] (2) to (1);
\draw[->, bend left=10, thick,line width=0.25mm] (3) to (4);
\draw[->,  bend left=10, thick,line width=0.25mm] (4) to (3);
\draw[->, bend left=10, thick,line width=0.25mm] (13) to (14);
\draw[->,  bend left=10, thick,line width=0.25mm] (14) to (13);
\draw[->, bend left=10, thick,line width=0.25mm] (11) to (12);
\draw[->,  bend left=10, thick,line width=0.25mm] (12) to (11);

\draw[->, bend left=10, dashed, thick,line width=0.25mm] (1) to (4);
\draw[->, bend left=10, dashed, thick,line width=0.25mm] (4) to (1);
\draw[->, bend left=10, dashed, thick,line width=0.25mm] (2) to (3);
\draw[->, bend left=10, dashed, thick,line width=0.25mm] (3) to (2);
\draw[->, bend left=10, dashed, thick,line width=0.25mm] (11) to (14);
\draw[->, bend left=10, dashed, thick,line width=0.25mm] (14) to (11);
\draw[->, bend left=10, dashed, thick,line width=0.25mm] (12) to (13);
\draw[->, bend left=10, dashed, thick,line width=0.25mm] (13) to (12);

\draw[->, thick,line width=0.25mm] (9) to (8);
\draw[->, thick,line width=0.25mm] (7) to (9);
\draw[->, thick,line width=0.25mm] (6) to (7);
\draw[->, thick,line width=0.25mm] (8) to (6);
\draw[->, thick,line width=0.25mm] (17) to (19);
\draw[->, thick,line width=0.25mm] (16) to (17);
\draw[->, thick,line width=0.25mm] (18) to (16);
\draw[->, thick,line width=0.25mm] (19) to (18);
\draw[->, bend left=10, dashed, thick,line width=0.25mm] (6) to (9);
\draw[->, bend left=10, dashed, thick,line width=0.25mm] (9) to (6);
\draw[->, bend left=10, dashed, thick,line width=0.25mm] (7) to (8);
\draw[->, bend left=10, dashed, thick,line width=0.25mm] (8) to (7);
\draw[->, bend left=10, dashed, thick,line width=0.25mm] (16) to (19);
\draw[->, bend left=10, dashed, thick,line width=0.25mm] (19) to (16);
\draw[->, bend left=10, dashed, thick,line width=0.25mm] (17) to (18);
\draw[->, bend left=10, dashed, thick,line width=0.25mm] (18) to (17);

\draw[->, bend left=10, thick,line width=0.25mm] (21) to (23);
\draw[->, bend left=10, thick,line width=0.25mm]  (23) to (21);
\draw[->, bend left=10, thick,line width=0.25mm] (31) to (33);
\draw[->, bend left=10, thick,line width=0.25mm] (33) to (31);
\draw[->, bend left=10, thick,line width=0.25mm] (22) to (24);
\draw[->, bend left=10, thick,line width=0.25mm]  (24) to (22);
\draw[->, bend left=10, thick,line width=0.25mm]  (32) to (34);
\draw[->, bend left=10, thick,line width=0.25mm]  (34) to (32);
\draw[->, bend left=10, dashed, thick,line width=0.25mm] (21) to (24);
\draw[->, bend left=10, dashed, thick,line width=0.25mm] (24) to (21);
\draw[->, bend left=10, dashed, thick,line width=0.25mm] (22) to (23);
\draw[->, bend left=10, dashed, thick,line width=0.25mm] (23) to (22);
\draw[->, bend left=10, dashed, thick,line width=0.25mm] (31) to (34);
\draw[->, bend left=10, dashed, thick,line width=0.25mm] (34) to (31);
\draw[->, bend left=10, dashed, thick,line width=0.25mm] (32) to (33);
\draw[->, bend left=10, dashed, thick,line width=0.25mm] (33) to (32);

\end{tikzpicture}
\caption{Illustration of \textbf{(Case 2.1)}: when $S_{ijk_1l_1} = S_{ijk_2l_2} = 0$ (left), $S_{ijk_1l_1} = S_{ijk_2l_2} = 1$ (middle), and $S_{ijk_1l_1} = S_{ijk_2l_2} = -1$ (right).}
\label{Case 2.1}
\end{figure}
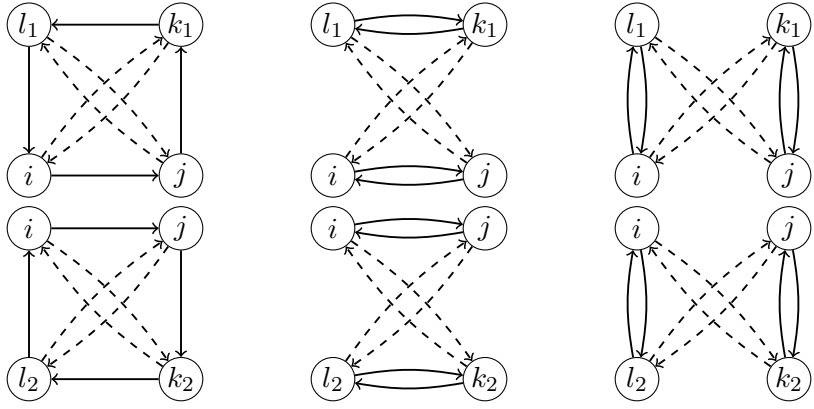

Next, we compute the joint probabilities:
\begin{align*}
\mathbb{P}(S_{ijk_1l_1} = 1, S_{ijk_2l_2} = 1) &= \mathbb{P}(A_{ij} A_{ji}) \mathbb{P}(A_{l_1k_1} A_{k_1l_1}) \mathbb{P}(A_{l_2k_2} A_{k_2l_2}) \\
&\quad \cdot \mathbb{P}((1 - A_{jk_1})(1 - A_{k_1j})) \mathbb{P}((1 - A_{jk_2})(1 - A_{k_2j})) \\
&\quad \cdot \mathbb{P}((1 - A_{il_1})(1 - A_{l_1i})) \mathbb{P}((1 - A_{il_2})(1 - A_{l_2i})) \\
&= O(n^{-6a + 3b}),
\end{align*}
by Assumptions \ref{covass}, \ref{nodeass} and \ref{sparseass}. Similarly,
\[
\mathbb{P}(S_{ijk_1l_1} = -1, S_{ijk_2l_2} = -1) = O(n^{-8a + 4b}), \quad \mathbb{P}(S_{ijk_1l_1} = 0, S_{ijk_2l_2} = 0) = O(n^{-7a}).
\]
By Lemma~\ref{Lem_score},
\[
\mathbb{E}\left( \nabla_{\rho} \ell_{ijk_1l_1}(\vartheta) \nabla_{\rho} \ell_{ijk_2l_2}(\vartheta) \right) = O(n^{-4\max\{b, 0\}}) \cdot O(n^{-6a + 3b}) + O(n^{4\min\{b, 0\}}) \cdot O(n^{-7a}).
\]
We also compute the mean:
\begin{align*}
\mathbb{E}(\nabla_{\rho} \ell_{ijk_1l_1}(\vartheta)) &= \mathbb{E}(\nabla_{\rho} \ell_{ijk_1l_1}(\vartheta) \mid S_{ijk_1l_1} = 1) \mathbb{P}(S_{ijk_1l_1} = 1) \\
&\quad + \mathbb{E}(\nabla_{\rho} \ell_{ijk_1l_1}(\vartheta) \mid S_{ijk_1l_1} = -1) \mathbb{P}(S_{ijk_1l_1} = -1) \\
&\quad + \mathbb{E}(\nabla_{\rho} \ell_{ijk_1l_1}(\vartheta) \mid S_{ijk_1l_1} = 0) \mathbb{P}(S_{ijk_1l_1} = 0) \\
&= O(n^{-2\max\{b, 0\}}) \cdot O(n^{-4a + 2b}) + O(n^{2\min\{b, 0\}}) \cdot O(n^{-4a}).
\end{align*}
Therefore, the covariance satisfies:
\[
\operatorname{Cov}(\nabla_{\rho} \ell_{ijk_1l_1}(\vartheta), \nabla_{\rho} \ell_{ijk_2l_2}(\vartheta)) = 
\begin{cases}
O(n^{-6a - \min\{a, b\}}), & \text{if } b > 0, \\
O(n^{-6a + 3b}), & \text{if } b \leq 0.
\end{cases}
\]

Similarly, by Lemma~\ref{Lem_score},
\[
\mathbb{E}(\nabla_{\gamma} \ell_{ijk_1l_1}(\vartheta) \nabla_{\gamma}^\top \ell_{ijk_2l_2}(\vartheta)) = O(1) \cdot O(n^{-6a + 3b}) + O(n^{4\min\{b, 0\}}) \cdot O(n^{-7a}),
\]
and
\[
\mathbb{E}(\nabla_{\gamma} \ell_{ijk_1l_1}(\vartheta)) = O(1) \cdot O(n^{-4a + 2b}) + O(n^{2\min\{b, 0\}}) \cdot O(n^{-4a}).
\]
Therefore,
\[
\operatorname{Cov}(\nabla_{\gamma} \ell_{ijk_1l_1}(\vartheta), \nabla_{\gamma} \ell_{ijk_2l_2}(\vartheta)) = O(n^{-6a + 3b}), \quad \text{for any } b.
\]
Lastly,
\[
\operatorname{Cov}(\nabla_{\rho} \ell_{ijk_1l_1}(\vartheta), \nabla_{\gamma} \ell_{ijk_2l_2}(\vartheta)) = 
\begin{cases}
O(n^{-6a + b}), & \text{if } b > 0, \\
O(n^{-6a + 3b}), & \text{if } b \leq 0.
\end{cases}
\]

\textbf{(Case 2.2):} $\pi^{\prime}(i,j,k_2,l_2) = (i,j,l_2,k_2)$. Due to the symmetry between $k_2$ and $l_2$, this case is equivalent to \textbf{(Case 2.1)}.

\textbf{(Case 2.3):} $\pi^{\prime}(i,j,k_2,l_2) = (i,k_2,l_2,j)$. For this case, as shown in Figure~\ref{Case 2.3}, we have: when $S_{ijk_1l_1} = 0$,
\[
\mathbb{P}(S_{ijk_1l_1} = 0, S_{ik_2l_2j} = 0 \text{ or } \pm 1) = 0.
\]
When $S_{ijk_1l_1} = 1$
\[
    \mathbb{P}(S_{ijk_1l_1} = 1, S_{ik_2l_2j} = 0 \text{ or } 1) = 0, \quad \mathbb{P}(S_{ijk_1l_1} = 1, S_{ik_2l_2j} = -1) = O(n^{-6a+3b}).
\]
When $S_{ijk_1l_1} = -1$, 
\[
    \mathbb{P}(S_{ijk_1l_1} = -1, S_{ik_2l_2j} = 0 \text{ or } -1) = 0, \quad \mathbb{P}(S_{ijk_1l_1} = -1, S_{ik_2l_2j} = 1) = O(n^{-8a+4b}).
\]

\begin{figure}[htbp]
\centering
\begin{tikzpicture}
\node[vertex] (6) at (2,0) {$i$};
\node[vertex] (7) at (4,0) {$k_2$};
\node[vertex] (8) at (2,-2) {$j$};
\node[vertex] (9) at (4,-2) {$l_2$};
\node[vertex] (16) at (2,0.7) {$i$};
\node[vertex] (17) at (4,0.7) {$j$};
\node[vertex] (18) at (2,2.7) {$l_1$};
\node[vertex] (19) at (4,2.7) {$k_1$};

\node[vertex] (1) at (6,0) {$i$};
\node[vertex] (2) at (8,0) {$k_2$};
\node[vertex] (3) at (6,-2) {$j$};
\node[vertex] (4) at (8,-2) {$l_2$};
\node[vertex] (11) at (6,0.7) {$i$};
\node[vertex] (12) at (8,0.7) {$j$};
\node[vertex] (13) at (6,2.7) {$l_1$};
\node[vertex] (14) at (8,2.7) {$k_1$};

\node[vertex] (21) at (10,0) {$i$};
\node[vertex] (22) at (12,0) {$k_2$};
\node[vertex] (23) at (10,-2) {$j$};
\node[vertex] (24) at (12,-2) {$l_2$};
\node[vertex] (31) at (10,0.7) {$i$};
\node[vertex] (32) at (12,0.7) {$j$};
\node[vertex] (33) at (10,2.7) {$l_1$};
\node[vertex] (34) at (12,2.7) {$k_1$};

\draw[->, bend left=10, thick,line width=0.25mm] (1) to (2);
\draw[->,  bend left=10, thick,line width=0.25mm] (2) to (1);
\draw[->, bend left=10, thick,line width=0.25mm] (3) to (4);
\draw[->,  bend left=10, thick,line width=0.25mm] (4) to (3);
\draw[->, bend left=10, thick,line width=0.25mm] (13) to (14);
\draw[->,  bend left=10, thick,line width=0.25mm] (14) to (13);
\draw[->, bend left=10, thick,line width=0.25mm] (11) to (12);
\draw[->,  bend left=10, thick,line width=0.25mm] (12) to (11);

\draw[->, bend left=10, dashed, thick,line width=0.25mm] (1) to (4);
\draw[->, bend left=10, dashed, thick,line width=0.25mm] (4) to (1);
\draw[->, bend left=10, dashed, thick,line width=0.25mm] (2) to (3);
\draw[->, bend left=10, dashed, thick,line width=0.25mm] (3) to (2);
\draw[->, bend left=10, dashed, thick,line width=0.25mm] (11) to (14);
\draw[->, bend left=10, dashed, thick,line width=0.25mm] (14) to (11);
\draw[->, bend left=10, dashed, thick,line width=0.25mm] (12) to (13);
\draw[->, bend left=10, dashed, thick,line width=0.25mm] (13) to (12);

\draw[->, thick,line width=0.25mm] (9) to (8);
\draw[->, thick,line width=0.25mm] (7) to (9);
\draw[->, thick,line width=0.25mm] (6) to (7);
\draw[->, thick,line width=0.25mm] (8) to (6);
\draw[->, thick,line width=0.25mm] (17) to (19);
\draw[->, thick,line width=0.25mm] (16) to (17);
\draw[->, thick,line width=0.25mm] (18) to (16);
\draw[->, thick,line width=0.25mm] (19) to (18);
\draw[->, bend left=10, dashed, thick,line width=0.25mm] (6) to (9);
\draw[->, bend left=10, dashed, thick,line width=0.25mm] (9) to (6);
\draw[->, bend left=10, dashed, thick,line width=0.25mm] (7) to (8);
\draw[->, bend left=10, dashed, thick,line width=0.25mm] (8) to (7);
\draw[->, bend left=10, dashed, thick,line width=0.25mm] (16) to (19);
\draw[->, bend left=10, dashed, thick,line width=0.25mm] (19) to (16);
\draw[->, bend left=10, dashed, thick,line width=0.25mm] (17) to (18);
\draw[->, bend left=10, dashed, thick,line width=0.25mm] (18) to (17);

\draw[->, bend left=10, thick,line width=0.25mm] (21) to (23);
\draw[->, bend left=10, thick,line width=0.25mm]  (23) to (21);
\draw[->, bend left=10, thick,line width=0.25mm] (31) to (33);
\draw[->, bend left=10, thick,line width=0.25mm] (33) to (31);
\draw[->, bend left=10, thick,line width=0.25mm] (22) to (24);
\draw[->, bend left=10, thick,line width=0.25mm]  (24) to (22);
\draw[->, bend left=10, thick,line width=0.25mm]  (32) to (34);
\draw[->, bend left=10, thick,line width=0.25mm]  (34) to (32);
\draw[->, bend left=10, dashed, thick,line width=0.25mm] (21) to (24);
\draw[->, bend left=10, dashed, thick,line width=0.25mm] (24) to (21);
\draw[->, bend left=10, dashed, thick,line width=0.25mm] (22) to (23);
\draw[->, bend left=10, dashed, thick,line width=0.25mm] (23) to (22);
\draw[->, bend left=10, dashed, thick,line width=0.25mm] (31) to (34);
\draw[->, bend left=10, dashed, thick,line width=0.25mm] (34) to (31);
\draw[->, bend left=10, dashed, thick,line width=0.25mm] (32) to (33);
\draw[->, bend left=10, dashed, thick,line width=0.25mm] (33) to (32);

\end{tikzpicture}
\caption{Illustration of \textbf{(Case 2.3)}: when $S_{i j k_1 l_1} = S_{i, k_2, l_2, j} = 0$ (left), $S_{i j k_1 l_1} = S_{i, k_2, l_2, j} = 1$ (middle), and $S_{i j k_1 l_1} = S_{i, k_2, l_2, j} = -1$ (right).}
\label{Case 2.3}
\end{figure}

Therefore,
\begin{align*} 
&\mathbb{E}(\nabla_{\vartheta}  \ell_{ijk_1l_1}(\vartheta)\nabla^{\top}_{\vartheta}  \ell_{ik_2l_2j}(\vartheta)) \\
= &\;\mathbb{E}(\nabla_{\vartheta}  \ell_{ijk_1l_1}(\vartheta)\nabla^{\top}_{\vartheta}  \ell_{ik_2l_2j}(\vartheta) \mid S_{ijk_1l_1} = 1, S_{ik_2l_2j} = -1)\; \mathbb{P}(S_{ijk_1l_1} = 1, S_{ik_2l_2j} = -1)\\
&+ \mathbb{E}(\nabla_{\vartheta}  \ell_{ijk_1l_1}(\vartheta)\nabla^{\top}_{\vartheta}  \ell_{ik_2l_2j}(\vartheta) \mid S_{ijk_1l_1} = -1, S_{ik_2l_2j} = 1)\; \mathbb{P}(S_{ijk_1l_1} = -1, S_{ik_2l_2j} = 1).
\end{align*}

Using Lemma~\ref{Lem_score}, we obtain
\[
\operatorname{Cov}\left( \nabla_{\gamma}  \ell_{ijk_1l_1}(\vartheta),  \nabla_{\gamma}  \ell_{ik_2l_2j}(\vartheta)\right) = O(n^{-6a+3b}),
\]
for any $b$. Meanwhile, when $b > 0$, we have
\begin{align*}
&\operatorname{Cov}\left( \nabla_{\rho}  \ell_{ijk_1l_1}(\vartheta),  \nabla_{\rho}  \ell_{ik_2l_2j}(\vartheta) \right) = O(n^{-6a-b}),\\
&\operatorname{Cov}\left( \nabla_{\rho}  \ell_{ijk_1l_1}(\vartheta),  \nabla_{\gamma}  \ell_{ik_2l_2j}(\vartheta) \right) = O(n^{-6a+b}),
\end{align*}
and when $b \leq 0$,
\begin{align*}
&\operatorname{Cov}\left( \nabla_{\rho}  \ell_{ijk_1l_1}(\vartheta),  \nabla_{\rho}  \ell_{ik_2l_2j}(\vartheta) \right) = O(n^{-6a+3b}),\\
&\operatorname{Cov}\left( \nabla_{\rho}  \ell_{ijk_1l_1}(\vartheta),  \nabla_{\gamma}  \ell_{ik_2l_2j}(\vartheta) \right) = O(n^{-6a+3b}).
\end{align*}

\textbf{(Case 2.4):} \(\pi^{\prime}(i,j,k_2,l_2) = (i, k_2, j, l_2)\). From Figure~\ref{Case 2.4}, when \(S_{i j k_1 l_1} = 0\),
\[
\mathbb{P}(S_{i j k_1 l_1} = 0, S_{i, k_2, j, l_2} = 0) = O(n^{-8a}), \quad 
\mathbb{P}(S_{i j k_1 l_1} = 0, S_{i, k_2, j, l_2} = \pm 1) = O(n^{-8a+2b});
\]
when \(S_{i j k_1 l_1} = \pm 1\),
\[
\mathbb{P}(S_{i j k_1 l_1} = \pm 1, S_{i, k_2, j, l_2} = 0) = O(n^{-8a+2b}), \quad 
\mathbb{P}(S_{i j k_1 l_1} = \pm 1, S_{i, k_2, j, l_2} = \pm 1) = O(n^{-8a+4b}).
\]

\begin{figure}[htbp]
\centering
\begin{tikzpicture}
\node[vertex] (6) at (2,0) {$i$};
\node[vertex] (7) at (4,0) {$k_2$};
\node[vertex] (8) at (2,-2) {$l_2$};
\node[vertex] (9) at (4,-2) {$j$};
\node[vertex] (16) at (2,0.7) {$i$};
\node[vertex] (17) at (4,0.7) {$j$};
\node[vertex] (18) at (2,2.7) {$l_1$};
\node[vertex] (19) at (4,2.7) {$k_1$};

\node[vertex] (1) at (6,0) {$i$};
\node[vertex] (2) at (8,0) {$k_2$};
\node[vertex] (3) at (6,-2) {$l_2$};
\node[vertex] (4) at (8,-2) {$j$};
\node[vertex] (11) at (6,0.7) {$i$};
\node[vertex] (12) at (8,0.7) {$j$};
\node[vertex] (13) at (6,2.7) {$l_1$};
\node[vertex] (14) at (8,2.7) {$k_1$};

\node[vertex] (21) at (10,0) {$i$};
\node[vertex] (22) at (12,0) {$k_2$};
\node[vertex] (23) at (10,-2) {$l_2$};
\node[vertex] (24) at (12,-2) {$j$};
\node[vertex] (31) at (10,0.7) {$i$};
\node[vertex] (32) at (12,0.7) {$j$};
\node[vertex] (33) at (10,2.7) {$l_1$};
\node[vertex] (34) at (12,2.7) {$k_1$};

\draw[->, bend left=10, thick,line width=0.25mm] (1) to (2);
\draw[->,  bend left=10, thick,line width=0.25mm] (2) to (1);
\draw[->, bend left=10, thick,line width=0.25mm] (3) to (4);
\draw[->,  bend left=10, thick,line width=0.25mm] (4) to (3);
\draw[->, bend left=10, thick,line width=0.25mm] (13) to (14);
\draw[->,  bend left=10, thick,line width=0.25mm] (14) to (13);
\draw[->, bend left=10, thick,line width=0.25mm] (11) to (12);
\draw[->,  bend left=10, thick,line width=0.25mm] (12) to (11);

\draw[->, bend left=10, dashed, thick,line width=0.25mm] (1) to (4);
\draw[->, bend left=10, dashed, thick,line width=0.25mm] (4) to (1);
\draw[->, bend left=10, dashed, thick,line width=0.25mm] (2) to (3);
\draw[->, bend left=10, dashed, thick,line width=0.25mm] (3) to (2);
\draw[->, bend left=10, dashed, thick,line width=0.25mm] (11) to (14);
\draw[->, bend left=10, dashed, thick,line width=0.25mm] (14) to (11);
\draw[->, bend left=10, dashed, thick,line width=0.25mm] (12) to (13);
\draw[->, bend left=10, dashed, thick,line width=0.25mm] (13) to (12);

\draw[->, thick,line width=0.25mm] (9) to (8);
\draw[->, thick,line width=0.25mm] (7) to (9);
\draw[->, thick,line width=0.25mm] (6) to (7);
\draw[->, thick,line width=0.25mm] (8) to (6);
\draw[->, thick,line width=0.25mm] (17) to (19);
\draw[->, thick,line width=0.25mm] (16) to (17);
\draw[->, thick,line width=0.25mm] (18) to (16);
\draw[->, thick,line width=0.25mm] (19) to (18);
\draw[->, bend left=10, dashed, thick,line width=0.25mm] (6) to (9);
\draw[->, bend left=10, dashed, thick,line width=0.25mm] (9) to (6);
\draw[->, bend left=10, dashed, thick,line width=0.25mm] (7) to (8);
\draw[->, bend left=10, dashed, thick,line width=0.25mm] (8) to (7);
\draw[->, bend left=10, dashed, thick,line width=0.25mm] (16) to (19);
\draw[->, bend left=10, dashed, thick,line width=0.25mm] (19) to (16);
\draw[->, bend left=10, dashed, thick,line width=0.25mm] (17) to (18);
\draw[->, bend left=10, dashed, thick,line width=0.25mm] (18) to (17);

\draw[->, bend left=10, thick,line width=0.25mm] (21) to (23);
\draw[->, bend left=10, thick,line width=0.25mm]  (23) to (21);
\draw[->, bend left=10, thick,line width=0.25mm] (31) to (33);
\draw[->, bend left=10, thick,line width=0.25mm] (33) to (31);
\draw[->, bend left=10, thick,line width=0.25mm] (22) to (24);
\draw[->, bend left=10, thick,line width=0.25mm]  (24) to (22);
\draw[->, bend left=10, thick,line width=0.25mm]  (32) to (34);
\draw[->, bend left=10, thick,line width=0.25mm]  (34) to (32);
\draw[->, bend left=10, dashed, thick,line width=0.25mm] (21) to (24);
\draw[->, bend left=10, dashed, thick,line width=0.25mm] (24) to (21);
\draw[->, bend left=10, dashed, thick,line width=0.25mm] (22) to (23);
\draw[->, bend left=10, dashed, thick,line width=0.25mm] (23) to (22);
\draw[->, bend left=10, dashed, thick,line width=0.25mm] (31) to (34);
\draw[->, bend left=10, dashed, thick,line width=0.25mm] (34) to (31);
\draw[->, bend left=10, dashed, thick,line width=0.25mm] (32) to (33);
\draw[->, bend left=10, dashed, thick,line width=0.25mm] (33) to (32);

\end{tikzpicture}
\caption{Illustration of \textbf{(Case 2.4)}:  when 
\(S_{i j k_1 l_1} = S_{i, k_2, j, l_2} = 0\) (left), 
\(S_{i j k_1 l_1} = S_{i, k_2, j, l_2} = 1\) (middle), and 
\(S_{i j k_1 l_1} = S_{i, k_2, j, l_2} = -1\) (right).}
\label{Case 2.4}
\end{figure}

Then
\begin{align*} 
&\mathbb{E}\big(\nabla_{\vartheta} \ell_{i j k_1 l_1}(\vartheta) \nabla_{\vartheta}^{\top} \ell_{i, k_2, j, l_2}(\vartheta)\big) \\
= \, &\mathbb{E}\big(\nabla_{\vartheta} \ell_{i j k_1 l_1}(\vartheta) \nabla_{\vartheta}^{\top} \ell_{i, k_2, j, l_2}(\vartheta) \mid S_{i j k_1 l_1} = 0, S_{i, k_2, j, l_2} = 0 \big) \mathbb{P}(S_{i j k_1 l_1} = 0, S_{i, k_2, j, l_2} = 0) \\
&+ \mathbb{E}\big(\nabla_{\vartheta} \ell_{i j k_1 l_1}(\vartheta) \nabla_{\vartheta}^{\top} \ell_{i, k_2, j, l_2}(\vartheta) \mid S_{i j k_1 l_1} = 0, S_{i, k_2, j, l_2} = \pm 1 \big) \mathbb{P}(S_{i j k_1 l_1} = 0, S_{i, k_2, j, l_2} = \pm 1) \\
&+ \mathbb{E}\big(\nabla_{\vartheta} \ell_{i j k_1 l_1}(\vartheta) \nabla_{\vartheta}^{\top} \ell_{i, k_2, j, l_2}(\vartheta) \mid S_{i j k_1 l_1} = \pm 1, S_{i, k_2, j, l_2} = 0 \big) \mathbb{P}(S_{i j k_1 l_1} = \pm 1, S_{i, k_2, j, l_2} = 0) \\
&+ \mathbb{E}\big(\nabla_{\vartheta} \ell_{i j k_1 l_1}(\vartheta) \nabla_{\vartheta}^{\top} \ell_{i, k_2, j, l_2}(\vartheta) \mid S_{i j k_1 l_1} = \pm 1, S_{i, k_2, j, l_2} = \pm 1 \big) \mathbb{P}(S_{i j k_1 l_1} = \pm 1, S_{i, k_2, j, l_2} = \pm 1).
\end{align*}

By Lemma~\ref{Lem_score}, when \(b > 0\),
\begin{align*}
&\operatorname{Cov}\big( \nabla_{\rho} \ell_{i j k_1 l_1}(\vartheta), \nabla_{\rho} \ell_{i, k_2, j, l_2}(\vartheta) \big) = O(n^{-8a}), \\
&\operatorname{Cov}\big( \nabla_{\gamma} \ell_{i j k_1 l_1}(\vartheta), \nabla_{\gamma} \ell_{i, k_2, j, l_2}(\vartheta) \big) = O(n^{-8a+4b}), \\
&\operatorname{Cov}\big( \nabla_{\rho} \ell_{i j k_1 l_1}(\vartheta), \nabla_{\gamma} \ell_{i, k_2, j, l_2}(\vartheta) \big) = O(n^{-8a+2b}),
\end{align*}
and when \(b < 0\),
\begin{align*}
&\operatorname{Cov}\big( \nabla_{\rho} \ell_{i j k_1 l_1}(\vartheta), \nabla_{\rho} \ell_{i, k_2, j, l_2}(\vartheta) \big) = O(n^{-8a+4b}), \\
&\operatorname{Cov}\big( \nabla_{\gamma} \ell_{i j k_1 l_1}(\vartheta), \nabla_{\gamma} \ell_{i, k_2, j, l_2}(\vartheta) \big) = O(n^{-8a+4b}), \\
&\operatorname{Cov}\big( \nabla_{\rho} \ell_{i j k_1 l_1}(\vartheta), \nabla_{\gamma} \ell_{i, k_2, j, l_2}(\vartheta) \big) = O(n^{-8a+4b}).
\end{align*}

\textbf{(Case 2.5):} \(\pi^{\prime}(i,j,k_2,l_2) = (i, l_2, k_2, j)\). Due to the symmetry between \(j_2\) and \(k_2\), this case is the same as \textbf{(Case 2.3)}.

\textbf{(Case 2.6):} \(\pi^{\prime}(i,j,k_2,l_2) = (i, l_2, j, k_2)\). Due to the symmetry between \(j_2\) and \(k_2\), this case is the same as \textbf{(Case 2.4)}.

\textbf{(Case 2.7):} \(\pi^{\prime}(i,j,k_2,l_2) = (j, i, k_2, l_2)\).

For this case, from Figure~\ref{Case 2.7}, when \(S_{i j k_1 l_1} = 0\),
\[
\mathbb{P}\big(S_{i j k_1 l_1} = 0, S_{j, i, k_2, l_2} = 0 \text{ or } \pm 1 \big) = 0;
\]
when \(S_{i j k_1 l_1} = 1\),
\[
\mathbb{P}\big(S_{i j k_1 l_1} = 1, S_{j, i, k_2, l_2} = 0 \text{ or } -1 \big) = 0, \quad
\mathbb{P}\big(S_{i j k_1 l_1} = 1, S_{j, i, k_2, l_2} = 1 \big) = O(n^{-6a + 3b});
\]
when \(S_{i j k_1 l_1} = -1\),
\[
\mathbb{P}\big(S_{i j k_1 l_1} = -1, S_{j, i, k_2, l_2} = 0\text{ or }  1 \big) = 0, \quad
\mathbb{P}\big(S_{i j k_1 l_1} = -1, S_{j, i, k_2, l_2} = -1 \big) = O(n^{-8a + 4b}).
\]

\begin{figure}[htbp]
\centering
\begin{tikzpicture}
\node[vertex] (6) at (2,0) {$j$};
\node[vertex] (7) at (4,0) {$i$};
\node[vertex] (8) at (2,-2) {$l_2$};
\node[vertex] (9) at (4,-2) {$k_2$};
\node[vertex] (16) at (2,0.7) {$i$};
\node[vertex] (17) at (4,0.7) {$j$};
\node[vertex] (18) at (2,2.7) {$l_1$};
\node[vertex] (19) at (4,2.7) {$k_1$};

\node[vertex] (1) at (6,0) {$j$};
\node[vertex] (2) at (8,0) {$i$};
\node[vertex] (3) at (6,-2) {$l_2$};
\node[vertex] (4) at (8,-2) {$k_2$};
\node[vertex] (11) at (6,0.7) {$i$};
\node[vertex] (12) at (8,0.7) {$j$};
\node[vertex] (13) at (6,2.7) {$l_1$};
\node[vertex] (14) at (8,2.7) {$k_1$};

\node[vertex] (21) at (10,0) {$j$};
\node[vertex] (22) at (12,0) {$i$};
\node[vertex] (23) at (10,-2) {$l_2$};
\node[vertex] (24) at (12,-2) {$k_2$};
\node[vertex] (31) at (10,0.7) {$i$};
\node[vertex] (32) at (12,0.7) {$j$};
\node[vertex] (33) at (10,2.7) {$l_1$};
\node[vertex] (34) at (12,2.7) {$k_1$};

\draw[->, bend left=10, thick,line width=0.25mm] (1) to (2);
\draw[->,  bend left=10, thick,line width=0.25mm] (2) to (1);
\draw[->, bend left=10, thick,line width=0.25mm] (3) to (4);
\draw[->,  bend left=10, thick,line width=0.25mm] (4) to (3);
\draw[->, bend left=10, thick,line width=0.25mm] (13) to (14);
\draw[->,  bend left=10, thick,line width=0.25mm] (14) to (13);
\draw[->, bend left=10, thick,line width=0.25mm] (11) to (12);
\draw[->,  bend left=10, thick,line width=0.25mm] (12) to (11);

\draw[->, bend left=10, dashed, thick,line width=0.25mm] (1) to (4);
\draw[->, bend left=10, dashed, thick,line width=0.25mm] (4) to (1);
\draw[->, bend left=10, dashed, thick,line width=0.25mm] (2) to (3);
\draw[->, bend left=10, dashed, thick,line width=0.25mm] (3) to (2);
\draw[->, bend left=10, dashed, thick,line width=0.25mm] (11) to (14);
\draw[->, bend left=10, dashed, thick,line width=0.25mm] (14) to (11);
\draw[->, bend left=10, dashed, thick,line width=0.25mm] (12) to (13);
\draw[->, bend left=10, dashed, thick,line width=0.25mm] (13) to (12);

\draw[->, thick,line width=0.25mm] (9) to (8);
\draw[->, thick,line width=0.25mm] (7) to (9);
\draw[->, thick,line width=0.25mm] (6) to (7);
\draw[->, thick,line width=0.25mm] (8) to (6);
\draw[->, thick,line width=0.25mm] (17) to (19);
\draw[->, thick,line width=0.25mm] (16) to (17);
\draw[->, thick,line width=0.25mm] (18) to (16);
\draw[->, thick,line width=0.25mm] (19) to (18);
\draw[->, bend left=10, dashed, thick,line width=0.25mm] (6) to (9);
\draw[->, bend left=10, dashed, thick,line width=0.25mm] (9) to (6);
\draw[->, bend left=10, dashed, thick,line width=0.25mm] (7) to (8);
\draw[->, bend left=10, dashed, thick,line width=0.25mm] (8) to (7);
\draw[->, bend left=10, dashed, thick,line width=0.25mm] (16) to (19);
\draw[->, bend left=10, dashed, thick,line width=0.25mm] (19) to (16);
\draw[->, bend left=10, dashed, thick,line width=0.25mm] (17) to (18);
\draw[->, bend left=10, dashed, thick,line width=0.25mm] (18) to (17);

\draw[->, bend left=10, thick,line width=0.25mm] (21) to (23);
\draw[->, bend left=10, thick,line width=0.25mm]  (23) to (21);
\draw[->, bend left=10, thick,line width=0.25mm] (31) to (33);
\draw[->, bend left=10, thick,line width=0.25mm] (33) to (31);
\draw[->, bend left=10, thick,line width=0.25mm] (22) to (24);
\draw[->, bend left=10, thick,line width=0.25mm]  (24) to (22);
\draw[->, bend left=10, thick,line width=0.25mm]  (32) to (34);
\draw[->, bend left=10, thick,line width=0.25mm]  (34) to (32);
\draw[->, bend left=10, dashed, thick,line width=0.25mm] (21) to (24);
\draw[->, bend left=10, dashed, thick,line width=0.25mm] (24) to (21);
\draw[->, bend left=10, dashed, thick,line width=0.25mm] (22) to (23);
\draw[->, bend left=10, dashed, thick,line width=0.25mm] (23) to (22);
\draw[->, bend left=10, dashed, thick,line width=0.25mm] (31) to (34);
\draw[->, bend left=10, dashed, thick,line width=0.25mm] (34) to (31);
\draw[->, bend left=10, dashed, thick,line width=0.25mm] (32) to (33);
\draw[->, bend left=10, dashed, thick,line width=0.25mm] (33) to (32);

\end{tikzpicture}
\caption{Illustration of \textbf{(Case 2.7)}: when \(S_{i j k_1 l_1} = S_{j, i, k_2, l_2} = 0\) (left), \(S_{i j k_1 l_1} = S_{j, i, k_2, l_2} = 1\) (middle), and \(S_{i j k_1 l_1} = S_{j, i, k_2, l_2} = -1\) (right).}
\label{Case 2.7}
\end{figure}

Then
\begin{align*} 
& \mathbb{E}\big(\nabla_{\vartheta} \ell_{i j k_1 l_1}(\vartheta) \nabla_{\vartheta}^{\top} \ell_{j, i, k_2, l_2}(\vartheta)\big) \\
= \; & \mathbb{E}\big(\nabla_{\vartheta} \ell_{i j k_1 l_1}(\vartheta) \nabla_{\vartheta}^{\top} \ell_{j, i, k_2, l_2}(\vartheta) \mid S_{i j k_1 l_1} = 1, S_{j, i, k_2, l_2} = 1 \big) \mathbb{P}(S_{i j k_1 l_1} = 1, S_{j, i, k_2, l_2} = 1) \\
& + \mathbb{E}\big(\nabla_{\vartheta} \ell_{i j k_1 l_1}(\vartheta) \nabla_{\vartheta}^{\top} \ell_{j, i, k_2, l_2}(\vartheta) \mid S_{i j k_1 l_1} = -1, S_{j, i, k_2, l_2} = -1 \big) \mathbb{P}(S_{i j k_1 l_1} = -1, S_{j, i, k_2, l_2} = -1).
\end{align*}

Using Lemma~\ref{Lem_score}, we have
\[
\operatorname{Cov}\big( \nabla_{\gamma} \ell_{i j k_1 l_1}(\vartheta), \nabla_{\gamma} \ell_{j, i, k_2, l_2}(\vartheta) \big) = O(n^{-6a + 3b}),
\]
for any \(b\). Meanwhile, 
when \(b > 0\),
\[
\operatorname{Cov}\big( \nabla_{\rho} \ell_{i j k_1 l_1}(\vartheta), \nabla_{\rho} \ell_{j, i, k_2, l_2}(\vartheta) \big) = O(n^{-6a - b}), \quad
\operatorname{Cov}\big( \nabla_{\rho} \ell_{i j k_1 l_1}(\vartheta), \nabla_{\gamma} \ell_{j, i, k_2, l_2}(\vartheta) \big) = O(n^{-6a + b}),
\]
and when \(b \leq 0\),
\[
\operatorname{Cov}\big( \nabla_{\rho} \ell_{i j k_1 l_1}(\vartheta), \nabla_{\rho} \ell_{j, i, k_2, l_2}(\vartheta) \big) = O(n^{-6a + 3b}), \quad
\operatorname{Cov}\big( \nabla_{\rho} \ell_{i j k_1 l_1}(\vartheta), \nabla_{\gamma} \ell_{j, i, k_2, l_2}(\vartheta) \big) = O(n^{-6a + 3b}).
\]

\textbf{(Case 2.8):} \(\pi^{\prime}(i,j,k_2,l_2) = (j, i, l_2, k_2)\). Due to the symmetry between \(j_2\) and \(k_2\), this case is the same as \textbf{(Case 2.7)}.

\textbf{(Case 2.9):} \(\pi^{\prime}(i,j,k_2,l_2) = (j, k_2, l_2, i)\). From Figure~\ref{Case 2.9}, when \(S_{i j k_1 l_1} = O(n^{-7a})\),
\[
\mathbb{P}(S_{i j k_1 l_1} = 0, S_{j, k_2, l_2, i} = 0) = 0, \quad \mathbb{P}(S_{i j k_1 l_1} = 0, S_{j, k_2, l_2, i} = \pm 1) = 0;
\]
when \(S_{i j k_1 l_1} = 1\),
\[
\mathbb{P}(S_{i j k_1 l_1} = 1, S_{j, k_2, l_2, i} = 0\text{ or } 1) = 0, \quad \mathbb{P}(S_{i j k_1 l_1} = 1, S_{j, k_2, l_2, i} = -1) = O(n^{-6a+3b});
\]
when \(S_{i j k_1 l_1} = -1\),
\[
\mathbb{P}(S_{i j k_1 l_1} = -1, S_{j, k_2, l_2, i} = 0\text{ or } -1) = 0, \quad \mathbb{P}(S_{i j k_1 l_1} = -1, S_{j, k_2, l_2, i} = 1) = O(n^{-8a+4b}).
\]

\begin{figure}[htbp]
\centering
\begin{tikzpicture}
\node[vertex] (6) at (2,0) {$j$};
\node[vertex] (7) at (4,0) {$k_2$};
\node[vertex] (8) at (2,-2) {$i$};
\node[vertex] (9) at (4,-2) {$l_2$};
\node[vertex] (16) at (2,0.7) {$i$};
\node[vertex] (17) at (4,0.7) {$j$};
\node[vertex] (18) at (2,2.7) {$l_1$};
\node[vertex] (19) at (4,2.7) {$k_1$};

\node[vertex] (1) at (6,0) {$j$};
\node[vertex] (2) at (8,0) {$k_2$};
\node[vertex] (3) at (6,-2) {$i$};
\node[vertex] (4) at (8,-2) {$l_2$};
\node[vertex] (11) at (6,0.7) {$i$};
\node[vertex] (12) at (8,0.7) {$j$};
\node[vertex] (13) at (6,2.7) {$l_1$};
\node[vertex] (14) at (8,2.7) {$k_1$};

\node[vertex] (21) at (10,0) {$j$};
\node[vertex] (22) at (12,0) {$k_2$};
\node[vertex] (23) at (10,-2) {$i$};
\node[vertex] (24) at (12,-2) {$l_2$};
\node[vertex] (31) at (10,0.7) {$i$};
\node[vertex] (32) at (12,0.7) {$j$};
\node[vertex] (33) at (10,2.7) {$l_1$};
\node[vertex] (34) at (12,2.7) {$k_1$};

\draw[->, bend left=10, thick,line width=0.25mm] (1) to (2);
\draw[->,  bend left=10, thick,line width=0.25mm] (2) to (1);
\draw[->, bend left=10, thick,line width=0.25mm] (3) to (4);
\draw[->,  bend left=10, thick,line width=0.25mm] (4) to (3);
\draw[->, bend left=10, thick,line width=0.25mm] (13) to (14);
\draw[->,  bend left=10, thick,line width=0.25mm] (14) to (13);
\draw[->, bend left=10, thick,line width=0.25mm] (11) to (12);
\draw[->,  bend left=10, thick,line width=0.25mm] (12) to (11);

\draw[->, bend left=10, dashed, thick,line width=0.25mm] (1) to (4);
\draw[->, bend left=10, dashed, thick,line width=0.25mm] (4) to (1);
\draw[->, bend left=10, dashed, thick,line width=0.25mm] (2) to (3);
\draw[->, bend left=10, dashed, thick,line width=0.25mm] (3) to (2);
\draw[->, bend left=10, dashed, thick,line width=0.25mm] (11) to (14);
\draw[->, bend left=10, dashed, thick,line width=0.25mm] (14) to (11);
\draw[->, bend left=10, dashed, thick,line width=0.25mm] (12) to (13);
\draw[->, bend left=10, dashed, thick,line width=0.25mm] (13) to (12);

\draw[->, thick,line width=0.25mm] (9) to (8);
\draw[->, thick,line width=0.25mm] (7) to (9);
\draw[->, thick,line width=0.25mm] (6) to (7);
\draw[->, thick,line width=0.25mm] (8) to (6);
\draw[->, thick,line width=0.25mm] (17) to (19);
\draw[->, thick,line width=0.25mm] (16) to (17);
\draw[->, thick,line width=0.25mm] (18) to (16);
\draw[->, thick,line width=0.25mm] (19) to (18);
\draw[->, bend left=10, dashed, thick,line width=0.25mm] (6) to (9);
\draw[->, bend left=10, dashed, thick,line width=0.25mm] (9) to (6);
\draw[->, bend left=10, dashed, thick,line width=0.25mm] (7) to (8);
\draw[->, bend left=10, dashed, thick,line width=0.25mm] (8) to (7);
\draw[->, bend left=10, dashed, thick,line width=0.25mm] (16) to (19);
\draw[->, bend left=10, dashed, thick,line width=0.25mm] (19) to (16);
\draw[->, bend left=10, dashed, thick,line width=0.25mm] (17) to (18);
\draw[->, bend left=10, dashed, thick,line width=0.25mm] (18) to (17);

\draw[->, bend left=10, thick,line width=0.25mm] (21) to (23);
\draw[->, bend left=10, thick,line width=0.25mm] (23) to (21);
\draw[->, bend left=10, thick,line width=0.25mm] (31) to (33);
\draw[->, bend left=10, thick,line width=0.25mm] (33) to (31);
\draw[->, bend left=10, thick,line width=0.25mm] (22) to (24);
\draw[->, bend left=10, thick,line width=0.25mm]  (24) to (22);
\draw[->, bend left=10, thick,line width=0.25mm]  (32) to (34);
\draw[->, bend left=10, thick,line width=0.25mm]  (34) to (32);
\draw[->, bend left=10, dashed, thick,line width=0.25mm] (21) to (24);
\draw[->, bend left=10, dashed, thick,line width=0.25mm] (24) to (21);
\draw[->, bend left=10, dashed, thick,line width=0.25mm] (22) to (23);
\draw[->, bend left=10, dashed, thick,line width=0.25mm] (23) to (22);
\draw[->, bend left=10, dashed, thick,line width=0.25mm] (31) to (34);
\draw[->, bend left=10, dashed, thick,line width=0.25mm] (34) to (31);
\draw[->, bend left=10, dashed, thick,line width=0.25mm] (32) to (33);
\draw[->, bend left=10, dashed, thick,line width=0.25mm] (33) to (32);

\end{tikzpicture}
\caption{Illustration of \textbf{(Case 2.9)}: when \(S_{i j k_1 l_1} = S_{j, k_2, l_2, i} = 0\) (left), \(S_{i j k_1 l_1} = S_{j, k_2, l_2, i} = 1\) (middle), and \(S_{i j k_1 l_1} = S_{j, k_2, l_2, i} = -1\) (right).}
\label{Case 2.9}
\end{figure}

Similarly to \textbf{(Case 2.1)}, when \(b > 0\),
\begin{align*}
& \operatorname{Cov}\big( \nabla_{\rho} \ell_{i j k_1 l_1}(\vartheta), \nabla_{\rho} \ell_{j, k_2, l_2, i}(\vartheta) \big) = O(n^{-6a - \min\{a,b\}}), \\
& \operatorname{Cov}\big( \nabla_{\gamma} \ell_{i j k_1 l_1}(\vartheta), \nabla_{\gamma} \ell_{j, k_2, l_2, i}(\vartheta) \big) = O(n^{-6a + 3b}), \\
& \operatorname{Cov}\big( \nabla_{\rho} \ell_{i j k_1 l_1}(\vartheta), \nabla_{\gamma} \ell_{j, k_2, l_2, i}(\vartheta) \big) = O(n^{-6a + b}),
\end{align*}
and when \(b \leq 0\),
\begin{align*}
& \operatorname{Cov}\big( \nabla_{\rho} \ell_{i j k_1 l_1}(\vartheta), \nabla_{\rho} \ell_{j, k_2, l_2, i}(\vartheta) \big) = O(n^{-6a + 3b}), \\
& \operatorname{Cov}\big( \nabla_{\gamma} \ell_{i j k_1 l_1}(\vartheta), \nabla_{\gamma} \ell_{j, k_2, l_2, i}(\vartheta) \big) = O(n^{-6a + 3b}), \\
& \operatorname{Cov}\big( \nabla_{\rho} \ell_{i j k_1 l_1}(\vartheta), \nabla_{\gamma} \ell_{j, k_2, l_2, i}(\vartheta) \big) = O(n^{-6a + 3b}).
\end{align*}

\textbf{(Case 2.10):} \(\pi^{\prime}(i,j,k_2,l_2) = (j, k_2, i, l_2)\). From Figure~\ref{Case 2.10}, when \(S_{i j k_1 l_1} = 0\),
\[
\mathbb{P}(S_{i j k_1 l_1} = 0, S_{i, k_2, j, l_2} = 0) = O(n^{-8a}), \quad
\mathbb{P}(S_{i j k_1 l_1} = 0, S_{i, k_2, j, l_2} = \pm 1) = O(n^{-8a + 2b});
\]
when \(S_{i j k_1 l_1} = \pm 1\),
\[
\mathbb{P}(S_{i j k_1 l_1} = \pm 1, S_{i, k_2, j, l_2} = 0) = O(n^{-8a + 2b}), \quad
\mathbb{P}(S_{i j k_1 l_1} = \pm 1, S_{i, k_2, j, l_2} = \pm 1) = O(n^{-8a + 4b}).
\]

Therefore, in this case, the bound for 
\[
\operatorname{Cov}\big(\nabla_{\vartheta} \ell_{i j k_1 l_1}(\vartheta), \nabla_{\vartheta} \ell_{j, i, k_2, l_2}(\vartheta)\big)
\]
is the same as in \textbf{(Case 2.4)}.

\begin{figure}[htbp]
\centering
\begin{tikzpicture}
\node[vertex] (6) at (2,0) {$j$};
\node[vertex] (7) at (4,0) {$k_2$};
\node[vertex] (8) at (2,-2) {$i$};
\node[vertex] (9) at (4,-2) {$l_2$};
\node[vertex] (16) at (2,0.7) {$i$};
\node[vertex] (17) at (4,0.7) {$j$};
\node[vertex] (18) at (2,2.7) {$l_1$};
\node[vertex] (19) at (4,2.7) {$k_1$};

\node[vertex] (1) at (6,0) {$j$};
\node[vertex] (2) at (8,0) {$k_2$};
\node[vertex] (3) at (6,-2) {$i$};
\node[vertex] (4) at (8,-2) {$l_2$};
\node[vertex] (11) at (6,0.7) {$i$};
\node[vertex] (12) at (8,0.7) {$j$};
\node[vertex] (13) at (6,2.7) {$l_1$};
\node[vertex] (14) at (8,2.7) {$k_1$};

\node[vertex] (21) at (10,0) {$j$};
\node[vertex] (22) at (12,0) {$k_2$};
\node[vertex] (23) at (10,-2) {$i$};
\node[vertex] (24) at (12,-2) {$l_2$};
\node[vertex] (31) at (10,0.7) {$i$};
\node[vertex] (32) at (12,0.7) {$j$};
\node[vertex] (33) at (10,2.7) {$l_1$};
\node[vertex] (34) at (12,2.7) {$k_1$};

\draw[->, bend left=10, thick,line width=0.25mm] (1) to (2);
\draw[->,  bend left=10, thick,line width=0.25mm] (2) to (1);
\draw[->, bend left=10, thick,line width=0.25mm] (3) to (4);
\draw[->,  bend left=10, thick,line width=0.25mm] (4) to (3);
\draw[->, bend left=10, thick,line width=0.25mm] (13) to (14);
\draw[->,  bend left=10, thick,line width=0.25mm] (14) to (13);
\draw[->, bend left=10, thick,line width=0.25mm] (11) to (12);
\draw[->,  bend left=10, thick,line width=0.25mm] (12) to (11);

\draw[->, bend left=10, dashed, thick,line width=0.25mm] (1) to (4);
\draw[->, bend left=10, dashed, thick,line width=0.25mm] (4) to (1);
\draw[->, bend left=10, dashed, thick,line width=0.25mm] (2) to (3);
\draw[->, bend left=10, dashed, thick,line width=0.25mm] (3) to (2);
\draw[->, bend left=10, dashed, thick,line width=0.25mm] (11) to (14);
\draw[->, bend left=10, dashed, thick,line width=0.25mm] (14) to (11);
\draw[->, bend left=10, dashed, thick,line width=0.25mm] (12) to (13);
\draw[->, bend left=10, dashed, thick,line width=0.25mm] (13) to (12);

\draw[->, thick,line width=0.25mm] (9) to (8);
\draw[->, thick,line width=0.25mm] (7) to (9);
\draw[->, thick,line width=0.25mm] (6) to (7);
\draw[->, thick,line width=0.25mm] (8) to (6);
\draw[->, thick,line width=0.25mm] (17) to (19);
\draw[->, thick,line width=0.25mm] (16) to (17);
\draw[->, thick,line width=0.25mm] (18) to (16);
\draw[->, thick,line width=0.25mm] (19) to (18);
\draw[->, bend left=10, dashed, thick,line width=0.25mm] (6) to (9);
\draw[->, bend left=10, dashed, thick,line width=0.25mm] (9) to (6);
\draw[->, bend left=10, dashed, thick,line width=0.25mm] (7) to (8);
\draw[->, bend left=10, dashed, thick,line width=0.25mm] (8) to (7);
\draw[->, bend left=10, dashed, thick,line width=0.25mm] (16) to (19);
\draw[->, bend left=10, dashed, thick,line width=0.25mm] (19) to (16);
\draw[->, bend left=10, dashed, thick,line width=0.25mm] (17) to (18);
\draw[->, bend left=10, dashed, thick,line width=0.25mm] (18) to (17);

\draw[->, bend left=10, thick,line width=0.25mm] (21) to (23);
\draw[->, bend left=10, thick,line width=0.25mm] (23) to (21);
\draw[->, bend left=10, thick,line width=0.25mm] (31) to (33);
\draw[->, bend left=10, thick,line width=0.25mm] (33) to (31);
\draw[->, bend left=10, thick,line width=0.25mm] (22) to (24);
\draw[->, bend left=10, thick,line width=0.25mm]  (24) to (22);
\draw[->, bend left=10, thick,line width=0.25mm]  (32) to (34);
\draw[->, bend left=10, thick,line width=0.25mm]  (34) to (32);
\draw[->, bend left=10, dashed, thick,line width=0.25mm] (21) to (24);
\draw[->, bend left=10, dashed, thick,line width=0.25mm] (24) to (21);
\draw[->, bend left=10, dashed, thick,line width=0.25mm] (22) to (23);
\draw[->, bend left=10, dashed, thick,line width=0.25mm] (23) to (22);
\draw[->, bend left=10, dashed, thick,line width=0.25mm] (31) to (34);
\draw[->, bend left=10, dashed, thick,line width=0.25mm] (34) to (31);
\draw[->, bend left=10, dashed, thick,line width=0.25mm] (32) to (33);
\draw[->, bend left=10, dashed, thick,line width=0.25mm] (33) to (32);

\end{tikzpicture}
\caption{Illustration of \textbf{(Case 2.10)}: when \(S_{i j k_1 l_1} = S_{j, k_2, i, l_2} = 0\) (left), \(S_{i j k_1 l_1} = S_{j, k_2, i, l_2} = 1\) (middle), and \(S_{i j k_1 l_1} = S_{j, k_2, i, l_2} = -1\) (right).}
\label{Case 2.10}
\end{figure}

\textbf{(Case 2.11):} \(\pi^{\prime}(i,j,k_2,l_2) = (j, l_2, k_2, i)\). By symmetry between \(j_2\) and \(k_2\), this case is the same as \textbf{(Case 2.9)}.

\textbf{(Case 2.12):} \(\pi^{\prime}(i,j,k_2,l_2) = (j, l_2, i, k_2)\). By symmetry between \(j_2\) and \(k_2\), this case is the same as \textbf{(Case 2.10)}.

\textbf{(Case 2.13):} \(\pi^{\prime}(i,j,k_2,l_2) = (k_2, i, j, l_2)\). From Figure~\ref{Case 2.13}, when \(S_{i j k_1 l_1} = 0\),
\[
\mathbb{P}(S_{i j k_1 l_1} = 0, S_{k_2, i, j, l_2} = 0) = 0, \quad
\mathbb{P}(S_{i j k_1 l_1} = 0, S_{k_2, i, j, l_2} = \pm 1) = 0;
\]
when \(S_{i j k_1 l_1} = 1\),
\[
\mathbb{P}(S_{i j k_1 l_1} = 1, S_{k_2, i, j, l_2} = 0 \text{ or } 1) = 0, \quad
\mathbb{P}(S_{i j k_1 l_1} = 1, S_{k_2, i, j, l_2} = -1) = O(n^{-6a + 3b});
\]
when \(S_{i j k_1 l_1} = -1\),
\[
\mathbb{P}(S_{i j k_1 l_1} = -1, S_{k_2, i, j, l_2} = 0\text{ or }-1) = 0, \quad
\mathbb{P}(S_{i j k_1 l_1} = -1, S_{k_2, i, j, l_2} = 1) = O(n^{-8a + 4b}).
\]

Therefore, in this case, the bound for 
\[
\operatorname{Cov} \big(\nabla_{\vartheta} \ell_{i j k_1 l_1}(\vartheta), \nabla_{\vartheta} \ell_{k_2, i, j, l_2}(\vartheta)\big)
\]
is the same as in \textbf{(Case 2.9)}.

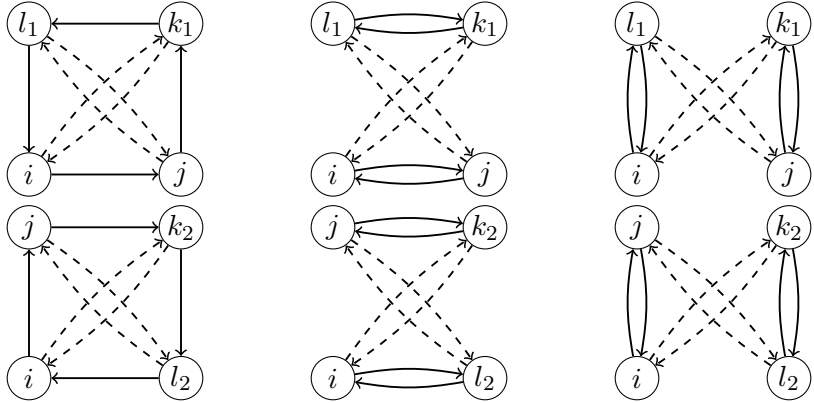
\begin{figure}[htbp]
\centering
\begin{tikzpicture}
\node[vertex] (6) at (2,0) {$j$};
\node[vertex] (7) at (4,0) {$k_2$};
\node[vertex] (8) at (2,-2) {$i$};
\node[vertex] (9) at (4,-2) {$l_2$};
\node[vertex] (16) at (2,0.7) {$i$};
\node[vertex] (17) at (4,0.7) {$j$};
\node[vertex] (18) at (2,2.7) {$l_1$};
\node[vertex] (19) at (4,2.7) {$k_1$};

\node[vertex] (1) at (6,0) {$j$};
\node[vertex] (2) at (8,0) {$k_2$};
\node[vertex] (3) at (6,-2) {$i$};
\node[vertex] (4) at (8,-2) {$l_2$};
\node[vertex] (11) at (6,0.7) {$i$};
\node[vertex] (12) at (8,0.7) {$j$};
\node[vertex] (13) at (6,2.7) {$l_1$};
\node[vertex] (14) at (8,2.7) {$k_1$};

\node[vertex] (21) at (10,0) {$j$};
\node[vertex] (22) at (12,0) {$k_2$};
\node[vertex] (23) at (10,-2) {$i$};
\node[vertex] (24) at (12,-2) {$l_2$};
\node[vertex] (31) at (10,0.7) {$i$};
\node[vertex] (32) at (12,0.7) {$j$};
\node[vertex] (33) at (10,2.7) {$l_1$};
\node[vertex] (34) at (12,2.7) {$k_1$};

\draw[->, bend left=10, thick,line width=0.25mm] (1) to (2);
\draw[->,  bend left=10, thick,line width=0.25mm] (2) to (1);
\draw[->, bend left=10, thick,line width=0.25mm] (3) to (4);
\draw[->,  bend left=10, thick,line width=0.25mm] (4) to (3);
\draw[->, bend left=10, thick,line width=0.25mm] (13) to (14);
\draw[->,  bend left=10, thick,line width=0.25mm] (14) to (13);
\draw[->, bend left=10, thick,line width=0.25mm] (11) to (12);
\draw[->,  bend left=10, thick,line width=0.25mm] (12) to (11);

\draw[->, bend left=10, dashed, thick,line width=0.25mm] (1) to (4);
\draw[->, bend left=10, dashed, thick,line width=0.25mm] (4) to (1);
\draw[->, bend left=10, dashed, thick,line width=0.25mm] (2) to (3);
\draw[->, bend left=10, dashed, thick,line width=0.25mm] (3) to (2);
\draw[->, bend left=10, dashed, thick,line width=0.25mm] (11) to (14);
\draw[->, bend left=10, dashed, thick,line width=0.25mm] (14) to (11);
\draw[->, bend left=10, dashed, thick,line width=0.25mm] (12) to (13);
\draw[->, bend left=10, dashed, thick,line width=0.25mm] (13) to (12);

\draw[->, thick,line width=0.25mm] (9) to (8);
\draw[->, thick,line width=0.25mm] (7) to (9);
\draw[->, thick,line width=0.25mm] (6) to (7);
\draw[->, thick,line width=0.25mm] (8) to (6);
\draw[->, thick,line width=0.25mm] (17) to (19);
\draw[->, thick,line width=0.25mm] (16) to (17);
\draw[->, thick,line width=0.25mm] (18) to (16);
\draw[->, thick,line width=0.25mm] (19) to (18);
\draw[->, bend left=10, dashed, thick,line width=0.25mm] (6) to (9);
\draw[->, bend left=10, dashed, thick,line width=0.25mm] (9) to (6);
\draw[->, bend left=10, dashed, thick,line width=0.25mm] (7) to (8);
\draw[->, bend left=10, dashed, thick,line width=0.25mm] (8) to (7);
\draw[->, bend left=10, dashed, thick,line width=0.25mm] (16) to (19);
\draw[->, bend left=10, dashed, thick,line width=0.25mm] (19) to (16);
\draw[->, bend left=10, dashed, thick,line width=0.25mm] (17) to (18);
\draw[->, bend left=10, dashed, thick,line width=0.25mm] (18) to (17);

\draw[->, bend left=10, thick,line width=0.25mm] (21) to (23);
\draw[->, bend left=10, thick,line width=0.25mm] (23) to (21);
\draw[->, bend left=10, thick,line width=0.25mm] (31) to (33);
\draw[->, bend left=10, thick,line width=0.25mm] (33) to (31);
\draw[->, bend left=10, thick,line width=0.25mm] (22) to (24);
\draw[->, bend left=10, thick,line width=0.25mm]  (24) to (22);
\draw[->, bend left=10, thick,line width=0.25mm]  (32) to (34);
\draw[->, bend left=10, thick,line width=0.25mm]  (34) to (32);
\draw[->, bend left=10, dashed, thick,line width=0.25mm] (21) to (24);
\draw[->, bend left=10, dashed, thick,line width=0.25mm] (24) to (21);
\draw[->, bend left=10, dashed, thick,line width=0.25mm] (22) to (23);
\draw[->, bend left=10, dashed, thick,line width=0.25mm] (23) to (22);
\draw[->, bend left=10, dashed, thick,line width=0.25mm] (31) to (34);
\draw[->, bend left=10, dashed, thick,line width=0.25mm] (34) to (31);
\draw[->, bend left=10, dashed, thick,line width=0.25mm] (32) to (33);
\draw[->, bend left=10, dashed, thick,line width=0.25mm] (33) to (32);

\end{tikzpicture}
\caption{Illustration of \textbf{(Case 2.13)}: when \(S_{i j k_1 l_1} = S_{k_2, i, j, l_2} = 0\) (left), \(S_{i j k_1 l_1} = S_{k_2, i, j, l_2} = 1\) (middle), and \(S_{i j k_1 l_1} = S_{k_2, i, j, l_2} = -1\) (right).}
\label{Case 2.13}
\end{figure}

\textbf{(Case 2.14):} $\pi^{\prime}(i,j,k_2,l_2) = (k_2, i, l_2, j)$. From Figure~\ref{Case 2.14}, when $S_{i j k_1 l_1} = 0$, we have 
\[
\mathbb{P}(S_{i j k_1 l_1} = 0, S_{k_2, i, l_2, j} = 0) = O(n^{-8a}), \quad 
\mathbb{P}(S_{i j k_1 l_1} = 0, S_{k_2, i, l_2, j} = \pm 1) = O(n^{-8a+2b}).
\]
When $S_{i j k_1 l_1} = \pm 1$, we have 
\[
\mathbb{P}(S_{i j k_1 l_1} = \pm 1, S_{k_2, i, l_2, j} = 0) = O(n^{-8a+2b}), \quad 
\mathbb{P}(S_{i j k_1 l_1} = \pm 1, S_{k_2, i, l_2, j} = \pm 1) = O(n^{-8a+4b}).
\]
Therefore, the bound for $\operatorname{Cov}(\nabla_{\vartheta}  \ell_{i j k_1 l_1}(\vartheta), \nabla_{\vartheta}  \ell_{k_2, i, l_2, j}(\vartheta))$ in this case is the same as in \textbf{(Case 2.4)}.

\begin{figure}[htbp]
\centering
\begin{tikzpicture}
\node[vertex] (6) at (2,0) {$j$};
\node[vertex] (7) at (4,0) {$k_2$};
\node[vertex] (8) at (2,-2) {$i$};
\node[vertex] (9) at (4,-2) {$l_2$};
\node[vertex] (16) at (2,0.7) {$i$};
\node[vertex] (17) at (4,0.7) {$j$};
\node[vertex] (18) at (2,2.7) {$l_1$};
\node[vertex] (19) at (4,2.7) {$k_1$};

\node[vertex] (1) at (6,0) {$j$};
\node[vertex] (2) at (8,0) {$k_2$};
\node[vertex] (3) at (6,-2) {$i$};
\node[vertex] (4) at (8,-2) {$l_2$};
\node[vertex] (11) at (6,0.7) {$i$};
\node[vertex] (12) at (8,0.7) {$j$};
\node[vertex] (13) at (6,2.7) {$l_1$};
\node[vertex] (14) at (8,2.7) {$k_1$};

\node[vertex] (21) at (10,0) {$j$};
\node[vertex] (22) at (12,0) {$k_2$};
\node[vertex] (23) at (10,-2) {$i$};
\node[vertex] (24) at (12,-2) {$l_2$};
\node[vertex] (31) at (10,0.7) {$i$};
\node[vertex] (32) at (12,0.7) {$j$};
\node[vertex] (33) at (10,2.7) {$l_1$};
\node[vertex] (34) at (12,2.7) {$k_1$};

\draw[->, bend left=10, thick,line width=0.25mm] (1) to (2);
\draw[->,  bend left=10, thick,line width=0.25mm] (2) to (1);
\draw[->, bend left=10, thick,line width=0.25mm] (3) to (4);
\draw[->,  bend left=10, thick,line width=0.25mm] (4) to (3);
\draw[->, bend left=10, thick,line width=0.25mm] (13) to (14);
\draw[->,  bend left=10, thick,line width=0.25mm] (14) to (13);
\draw[->, bend left=10, thick,line width=0.25mm] (11) to (12);
\draw[->,  bend left=10, thick,line width=0.25mm] (12) to (11);

\draw[->, bend left=10, dashed, thick,line width=0.25mm] (1) to (4);
\draw[->, bend left=10, dashed, thick,line width=0.25mm] (4) to (1);
\draw[->, bend left=10, dashed, thick,line width=0.25mm] (2) to (3);
\draw[->, bend left=10, dashed, thick,line width=0.25mm] (3) to (2);
\draw[->, bend left=10, dashed, thick,line width=0.25mm] (11) to (14);
\draw[->, bend left=10, dashed, thick,line width=0.25mm] (14) to (11);
\draw[->, bend left=10, dashed, thick,line width=0.25mm] (12) to (13);
\draw[->, bend left=10, dashed, thick,line width=0.25mm] (13) to (12);

\draw[->, thick,line width=0.25mm] (9) to (8);
\draw[->, thick,line width=0.25mm] (7) to (9);
\draw[->, thick,line width=0.25mm] (6) to (7);
\draw[->, thick,line width=0.25mm] (8) to (6);
\draw[->, thick,line width=0.25mm] (17) to (19);
\draw[->, thick,line width=0.25mm] (16) to (17);
\draw[->, thick,line width=0.25mm] (18) to (16);
\draw[->, thick,line width=0.25mm] (19) to (18);
\draw[->, bend left=10, dashed, thick,line width=0.25mm] (6) to (9);
\draw[->, bend left=10, dashed, thick,line width=0.25mm] (9) to (6);
\draw[->, bend left=10, dashed, thick,line width=0.25mm] (7) to (8);
\draw[->, bend left=10, dashed, thick,line width=0.25mm] (8) to (7);
\draw[->, bend left=10, dashed, thick,line width=0.25mm] (16) to (19);
\draw[->, bend left=10, dashed, thick,line width=0.25mm] (19) to (16);
\draw[->, bend left=10, dashed, thick,line width=0.25mm] (17) to (18);
\draw[->, bend left=10, dashed, thick,line width=0.25mm] (18) to (17);

\draw[->, bend left=10, thick,line width=0.25mm] (21) to (23);
\draw[->, bend left=10, thick,line width=0.25mm] (23) to (21);
\draw[->, bend left=10, thick,line width=0.25mm] (31) to (33);
\draw[->, bend left=10, thick,line width=0.25mm] (33) to (31);
\draw[->, bend left=10, thick,line width=0.25mm] (22) to (24);
\draw[->, bend left=10, thick,line width=0.25mm]  (24) to (22);
\draw[->, bend left=10, thick,line width=0.25mm]  (32) to (34);
\draw[->, bend left=10, thick,line width=0.25mm]  (34) to (32);
\draw[->, bend left=10, dashed, thick,line width=0.25mm] (21) to (24);
\draw[->, bend left=10, dashed, thick,line width=0.25mm] (24) to (21);
\draw[->, bend left=10, dashed, thick,line width=0.25mm] (22) to (23);
\draw[->, bend left=10, dashed, thick,line width=0.25mm] (23) to (22);
\draw[->, bend left=10, dashed, thick,line width=0.25mm] (31) to (34);
\draw[->, bend left=10, dashed, thick,line width=0.25mm] (34) to (31);
\draw[->, bend left=10, dashed, thick,line width=0.25mm] (32) to (33);
\draw[->, bend left=10, dashed, thick,line width=0.25mm] (33) to (32);

\end{tikzpicture}
\caption{Illustration of \textbf{(Case 2.14)}: when 
$S_{i j k_1 l_1} = S_{k_2, i, l_2, j} = 0$ (left), 
$S_{i j k_1 l_1} = S_{k_2, i, l_2, j} = 1$ (middle), and 
$S_{i j k_1 l_1} = S_{k_2, i, l_2, j} = -1$ (right).}
\label{Case 2.14}
\end{figure}
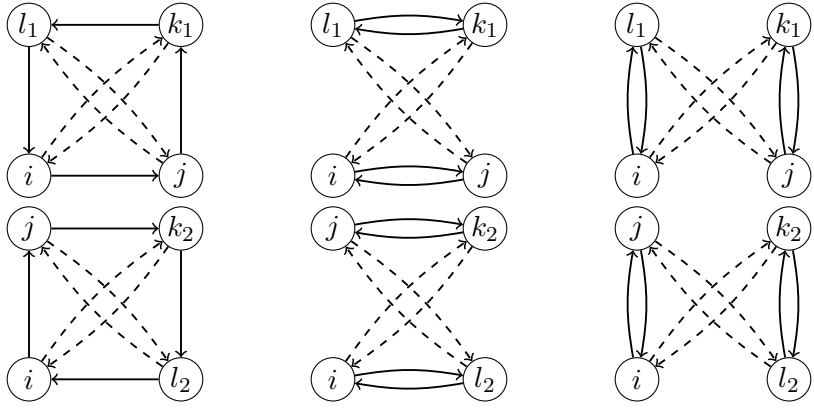
\textbf{(Case 2.15):} $\pi^{\prime}(i,j,k_2,l_2) = (k_2, j, i, l_2)$. From Figure~\ref{Case 2.15}, when $S_{i j k_1 l_1} = 0$, we have 
\[
\mathbb{P}(S_{i j k_1 l_1} = 0, S_{k_2, j, i, l_2} = 0 \text{ or } \pm 1) = 0.
\]
When $S_{i j k_1 l_1} = 1$, we have 
\[
\mathbb{P}(S_{i j k_1 l_1} = 1, S_{k_2, j, i, l_2} = 0 \text{ or } 1) = 0, \quad 
\mathbb{P}(S_{i j k_1 l_1} = 1, S_{k_2, j, i, l_2} = -1) = O(n^{-6a+3b}).
\]
When $S_{i j k_1 l_1} = -1$, we have 
\[
\mathbb{P}(S_{i j k_1 l_1} = -1, S_{k_2, j, i, l_2} = 0 \text{ or } -1) = 0, \quad 
\mathbb{P}(S_{i j k_1 l_1} = -1, S_{k_2, j, i, l_2} = 1) = O(n^{-8a+4b}).
\]
Therefore, the bound for 
\[
\operatorname{Cov}(\nabla_{\vartheta}  \ell_{i j k_1 l_1}(\vartheta), \nabla_{\vartheta}  \ell_{k_2, j, i, l_2}(\vartheta))
\]
in this case is the same as in \textbf{(Case 2.3)}.

\begin{figure}[htbp]
\centering
\begin{tikzpicture}
\node[vertex] (6) at (2,0) {$j$};
\node[vertex] (7) at (4,0) {$k_2$};
\node[vertex] (8) at (2,-2) {$i$};
\node[vertex] (9) at (4,-2) {$l_2$};
\node[vertex] (16) at (2,0.7) {$i$};
\node[vertex] (17) at (4,0.7) {$j$};
\node[vertex] (18) at (2,2.7) {$l_1$};
\node[vertex] (19) at (4,2.7) {$k_1$};

\node[vertex] (1) at (6,0) {$j$};
\node[vertex] (2) at (8,0) {$k_2$};
\node[vertex] (3) at (6,-2) {$i$};
\node[vertex] (4) at (8,-2) {$l_2$};
\node[vertex] (11) at (6,0.7) {$i$};
\node[vertex] (12) at (8,0.7) {$j$};
\node[vertex] (13) at (6,2.7) {$l_1$};
\node[vertex] (14) at (8,2.7) {$k_1$};

\node[vertex] (21) at (10,0) {$j$};
\node[vertex] (22) at (12,0) {$k_2$};
\node[vertex] (23) at (10,-2) {$i$};
\node[vertex] (24) at (12,-2) {$l_2$};
\node[vertex] (31) at (10,0.7) {$i$};
\node[vertex] (32) at (12,0.7) {$j$};
\node[vertex] (33) at (10,2.7) {$l_1$};
\node[vertex] (34) at (12,2.7) {$k_1$};

\draw[->, bend left=10, thick,line width=0.25mm] (1) to (2);
\draw[->,  bend left=10, thick,line width=0.25mm] (2) to (1);
\draw[->, bend left=10, thick,line width=0.25mm] (3) to (4);
\draw[->,  bend left=10, thick,line width=0.25mm] (4) to (3);
\draw[->, bend left=10, thick,line width=0.25mm] (13) to (14);
\draw[->,  bend left=10, thick,line width=0.25mm] (14) to (13);
\draw[->, bend left=10, thick,line width=0.25mm] (11) to (12);
\draw[->,  bend left=10, thick,line width=0.25mm] (12) to (11);

\draw[->, bend left=10, dashed, thick,line width=0.25mm] (1) to (4);
\draw[->, bend left=10, dashed, thick,line width=0.25mm] (4) to (1);
\draw[->, bend left=10, dashed, thick,line width=0.25mm] (2) to (3);
\draw[->, bend left=10, dashed, thick,line width=0.25mm] (3) to (2);
\draw[->, bend left=10, dashed, thick,line width=0.25mm] (11) to (14);
\draw[->, bend left=10, dashed, thick,line width=0.25mm] (14) to (11);
\draw[->, bend left=10, dashed, thick,line width=0.25mm] (12) to (13);
\draw[->, bend left=10, dashed, thick,line width=0.25mm] (13) to (12);

\draw[->, thick,line width=0.25mm] (9) to (8);
\draw[->, thick,line width=0.25mm] (7) to (9);
\draw[->, thick,line width=0.25mm] (6) to (7);
\draw[->, thick,line width=0.25mm] (8) to (6);
\draw[->, thick,line width=0.25mm] (17) to (19);
\draw[->, thick,line width=0.25mm] (16) to (17);
\draw[->, thick,line width=0.25mm] (18) to (16);
\draw[->, thick,line width=0.25mm] (19) to (18);
\draw[->, bend left=10, dashed, thick,line width=0.25mm] (6) to (9);
\draw[->, bend left=10, dashed, thick,line width=0.25mm] (9) to (6);
\draw[->, bend left=10, dashed, thick,line width=0.25mm] (7) to (8);
\draw[->, bend left=10, dashed, thick,line width=0.25mm] (8) to (7);
\draw[->, bend left=10, dashed, thick,line width=0.25mm] (16) to (19);
\draw[->, bend left=10, dashed, thick,line width=0.25mm] (19) to (16);
\draw[->, bend left=10, dashed, thick,line width=0.25mm] (17) to (18);
\draw[->, bend left=10, dashed, thick,line width=0.25mm] (18) to (17);

\draw[->, bend left=10, thick,line width=0.25mm] (21) to (23);
\draw[->, bend left=10, thick,line width=0.25mm] (23) to (21);
\draw[->, bend left=10, thick,line width=0.25mm] (31) to (33);
\draw[->, bend left=10, thick,line width=0.25mm] (33) to (31);
\draw[->, bend left=10, thick,line width=0.25mm] (22) to (24);
\draw[->, bend left=10, thick,line width=0.25mm]  (24) to (22);
\draw[->, bend left=10, thick,line width=0.25mm]  (32) to (34);
\draw[->, bend left=10, thick,line width=0.25mm]  (34) to (32);
\draw[->, bend left=10, dashed, thick,line width=0.25mm] (21) to (24);
\draw[->, bend left=10, dashed, thick,line width=0.25mm] (24) to (21);
\draw[->, bend left=10, dashed, thick,line width=0.25mm] (22) to (23);
\draw[->, bend left=10, dashed, thick,line width=0.25mm] (23) to (22);
\draw[->, bend left=10, dashed, thick,line width=0.25mm] (31) to (34);
\draw[->, bend left=10, dashed, thick,line width=0.25mm] (34) to (31);
\draw[->, bend left=10, dashed, thick,line width=0.25mm] (32) to (33);
\draw[->, bend left=10, dashed, thick,line width=0.25mm] (33) to (32);

\end{tikzpicture}
\caption{Illustration of \textbf{(Case 2.15)}: when $S_{i j k_1 l_1} = S_{k_2, j, i, l_2} = 0$ (left), $S_{i j k_1 l_1} = S_{k_2, j, i, l_2} = 1$ (middle), and $S_{i j k_1 l_1} = S_{k_2, j, i, l_2} = -1$ (right).}
\label{Case 2.15}
\end{figure}
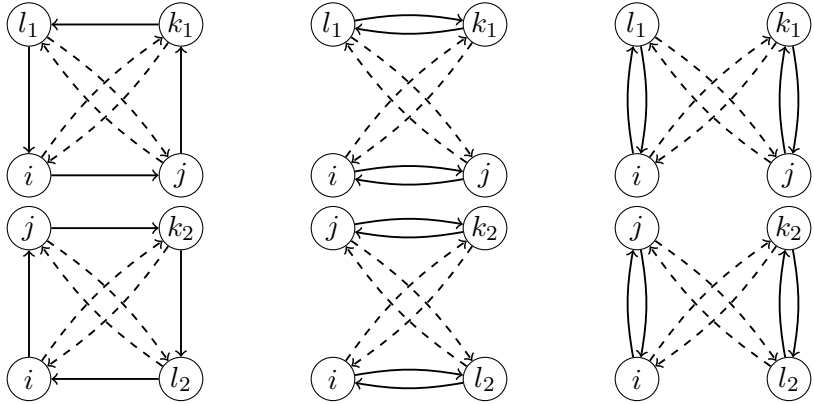

\textbf{(Case 2.16):} $\pi^{\prime}(i,j,k_2,l_2) = (k_2, j, l_2, i)$. From Figure~\ref{Case 2.16}, when $S_{i j k_1 l_1} = 0$, we have 
\[
\mathbb{P}(S_{i j k_1 l_1} = 0, S_{k_2, j, l_2, i} = 0) = O(n^{-8a}), \quad \mathbb{P}(S_{i j k_1 l_1} = 0, S_{k_2, j, l_2, i} = \pm 1) = O(n^{-8a+2b}).
\]
When $S_{i j k_1 l_1} = \pm 1$, 
\[
\mathbb{P}(S_{i j k_1 l_1} = \pm 1, S_{k_2, j, l_2, i} = 0) = O(n^{-8a+2b}), \quad \mathbb{P}(S_{i j k_1 l_1} = \pm 1, S_{k_2, j, l_2, i} = \pm 1) = O(n^{-8a+4b}).
\]
Therefore, in this case, the bound for $\operatorname{Cov}(\nabla_{\vartheta} \ell_{i j k_1 l_1}(\vartheta), \nabla_{\vartheta} \ell_{k_2, j, l_2, i}(\vartheta))$ is the same as in \textbf{(Case 2.4)}.

\begin{figure}[htbp]
\centering
\begin{tikzpicture}
\node[vertex] (6) at (2,0) {$j$};
\node[vertex] (7) at (4,0) {$k_2$};
\node[vertex] (8) at (2,-2) {$i$};
\node[vertex] (9) at (4,-2) {$l_2$};
\node[vertex] (16) at (2,0.7) {$i$};
\node[vertex] (17) at (4,0.7) {$j$};
\node[vertex] (18) at (2,2.7) {$l_1$};
\node[vertex] (19) at (4,2.7) {$k_1$};

\node[vertex] (1) at (6,0) {$j$};
\node[vertex] (2) at (8,0) {$k_2$};
\node[vertex] (3) at (6,-2) {$i$};
\node[vertex] (4) at (8,-2) {$l_2$};
\node[vertex] (11) at (6,0.7) {$i$};
\node[vertex] (12) at (8,0.7) {$j$};
\node[vertex] (13) at (6,2.7) {$l_1$};
\node[vertex] (14) at (8,2.7) {$k_1$};

\node[vertex] (21) at (10,0) {$j$};
\node[vertex] (22) at (12,0) {$k_2$};
\node[vertex] (23) at (10,-2) {$i$};
\node[vertex] (24) at (12,-2) {$l_2$};
\node[vertex] (31) at (10,0.7) {$i$};
\node[vertex] (32) at (12,0.7) {$j$};
\node[vertex] (33) at (10,2.7) {$l_1$};
\node[vertex] (34) at (12,2.7) {$k_1$};

\draw[->, bend left=10, thick,line width=0.25mm] (1) to (2);
\draw[->,  bend left=10, thick,line width=0.25mm] (2) to (1);
\draw[->, bend left=10, thick,line width=0.25mm] (3) to (4);
\draw[->,  bend left=10, thick,line width=0.25mm] (4) to (3);
\draw[->, bend left=10, thick,line width=0.25mm] (13) to (14);
\draw[->,  bend left=10, thick,line width=0.25mm] (14) to (13);
\draw[->, bend left=10, thick,line width=0.25mm] (11) to (12);
\draw[->,  bend left=10, thick,line width=0.25mm] (12) to (11);

\draw[->, bend left=10, dashed, thick,line width=0.25mm] (1) to (4);
\draw[->, bend left=10, dashed, thick,line width=0.25mm] (4) to (1);
\draw[->, bend left=10, dashed, thick,line width=0.25mm] (2) to (3);
\draw[->, bend left=10, dashed, thick,line width=0.25mm] (3) to (2);
\draw[->, bend left=10, dashed, thick,line width=0.25mm] (11) to (14);
\draw[->, bend left=10, dashed, thick,line width=0.25mm] (14) to (11);
\draw[->, bend left=10, dashed, thick,line width=0.25mm] (12) to (13);
\draw[->, bend left=10, dashed, thick,line width=0.25mm] (13) to (12);

\draw[->, thick,line width=0.25mm] (9) to (8);
\draw[->, thick,line width=0.25mm] (7) to (9);
\draw[->, thick,line width=0.25mm] (6) to (7);
\draw[->, thick,line width=0.25mm] (8) to (6);
\draw[->, thick,line width=0.25mm] (17) to (19);
\draw[->, thick,line width=0.25mm] (16) to (17);
\draw[->, thick,line width=0.25mm] (18) to (16);
\draw[->, thick,line width=0.25mm] (19) to (18);
\draw[->, bend left=10, dashed, thick,line width=0.25mm] (6) to (9);
\draw[->, bend left=10, dashed, thick,line width=0.25mm] (9) to (6);
\draw[->, bend left=10, dashed, thick,line width=0.25mm] (7) to (8);
\draw[->, bend left=10, dashed, thick,line width=0.25mm] (8) to (7);
\draw[->, bend left=10, dashed, thick,line width=0.25mm] (16) to (19);
\draw[->, bend left=10, dashed, thick,line width=0.25mm] (19) to (16);
\draw[->, bend left=10, dashed, thick,line width=0.25mm] (17) to (18);
\draw[->, bend left=10, dashed, thick,line width=0.25mm] (18) to (17);

\draw[->, bend left=10, thick,line width=0.25mm] (21) to (23);
\draw[->, bend left=10, thick,line width=0.25mm] (23) to (21);
\draw[->, bend left=10, thick,line width=0.25mm] (31) to (33);
\draw[->, bend left=10, thick,line width=0.25mm] (33) to (31);
\draw[->, bend left=10, thick,line width=0.25mm] (22) to (24);
\draw[->, bend left=10, thick,line width=0.25mm]  (24) to (22);
\draw[->, bend left=10, thick,line width=0.25mm]  (32) to (34);
\draw[->, bend left=10, thick,line width=0.25mm]  (34) to (32);
\draw[->, bend left=10, dashed, thick,line width=0.25mm] (21) to (24);
\draw[->, bend left=10, dashed, thick,line width=0.25mm] (24) to (21);
\draw[->, bend left=10, dashed, thick,line width=0.25mm] (22) to (23);
\draw[->, bend left=10, dashed, thick,line width=0.25mm] (23) to (22);
\draw[->, bend left=10, dashed, thick,line width=0.25mm] (31) to (34);
\draw[->, bend left=10, dashed, thick,line width=0.25mm] (34) to (31);
\draw[->, bend left=10, dashed, thick,line width=0.25mm] (32) to (33);
\draw[->, bend left=10, dashed, thick,line width=0.25mm] (33) to (32);

\end{tikzpicture}
\caption{Illustration of \textbf{(Case 2.16)}: when  $S_{i j k_1 l_1} = S_{k_2, j, l_2, i} = 0$ (left), $S_{i j k_1 l_1} = S_{k_2, j, l_2, i} = 1$ (middle), and $S_{i j k_1 l_1} = S_{k_2, j, l_2, i} = -1$ (right).}
\label{Case 2.16}
\end{figure}
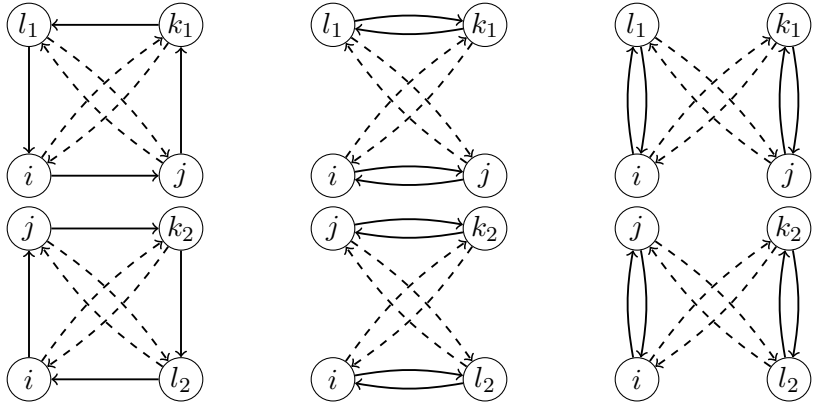

\textbf{(Case 2.17):} $\pi^{\prime}(i,j,k_2,l_2) = (k_2, l_2, i, j)$. From Figure~\ref{Case 2.17}, when $S_{i j k_1 l_1} = 0$, we have 
\[
\mathbb{P}(S_{i j k_1 l_1} = 0, S_{k_2, l_2, i, j} = 0) = O(n^{-7a}),\quad \mathbb{P}(S_{i j k_1 l_1} = 0, S_{k_2, l_2, i, j} = \pm 1) = 0.
\]
When $S_{i j k_1 l_1} = 1$, 
\[
\mathbb{P}(S_{i j k_1 l_1} = 1, S_{k_2, l_2, i, j} = 0\text{ or } -1) = 0,\quad \mathbb{P}(S_{i j k_1 l_1} = 1, S_{k_2, l_2, i, j} = 1) = O(n^{-6a+3b}).
\]
When $S_{i j k_1 l_1} = -1$, 
\[
\mathbb{P}(S_{i j k_1 l_1} = -1, S_{k_2, l_2, i, j} = 0\text{ or } 1) = 0,\quad \mathbb{P}(S_{i j k_1 l_1} = -1, S_{k_2, l_2, i, j} = -1) = O(n^{-8a+4b}).
\] 
Therefore, in this case, the bound for $\operatorname{Cov}(\nabla_{\vartheta} \ell_{i j k_1 l_1}(\vartheta), \nabla_{\vartheta} \ell_{k_2, l_2, i, j}(\vartheta))$ is the same as in \textbf{(Case 2.1)}.

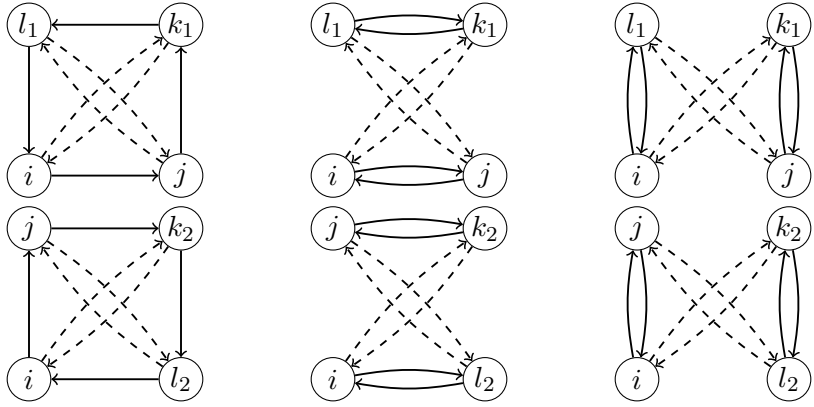
\begin{figure}[htbp]
\centering
\begin{tikzpicture}
\node[vertex] (6) at (2,0) {$j$};
\node[vertex] (7) at (4,0) {$k_2$};
\node[vertex] (8) at (2,-2) {$i$};
\node[vertex] (9) at (4,-2) {$l_2$};
\node[vertex] (16) at (2,0.7) {$i$};
\node[vertex] (17) at (4,0.7) {$j$};
\node[vertex] (18) at (2,2.7) {$l_1$};
\node[vertex] (19) at (4,2.7) {$k_1$};

\node[vertex] (1) at (6,0) {$j$};
\node[vertex] (2) at (8,0) {$k_2$};
\node[vertex] (3) at (6,-2) {$i$};
\node[vertex] (4) at (8,-2) {$l_2$};
\node[vertex] (11) at (6,0.7) {$i$};
\node[vertex] (12) at (8,0.7) {$j$};
\node[vertex] (13) at (6,2.7) {$l_1$};
\node[vertex] (14) at (8,2.7) {$k_1$};

\node[vertex] (21) at (10,0) {$j$};
\node[vertex] (22) at (12,0) {$k_2$};
\node[vertex] (23) at (10,-2) {$i$};
\node[vertex] (24) at (12,-2) {$l_2$};
\node[vertex] (31) at (10,0.7) {$i$};
\node[vertex] (32) at (12,0.7) {$j$};
\node[vertex] (33) at (10,2.7) {$l_1$};
\node[vertex] (34) at (12,2.7) {$k_1$};

\draw[->, bend left=10, thick,line width=0.25mm] (1) to (2);
\draw[->,  bend left=10, thick,line width=0.25mm] (2) to (1);
\draw[->, bend left=10, thick,line width=0.25mm] (3) to (4);
\draw[->,  bend left=10, thick,line width=0.25mm] (4) to (3);
\draw[->, bend left=10, thick,line width=0.25mm] (13) to (14);
\draw[->,  bend left=10, thick,line width=0.25mm] (14) to (13);
\draw[->, bend left=10, thick,line width=0.25mm] (11) to (12);
\draw[->,  bend left=10, thick,line width=0.25mm] (12) to (11);

\draw[->, bend left=10, dashed, thick,line width=0.25mm] (1) to (4);
\draw[->, bend left=10, dashed, thick,line width=0.25mm] (4) to (1);
\draw[->, bend left=10, dashed, thick,line width=0.25mm] (2) to (3);
\draw[->, bend left=10, dashed, thick,line width=0.25mm] (3) to (2);
\draw[->, bend left=10, dashed, thick,line width=0.25mm] (11) to (14);
\draw[->, bend left=10, dashed, thick,line width=0.25mm] (14) to (11);
\draw[->, bend left=10, dashed, thick,line width=0.25mm] (12) to (13);
\draw[->, bend left=10, dashed, thick,line width=0.25mm] (13) to (12);

\draw[->, thick,line width=0.25mm] (9) to (8);
\draw[->, thick,line width=0.25mm] (7) to (9);
\draw[->, thick,line width=0.25mm] (6) to (7);
\draw[->, thick,line width=0.25mm] (8) to (6);
\draw[->, thick,line width=0.25mm] (17) to (19);
\draw[->, thick,line width=0.25mm] (16) to (17);
\draw[->, thick,line width=0.25mm] (18) to (16);
\draw[->, thick,line width=0.25mm] (19) to (18);
\draw[->, bend left=10, dashed, thick,line width=0.25mm] (6) to (9);
\draw[->, bend left=10, dashed, thick,line width=0.25mm] (9) to (6);
\draw[->, bend left=10, dashed, thick,line width=0.25mm] (7) to (8);
\draw[->, bend left=10, dashed, thick,line width=0.25mm] (8) to (7);
\draw[->, bend left=10, dashed, thick,line width=0.25mm] (16) to (19);
\draw[->, bend left=10, dashed, thick,line width=0.25mm] (19) to (16);
\draw[->, bend left=10, dashed, thick,line width=0.25mm] (17) to (18);
\draw[->, bend left=10, dashed, thick,line width=0.25mm] (18) to (17);

\draw[->, bend left=10, thick,line width=0.25mm] (21) to (23);
\draw[->, bend left=10, thick,line width=0.25mm] (23) to (21);
\draw[->, bend left=10, thick,line width=0.25mm] (31) to (33);
\draw[->, bend left=10, thick,line width=0.25mm] (33) to (31);
\draw[->, bend left=10, thick,line width=0.25mm] (22) to (24);
\draw[->, bend left=10, thick,line width=0.25mm]  (24) to (22);
\draw[->, bend left=10, thick,line width=0.25mm]  (32) to (34);
\draw[->, bend left=10, thick,line width=0.25mm]  (34) to (32);
\draw[->, bend left=10, dashed, thick,line width=0.25mm] (21) to (24);
\draw[->, bend left=10, dashed, thick,line width=0.25mm] (24) to (21);
\draw[->, bend left=10, dashed, thick,line width=0.25mm] (22) to (23);
\draw[->, bend left=10, dashed, thick,line width=0.25mm] (23) to (22);
\draw[->, bend left=10, dashed, thick,line width=0.25mm] (31) to (34);
\draw[->, bend left=10, dashed, thick,line width=0.25mm] (34) to (31);
\draw[->, bend left=10, dashed, thick,line width=0.25mm] (32) to (33);
\draw[->, bend left=10, dashed, thick,line width=0.25mm] (33) to (32);

\end{tikzpicture}
\caption{Illustration of \textbf{(Case 2.17)}: when  $S_{i j k_1 l_1} = S_{k_2, l_2, i, j} = 0$ (left), $S_{i j k_1 l_1} = S_{k_2, l_2, i, j} = 1$ (middle), and $S_{i j k_1 l_1} = S_{k_2, l_2, i, j} = -1$ (right).}
\label{Case 2.17}
\end{figure}

\textbf{(Case 2.18):} $\pi^{\prime}(i,j,k_2,l_2) = (k_2, l_2, j, i)$. From Figure~\ref{Case 2.18}, when $S_{i j k_1 l_1} = 0$, we have 
\[
\mathbb{P}(S_{i j k_1 l_1} = 0, S_{k_2, l_2, j, i} = 0 \text{ or } \pm 1) = 0.
\]
When $S_{i j k_1 l_1} = 1$, 
\[
\mathbb{P}(S_{i j k_1 l_1} = 1, S_{k_2, l_2, j, i} = 0\text{ or } -1) = 0, \quad \mathbb{P}(S_{i j k_1 l_1} = 1, S_{k_2, l_2, j, i} = 1) = O(n^{-6a+3b}).
\] 
When $S_{i j k_1 l_1} = -1$, 
\[\mathbb{P}(S_{i j k_1 l_1} = -1, S_{k_2, l_2, j, i} = 0\text{ or } 1) = 0, \quad \mathbb{P}(S_{i j k_1 l_1} = -1, S_{k_2, l_2, j, i} = -1) = O(n^{-8a+4b}).
\]
Therefore, in this case, the bound for the covariance 
\[
\operatorname{Cov}\big(\nabla_{\vartheta} \ell_{i j k_1 l_1}(\vartheta), \nabla_{\vartheta} \ell_{k_2, l_2, j, i}(\vartheta)\big)
\]
is the same as in \textbf{(Case 2.7)}.

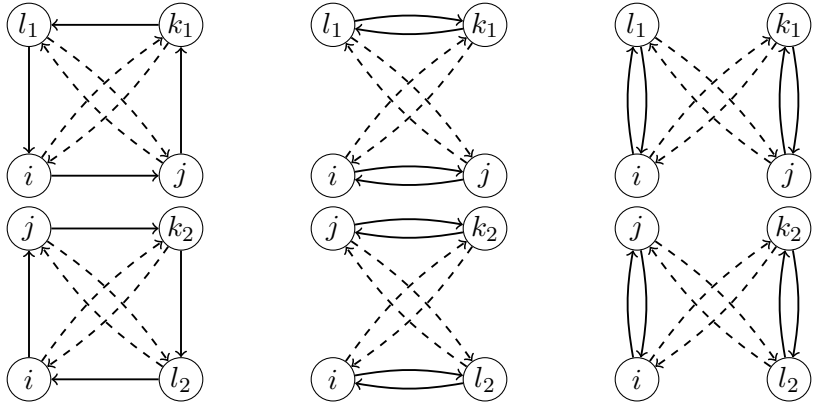
\begin{figure}[htbp]
\centering
\begin{tikzpicture}
\node[vertex] (6) at (2,0) {$j$};
\node[vertex] (7) at (4,0) {$k_2$};
\node[vertex] (8) at (2,-2) {$i$};
\node[vertex] (9) at (4,-2) {$l_2$};
\node[vertex] (16) at (2,0.7) {$i$};
\node[vertex] (17) at (4,0.7) {$j$};
\node[vertex] (18) at (2,2.7) {$l_1$};
\node[vertex] (19) at (4,2.7) {$k_1$};

\node[vertex] (1) at (6,0) {$j$};
\node[vertex] (2) at (8,0) {$k_2$};
\node[vertex] (3) at (6,-2) {$i$};
\node[vertex] (4) at (8,-2) {$l_2$};
\node[vertex] (11) at (6,0.7) {$i$};
\node[vertex] (12) at (8,0.7) {$j$};
\node[vertex] (13) at (6,2.7) {$l_1$};
\node[vertex] (14) at (8,2.7) {$k_1$};

\node[vertex] (21) at (10,0) {$j$};
\node[vertex] (22) at (12,0) {$k_2$};
\node[vertex] (23) at (10,-2) {$i$};
\node[vertex] (24) at (12,-2) {$l_2$};
\node[vertex] (31) at (10,0.7) {$i$};
\node[vertex] (32) at (12,0.7) {$j$};
\node[vertex] (33) at (10,2.7) {$l_1$};
\node[vertex] (34) at (12,2.7) {$k_1$};

\draw[->, bend left=10, thick,line width=0.25mm] (1) to (2);
\draw[->,  bend left=10, thick,line width=0.25mm] (2) to (1);
\draw[->, bend left=10, thick,line width=0.25mm] (3) to (4);
\draw[->,  bend left=10, thick,line width=0.25mm] (4) to (3);
\draw[->, bend left=10, thick,line width=0.25mm] (13) to (14);
\draw[->,  bend left=10, thick,line width=0.25mm] (14) to (13);
\draw[->, bend left=10, thick,line width=0.25mm] (11) to (12);
\draw[->,  bend left=10, thick,line width=0.25mm] (12) to (11);

\draw[->, bend left=10, dashed, thick,line width=0.25mm] (1) to (4);
\draw[->, bend left=10, dashed, thick,line width=0.25mm] (4) to (1);
\draw[->, bend left=10, dashed, thick,line width=0.25mm] (2) to (3);
\draw[->, bend left=10, dashed, thick,line width=0.25mm] (3) to (2);
\draw[->, bend left=10, dashed, thick,line width=0.25mm] (11) to (14);
\draw[->, bend left=10, dashed, thick,line width=0.25mm] (14) to (11);
\draw[->, bend left=10, dashed, thick,line width=0.25mm] (12) to (13);
\draw[->, bend left=10, dashed, thick,line width=0.25mm] (13) to (12);

\draw[->, thick,line width=0.25mm] (9) to (8);
\draw[->, thick,line width=0.25mm] (7) to (9);
\draw[->, thick,line width=0.25mm] (6) to (7);
\draw[->, thick,line width=0.25mm] (8) to (6);
\draw[->, thick,line width=0.25mm] (17) to (19);
\draw[->, thick,line width=0.25mm] (16) to (17);
\draw[->, thick,line width=0.25mm] (18) to (16);
\draw[->, thick,line width=0.25mm] (19) to (18);
\draw[->, bend left=10, dashed, thick,line width=0.25mm] (6) to (9);
\draw[->, bend left=10, dashed, thick,line width=0.25mm] (9) to (6);
\draw[->, bend left=10, dashed, thick,line width=0.25mm] (7) to (8);
\draw[->, bend left=10, dashed, thick,line width=0.25mm] (8) to (7);
\draw[->, bend left=10, dashed, thick,line width=0.25mm] (16) to (19);
\draw[->, bend left=10, dashed, thick,line width=0.25mm] (19) to (16);
\draw[->, bend left=10, dashed, thick,line width=0.25mm] (17) to (18);
\draw[->, bend left=10, dashed, thick,line width=0.25mm] (18) to (17);

\draw[->, bend left=10, thick,line width=0.25mm] (21) to (23);
\draw[->, bend left=10, thick,line width=0.25mm] (23) to (21);
\draw[->, bend left=10, thick,line width=0.25mm] (31) to (33);
\draw[->, bend left=10, thick,line width=0.25mm] (33) to (31);
\draw[->, bend left=10, thick,line width=0.25mm] (22) to (24);
\draw[->, bend left=10, thick,line width=0.25mm]  (24) to (22);
\draw[->, bend left=10, thick,line width=0.25mm]  (32) to (34);
\draw[->, bend left=10, thick,line width=0.25mm]  (34) to (32);
\draw[->, bend left=10, dashed, thick,line width=0.25mm] (21) to (24);
\draw[->, bend left=10, dashed, thick,line width=0.25mm] (24) to (21);
\draw[->, bend left=10, dashed, thick,line width=0.25mm] (22) to (23);
\draw[->, bend left=10, dashed, thick,line width=0.25mm] (23) to (22);
\draw[->, bend left=10, dashed, thick,line width=0.25mm] (31) to (34);
\draw[->, bend left=10, dashed, thick,line width=0.25mm] (34) to (31);
\draw[->, bend left=10, dashed, thick,line width=0.25mm] (32) to (33);
\draw[->, bend left=10, dashed, thick,line width=0.25mm] (33) to (32);

\end{tikzpicture}
\caption{Illustration of \textbf{(Case 2.18)}: when $S_{i j k_1 l_1} = S_{k_2, l_2, j, i} = 0$ (left), $1$ (middle), and $-1$ (right).}
\label{Case 2.18}
\end{figure}

\textbf{(Case 2.19):} $\pi^{\prime}(i,j,k_2,l_2) = (l_2, i, j, k_2)$. Due to the symmetry between $j_2$ and $k_2$, this case is the same as \textbf{(Case 2.13)}.

\textbf{(Case 2.20):} $\pi^{\prime}(i,j,k_2,l_2) = (l_2, i, k_2, j)$. Due to the symmetry between $j_2$ and $k_2$, this case is the same as \textbf{(Case 2.14)}.

\textbf{(Case 2.21):} $\pi^{\prime}(i,j,k_2,l_2) = (l_2, j, i, k_2)$. Due to the symmetry between $j_2$ and $k_2$, this case is the same as \textbf{(Case 2.15)}.

\textbf{(Case 2.22):} $\pi^{\prime}(i,j,k_2,l_2) = (l_2, j, k_2, i)$. Due to the symmetry between $j_2$ and $k_2$, this case is the same as \textbf{(Case 2.16)}.

\textbf{(Case 2.23):} $\pi^{\prime}(i,j,k_2,l_2) = (l_2, k_2, i, j)$. Due to the symmetry between $j_2$ and $k_2$, this case is the same as \textbf{(Case 2.17)}.

\textbf{(Case 2.24):} $\pi^{\prime}(i,j,k_2,l_2) = (l_2, k_2, j, i)$. Due to the symmetry between $j_2$ and $k_2$, this case is the same as \textbf{(Case 2.18)}.

Combining all the 24 cases, we have when $b > 0$,
\begin{align*}
    &\operatorname{Cov}(\psi_{\rho, ijk_1l_1}(\vartheta), \psi_{\rho, ijk_2l_2}(\vartheta)) = O\bigl(n^{-6a - \min\{a,b\}}\bigr), \\
    &\operatorname{Cov}(\psi_{\rho, ijk_1l_1}(\vartheta), \psi_{\gamma, ijk_2l_2}(\vartheta)) = O(n^{-6a + b}), \\
    &\operatorname{Cov}(\psi_{\gamma, ijk_1l_1}(\vartheta), \psi_{\gamma, ijk_2l_2}(\vartheta)) = O(n^{-6a + 3b}),
\end{align*}
and when $b \leq 0$,
\begin{align*}
    &\operatorname{Cov}(\psi_{\rho, ijk_1l_1}(\vartheta), \psi_{\rho, ijk_2l_2}(\vartheta)) = O(n^{-6a + 3b}), \qquad \\
    &\operatorname{Cov}(\psi_{\rho, ijk_1l_1}(\vartheta), \psi_{\gamma, ijk_2l_2}(\vartheta)) = O(n^{-6a + 3b}), \\
    &\operatorname{Cov}(\psi_{\gamma, ijk_1l_1}(\vartheta), \psi_{\gamma, ijk_2l_2}(\vartheta)) = O(n^{-6a + 3b}).
\end{align*}

\textbf{Case 3:} When $|\{i, j, k, l\} \cap \{p, q, t, r\}| = 3$, we consider 
\[
\operatorname{Cov}(\psi_{ijkl_1}(\vartheta), \psi_{ijkl_2}(\vartheta))
\]
where $l_1 \neq l_2$. To analyze this, we study 
\[
\operatorname{Cov}\bigl(\nabla_{\vartheta} \ell_{ijkl_1}(\vartheta), \nabla_{\vartheta} \ell_{\pi^{\prime}(i,j,k,l_2)}(\vartheta)\bigr)
\]
for any permutation $\pi^{\prime} \in \Pi_4$.

\textbf{(Case 3.1):} $\pi^{\prime}(i,j,k,l_2) = (i,j,k,l_2)$. From Figure \ref{Case 3.1}, we have $\mathbb{P}(S_{i j k l_1} \neq S_{i j k l_2})= 0$. Using the law of total expectation,
\begin{align*} 
&\mathbb{E}\bigl(\nabla_{\rho} \ell_{i j k l_1}(\vartheta) \nabla_{\rho} \ell_{i j k l_2}(\vartheta)\bigr) \\
=\, &\mathbb{E}\bigl(\nabla_{\rho} \ell_{i j k l_1}(\vartheta) \nabla_{\rho} \ell_{i j k l_2}(\vartheta) \mid S_{i j k l_1} = 1, S_{i j k l_2} = 1 \bigr) \mathbb{P}(S_{i j k l_1} = 1, S_{i j k l_2} = 1) \\
&+ \mathbb{E}\bigl(\nabla_{\rho} \ell_{i j k l_1}(\vartheta) \nabla_{\rho} \ell_{i j k l_2}(\vartheta) \mid S_{i j k l_1} = -1, S_{i j k l_2} = -1 \bigr) \mathbb{P}(S_{i j k l_1} = -1, S_{i j k l_2} = -1) \\
&+ \mathbb{E}\bigl(\nabla_{\rho} \ell_{i j k l_1}(\vartheta) \nabla_{\rho} \ell_{i j k l_2}(\vartheta) \mid S_{i j k l_1} = 0, S_{i j k l_2} = 0 \bigr) \mathbb{P}(S_{i j k l_1} = 0, S_{i j k l_2} = 0).
\end{align*}

\begin{figure}[htbp]
\centering
\begin{tikzpicture}
\node[vertex] (6) at (2,0) {$i$};
\node[vertex] (7) at (4,0) {$j$};
\node[vertex] (8) at (2,-2) {$l_2$};
\node[vertex] (9) at (4,-2) {$k$};
\node[vertex] (16) at (2,0.7) {$i$};
\node[vertex] (17) at (4,0.7) {$j$};
\node[vertex] (18) at (2,2.7) {$l_1$};
\node[vertex] (19) at (4,2.7) {$k$};

\node[vertex] (1) at (6,0) {$i$};
\node[vertex] (2) at (8,0) {$j$};
\node[vertex] (3) at (6,-2) {$l_2$};
\node[vertex] (4) at (8,-2) {$k$};
\node[vertex] (11) at (6,0.7) {$i$};
\node[vertex] (12) at (8,0.7) {$j$};
\node[vertex] (13) at (6,2.7) {$l_1$};
\node[vertex] (14) at (8,2.7) {$k$};

\node[vertex] (21) at (10,0) {$i$};
\node[vertex] (22) at (12,0) {$j$};
\node[vertex] (23) at (10,-2) {$l_2$};
\node[vertex] (24) at (12,-2) {$k$};
\node[vertex] (31) at (10,0.7) {$i$};
\node[vertex] (32) at (12,0.7) {$j$};
\node[vertex] (33) at (10,2.7) {$l_1$};
\node[vertex] (34) at (12,2.7) {$k$};

\draw[->, bend left=10, thick,line width=0.25mm] (1) to (2);
\draw[->,  bend left=10, thick,line width=0.25mm] (2) to (1);
\draw[->, bend left=10, thick,line width=0.25mm] (3) to (4);
\draw[->,  bend left=10, thick,line width=0.25mm] (4) to (3);
\draw[->, bend left=10, thick,line width=0.25mm] (13) to (14);
\draw[->,  bend left=10, thick,line width=0.25mm] (14) to (13);
\draw[->, bend left=10, thick,line width=0.25mm] (11) to (12);
\draw[->,  bend left=10, thick,line width=0.25mm] (12) to (11);

\draw[->, bend left=10, dashed, thick,line width=0.25mm] (1) to (4);
\draw[->, bend left=10, dashed, thick,line width=0.25mm] (4) to (1);
\draw[->, bend left=10, dashed, thick,line width=0.25mm] (2) to (3);
\draw[->, bend left=10, dashed, thick,line width=0.25mm] (3) to (2);
\draw[->, bend left=10, dashed, thick,line width=0.25mm] (11) to (14);
\draw[->, bend left=10, dashed, thick,line width=0.25mm] (14) to (11);
\draw[->, bend left=10, dashed, thick,line width=0.25mm] (12) to (13);
\draw[->, bend left=10, dashed, thick,line width=0.25mm] (13) to (12);

\draw[->, thick,line width=0.25mm] (9) to (8);
\draw[->, thick,line width=0.25mm] (7) to (9);
\draw[->, thick,line width=0.25mm] (6) to (7);
\draw[->, thick,line width=0.25mm] (8) to (6);
\draw[->, thick,line width=0.25mm] (17) to (19);
\draw[->, thick,line width=0.25mm] (16) to (17);
\draw[->, thick,line width=0.25mm] (18) to (16);
\draw[->, thick,line width=0.25mm] (19) to (18);
\draw[->, bend left=10, dashed, thick,line width=0.25mm] (6) to (9);
\draw[->, bend left=10, dashed, thick,line width=0.25mm] (9) to (6);
\draw[->, bend left=10, dashed, thick,line width=0.25mm] (7) to (8);
\draw[->, bend left=10, dashed, thick,line width=0.25mm] (8) to (7);
\draw[->, bend left=10, dashed, thick,line width=0.25mm] (16) to (19);
\draw[->, bend left=10, dashed, thick,line width=0.25mm] (19) to (16);
\draw[->, bend left=10, dashed, thick,line width=0.25mm] (17) to (18);
\draw[->, bend left=10, dashed, thick,line width=0.25mm] (18) to (17);

\draw[->, bend left=10, thick,line width=0.25mm] (21) to (23);
\draw[->, bend left=10, thick,line width=0.25mm]  (23) to (21);
\draw[->, bend left=10, thick,line width=0.25mm] (31) to (33);
\draw[->, bend left=10, thick,line width=0.25mm] (33) to (31);
\draw[->, bend left=10, thick,line width=0.25mm] (22) to (24);
\draw[->, bend left=10, thick,line width=0.25mm]  (24) to (22);
\draw[->, bend left=10, thick,line width=0.25mm]  (32) to (34);
\draw[->, bend left=10, thick,line width=0.25mm]  (34) to (32);
\draw[->, bend left=10, dashed, thick,line width=0.25mm] (21) to (24);
\draw[->, bend left=10, dashed, thick,line width=0.25mm] (24) to (21);
\draw[->, bend left=10, dashed, thick,line width=0.25mm] (22) to (23);
\draw[->, bend left=10, dashed, thick,line width=0.25mm] (23) to (22);
\draw[->, bend left=10, dashed, thick,line width=0.25mm] (31) to (34);
\draw[->, bend left=10, dashed, thick,line width=0.25mm] (34) to (31);
\draw[->, bend left=10, dashed, thick,line width=0.25mm] (32) to (33);
\draw[->, bend left=10, dashed, thick,line width=0.25mm] (33) to (32);

\end{tikzpicture}
\caption{Illustration of \textbf{(Case 3.1)}: when  $S_{i j k l_1} = S_{i j k l_2} = 0$ (left), $S_{i j k l_1} = S_{i j k l_2} = 1$ (middle), and $S_{i j k l_1} = S_{i j k l_2} = -1$ (right).}
\label{Case 3.1}
\end{figure}

By Assumptions \ref{covass}, \ref{nodeass}, and \ref{sparseass}, we have
\[
\mathbb{P}(S_{i j k l_1} = 0, S_{i j k l_2} = 0) = O(n^{-6a}), \quad
\]
and
\[
\mathbb{P}(S_{i j k l_1} = -1, S_{i j k l_2} = -1) = O(n^{-6a+3b}), \quad
\mathbb{P}(S_{i j k l_1} = 1, S_{i j k l_2} = 1) = O(n^{-6a+3b}).
\]
Then by Lemma \ref{Lem_score},
\[
\mathbb{E}\bigl(\nabla_{\rho} \ell_{i j k l_1}(\vartheta) \nabla_{\rho} \ell_{i j k l_2}(\vartheta)\bigr)
= O(n^{-4\max\{b,0\}}) \cdot O(n^{-6a+3b}) + O(n^{4\min\{b,0\}}) \cdot O(n^{-6a}).
\]
Also,
\[
\mathbb{E}(\nabla_{\rho} \ell_{i j k l_1}(\vartheta)) = O(n^{-2\max\{b,0\}}) \cdot O(n^{-4a+2b}) + O(n^{2\min\{b,0\}}) \cdot O(n^{-4a}).
\]
Therefore,
\[
\operatorname{Cov}\bigl(\nabla_{\rho} \ell_{i j k l_1}(\vartheta), \nabla_{\rho} \ell_{i j k l_2}(\vartheta)\bigr) = 
\begin{cases}
O(n^{-6a}), & \text{if } b > 0, \\
O(n^{-6a+3b}), & \text{if } b \leq 0.
\end{cases}
\]

Similarly, by Lemma \ref{Lem_score},
\[
\mathbb{E}\bigl(\nabla_{\gamma} \ell_{i j k l_1}(\vartheta) \nabla_{\gamma}^\top \ell_{i j k l_2}(\vartheta)\bigr) 
= O(1) \cdot O(n^{-6a+3b}) + O(n^{4\min\{b,0\}}) \cdot O(n^{-6a}).
\]
Combining this with 
\[
\mathbb{E}(\nabla_{\gamma} \ell_{i j k l_1}(\vartheta)) = O(n^{-4a+2b}),
\]
we get
\[
\operatorname{Cov}\bigl(\nabla_{\gamma} \ell_{i j k l_1}(\vartheta), \nabla_{\gamma} \ell_{i j k l_2}(\vartheta)\bigr) = O(n^{-6a+3b}),
\]
for any value of $b$. Following a similar argument,
\[
\operatorname{Cov}\bigl(\nabla_{\rho} \ell_{i j k l_1}(\vartheta), \nabla_{\gamma} \ell_{i j k l_2}(\vartheta)\bigr) = 
\begin{cases}
O(n^{-6a + b}), & \text{if } b > 0, \\
O(n^{-6a + 3b}), & \text{if } b \leq 0.
\end{cases}
\]

For the other 23 cases, we always have 
\[
\mathbb{P}\bigl(S_{i j k l_1} \neq S_{\pi^{\prime}(i,j,k,l_2)}\bigr) = 0,
\]
for any permutation $\pi^{\prime}$, and
\[
\mathbb{P}\bigl(S_{i j k l_1} = 0, S_{\pi^{\prime}(i,j,k,l_2)} = 0\bigr) = O(n^{-6a}),
\quad
\mathbb{P}\bigl(S_{i j k l_1} = \pm 1, S_{\pi^{\prime}(i,j,k,l_2)} = \pm 1\bigr) = O(n^{-6a+3b}).
\]

Therefore, following a similar argument as in \textbf{(Case 3.1)} and combining all 24 cases together, we conclude: when $b > 0$,
\begin{align*}
    \operatorname{Cov}(\psi_{\rho, i j k l_1}(\vartheta), \psi_{\rho, i j k l_2}(\vartheta)) &= O(n^{-6a}), \\
    \operatorname{Cov}(\psi_{\rho, i j k l_1}(\vartheta), \psi_{\gamma, i j k l_2}(\vartheta)) &= O(n^{-6a + b}), \\
    \operatorname{Cov}(\psi_{\gamma, i j k l_1}(\vartheta), \psi_{\gamma, i j k l_2}(\vartheta)) &= O(n^{-6a + 3b}),
\end{align*}
and when $b \leq 0$,
\begin{align*}
    \operatorname{Cov}(\psi_{\rho, i j k l_1}(\vartheta), \psi_{\rho, i j k l_2}(\vartheta)) &= O(n^{-6a + 3b}), \\
    \operatorname{Cov}(\psi_{\rho, i j l_1}(\vartheta), \psi_{\gamma, i j k l_2}(\vartheta)) &= O(n^{-6a + 3b}), \\
    \operatorname{Cov}(\psi_{\gamma, i j k l_1}(\vartheta), \psi_{\gamma, i j k l_2}(\vartheta)) &= O(n^{-6a + 3b}).
\end{align*}

\textbf{Case 4:} When $|\{i, j, k, l\} \cap \{p, q, t, r\}| = 4$, we consider
\[
\operatorname{Cov}\bigl(\nabla_{\vartheta} \ell_{i j k l}(\vartheta), \nabla_{\vartheta} \ell_{\pi^{\prime}(i,j,k,l)}(\vartheta)\bigr)
\]
for any permutation $\pi^{\prime} \in \Pi_4$.

\textbf{(Case 4.1):} $\pi^{\prime}(i,j,k,l) = (i,j,k,l)$. From Figure \ref{Case 4.1}, we have $\mathbb{P}(S_{i j k l} \neq S_{i j k l}) = 0$. Using the law of total expectation,
\begin{align*} 
&\mathbb{E}\bigl(\nabla_{\rho} \ell_{i j k l}(\vartheta) \nabla_{\rho} \ell_{i j k l}(\vartheta)\bigr) \\
=\, & \mathbb{E}\bigl(\nabla_{\rho} \ell_{i j k l}(\vartheta) \nabla_{\rho} \ell_{i j k l}(\vartheta) \mid S_{i j k l} = 1\bigr) \mathbb{P}(S_{i j k l} = 1) \\
& + \mathbb{E}\bigl(\nabla_{\rho} \ell_{i j k l}(\vartheta) \nabla_{\rho} \ell_{i j k l}(\vartheta) \mid S_{i j k l} = -1\bigr) \mathbb{P}(S_{i j k l} = -1) \\
& + \mathbb{E}\bigl(\nabla_{\rho} \ell_{i j k l}(\vartheta) \nabla_{\rho} \ell_{i j k l}(\vartheta) \mid S_{i j k l} = 0\bigr) \mathbb{P}(S_{i j k l} = 0).
\end{align*}

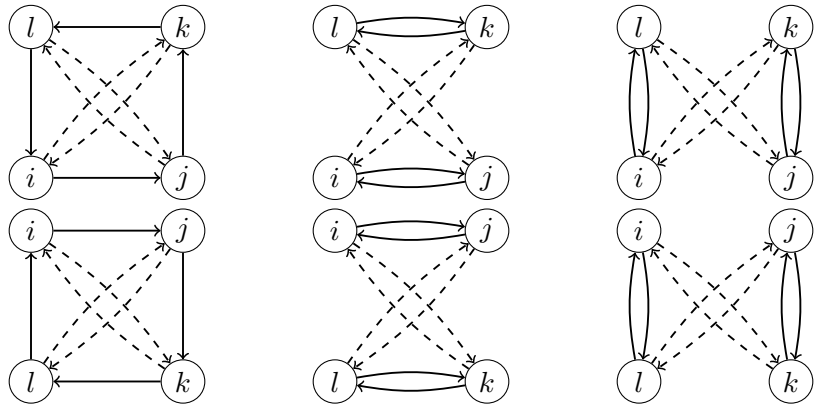
\begin{figure}[htbp]
\centering
\begin{tikzpicture}
\node[vertex] (6) at (2,0) {$i$};
\node[vertex] (7) at (4,0) {$j$};
\node[vertex] (8) at (2,-2) {$l$};
\node[vertex] (9) at (4,-2) {$k$};
\node[vertex] (16) at (2,0.7) {$i$};
\node[vertex] (17) at (4,0.7) {$j$};
\node[vertex] (18) at (2,2.7) {$l$};
\node[vertex] (19) at (4,2.7) {$k$};

\node[vertex] (1) at (6,0) {$i$};
\node[vertex] (2) at (8,0) {$j$};
\node[vertex] (3) at (6,-2) {$l$};
\node[vertex] (4) at (8,-2) {$k$};
\node[vertex] (11) at (6,0.7) {$i$};
\node[vertex] (12) at (8,0.7) {$j$};
\node[vertex] (13) at (6,2.7) {$l$};
\node[vertex] (14) at (8,2.7) {$k$};

\node[vertex] (21) at (10,0) {$i$};
\node[vertex] (22) at (12,0) {$j$};
\node[vertex] (23) at (10,-2) {$l$};
\node[vertex] (24) at (12,-2) {$k$};
\node[vertex] (31) at (10,0.7) {$i$};
\node[vertex] (32) at (12,0.7) {$j$};
\node[vertex] (33) at (10,2.7) {$l$};
\node[vertex] (34) at (12,2.7) {$k$};

\draw[->, bend left=10, thick,line width=0.25mm] (1) to (2);
\draw[->,  bend left=10, thick,line width=0.25mm] (2) to (1);
\draw[->, bend left=10, thick,line width=0.25mm] (3) to (4);
\draw[->,  bend left=10, thick,line width=0.25mm] (4) to (3);
\draw[->, bend left=10, thick,line width=0.25mm] (13) to (14);
\draw[->,  bend left=10, thick,line width=0.25mm] (14) to (13);
\draw[->, bend left=10, thick,line width=0.25mm] (11) to (12);
\draw[->,  bend left=10, thick,line width=0.25mm] (12) to (11);

\draw[->, bend left=10, dashed, thick,line width=0.25mm] (1) to (4);
\draw[->, bend left=10, dashed, thick,line width=0.25mm] (4) to (1);
\draw[->, bend left=10, dashed, thick,line width=0.25mm] (2) to (3);
\draw[->, bend left=10, dashed, thick,line width=0.25mm] (3) to (2);
\draw[->, bend left=10, dashed, thick,line width=0.25mm] (11) to (14);
\draw[->, bend left=10, dashed, thick,line width=0.25mm] (14) to (11);
\draw[->, bend left=10, dashed, thick,line width=0.25mm] (12) to (13);
\draw[->, bend left=10, dashed, thick,line width=0.25mm] (13) to (12);

\draw[->, thick,line width=0.25mm] (9) to (8);
\draw[->, thick,line width=0.25mm] (7) to (9);
\draw[->, thick,line width=0.25mm] (6) to (7);
\draw[->, thick,line width=0.25mm] (8) to (6);
\draw[->, thick,line width=0.25mm] (17) to (19);
\draw[->, thick,line width=0.25mm] (16) to (17);
\draw[->, thick,line width=0.25mm] (18) to (16);
\draw[->, thick,line width=0.25mm] (19) to (18);
\draw[->, bend left=10, dashed, thick,line width=0.25mm] (6) to (9);
\draw[->, bend left=10, dashed, thick,line width=0.25mm] (9) to (6);
\draw[->, bend left=10, dashed, thick,line width=0.25mm] (7) to (8);
\draw[->, bend left=10, dashed, thick,line width=0.25mm] (8) to (7);
\draw[->, bend left=10, dashed, thick,line width=0.25mm] (16) to (19);
\draw[->, bend left=10, dashed, thick,line width=0.25mm] (19) to (16);
\draw[->, bend left=10, dashed, thick,line width=0.25mm] (17) to (18);
\draw[->, bend left=10, dashed, thick,line width=0.25mm] (18) to (17);

\draw[->, bend left=10, thick,line width=0.25mm] (21) to (23);
\draw[->, bend left=10, thick,line width=0.25mm]  (23) to (21);
\draw[->, bend left=10, thick,line width=0.25mm] (31) to (33);
\draw[->, bend left=10, thick,line width=0.25mm] (33) to (31);
\draw[->, bend left=10, thick,line width=0.25mm] (22) to (24);
\draw[->, bend left=10, thick,line width=0.25mm]  (24) to (22);
\draw[->, bend left=10, thick,line width=0.25mm]  (32) to (34);
\draw[->, bend left=10, thick,line width=0.25mm]  (34) to (32);
\draw[->, bend left=10, dashed, thick,line width=0.25mm] (21) to (24);
\draw[->, bend left=10, dashed, thick,line width=0.25mm] (24) to (21);
\draw[->, bend left=10, dashed, thick,line width=0.25mm] (22) to (23);
\draw[->, bend left=10, dashed, thick,line width=0.25mm] (23) to (22);
\draw[->, bend left=10, dashed, thick,line width=0.25mm] (31) to (34);
\draw[->, bend left=10, dashed, thick,line width=0.25mm] (34) to (31);
\draw[->, bend left=10, dashed, thick,line width=0.25mm] (32) to (33);
\draw[->, bend left=10, dashed, thick,line width=0.25mm] (33) to (32);

\end{tikzpicture}
\caption{Illustration of \textbf{(Case 4.1)}: when  $S_{i j k l} = 0$ (left), $S_{i j k l} = 1$ (middle), and $S_{i j k l} = -1$ (right).}
\label{Case 4.1}
\end{figure}

By Assumptions \ref{covass}, \ref{nodeass}, and \ref{sparseass}, we have
\[
\mathbb{P}(S_{i j k l} = 1) = O(n^{-4a + 2b}), \quad \mathbb{P}(S_{i j k l} = -1) = O(n^{-4a + 2b}), \quad \mathbb{P}(S_{i j k l} = 0) = O(n^{-4a}).
\]
Then by Lemma \ref{Lem_score},
\[
\mathbb{E}\bigl(\nabla_{\rho} \ell_{i j k l}(\vartheta) \nabla_{\rho} \ell_{i j k l}(\vartheta)\bigr) = O(n^{-4\max\{b,0\}}) \cdot O(n^{-4a + 2b}) + O(n^{4\min\{b,0\}}) \cdot O(n^{-4a}).
\]

Similarly,
\[
\mathbb{E}(\nabla_{\rho} \ell_{i j k l}(\vartheta)) = O(n^{-2\max\{b,0\}}) \cdot O(n^{-4a + 2b}) + O(n^{2\min\{b,0\}}) \cdot O(n^{-4a}).
\]
Therefore,
\[
\operatorname{Cov}\bigl(\nabla_{\rho} \ell_{i j k l}(\vartheta), \nabla_{\rho} \ell_{i j k l}(\vartheta)\bigr) = 
\begin{cases}
O(n^{-4a}), & b > 0, \\[6pt]
O(n^{-4a + 2b}), & b \leq 0.
\end{cases}
\]
By Lemma \ref{Lem_score}, we also have
\[
\mathbb{E}\bigl(\nabla_{\gamma} \ell_{i j k l}(\vartheta) \nabla_{\gamma}^\top \ell_{i j k l}(\vartheta)\bigr) = O(1) \cdot O(n^{-4a + 2b}) + O(n^{4\min\{b,0\}}) \cdot O(n^{-4a}).
\]
Combining this with
\[
\mathbb{E}(\nabla_{\gamma} \ell_{i j k l}(\vartheta)) = O(n^{-4a + 2b}),
\]
we get
\[
\operatorname{Cov}\bigl(\nabla_{\gamma} \ell_{i j k l}(\vartheta), \nabla_{\gamma} \ell_{i j k l}(\vartheta)\bigr) = O(n^{-4a + 2b}),
\]
for any value of $b$. Following a similar argument,
\[
\operatorname{Cov}\bigl(\nabla_{\rho} \ell_{i j k l}(\vartheta), \nabla_{\gamma} \ell_{i j k l}(\vartheta)\bigr) =
\begin{cases}
O(n^{-4a}), & b > 0, \\[6pt]
O(n^{-4a + 2b}), & b \leq 0.
\end{cases}
\]

For the other 23 cases, we always have 
\[
\mathbb{P}(S_{i j k l} \neq S_{\pi^{\prime}(i,j,k,l)}) = 0,
\]
for any permutation $\pi'$, and 
\[
\mathbb{P}(S_{i j k l} = 0, S_{\pi^{\prime}(i,j,k,l)} = 0) = O(n^{-4a}), \quad \mathbb{P}(S_{i j k l} = \pm 1, S_{\pi^{\prime}(i,j,k,l)} = \pm 1) = O(n^{-4a+2b}).
\]

Therefore, following a similar argument as in \textbf{(Case 4.1)} and combining all 24 cases, we conclude that when $b > 0$,
\begin{align*}
&\operatorname{Cov}(\psi_{\rho, ijkl}(\vartheta), \psi_{\rho, ijkl}(\vartheta)) = O(n^{-4a}), \\
&\operatorname{Cov}(\psi_{\rho, ijkl}(\vartheta), \psi_{\gamma, ijkl}(\vartheta)) = O(n^{-4a}), \\
&\operatorname{Cov}(\psi_{\gamma, ijkl}(\vartheta), \psi_{\gamma, ijkl}(\vartheta)) = O(n^{-4a + 2b}),
\end{align*}
and when $b \leq 0$,
\begin{align*}
&\operatorname{Cov}(\psi_{\rho, ijkl}(\vartheta), \psi_{\rho, ijkl}(\vartheta)) = O(n^{-4a + 2b}), \\
&\operatorname{Cov}(\psi_{\rho, ijkl}(\vartheta), \psi_{\gamma, ijkl}(\vartheta)) = O(n^{-4a + 2b}), \\
&\operatorname{Cov}(\psi_{\gamma, ijkl}(\vartheta), \psi_{\gamma, ijkl}(\vartheta)) = O(n^{-4a + 2b}).
\end{align*}

For any fixed set $\{i,j,k,l\}$, the number of quadruples $\{p,q,t,r\}$ such that 
\[
|\{i,j,k,l\} \cap \{p,q,t,r\}| = 2
\]
is at most 
\[
6 \cdot \binom{n-2}{2}.
\]
The number of quadruples $(p,q,t,r)$ such that 
\[
|\{i,j,k,l\} \cap \{p,q,t,r\}| = 3
\]
is at most 
\[
4 \cdot (n-3),
\]
and only when 
\[
\{p,q,t,r\} = \{i,j,k,l\},
\]
do we have 
\[
|\{i,j,k,l\} \cap \{p,q,t,r\}| = 4.
\]

Recall that 
\[
\Psi_n(\vartheta) = \binom{n}{4}^{-1} \sum_{i<j<k<l} \psi_{ijkl}(\vartheta).
\]
Utilizing the results on $\operatorname{Cov}(\psi_{ijkl}(\vartheta), \psi_{pqtr}(\vartheta))$ obtained before and by Assumption \ref{sparseass}, we have: when $b \leq 0$,
\begin{align*}
\operatorname{Var}\big(n^{4a - 2b} \Psi_{n, \rho}(\vartheta)\big) &= O\left(\frac{n^{2a - b}}{n^2} + \frac{n^{2a - b}}{n^3} + \frac{n^{4a - 2b}}{n^4}\right) = o(1), \\
\operatorname{Var}\big(n^{4a - 2b} \Psi_{n, \gamma}(\vartheta)\big) &= O\left(\frac{n^{2a - b}}{n^2} + \frac{n^{2a - b}}{n^3} + \frac{n^{4a - 2b}}{n^4}\right) = o(1), \\
\operatorname{Cov}\big(n^{4a - 2b} \Psi_{n, \rho}(\vartheta), n^{4a - 2b} \Psi_{n, \gamma}(\vartheta)\big) &= O\left(\frac{n^{2a - b}}{n^2} + \frac{n^{2a - b}}{n^3} + \frac{n^{4a - 2b}}{n^4}\right) = o(1),
\end{align*}
when $b > 0$,
\begin{align*}
\operatorname{Var}\big(n^{4a} \Psi_{n, \rho}(\vartheta)\big) &= O\left(\frac{n^{2a - \min\{a,b\}}}{n^2} + \frac{n^{2a}}{n^3} + \frac{n^{4a}}{n^4}\right) = o(1), \\
\operatorname{Var}\big(n^{4a - 2b} \Psi_{n, \gamma}(\vartheta)\big) &= O\left(\frac{n^{2a - b}}{n^2} + \frac{n^{2a - b}}{n^3} + \frac{n^{4a - 2b}}{n^4}\right) = o(1), \\
\operatorname{Cov}\big(n^{4a} \Psi_{n, \rho}(\vartheta), n^{4a - 2b} \Psi_{n, \gamma}(\vartheta)\big) &= O\left(\frac{n^{2a - b}}{n^2} + \frac{n^{2a - b}}{n^3} + \frac{n^{4a - 2b}}{n^4}\right) = o(1).
\end{align*}

\end{proof}

\begin{proof}[Proof of Theorem~\ref{consistency}]
We verify the conditions of Theorem 5.9 in \cite{van2000asymptotic}. First, consider the case when $b \leq 0$. By Assumption \ref{Hessian}, for any $\vartheta$,
\[
\lim_{n \rightarrow +\infty} \mathbb{E}\big(n^{4a - 2b} \Psi_n(\vartheta)\big) = \Psi(\vartheta).
\]
By Proposition \ref{variance}, 
\[
\operatorname{Var}\big(n^{4a - 2b} \Psi_n(\vartheta)\big) = o(1),
\]
hence
\[
\Vert n^{4a - 2b} \Psi_n(\vartheta) - \Psi(\vartheta) \Vert_2 \xrightarrow{\mathbb{P}} 0
\]
for any $\vartheta$. Since both $\Psi_n(\vartheta)$ and $\Psi(\vartheta)$ are continuous in $\vartheta$, and the parameter space is compact by Assumption \ref{covass}, the first condition of Theorem 5.9 in \cite{van2000asymptotic} holds.

Note that for any $i < j < k < l$, 
\[
\mathbb{E}(\psi_{ijkl}(\vartheta_0)) = 0,
\]
which implies $\vartheta_0$ is a root of $\Psi(\vartheta)$. By Assumption \ref{Hessian},
\[
\lim_{n \rightarrow +\infty} \mathbb{E}\big(n^{4a - 2b} H_n(\vartheta_0)\big) = H_0(\vartheta_0),
\]
where $H_0(\vartheta_0)$ is strictly positive definite. Thus, the second condition of Theorem 5.9 in \cite{van2000asymptotic} also holds. Therefore, 
\[
\hat{\vartheta} \xrightarrow{\mathbb{P}} \vartheta_0 \quad \text{when } b \leq 0.
\]

Now consider the case $b > 0$. By Assumption \ref{Hessian}, for any $\vartheta$,
\[
\lim_{n \rightarrow +\infty} \mathbb{E}\big(n^{4a} \Psi_{n,\rho}(\vartheta),\, n^{4a - 2b} \Psi_{n,\gamma}(\vartheta)\big) = \Psi(\vartheta).
\]
By Proposition \ref{variance}, 
\[
\operatorname{Var}\big(n^{4a} \Psi_{n,\rho}(\vartheta),\, n^{4a - 2b} \Psi_{n,\gamma}(\vartheta)\big) = o(1).
\]
Thus, the first condition of Theorem 5.9 in \cite{van2000asymptotic} follows by a similar argument as in the case $b \leq 0$. Using the mean value theorem, there exists some $\tilde{\vartheta}$ between $\vartheta$ and $\vartheta_0$ such that
\[
\Psi(\vartheta) - \Psi(\vartheta_0) = \bar{H}(\tilde{\vartheta})(\vartheta - \vartheta_0).
\]
Hence,
\[
\Vert \Psi(\vartheta) - \Psi(\vartheta_0) \Vert_2 \geq \lambda_{\min}\big(\bar{H}(\tilde{\vartheta}) \bar{H}^{\top}(\tilde{\vartheta})\big) \Vert \vartheta - \vartheta_0 \Vert_2.
\]
By Assumption \ref{Hessian}, 
\[
\lim_{n \rightarrow +\infty} \bar{H}_n(\vartheta_0) = H_0(\vartheta_0),
\]
and $H_0(\vartheta_0)$ is non-singular. Combined with the continuity of $\bar{H}(\vartheta)$, the second condition of Theorem 5.9 in \cite{van2000asymptotic} holds. Therefore,
\[
\hat{\vartheta} \xrightarrow{\mathbb{P}} \vartheta_0 \quad \text{when } b > 0,
\]
which completes the proof.
\end{proof}

\begin{proof}[Proof of Theorem~\ref{AN}]
We present the proof by considering three distinct cases based on the sparsity levels: \textbf{Case A} ($b \leq 0$), \textbf{Case B} ($0 < b \leq a$), and \textbf{Case C} ($b > a$). We will analyze each case in sequence. In the following, let $H_A(\vartheta) = H_0(\vartheta)$ when $b \leq 0$; $H_B(\vartheta) = H_0(\vartheta)$ when $0 < b \leq a$ and 
\[
H_C(\vartheta) = \begin{pmatrix}
H_{0,1,1}(\vartheta) & \mathbf{0}_q^\top \\
\mathbf{0}_q & H_{0,2:(q+1), 2:(q+1)}(\vartheta)
\end{pmatrix}.
\]
when $b > a$. $H_0(\vartheta)$ is defined in Assumption \ref{Hessian}.

\textbf{Case A}. Using the mean value theorem, we have
\[
\sqrt{n^{2 - 2a + b}} \, (\hat{\vartheta} - \vartheta_0) = \big(n^{4a - 2b} H_n(\tilde{\vartheta})\big)^{-1} \cdot \sqrt{n^{2 + 6a - 3b}} \, \Psi_n(\vartheta_0),
\]
where $\tilde{\vartheta}$ lies between $\hat{\vartheta}$ and $\vartheta_0$. We first prove that 
\[
n^{4a - 2b} H_n(\tilde{\vartheta}) = H_A(\vartheta_0) + o_P(1),
\]
where the $o_P(1)$ is understood element-wise for the matrix. Recall the second-order derivative of $\ell_{ijkl}(\vartheta)$ with respect to $\rho$ is
\[
\frac{\partial^2 \ell_{ijkl}(\vartheta)}{\partial \rho^2} = \frac{4 \, \mathbb{I}(S_{ijkl} = 0, \pm 1) \, r_{ijkl}(\vartheta)}{\big(1 + r_{ijkl}(\vartheta)\big)^2}.
\]
Define 
\[
f_{ijkl}^{(1)}(\vartheta) = 2b \log n + (W_{ij}^{\top} + W_{kl}^{\top}) \vartheta, \quad f_{ijkl}^{(2)}(\vartheta) = 2b \log n + (W_{ik}^{\top} + W_{jl}^{\top}) \vartheta.
\]
The third-order derivative with respect to $\vartheta$ is
\[
\nabla_{\vartheta} \frac{\partial^2 \ell_{ijkl}(\vartheta)}{\partial \rho^2} = \frac{4 \, \mathbb{I}(S_{ijkl} = 0, \pm 1) \, e_{ijkl}(\vartheta)}{\big(1 + r_{ijkl}(\vartheta)\big)^3},
\]
where
\[
e_{ijkl}(\vartheta) = (1 - r_{ijkl}(\vartheta)) \Big[ (W_{ij} + W_{kl}) \exp(f_{ijkl}^{(1)}(\vartheta)) + (W_{ik} + W_{jl}) \exp(f_{ijkl}^{(2)}(\vartheta)) \Big].
\]

When $b \leq 0$, note that $(1 + r_{ijkl}(\vartheta))^3 = \Theta(1)$, and
\[
\mathbb{P}(S_{ijkl} = 0, \pm 1) = O(n^{-4a}) + O(n^{-4a + 2b}) = O(n^{-4a}).
\]
Also, $e_{ijkl}(\vartheta) = O(n^{2b})$. Therefore, by the law of large numbers and Assumption~\ref{Hessian}, it follows that
\[
n^{4a - 2b} H_{n,1,1}(\tilde{\vartheta}) = H_{A,1,1}(\tilde{\vartheta}) + o_P(1).
\]
Further, applying the mean value theorem and noting that $\hat{\vartheta} \xrightarrow{P} \vartheta_0$, we have
\[
H_{A,1,1}(\tilde{\vartheta}) = H_{A,1,1}(\vartheta_0) + o_P(1).
\]
Therefore,
\[
n^{4a - 2b} H_{n,1,1}(\tilde{\vartheta}) = H_{A,1,1}(\vartheta_0) + o_P(1).
\]

Next, we consider the $(t,m)$-th element of the second-order derivative matrix $\nabla_{\gamma \gamma} \ell_{ijkl}$, which is
\[
\frac{\partial^2 \ell_{ijkl}(\vartheta)}{\partial \gamma_t \partial \gamma_m} = \frac{\mathbb{I}(S_{ijkl} = 0, \pm 1) \, h_{ijkl}(\vartheta)}{\big(1 + r_{ijkl}(\vartheta)\big)^2},
\]
where
\[
\begin{aligned}
h_{ijkl}(\vartheta) &= (V_{ij,t} + V_{kl,t})(V_{ij,m} + V_{kl,m}) \exp\big(f_{ijkl}^{(1)}(\vartheta)\big) \\
&\quad + (V_{ik,t} + V_{jl,t})(V_{ik,m} + V_{jl,m}) \exp\big(f_{ijkl}^{(2)}(\vartheta)\big) \\
&\quad + Z_{ijkl,t} Z_{ijkl,m} \exp\big(f_{ijkl}^{(1)}(\vartheta) + f_{ijkl}^{(2)}(\vartheta)\big).
\end{aligned}
\]
The corresponding third-order derivative is
\[
\nabla_{\vartheta} \frac{\partial^2 \ell_{ijkl}(\vartheta)}{\partial \gamma_t \partial \gamma_m} = \frac{\mathbb{I}(S_{ijkl} = 0, \pm 1) \, q_{ijkl}(\vartheta)}{\big(1 + r_{ijkl}(\vartheta)\big)^3},
\]
where
\[
\begin{aligned}
q_{ijkl}(\vartheta) &= (W_{ij} + W_{kl})(V_{ij,t} + V_{kl,t})(V_{ij,m} + V_{kl,m}) \exp\big(f_{ijkl}^{(1)}(\vartheta)\big) \big[1 - r_{ijkl}(\vartheta)\big] \\
&\quad + (W_{ik} + W_{jl})(V_{ik,t} + V_{jl,t})(V_{ik,m} + V_{jl,m}) \exp\big(f_{ijkl}^{(2)}(\vartheta)\big) \big[1 - r_{ijkl}(\vartheta)\big] \\
&\quad + (W_{ij} + W_{kl} + W_{ik} + W_{jl}) Z_{ijkl,t} Z_{ijkl,m} \exp\big(f_{ijkl}^{(1)}(\vartheta) + f_{ijkl}^{(2)}(\vartheta)\big) \\
&\quad - \bar{Z}_{ijkl} Z_{ijkl,t} Z_{ijkl,m} \exp\big(f_{ijkl}^{(1)}(\vartheta) + f_{ijkl}^{(2)}(\vartheta)\big) \big[\exp\big(f_{ijkl}^{(1)}(\vartheta)\big) - \exp\big(f_{ijkl}^{(2)}(\vartheta)\big)\big].
\end{aligned}
\]
Here, $\bar{Z}_{ijkl} = (0, Z_{ijkl}^{\top})^{\top}$. Since $q_{ijkl}(\vartheta) = O(n^{2b})$, by a similar argument as before, it follows that for any $t = 1, 2, \ldots, q$,
\[
n^{4a - 2b} H_{n,1+t,1+t}(\tilde{\vartheta}) = H_{A,1+t,1+t}(\vartheta_0) + o_P(1).
\]
Similarly, we have
\begin{align*}
\frac{\partial^{2} \ell_{ i j k l } }{\partial\rho  \partial\gamma_{m}} &= \frac{2\mathbb{I}\left(S_{ijkl}=0, \pm 1 \right)\big[(V_{ij,m}+V_{kl,m})\exp\big(f^{(1)}_{ijkl}(\vartheta) \big) + (V_{i k,m}+V_{j l,m})\exp\big(f^{(2)}_{ijkl}(\vartheta) \big)\big]}{(1+r_{ijkl}(\vartheta))^2},
\end{align*}
and 
\begin{align*}
\nabla_{\vartheta}\frac{\partial^{2} \ell_{ i j k l } }{\partial\rho  \partial\gamma_{m}} &= \frac{2\mathbb{I}\left(S_{ijkl}=0, \pm 1 \right)s_{ijkl}(\vartheta)}{(1+r_{ijkl}(\vartheta))^3},
\end{align*}
where 
\[
\begin{aligned}
s_{ijkl}(\vartheta) &=2 ((W_{ij}+W_{kl})(V_{ij,m}+V_{kl,m})\exp\big(f^{(1)}_{ijkl}(\vartheta) \big) \big(1+r_{ijkl}(\vartheta)\big)\\
&\quad + 2(W_{ik}+W_{jl})(V_{i k,m}+V_{j l,m})\exp\big(f^{(2)}_{ijkl}(\vartheta)\big) \big(1+r_{ijkl}(\vartheta)\big) \\
&\quad -2 (V_{ij,m}+V_{kl,m})\exp\big(f^{(1)}_{ijkl}(\vartheta) \big)\big[(W_{ij}+W_{kl})\exp\big(f^{(1)}_{ijkl}(\vartheta) \big) +(W_{ik}+W_{jl})\exp\big(f^{(2)}_{ijkl}(\vartheta) \big)\big] \\
&\quad -2(V_{i k,m}+V_{j l,m})\exp\big(f^{(2)}_{ijkl}(\vartheta) \big)\big[(W_{ij}+W_{kl})\exp\big(f^{(1)}_{ijkl}(\vartheta) \big) +(W_{ik}+W_{jl})\exp\big(f^{(2)}_{ijkl}(\vartheta) \big)\big].
\end{aligned}
\]
We also have $s_{ijkl}(\vartheta)=O(n^{2b})$ and $n^{4a-2b} H_{n,1,1+t}(\tilde{\vartheta}) = H_{A,1,1+t}(\vartheta_0)+o_{P}(1)$ for any $t =1,2, ..., q$. In conclusion, $n^{4a-2b} H_n(\tilde{\vartheta}) = H_A(\vartheta_0)+o_{P}(1)$. Next, we consider a decomposition of $\Psi_n(\vartheta_0)$, where the main term is proved to be a summation of independent random vectors. Let 
$$
\bar{\psi}_{i j}(\vartheta_0)= \frac{\sum_{(k,l) \in \mathcal{H}_{ij}}\mathbb{E}\left(\psi_{i j k l}(\vartheta_0) \vert A_{i j}, A_{j i}, W_{i j}\right)}{\binom{n-2}{2}}. 
$$
For ease of presentation, we no longer require the ordering $i < j < k < l$ for the indices of $\psi_{ijkl}$ as before. For example, if $(i,j) = (3,5)$ and $(k,l) = (2,4)$, we write $\psi_{3524}$ here instead of $\psi_{2345}$ used earlier. These two expressions are equivalent because $\psi_{ijkl}$ is permutation invariant with respect to the indices $(i,j,k,l)$. Note that if $\{i,j\} \not\subset \{k,l,p,q\}$, then
\[
\mathbb{E}\big(\psi_{klpq}(\vartheta_0) \mid A_{ij}, A_{ji}, W_{ij}\big) = 0.
\]
Therefore,
\[
\mathbb{E}\big(\Psi_n(\vartheta_0) \mid A_{ij}, A_{ji}, W_{ij}\big) = \frac{1}{\binom{n}{4}} \sum_{(k,l) \in \mathcal{H}_{ij}} \mathbb{E}\big(\psi_{ijkl}(\vartheta_0) \mid A_{ij}, A_{ji}, W_{ij}\big) = \frac{6}{\binom{n}{2}} \bar{\psi}_{ij}(\vartheta_0).
\]
We define
\[
\bar{\Psi}_n(\vartheta_0):= \frac{6}{\binom{n}{2}} \sum_{i<j}\bar{\psi}_{ij}(\vartheta_0).
\]
In what follows, we always evaluate the functions $\Psi_n$, $\bar{\Psi}_n$, $\psi_{ijkl}$, and $\bar{\psi}_{ij}$ at $\vartheta_0$. For brevity, we write simply $\Psi_n$ instead of $\Psi_n(\vartheta_0)$ and similarly for $\bar{\Psi}_n$, $\psi_{ijkl}$, and $\bar{\psi}_{ij}$. Now, consider the variances $\operatorname{Var}(\Psi_n)$ and $\operatorname{Var}(\bar{\Psi}_n)$. We have
\[
\begin{aligned}
\operatorname{Var}(\Psi_n) &= \frac{\sum_{\{i_1,j_1,k_1,l_1\}, \{i_2,j_2,k_2,l_2\}} \operatorname{Cov}\big(\psi_{i_1 j_1 k_1 l_1}, \psi_{i_2 j_2 k_2 l_2}\big)}{\binom{n}{4}^2} \\
&= \frac{\sum_{i < j} \sum_{\substack{(k_1,l_1), (k_2,l_2) \in \mathcal{H}_{ij}, \{k_1,l_1\} \cap \{k_2,l_2\} = \emptyset}} \operatorname{Cov}\big(\psi_{i j k_1 l_1}, \psi_{i j k_2 l_2}\big)}{\binom{n}{4}^2} \\
&\quad + \frac{\sum_{i < j < k} \sum_{\substack{\{l_1,l_2\} \cap \{i,j,k\} = \emptyset, l_1 \neq l_2}} \operatorname{Cov}\big(\psi_{i j k l_1}, \psi_{i j k l_2}\big)}{\binom{n}{4}^2} \\
&\quad + \frac{\sum_{i < j < k < l} \operatorname{Cov}\big(\psi_{i j k l}, \psi_{i j k l}\big)}{\binom{n}{4}^2} \\
&= (A_1) + (A_2) + (A_3).
\end{aligned}
\]
From the proof of Proposition~\ref{variance}, when $b \leq 0$, we have
\[
(A_1) = O\big(n^{-2 - 6a + 3b}\big), \quad (A_2) = O\big(n^{-3 - 6a + 3b}\big), \quad \text{and} \quad (A_3) = O\big(n^{-4 - 4a + 2b}\big).
\]

Next, we consider $\operatorname{Var}(\bar{\Psi}_n)$. Since $\{\bar{\psi}_{ij}\}_{i < j}$ are independent random vectors, we have
\begin{align*}
\operatorname{Var}\left(\bar{\Psi}_n\right) 
&= \frac{36}{\binom{n}{2}^2} \sum_{i < j} \operatorname{Var}\left(\bar{\psi}_{ij}\right) \\
&= \frac{36 \sum_{i < j} \sum_{(k_1, l_1), (k_2, l_2) \in \mathcal{H}_{ij}} \operatorname{Cov}\left(
\mathbb{E}\left(\psi_{i j k_1 l_1} \mid A_{ij}, A_{ji}, W_{ij}\right),
\mathbb{E}\left(\psi_{i j k_2 l_2} \mid A_{ij}, A_{ji}, W_{ij}\right)
\right)}{\binom{n}{2}^2 \binom{n-2}{2}^2}.
\end{align*}

Here we may not have $(k_1, l_1) \cap (k_2, l_2) = \emptyset$. Specifically, for each $(k_1, l_1)$, there are $\binom{n-4}{2}$ different $(k_2, l_2)$ such that $(k_1, l_1) \cap (k_2, l_2) = \emptyset$, and for the remaining $\binom{n-2}{2} - \binom{n-4}{2}$ different $(k_2, l_2)$, $(k_1, l_1) \cap (k_2, l_2) \neq \emptyset$.  When $(k_1, l_1) \cap (k_2, l_2) = \emptyset$, conditional on $A_{ij}, A_{ji}, W_{ij}$, $\psi_{i j k_1 l_1}$ and $\psi_{i j k_2 l_2}$ are independent. Then
\begin{align*}
& \operatorname{Cov}\left(\mathbb{E}\left(\psi_{i j k_1 l_1} \mid A_{ij}, A_{ji}, W_{ij}\right), \mathbb{E}\left(\psi_{i j k_2 l_2} \mid A_{ij}, A_{ji}, W_{ij}\right)\right) \\
=& \mathbb{E}\left(
\mathbb{E}\left(\psi_{i j k_1 l_1} \mid A_{ij}, A_{ji}, W_{ij}\right) 
\mathbb{E}\left(\psi_{i j k_2 l_2} \mid A_{ij}, A_{ji}, W_{ij}\right)
\right) \\
=& \mathbb{E}\left(
\mathbb{E}\left(\psi_{i j k_1 l_1} \psi_{i j k_2 l_2} \mid A_{ij}, A_{ji}, W_{ij}\right)
\right) \\
=& \mathbb{E}\left(\psi_{i j k_1 l_1} \psi_{i j k_2 l_2}\right) \\
=& \operatorname{Cov}\left(\psi_{i j k_1 l_1}, \psi_{i j k_2 l_2}\right).
\end{align*}
Using a similar discussion as in the proof of Proposition \ref{variance}, we have the following results: when $\vert (k_1,l_1)\cap (k_2,l_2)\vert=1$, 
\[
\operatorname{Cov}\big(\mathbb{E}(\psi_{i j k_1 l_1} \mid A_{i j}, A_{j i}, W_{i j}), \mathbb{E}(\psi_{i j k_2 l_2} \mid A_{i j}, A_{j i}, W_{i j})\big) = O(n^{-6a+3b}),
\]
and if $\vert (k_1,l_1)\cap (k_2,l_2)\vert=2$,
\[
\operatorname{Cov}\big(\mathbb{E}(\psi_{i j k_1 l_1} \mid A_{i j}, A_{j i}, W_{i j}), \mathbb{E}(\psi_{i j k_2 l_2} \mid A_{i j}, A_{j i}, W_{i j})\big) = O(n^{-4a+2b}).
\]
Then, we conclude that
\[
\operatorname{Var}(\Psi_n) - \operatorname{Var}(\bar{\Psi}_n) = O(n^{-3 - 6a + 3b}) + O(n^{-4 - 4a + 2b}).
\]
Therefore,
\begin{align*}
& n^{2 + 6a - 3b} \, \mathbb{E}\big((\Psi_n - \bar{\Psi}_n)(\Psi_n - \bar{\Psi}_n)^{\top}\big) \\
=& \, n^{2 + 6a - 3b} \big(\operatorname{Var}(\Psi_n) + \operatorname{Var}(\bar{\Psi}_n) - 2 \mathbb{E}(\Psi_n \bar{\Psi}_n^{\top})\big) \\
=& \, n^{2 + 6a - 3b} \big(\operatorname{Var}(\Psi_n) + \operatorname{Var}(\bar{\Psi}_n) - 2 \mathbb{E}((\Psi_n - \bar{\Psi}_n) \bar{\Psi}_n) - 2 \mathbb{E}(\bar{\Psi}_n \bar{\Psi}_n^{\top})\big) \\
=& \, n^{2 + 6a - 3b} \big(\operatorname{Var}(\Psi_n) - \operatorname{Var}(\bar{\Psi}_n)\big) \\
=& \, O(n^{-1}) + O(n^{-2 + 2a - b}) \\
=& \, o(1),
\end{align*}
which implies
\[
\sqrt{n^{2 + 6a - 3b}} (\Psi_n - \bar{\Psi}_n) = o_P(1).
\]
Since $\bar{\Psi}_n$ is a summation of independent random vectors, by Assumption \ref{score},
\[
\sqrt{n^{2 + 6a - 3b}} \, \bar{\Psi}_n \xrightarrow{d} \Omega, \quad \text{as } n \to \infty.
\]

Finally, combining all these results,
\begin{align*}
\sqrt{n^{2 - 2a + b}} (\hat{\vartheta} - \vartheta_0) 
&= \big(n^{4a - 2b} H_n(\tilde{\vartheta})\big)^{-1} \cdot \sqrt{n^{2 + 6a - 3b}} \Psi_n(\vartheta_0) \\
&= H_A^{-1}(\vartheta_0) \sqrt{n^{2 + 6a - 3b}} \bar{\Psi}_n + o_P(1) \\
&\xrightarrow{d} N\big(0,\, H_A^{-1}(\vartheta_0) \, \Omega \, H_A^{-1}(\vartheta_0)\big),
\end{align*}
where the last step uses Assumptions \ref{Hessian} and \ref{score}.

\textbf{Case B}. Using the mean-value theorem, we obtain
\begin{align*}
n^{4a}\Psi_{n, \rho}(\vartheta_0) 
&= n^{4a}H_{n,1,1}(\tilde{\vartheta})(\hat{\rho} - \rho_0) + n^{4a}H_{n,1,2:(q+1)}(\tilde{\vartheta})(\hat{\gamma} - \gamma_0), \\
n^{4a - 2b}\Psi_{n, \gamma}(\vartheta_0) 
&= n^{4a - 2b}H_{n,2:(q+1),1}(\tilde{\vartheta})(\hat{\rho} - \rho_0) + n^{4a - 2b}H_{n,2:(q+1),2:(q+1)}(\tilde{\vartheta})(\hat{\gamma} - \gamma_0).
\end{align*}
Then, 
\[
\sqrt{n^{2 - 2a + b}} \big(\hat{\vartheta} - \vartheta_0\big) = \bar{H}_n(\tilde{\vartheta})^{-1} \cdot \begin{pmatrix}
\sqrt{n^{2 + 6a + b}} \, \Psi_{n, \rho}(\vartheta_0) \\
\sqrt{n^{2 + 6a - 3b}} \, \Psi_{n, \gamma}(\vartheta_0)
\end{pmatrix},
\]
where
\[
\bar{H}_n(\tilde{\vartheta}) = 
\begin{pmatrix}
n^{4a} H_{n,1,1}(\tilde{\vartheta}) & n^{4a} H_{n,1,2:(q+1)}(\tilde{\vartheta}) \\
n^{4a - 2b} H_{n,2:(q+1),1}(\tilde{\vartheta}) & n^{4a - 2b} H_{n,2:(q+1),2:(q+1)}(\tilde{\vartheta})
\end{pmatrix}.
\]
Through some straightforward calculations, we obtain that 
\[
\mathbb{P}(S_{i j k l} = 0, \pm 1) = O(n^{-4a}) + O(n^{-4a+2b}) = O(n^{-4a+2b}),
\]
\[
e_{ijkl}(\vartheta) = O(n^{-2b}), \quad q_{ijkl}(\vartheta) = O(1), \quad \text{and} \quad s_{ijkl}(\vartheta) = O(n^{-2b}).
\]
Then, using a similar argument as in \textbf{Case A}, we have 
\[
n^{4a} H_{n,1,1}(\tilde{\vartheta}) = H_{B,1,1}(\vartheta_0) + o_{P}(1),
\]
\[
n^{4a} H_{n,1,2:(q+1)}(\tilde{\vartheta}) = H_{B,1,2:(q+1)}(\vartheta_0) + o_{P}(1),
\]
\[
n^{4a - 2b} H_{n,2:(q+1),1}(\tilde{\vartheta}) = o_{P}(1),
\]
and
\[
n^{4a - 2b} H_{n,2:(q+1),2:(q+1)}(\tilde{\vartheta}) = H_{B,2:(q+1),2:(q+1)}(\vartheta_0) + o_{P}(1).
\]

Let 
\[
\psi^{s}_{i j k l} = \begin{pmatrix}
\sqrt{n^{2 + 6a + b}} \, \psi^{s}_{i j k l, \rho} \\
\sqrt{n^{2 + 6a - 3b}} \, \psi^{s}_{i j k l, \gamma}
\end{pmatrix}, \quad \Psi^{s}_{n} = \sum_{i<j<k<l} \psi^{s}_{i j k l}.
\]
Define $\bar{\psi}^{s}_{i j}$ and $\bar{\Psi}^{s}_{n}$ correspondingly as before. Then, by Proposition \ref{variance} and using a similar discussion as in \textbf{Case A}, we have 
\[
\Psi^{s}_{n} - \bar{\Psi}^{s}_{n} = o_{P}(1), \quad \bar{\Psi}^{s}_{n} \xrightarrow{d} \Omega,
\]
and
\[
\sqrt{n^{2 - 2a + b}} (\hat{\vartheta} - \vartheta_0) \xrightarrow{d} N\big(0, \, H_{B}^{-1}(\vartheta_0) \, \Omega \, H_{B}^{-1}(\vartheta_0)\big).
\]

\textbf{Case C}. As in \textbf{Case B}, 
\begin{align*}
n^{4a} \Psi_{n, \rho}(\vartheta_0) &= \bar{H}_{n,1,1}(\tilde{\vartheta})(\hat{\rho} - \rho_0) + \bar{H}_{n,1,2:(q+1)}(\tilde{\vartheta})(\hat{\gamma} - \gamma_0), \\
n^{4a - 2b} \Psi_{n, \gamma}(\vartheta_0) &= \bar{H}_{n,2:(q+1),1}(\tilde{\vartheta})(\hat{\rho} - \rho_0) + \bar{H}_{n,2:(q+1),2:(q+1)}(\tilde{\vartheta})(\hat{\gamma} - \gamma_0).
\end{align*}
Then,
\begin{align*}
& n^{4a - 2b} \Psi_{n, \gamma}(\vartheta_0) - \bar{H}_{n,2:(q+1),1}(\tilde{\vartheta}) \bar{H}^{-1}_{n,1,1}(\tilde{\vartheta}) n^{4a} \Psi_{n, \rho}(\vartheta_0) \\
= &\bigl(\bar{H}_{n,2:(q+1),2:(q+1)}(\tilde{\vartheta}) - \bar{H}_{n,2:(q+1),1}(\tilde{\vartheta}) \bar{H}^{-1}_{n,1,1}(\tilde{\vartheta}) \bar{H}_{n,1,2:(q+1)}(\tilde{\vartheta})\bigr) (\hat{\gamma} - \gamma_0).
\end{align*}
From Proposition \ref{variance}, we have
\[
n^{4a - 2b} \Psi_{n, \gamma}(\vartheta_0) = O_{P}\bigl(n^{(-2 + 2a - b)/2}\bigr) \quad \text{and} \quad
n^{4a} \Psi_{n, \rho}(\vartheta_0) = O_{P}\bigl(n^{(-2 + a)/2}\bigr).
\]
Also,
\begin{align*}
&\bar{H}_{n,1,1}(\tilde{\vartheta}) = \Theta(1), \quad \bar{H}_{n,2:(q+1),2:(q+1)}(\tilde{\vartheta}) = \Theta(1),\\
&\bar{H}_{n,1,2:(q+1)}(\tilde{\vartheta}) = O(1), \quad  \bar{H}_{n,2:(q+1),1}(\tilde{\vartheta}) = O(n^{-2b}).
\end{align*}
Therefore,
\[
\hat{\gamma} - \gamma_0 = O\bigl(n^{(-2 + 2a - b)/2}\bigr).
\]
Then,
\[
\hat{\rho} - \rho_0 = \bar{H}^{-1}_{n,1,1}(\tilde{\vartheta}) \Bigl(n^{4a} \Psi_{n, \rho}(\vartheta_0) - \bar{H}_{n,1,2:(q+1)}(\tilde{\vartheta})(\hat{\gamma} - \gamma_0) \Bigr) = O_{P}(1).
\]

Finally, we conclude that
\[
\begin{pmatrix}
\sqrt{n^{2 - a}}(\hat{\rho} - \rho_0) \\
\sqrt{n^{2 - 2a + b}}(\hat{\gamma} - \gamma_0)
\end{pmatrix}
= \bar{H}_{n}(\tilde{\vartheta})^{-1} \cdot 
\begin{pmatrix}
\sqrt{n^{2 + 7a}} \Psi_{n, \rho}(\vartheta_0) \\
\sqrt{n^{2 + 6a - 3b}} \Psi_{n, \gamma}(\vartheta_0)
\end{pmatrix} + o_{P}(1),
\]
where
\[
\bar{H}_{n}(\tilde{\vartheta}) = \begin{pmatrix}
n^{4a} H_{n,1,1}(\tilde{\vartheta}) & \mathbf{0} \\
\mathbf{0} & n^{4a - 2b} H_{n, 2:(q+1), 2:(q+1)}(\tilde{\vartheta})
\end{pmatrix}.
\]

Let 
\[
\psi^{s}_{ijkl} = \begin{pmatrix}
\sqrt{n^{2+7a}} \, \psi^{s}_{ijkl, \rho} \\
\sqrt{n^{2+6a-3b}} \, \psi^{s}_{ijkl, \gamma}
\end{pmatrix},
\quad
\Psi^{s}_{n} = \sum_{i<j<k<l} \psi^{s}_{ijkl}.
\]
Define $\bar{\psi}^{s}_{ij}$ and $\bar{\Psi}_{n}$ correspondingly as before. Then, by Proposition \ref{variance} and Assumption \ref{sparseass2}, using a similar discussion as in \textbf{Case A}, we have
\[
\Psi^{s}_{n} - \bar{\Psi}^{s}_{n} = o_{P}(1), \quad \bar{\Psi}^{s}_{n} \overset{d}{\longrightarrow} \Omega,
\]
and
\[
\big(\sqrt{n^{2 - a}}(\hat{\rho} - \rho_0), \sqrt{n^{2 - 2a + b}}(\hat{\gamma} - \gamma_0)^{\top}\big)^{\top} \overset{d}{\longrightarrow} N\big(0, H_{C}^{-1}(\vartheta_0) \, \Omega \, H_{C}^{-1}(\vartheta_0)\big).
\]
\end{proof}

\begin{proof}[Proof of Theorem \ref{Inference}]
We discuss the proof by considering \textbf{Case A} ($b \leq 0$), \textbf{Case B} ($0 < b \leq a$), and \textbf{Case C} separately.

\textbf{Case A:}  
As in the proof of Theorem \ref{AN}, we have
\[
n^{4a - 2b} H_{n}(\hat{\vartheta}_n) = H_{A}(\vartheta_0) + o_{P}(1)
\]
using the fact that $\hat{\vartheta}_n \xrightarrow{\mathbb{P}} \vartheta_0$. Next, we consider
\[
n^{6a - 3b} \Omega_n(\hat{\vartheta}_n) = \Omega(\vartheta_{n0}) + o_{P}(1).
\]
To simplify, we prove
\[
n^{6a - 3b} \Omega_{n,1,1}(\hat{\vartheta}_n) = \Omega_{1,1}(\vartheta_{n0}) + o_{P}(1),
\]
with the other entries following by similar arguments.

We first show that
\[
n^{6a - 3b} \Omega_{n,1,1}(\vartheta_{n0}) = \Omega_{1,1}(\vartheta_{n0}) + o_{P}(1),
\]
which requires
\[
\operatorname{Var}\bigl(n^{6a - 3b} \Omega_{n,1,1}(\vartheta_{n0})\bigr) = o(1).
\]

To do so, we evaluate
\[
\operatorname{Cov}\bigl(
\psi_{i_1 j_1 k_1 l_1, \rho_n}(\vartheta_{n0}) \psi_{i_1 j_1 k_2 l_2, \rho_n}(\vartheta_{n0}),
\quad
\psi_{i_2 j_2 k_3 l_3, \rho_n}(\vartheta_{n0}) \psi_{i_2 j_2 k_4 l_4, \rho_n}(\vartheta_{n0})
\bigr)
\]
for any $i_1, j_1, i_2, j_2$ and $(k_1, l_1), (k_2, l_2) \in \mathcal{H}_{i_1 j_1}$, $(k_3, l_3), (k_4, l_4) \in \mathcal{H}_{i_2 j_2}$. We now consider the case when $(i_1, j_1) = (i_2, j_2)$ and 
\[
(k_1, l_1) \cap (k_2, l_2) = (k_3, l_3) \cap (k_4, l_4) = \emptyset.
\]

In the following, we omit the dependence on $\vartheta_{n0}$ in the notation $\psi_{ijkl, \rho_n}(\vartheta_{n0})$ and simply write $\psi_{ijkl, \rho_n}$ when it is clear from the context.

\textbf{(Case A.1):} $(k_1, l_1, k_2, l_2) \cap (k_3, l_3, k_4, l_4) = \emptyset$. We have
\begin{align*} 
&\mathbb{E}\big(\textstyle \prod^{4}_{t=1}\nabla_{\rho_n}  \ell_{ i j k_{t} l_{t}}(\vartheta_{n0})\big) \\
 = &\mathbb{E}\big(\textstyle \prod^{4}_{t=1}\nabla_{\rho_n}  \ell_{ i j k_{t} l_{t}}(\vartheta_{n0})\mid S_{i j k_t l_t} = -1 \text{ for all } t=1,2,3,4) \mathbb{P}(S_{i j k_t l_t} = -1 \text{ for all } t=1,2,3,4\big)\\
  &+ \mathbb{E}\big(\textstyle \prod^{4}_{t=1}\nabla_{\rho_n}  \ell_{ i j k_{t} l_{t}}(\vartheta_{n0})\mid S_{i j k_t l_t} = 1 \text{ for all } t=1,2,3,4) \mathbb{P}(S_{i j k_t l_t} = 1 \text{ for all } t=1,2,3,4\big)\\
    &+ \mathbb{E}\big(\textstyle \prod^{4}_{t=1}\nabla_{\rho_n}  \ell_{ i j k_{t} l_{t}}(\vartheta_{n0})\mid S_{i j k_t l_t} =0 \text{ for all } t=1,2,3,4) \mathbb{P}(S_{i j k_t l_t} = 0 \text{ for all } t=1,2,3,4\big),
\end{align*}
where this decomposition holds since 
\[
\mathbb{P}\bigl(S_{i j k_t l_t} \neq S_{i j k_m l_m}\bigr) = 0 \quad \text{for any } t \neq m.
\]
Hereafter, `for all $t$' abbreviates `for all $t=1,2,3,4$'. To evaluate $\mathbb{P}(S_{i j k_t l_t} = 1 \text{ for all } t)$, note that
\begin{align*}
\mathbb{P}\bigl(S_{i j k_t l_t} = 1 \text{ for all } t\bigr) 
=& \mathbb{P}(A_{ij} A_{ji} = 1) \cdot \textstyle \prod^{4}_{t=1}\mathbb{P}\bigl((1 - A_{j k_t})(1 - A_{k_t j}) = 1 \bigr) \\
&\quad \cdot \textstyle \prod^{4}_{t=1} \mathbb{P}\bigl((1 - A_{i l_t})(1 - A_{l_t i}) = 1 \bigr) \cdot \textstyle \prod^{4}_{t=1} \mathbb{P}(A_{l_t k_t} A_{k_t l_t} = 1).
\end{align*}
This probability is of order $O(n^{-10a + 5b})$. Similarly, we have 
\[
\mathbb{P}(S_{i j k_t l_t} = -1 \text{ for all } t) = O(n^{-10a + 5b}) \quad \text{and} \quad \mathbb{P}(S_{i j k_t l_t} = 0 \text{ for all } t) = O(n^{-13a}).
\]

By Lemma \ref{Lem_score}, it follows that
\[
\mathbb{E}\left(\textstyle \prod^{4}_{t=1} \nabla_{\rho_n} \ell_{i j k_t l_t}(\vartheta_{n0}) \right) = O(n^{-10a + 5b}) + O(n^{-13a + 8b}) = O(n^{-10a + 5b}).
\]
Further, by considering all permutations $\pi_t$ of $(i,j,k_t,l_t)$, we have
\[
\mathbb{E}\left(\textstyle \prod^{4}_{t=1} \nabla_{\rho_n} \ell_{\pi_t(i j k_t l_t)}(\vartheta_{n0}) \right) = O(n^{-10a + 5b}).
\]
From Proposition \ref{variance}, we also know
\[
\mathbb{E}\left( \psi_{i j k_1 l_1, \rho_n} \psi_{i j k_2 l_2, \rho_n} \right) = O(n^{-6a + 3b}).
\]
Therefore,
\[
\operatorname{Cov}\left( \psi_{i j k_1 l_1, \rho_n} \psi_{i j k_2 l_2, \rho_n}, \psi_{i j k_3 l_3, \rho_n} \psi_{i j k_4 l_4, \rho_n} \right) = O(n^{-10a + 5b}).
\]

\textbf{(Case A.2):} Suppose $k_3 = k_1$ and $(l_1, k_2, l_2) \cap (l_3, k_4, l_4) = \emptyset$. In this case, we still have
\[
\mathbb{P}(S_{i j k_t l_t} = 1 \text{ for all } t) = O(n^{-10a + 5b}), \quad
\mathbb{P}(S_{i j k_t l_t} = -1 \text{ for all } t) = O(n^{-10a + 5b}),
\]
but
\begin{align*}
\mathbb{P}(S_{i j k_t l_t} = 0 \text{ for all } t) 
= & \mathbb{P}(A_{ij}(1 - A_{ji})) \cdot \textstyle \prod_{t=1,2,4} \mathbb{P}\bigl(A_{j k_t} (1 - A_{k_t j})\bigr) \\
& \cdot \textstyle \prod_{t=1}^4 \mathbb{P}\bigl((1 - A_{i l_t}) A_{l_t i}\bigr) \cdot \textstyle \prod_{t=1}^4 \mathbb{P}\bigl((1 - A_{l_t k_t}) A_{k_t l_t}\bigr),
\end{align*}
which is of order $O(n^{-12a})$. Then, by Lemma \ref{Lem_score}, we have
\[
\mathbb{E}\left(\textstyle \prod_{t=1}^4 \nabla_{\rho_n} \ell_{i j k_t l_t}(\vartheta_{n0})\right) = O(n^{-10a + 5b}) + O(n^{-12a + 8b}) = O(n^{-10a + 5b}),
\]
and consequently,
\[
\operatorname{Cov}\left( \psi_{i j k_1 l_1, \rho_n} \psi_{i j k_2 l_2, \rho_n}, \psi_{i j k_3 l_3, \rho_n} \psi_{i j k_4 l_4, \rho_n} \right) = O(n^{-10a + 5b}).
\]

\textbf{(Case A.3):} Suppose $k_3 = k_1$, $l_3 = l_1$, and $(k_2, l_2) \cap (k_4, l_4) = \emptyset$. In this case, we have
\begin{align*}
\mathbb{P}\bigl(S_{i j k_t l_t} = 1 \text{ for all } t=1,2,3,4\bigr) 
= &\; \mathbb{P}(A_{ij} A_{ji}) \cdot \textstyle \prod_{t =1,2,4} \mathbb{P}\bigl((1 - A_{j k_t})(1 - A_{k_t j})\bigr) \\
& \cdot \textstyle \prod_{t =1,2,4} \mathbb{P}\bigl((1 - A_{i l_t})(1 - A_{l_t i})\bigr) \cdot \textstyle \prod_{t =1,2,4} \mathbb{P}(A_{l_t k_t} A_{k_t l_t}),
\end{align*}
which is of order $O(n^{-8a + 4b})$. Similarly,
\[
\mathbb{P}\bigl(S_{i j k_t l_t} = -1 \text{ for all } t\bigr) = O(n^{-8a + 4b}), \quad \text{and} \quad \mathbb{P}\bigl(S_{i j k_t l_t} = 0 \text{ for all } t\bigr) = O(n^{-10a}).
\]
Then, by Lemma \ref{Lem_score}, 
\[
\mathbb{E}\left(\textstyle \prod_{t=1}^4 \nabla_{\rho_n} \ell_{i j k_t l_t}(\vartheta_{n0}) \right) = O(n^{-8a + 4b}) + O(n^{-10a + 8b}) = O(n^{-8a + 4b}),
\]
and consequently,
\[
\operatorname{Cov}\bigl( \psi_{i j k_1 l_1, \rho_n} \psi_{i j k_2 l_2, \rho_n},  \psi_{i j k_3 l_3, \rho_n} \psi_{i j k_4 l_4, \rho_n} \bigr) = O(n^{-8a + 4b}).
\]

\textbf{(Case A.4):} Suppose $k_3 = k_1$, $k_4 = k_2$, and $(l_1, l_2) \cap (l_3, l_4) = \emptyset$. In this case, we have
\[
\mathbb{P}\bigl(S_{i j k_t l_t} = 1 \text{ for all } t\bigr) = O(n^{-10a + 5b}),
\]
\[
\mathbb{P}\bigl(S_{i j k_t l_t} = -1 \text{ for all } t\bigr) = O(n^{-10a + 5b}),
\]
and
\begin{align*}
\mathbb{P}\bigl(S_{i j k_t l_t} = 0 \text{ for all } t4\bigr) 
= &\; \mathbb{P}\bigl(A_{ij} (1 - A_{ji})\bigr) \cdot \textstyle \prod_{t=1,2} \mathbb{P}\bigl(A_{j k_t} (1 - A_{k_t j})\bigr) \\
& \cdot \textstyle \prod_{t=1}^4 \mathbb{P}\bigl((1 - A_{i l_t}) A_{l_t i}\bigr) \cdot \textstyle \prod_{t=1}^4 \mathbb{P}\bigl((1 - A_{l_t k_t}) A_{k_t l_t}\bigr),
\end{align*}
which is of order $O(n^{-11a})$.

Then, by Lemma \ref{Lem_score}, 
\[
\mathbb{E}\left(\textstyle \prod_{t=1}^4 \nabla_{\rho_n} \ell_{i j k_t l_t}(\vartheta_{n0})\right) = O(n^{-10a + 5b}) + O(n^{-11a + 8b}) = O(n^{-10a + 5b}),
\]
and therefore,
\[
\operatorname{Cov}\bigl(\psi_{i j k_1 l_1, \rho_n} \psi_{i j k_2 l_2, \rho_n}, \psi_{i j k_3 l_3, \rho_n} \psi_{i j k_4 l_4, \rho_n} \bigr) = O(n^{-10a + 5b}).
\]

\textbf{(Case A.5):} Suppose $k_3 = k_1$, $l_3 = l_1$, $k_4 = k_2$, and $l_4 \neq l_2$. In this case, we have
\[
\mathbb{P}\bigl(S_{i j k_t l_t} = 1 \text{ for all } t\bigr) = O(n^{-8a + 4b}),
\]
\[
\mathbb{P}\bigl(S_{i j k_t l_t} = -1 \text{ for all } t\bigr) = O(n^{-8a + 4b}),
\]
and
\[
\mathbb{P}\bigl(S_{i j k_t l_t} = 0 \text{ for all } t\bigr) = O(n^{-9a}).
\]
Therefore, by Lemma \ref{Lem_score},
\[
\mathbb{E}\left(\textstyle \prod_{t=1}^4 \nabla_{\rho_n} \ell_{i j k_t l_t}(\vartheta_{n0})\right) = O(n^{-8a + 4b}) + O(n^{-9a + 8b}) = O(n^{-8a + 4b}),
\]
and hence,
\[
\operatorname{Cov}\bigl(\psi_{i j k_1 l_1, \rho_n} \psi_{i j k_2 l_2, \rho_n}, \psi_{i j k_3 l_3, \rho_n} \psi_{i j k_4 l_4, \rho_n} \bigr) = O(n^{-8a + 4b}).
\]
\textbf{(Case A.6):} Suppose $k_3 = k_1$, $l_3 = l_1$, $k_4 = k_2$, and $l_4 = l_2$. In this case, we have
\begin{align*}
\mathbb{P}\bigl(S_{i j k_t l_t} = 1 \text{ for all } t\bigr) 
&= \mathbb{P}(A_{ij} A_{ji}) \cdot \textstyle \prod_{t=1}^4 \mathbb{P}((1 - A_{j k_t})(1 - A_{k_t j})) \\
&\quad \cdot \textstyle \prod_{t=1,2} \mathbb{P}((1 - A_{i l_t})(1 - A_{l_t i})) \cdot\textstyle \prod_{t=1}^4 \mathbb{P}(A_{l_t k_t} A_{k_t l_t}) \\
&= O(n^{-6a + 3b}).
\end{align*}
Similarly,
\[
\mathbb{P}\bigl(S_{i j k_t l_t} = -1 \text{ for all } t\bigr) = O(n^{-6a + 3b}),
\]
and
\begin{align*}
\mathbb{P}\bigl(S_{i j k_t l_t} = 0 \text{ for all } t\bigr) 
&= \mathbb{P}(A_{ij} (1 - A_{ji})) \cdot \textstyle \prod_{t=1,2} \mathbb{P}(A_{j k_t} (1 - A_{k_t j})) \\
&\quad \cdot \textstyle \prod_{t=1,2} \mathbb{P}((1 - A_{i l_t}) A_{l_t i}) \cdot \textstyle \prod_{t=1,2} \mathbb{P}((1 - A_{l_t k_t}) A_{k_t l_t}) \\
&= O(n^{-7a}).
\end{align*}
By Lemma \ref{Lem_score}, it follows that
\[
\mathbb{E}\left(\textstyle \prod_{t=1}^4 \nabla_{\rho_n} \ell_{i j k_t l_t}(\vartheta_{n0})\right) = O(n^{-6a + 3b}),
\]
and consequently,
\[
\operatorname{Cov}\bigl(\psi_{i j k_1 l_1, \rho_n} \psi_{i j k_2 l_2, \rho_n}, \ \psi_{i j k_3 l_3, \rho_n} \psi_{i j k_4 l_4, \rho_n}\bigr) = O(n^{-6a + 3b}).
\]

From the proof of Theorem \ref{AN}, we have
\begin{align*}
n^{6a - 3b} \, \Omega_{n,1,1}(\vartheta_{n0}) 
&= \frac{36 n^{6a - 3b}}{\binom{n}{2} \binom{n-2}{2}^2} \sum_{i < j} \sum_{(k_1, l_1), (k_2, l_2) \in \mathcal{H}_{ij}} \psi_{i j k_1 l_1, \rho_n} \, \psi_{i j k_2 l_2, \rho_n} \\
&= \frac{36 n^{6a - 3b}}{\binom{n}{2} \binom{n-2}{2}^2} \sum_{i < j} \sum_{\substack{(k_1, l_1), (k_2, l_2) \in \mathcal{H}_{ij} \\ (k_1, l_1) \cap (k_2, l_2) = \emptyset}} \psi_{i j k_1 l_1, \rho_n} \, \psi_{i j k_2 l_2, \rho_n} + o_P(1).
\end{align*}

For any fixed index tuple $(i_1, j_1, k_1, l_1, k_2, l_2)$, consider another tuple $(i_2, j_2, k_3, l_3, k_4, l_4)$.

\textbf{(Case A.01):} If
\[
\left| \{i_2, j_2, k_3, l_3, k_4, l_4\} \cap \{i_1, j_1, k_1, l_1, k_2, l_2\} \right| = 0 \text{ or } 1,
\]
then the random variables
\[
\psi_{i_1 j_1 k_1 l_1, \rho_n} \psi_{i_1 j_1 k_2 l_2, \rho_n} \quad \text{and} \quad \psi_{i_2 j_2 k_3 l_3, \rho_n} \psi_{i_2 j_2 k_4 l_4, \rho_n}
\]
are independent. Hence,
\[
\operatorname{Cov}\left( \psi_{i_1 j_1 k_1 l_1, \rho_n} \psi_{i_1 j_1 k_2 l_2, \rho_n}, \; \psi_{i_2 j_2 k_3 l_3, \rho_n} \psi_{i_2 j_2 k_4 l_4, \rho_n} \right) = 0.
\]

\textbf{(Case A.02):} 
\[
\left| \{i_2, j_2, k_3, l_3, k_4, l_4\} \cap \{i_1, j_1, k_1, l_1, k_2, l_2\} \right| = 2.
\]
Even if we may not have $(i_1, j_1) = (i_2, j_2)$ as in \textbf{(Case A.1)}, we still have 
\[
\operatorname{Cov}\big( \psi_{i_1 j_1 k_1 l_1, \rho_n} \psi_{i_1 j_1 k_2 l_2, \rho_n}, \; \psi_{i_2 j_2 k_3 l_3, \rho_n} \psi_{i_2 j_2 k_4 l_4, \rho_n} \big) = O(n^{-10a + 5b})
\]
in this case after some careful considerations. 

For any fixed $(i_1, j_1, k_1, l_1, k_2, l_2)$, there are at most 
\[
6! \cdot \binom{6}{2} \cdot \binom{n-6}{4}
\]
distinct $(i_2, j_2, k_3, l_3, k_4, l_4)$ such that
\[
\left| \{i_2, j_2, k_3, l_3, k_4, l_4\} \cap \{i_1, j_1, k_1, l_1, k_2, l_2\} \right| = 2.
\]

\vspace{1em}

\textbf{(Case A.03):} 
\[
\left| \{i_2, j_2, k_3, l_3, k_4, l_4\} \cap \{i_1, j_1, k_1, l_1, k_2, l_2\} \right| = 3.
\]
For any $(i_1, j_1, k_1, l_1, k_2, l_2)$, there are at most
\[
6! \cdot \binom{6}{3} \cdot \binom{n-6}{3}
\]
distinct $(i_2, j_2, k_3, l_3, k_4, l_4)$ such that 
\[
\left| \{i_2, j_2, k_3, l_3, k_4, l_4\} \cap \{i_1, j_1, k_1, l_1, k_2, l_2\} \right| = 3,
\]
and in this case, 
\[
\operatorname{Cov}\big( \psi_{i_1 j_1 k_1 l_1, \rho_n} \psi_{i_1 j_1 k_2 l_2, \rho_n}, \; \psi_{i_2 j_2 k_3 l_3, \rho_n} \psi_{i_2 j_2 k_4 l_4, \rho_n} \big) = O(n^{-10a + 5b}),
\]
as in \textbf{(Case A.02)}.

\textbf{(Case A.04):} 
\[
\left| \{i_2, j_2, k_3, l_3, k_4, l_4\} \cap \{i_1, j_1, k_1, l_1, k_2, l_2\} \right| = 4.
\]
For any $(i_1, j_1, k_1, l_1, k_2, l_2)$, there are at most 
\[
6! \cdot \binom{6}{4} \cdot \binom{n-6}{2}
\]
distinct $(i_2, j_2, k_3, l_3, k_4, l_4)$ such that 
\[
\left| \{i_2, j_2, k_3, l_3, k_4, l_4\} \cap \{i_1, j_1, k_1, l_1, k_2, l_2\} \right| = 4,
\]
and in this case,
\[
\operatorname{Cov}\big( \psi_{i_1 j_1 k_1 l_1, \rho_n} \psi_{i_1 j_1 k_2 l_2, \rho_n}, \; \psi_{i_2 j_2 k_3 l_3, \rho_n} \psi_{i_2 j_2 k_4 l_4, \rho_n} \big) = O(n^{-8a + 4b}),
\]
as in \textbf{(Case A.3)} and \textbf{(Case A.4)}.

\vspace{1em}

\textbf{(Case A.05):}  
\[
\left| \{i_2, j_2, k_3, l_3, k_4, l_4\} \cap \{i_1, j_1, k_1, l_1, k_2, l_2\} \right| = 5.
\]
For any $(i_1, j_1, k_1, l_1, k_2, l_2)$, there are at most
\[
6! \cdot \binom{6}{5} \cdot \binom{n-6}{1}
\]
distinct $(i_2, j_2, k_3, l_3, k_4, l_4)$ such that
\[
\left| \{i_2, j_2, k_3, l_3, k_4, l_4\} \cap \{i_1, j_1, k_1, l_1, k_2, l_2\} \right| = 5,
\]
and in this case,
\[
\operatorname{Cov}\big( \psi_{i_1 j_1 k_1 l_1, \rho_n} \psi_{i_1 j_1 k_2 l_2, \rho_n}, \; \psi_{i_2 j_2 k_3 l_3, \rho_n} \psi_{i_2 j_2 k_4 l_4, \rho_n} \big) = O(n^{-8a + 4b}),
\]
as in \textbf{(Case A.5)}.

\vspace{1em}

\textbf{(Case A.06):}  
\[
\left| \{i_2, j_2, k_3, l_3, k_4, l_4\} \cap \{i_1, j_1, k_1, l_1, k_2, l_2\} \right| = 6.
\]
For any $(i_1, j_1, k_1, l_1, k_2, l_2)$, there are at most $6!$ distinct $(i_2, j_2, k_3, l_3, k_4, l_4)$ such that
\[
\left| \{i_2, j_2, k_3, l_3, k_4, l_4\} \cap \{i_1, j_1, k_1, l_1, k_2, l_2\} \right| = 6,
\]
and in this case,
\[
\operatorname{Cov}\big( \psi_{i_1 j_1 k_1 l_1, \rho_n} \psi_{i_1 j_1 k_2 l_2, \rho_n}, \; \psi_{i_2 j_2 k_3 l_3, \rho_n} \psi_{i_2 j_2 k_4 l_4, \rho_n} \big) = O(n^{-6a + 3b}),
\]
as in \textbf{(Case A.6)}.

To summarize all the discussions above, we have
\begin{align*}
\operatorname{Var}\bigl(n^{6a-3b}\Omega_{n}(\vartheta_{n0})\bigr) 
= O\biggl(
\frac{n^{10 - 10a + 5b} + n^{9 - 10a + 5b} + n^{8 - 8a + 4b} + n^{7 - 8a + 4b} + n^{6 - 6a + 3b}}{n^{2(6 - 6a + 3b)}}
\biggr) = o(1),
\end{align*}
by Assumption \ref{sparseass}. Then,
\[
n^{6a - 3b} \Omega_{n,1,1}(\vartheta_{n0}) = \Omega_{1,1}(\vartheta_{n0}) + o_{P}(1)
\]
follows. Next, we consider
\[
n^{6a - 3b} \Omega_{n,1,1}(\hat{\vartheta}_{n}) - n^{6a - 3b} \Omega_{n,1,1}(\vartheta_{n0}).
\]
Notice that 
\begin{align*}
&\nabla_{\rho_n} \ell_{i j k_1 l_1}(\hat{\vartheta}_{n}) \nabla_{\rho_n} \ell_{i j k_2 l_2}(\hat{\vartheta}_{n})
- \nabla_{\rho_n} \ell_{i j k_1 l_1}(\vartheta_{n0}) \nabla_{\rho_n} \ell_{i j k_2 l_2}(\vartheta_{n0}) \\
=& \; 2 \mathbb{I}(S_{i j k_1 l_1} = S_{i j k_2 l_2} = \pm 1) 
\biggl\{ \frac{1}{(1 + r_{i j k_1 l_1}(\hat{\vartheta}_{n}))(1 + r_{i j k_2 l_2}(\hat{\vartheta}_{n}))} 
- \frac{1}{(1 + r_{i j k_1 l_1}(\vartheta_{n0}))(1 + r_{i j k_2 l_2}(\vartheta_{n0}))} \biggr\} \\
+& \; 2 \mathbb{I}(S_{i j k_1 l_1} = S_{i j k_2 l_2} = 0) 
\biggl\{ \frac{r_{i j k_1 l_1}(\hat{\vartheta}_{n}) r_{i j k_2 l_2}(\hat{\vartheta}_{n})}{(1 + r_{i j k_1 l_1}(\hat{\vartheta}_{n}))(1 + r_{i j k_2 l_2}(\hat{\vartheta}_{n}))} 
- \frac{r_{i j k_1 l_1}(\vartheta_{n0}) r_{i j k_2 l_2}(\vartheta_{n0})}{(1 + r_{i j k_1 l_1}(\vartheta_{n0}))(1 + r_{i j k_2 l_2}(\vartheta_{n0}))} \biggr\}.
\end{align*}
By the mean-value theorem and the consistency of $\hat{\vartheta}_n$, we have 
\[
r_{i j k_1 l_1}(\hat{\vartheta}_n) = r_{i j k_1 l_1}(\vartheta_{n0})(1 + o_P(1)) \quad \text{and} \quad r_{i j k_2 l_2}(\hat{\vartheta}_n) = r_{i j k_2 l_2}(\vartheta_{n0})(1 + o_P(1)),
\]
then 
\[
\nabla_{\rho_n} \ell_{i j k_1 l_1}(\hat{\vartheta}_n) \nabla_{\rho_n} \ell_{i j k_2 l_2}(\hat{\vartheta}_n) = \nabla_{\rho_n} \ell_{i j k_1 l_1}(\vartheta_{n0}) \nabla_{\rho_n} \ell_{i j k_2 l_2}(\vartheta_{n0}) (1 + o_P(1)).
\]
Therefore, 
\begin{align*}
n^{6a - 3b} \Omega_{n,1,1}(\hat{\vartheta}_n) - \Omega_{1,1}(\vartheta_{n0}) 
&= n^{6a - 3b} \Omega_{n,1,1}(\vartheta_{n0})(1 + o_P(1)) - \Omega_{1,1}(\vartheta_{n0}) \\
&= \bigl(\Omega_{1,1}(\vartheta_{n0}) + o_P(1)\bigr)(1 + o_P(1)) - \Omega_{1,1}(\vartheta_{n0}) \\
&= o_P(1).
\end{align*}
Similar results hold for other components of $\Omega_{n}(\hat{\vartheta}_{n})$. Hence, we have
\[
n^{6a - 3b} \Omega_{n}(\hat{\vartheta}_{n}) = \Omega(\vartheta_{n0}) + o_{P}(1)
\]
and
\[
\bigl(n^{4a - 2b} H_{n}(\hat{\vartheta}_{n})\bigr)^{-1}
\bigl(n^{6a - 3b} \Omega_{n}(\hat{\vartheta}_{n})\bigr)
\bigl(n^{4a - 2b} H^{\top}_{n}(\hat{\vartheta}_{n})\bigr)^{-1}
= H_{A}^{-1}(\vartheta_{0}) \, \Omega(\vartheta_{0}) \, H_{A}^{\top,-1}(\vartheta_{0}) + o_{P}(1).
\]
Therefore,
\begin{align*}
n \cdot \frac{\hat{\rho}_n - \rho_{n0}}{\sqrt{\hat{V}_{1,1}}}
&= \sqrt{n^{2 - 2a + b}} \cdot \frac{\hat{\rho}_n - \rho_{n0}}{
\sqrt{
\left(
\bigl(n^{4a - 2b} H_{n}(\hat{\vartheta}_{n})\bigr)^{-1}
\bigl(n^{6a - 3b} \Omega_{n}(\hat{\vartheta}_{n})\bigr)
\bigl(n^{4a - 2b} H_{n}^{\top}(\hat{\vartheta}_{n})\bigr)^{-1}
\right)_{1,1}
}
} \\
&= \sqrt{n^{2 - 2a + b}} \cdot \frac{\hat{\rho}_n - \rho_{n0}}{
\sqrt{
\bigl(H_{A}^{-1}(\vartheta_0) \, \Omega(\vartheta_0) \, H_{A}^{\top,-1}(\vartheta_0)\bigr)_{1,1} + o_P(1)
}
} \\
&\xrightarrow{d} N(0,1).
\end{align*}

Similarly,
\[
n \cdot \frac{\hat{\gamma}_k - \gamma_{k0}}{\sqrt{\hat{V}_{1+k,1+k}}} \xrightarrow{d} N(0,1).
\]

\textbf{Case B:} To simplify the notation, for a matrix $M$, we use $M_{1,2}$ for $M_{1,2:(q+1)}$, $M_{2,1}$ for $M_{2:(q+1),1}$ and $M_{2,2}$ for $M_{2:(q+1),2:(q+1)}$ in the following. As in \textbf{Case A}, now we have  
\begin{align*}
&n^{4a}H_{n,1,1}(\hat{\vartheta}_{n}) = H_{B,1,1}(\vartheta_0) + o_P(1), \quad n^{6a + b} \Omega_{n,1,1}(\hat{\vartheta}_{n}) = \Omega_{1,1}(\vartheta_0) + o_P(1),\\
&n^{4a}H_{n,1,2}(\hat{\vartheta}_{n}) = H_{B,1,2}(\vartheta_0) + o_P(1), \quad n^{6a - b} \Omega_{n,1,2}(\hat{\vartheta}_{n}) = \Omega_{1,2}(\vartheta_0) + o_P(1), \\
& n^{4a - 2b}H_{n,2,2}(\hat{\vartheta}_{n}) = H_{B,2,2}(\vartheta_0) + o_P(1),\quad n^{6a - 3b} \Omega_{n,2,2}(\hat{\vartheta}_{n}) = \Omega_{2,2}(\vartheta_0) + o_P(1).
\end{align*}
In the following, we write $\hat{G}_n$ and $\hat{\Omega}_n$ instead of $H_n^{-1}(\hat{\vartheta}_n)$ and $\Omega_n(\hat{\vartheta}_n)$ for brevity. Notice that we use $G_{n,2,2}$ for the submatrix of $H_n^{-1}$ formed by its $2$ to $(q+1)$ rows and $2$ to $(q+1)$ columns, whereas $H_{n,2,2}^{-1}$ represents the inverse of $H_{n,2,2}$.

We define $G_B$ as the inverse of the matrix $H_B$. Since $H_B$ is a block upper diagonal matrix, we have
\[
G_{B,1,1} = H_{B,1,1}^{-1}, \quad
G_{B,2,2} = H_{B,2,2}^{-1}, \quad
G_{B,1,2} = H_{B,1,1}^{-1} H_{B,1,2} H_{B,2,2}^{-1}, \quad
G_{B,2,1} = 0.
\]
Using the Schur complement (Chapter 0.8.5 in \cite{horn2012matrix}),
\[
\hat{G}_{n,1,1} = (\hat{H}_{n,1,1} - \hat{H}_{n,1,2} \hat{H}_{n,2,2}^{-1} \hat{H}_{n,2,1})^{-1},
\]
then
\[
n^{-4a} \hat{G}_{n,1,1} = ( n^{4a} \hat{H}_{n,1,1} - n^{4a} \hat{H}_{n,1,2} \hat{H}_{n,2,2}^{-1} \hat{H}_{n,2,1})^{-1} = \left( H_{B,1,1} + o_P(1) \right)^{-1} = G_{B,1,1} + o_P(1).
\]
Similarly, we have
\[
n^{-4a + 2b} \hat{G}_{n,2,2} = G_{B,2,2} + o_P(1), \quad
n^{-4a + 2b} \hat{G}_{n,1,2} = G_{B,1,2} + o_P(1).
\]
Therefore, we have the following:
\begin{align*}
& n^{-2a + b} \hat{G}_{n,1,1} \hat{\Omega}_{n,1,1} \hat{G}_{n,1,1}^{\top}
= (n^{-4a} \hat{G}_{n,1,1})(n^{6a + b} \hat{\Omega}_{n,1,1})(n^{-4a} \hat{G}_{n,1,1}^{\top})
= G_{B,1,1} \Omega_{1,1} G_{B,1,1}^{\top} + o_P(1), \\
& n^{-2a + b} \hat{G}_{n,1,2} \hat{\Omega}_{n,2,2} \hat{G}_{n,1,2}^{\top}
= (n^{-4a + 2b} \hat{G}_{n,1,2})(n^{6a - 3b} \hat{\Omega}_{n,2,2})(n^{-4a + 2b} \hat{G}_{n,1,2}^{\top})
= G_{B,1,2} \Omega_{2,2} G_{B,1,2}^{\top} + o_P(1), \\
& n^{-2a + b} \hat{G}_{n,1,2} \hat{\Omega}_{n,2,1} \hat{G}_{n,1,1}^{\top}
= (n^{-4a + 2b} \hat{G}_{n,1,2})(n^{6a - b} \hat{\Omega}_{n,2,1})(n^{-4a} \hat{G}_{n,1,1}^{\top})
= G_{B,1,2} \Omega_{2,1} G_{B,1,1}^{\top} + o_P(1), \\
& n^{-2a + b} \hat{G}_{n,1,1} \hat{\Omega}_{n,1,2} \hat{G}_{n,1,2}^{\top}
= (n^{-4a} \hat{G}_{n,1,1})(n^{6a - b} \hat{\Omega}_{n,1,2})(n^{-4a + 2b} \hat{G}_{n,1,2}^{\top})
= G_{B,1,1} \Omega_{1,2} G_{B,1,2}^{\top} + o_P(1).
\end{align*}
Using the identity
\[
\big(\hat{G}_n \hat{\Omega}_n \hat{G}_n^{\top}\big)_{1,1} = \hat{G}_{n,1,1} \hat{\Omega}_{n,1,1} \hat{G}_{n,1,1}^{\top} + \hat{G}_{n,1,2} \hat{\Omega}_{n,2,2} \hat{G}_{n,1,2}^{\top} + \hat{G}_{n,1,2} \hat{\Omega}_{n,2,1} \hat{G}_{n,1,1}^{\top} + \hat{G}_{n,1,1} \hat{\Omega}_{n,1,2} \hat{G}_{n,1,2}^{\top},
\]
we have
\[
n^{-2a + b} \big(\hat{G}_n \hat{\Omega}_n \hat{G}_n^{\top}\big)_{1,1} = (G_B \Omega G_B^{\top})_{1,1} + o_P(1).
\]
Therefore,
\begin{align*}
n \cdot \frac{\hat{\rho}_n - \rho_{n0}}{\sqrt{\hat{V}_{1,1}}}
&= \sqrt{n^{2 - 2a + b}} \cdot \frac{\hat{\rho} - \rho_0}{\sqrt{n^{-2a + b} \big(\hat{G}_n \hat{\Omega}_n \hat{G}_n^{\top}\big)_{1,1}}} \\
&= \sqrt{n^{2 - 2a + b}} \cdot \frac{\hat{\rho} - \rho_0}{\sqrt{(G_B \Omega G_B^{\top})_{1,1} + o_P(1)}} \\
&\xrightarrow{d} N(0,1).
\end{align*}

Similarly, we have
\begin{align*}
& n^{-2a + b} \hat{G}_{n,2,1} \hat{\Omega}_{n,1,1} \hat{G}_{n,2,1}^{\top}
= (n^{-4a} \hat{G}_{n,2,1})(n^{6a + b} \hat{\Omega}_{n,1,1})(n^{-4a} \hat{G}_{n,2,1}^{\top}) = o_P(1), \\
& n^{-2a + b} \hat{G}_{n,2,2} \hat{\Omega}_{n,2,1} \hat{G}_{n,2,1}^{\top}
= (n^{-4a + 2b} \hat{G}_{n,2,2})(n^{6a - b} \hat{\Omega}_{n,2,1})(n^{-4a} \hat{G}_{n,2,1}^{\top}) = o_P(1), \\
& n^{-2a + b} \hat{G}_{n,2,1} \hat{\Omega}_{n,1,2} \hat{G}_{n,2,2}^{\top}
= (n^{-4a + 2b} \hat{G}_{n,2,1})(n^{6a - b} \hat{\Omega}_{n,1,2})(n^{-4a} \hat{G}_{n,2,2}^{\top}) = o_P(1), \\
& n^{-2a + b} \hat{G}_{n,2,2} \hat{\Omega}_{n,2,2} \hat{G}_{n,2,2}^{\top}
= (n^{-4a + 2b} \hat{G}_{n,2,2})(n^{6a - 3b} \hat{\Omega}_{n,2,2})(n^{-4a + 2b} \hat{G}_{n,2,2}^{\top})
= G_{B,2,2} \Omega_{2,2} G_{B,2,2}^{\top} + o_P(1).
\end{align*}
Then,
\[
n^{-2a + b} \big(\hat{G}_n \hat{\Omega}_n \hat{G}_n^{\top}\big)_{2,2} = G_{B,2,2} \Omega_{2,2} G_{B,2,2}^{\top} + o_P(1) = (G_B \Omega G_B^{\top})_{2,2} + o_P(1).
\]
Therefore,
\begin{align*}
n \cdot \frac{\hat{\gamma}_k - \gamma_{k0}}{\sqrt{\hat{V}_{1+k,1+k}}}
&= \sqrt{n^{2 - 2a + b}} \cdot \frac{\hat{\gamma}_k - \gamma_{k0}}{\sqrt{n^{-2a + b} \big(\hat{G}_n \hat{\Omega}_n \hat{G}_n^{\top}\big)_{1+k,1+k}}} \\
&= \sqrt{n^{2 - 2a + b}} \cdot \frac{\hat{\gamma}_k - \gamma_{k0}}{\sqrt{(G_B \Omega G_B^{\top})_{1+k,1+k} + o_P(1)}} \\
&\xrightarrow{d} N(0,1).
\end{align*}

\textbf{Case C:} Using similar argument as in \textbf{Case A} and \textbf{B}, in this case, we have 
\[
n^{4a}H_{n, 1,1}(\hat{\vartheta}_{n})=H_{C,1,1}(\vartheta_{0}) + o_{P}(1) \quad \text{and} \quad n^{4a-2b}H_{n, 2,2}(\hat{\vartheta}_{n})=H_{C,2,2}(\vartheta_{0}) + o_{P}(1).
\] 
Also, 
\begin{align*}
n^{7a}\Omega_{n, 1,1}(\hat{\vartheta}_{n})&=\Omega_{1,1}(\vartheta_{0})+o_{P}(1), \quad n^{6a-b}\Omega_{n, 1,2}(\hat{\vartheta}_{n})=\Omega_{1,2}(\vartheta_{0})+o_{P}(1), \\
&n^{6a-3b}\Omega_{n, 2,2}(\hat{\vartheta}_{n})=\Omega_{2,2}(\vartheta_{0})+o_{P}(1).
\end{align*}
Since $H_{C}$ is a block diagonal matrix, we have
\[
G_{B,1,1} = H^{-1}_{B,1,1}, \quad G_{B,2,2} = H^{-1}_{B,2,2}, \quad \text{and} \quad G_{B,1,2}=G^{\top}_{B,2,1} = 0.
\]
Using the Schur complement, 
\[
n^{-4a}\hat{G}_{n,1,1} = G_{C,1,1}+o_{P}(1).
\]
Similarly, 
\[
n^{-4a+2b}\hat{G}_{n,2,2} = G_{C,2,2}+o_{P}(1), \quad \text{and} \quad n^{-4a+2b}\hat{G}_{n,1,2} = O_{P}(1).
\]
Then,
\begin{align*}
& n^{-a}\hat{G}_{n,1,1}\hat{\Omega}_{n,1,1}\hat{G}^{\top}_{n,1,1} 
= (n^{-4a}\hat{G}_{n,1,1}) (n^{7a}\hat{\Omega}_{n,1,1})(n^{-4a}\hat{G}^{\top}_{n,1,1}) 
= G_{C,1,1}\Omega_{1,1}G^{\top}_{C,1,1}+o_{P}(1), \\
& n^{-a}\hat{G}_{n,1,2}\hat{\Omega}_{n,2,2}\hat{G}^{\top}_{n,1,2} 
= (n^{-4a+2b}\hat{G}_{n,1,2})(n^{(6a-3b)+(a-b)}\hat{\Omega}_{n, 2,2})(n^{-4a+2b}\hat{G}^{\top}_{n,1,2})
= o_{P}(1), \\
& n^{-a}\hat{G}_{n,1,2}\hat{\Omega}_{n,2,1}\hat{G}^{\top}_{n,1,1} 
= (n^{-4a+2b}\hat{G}_{n,1,2})(n^{(6a-b)+(a-b)}\hat{\Omega}_{n, 2,1})(n^{-4a}\hat{G}^{\top}_{n,1,1}) 
= o_{P}(1), \\
& n^{-a}\hat{G}_{n,1,1}\hat{\Omega}_{n,1,2}\hat{G}^{\top}_{n,1,2} 
= (n^{-4a}\hat{G}_{n,1,1})(n^{(6a-b)+(a-b)}\hat{\Omega}_{n, 1,2})(n^{-4a+2b}\hat{G}^{\top}_{n,1,2}) 
= o_{P}(1).
\end{align*}
Thus, 
\[
n^{-a}\big(\hat{G}_{n}\hat{\Omega}_{n}\hat{G}^{\top}_{n}\big)_{1,1} = (G_{C}\Omega G_{C}^{\top})_{1,1}+o_{P}(1).
\]
Therefore,
\begin{align*}
n \cdot \frac{(\hat{\rho}_n-\rho_{n0})}{\sqrt{\hat{V}_{1,1}}}
&= \sqrt{n^{2-a}} \cdot \frac{(\hat{\rho}-\rho_{0})}{\sqrt{n^{-a}\big(\hat{G}_{n}\hat{\Omega}_{n}\hat{G}^{\top}_{n}\big)_{1,1} }} \\
&= \sqrt{n^{2-a}} \cdot \frac{(\hat{\rho}-\rho_{0})}{\sqrt{(G_C\Omega G_{C}^{\top})_{1,1}+o_{P}(1)}} \\
&\overset{d}{\longrightarrow} N(0, 1).
\end{align*}

Similarly, 
\begin{align*}
& n^{-2a+b}\hat{G}_{n,2,1}\hat{\Omega}_{n,1,1}\hat{G}^{\top}_{n,2,1} 
= (n^{-4a+2b}\hat{G}_{n,2,1})(n^{6a-3b}\hat{\Omega}_{n, 1,1})(n^{-4a+2b}\hat{G}^{\top}_{n,2,1}) 
= o_{P}(1), \\
& n^{-2a+b}\hat{G}_{n,2,2}\hat{\Omega}_{n,2,1}\hat{G}^{\top}_{n,2,1} 
= (n^{-4a+2b}\hat{G}_{n,2,2})(n^{6a-b}\hat{\Omega}_{n, 2,1})(n^{-4a}\hat{G}^{\top}_{n,2,1}) 
= o_{P}(1), \\
& n^{-2a+b}\hat{G}_{n,2,1}\hat{\Omega}_{n,1,2}\hat{G}^{\top}_{n,2,2} 
= (n^{-4a+2b}\hat{G}_{n,2,1})(n^{6a-b}\hat{\Omega}_{n, 1,2})(n^{-4a}\hat{G}^{\top}_{n,2,2}) 
= o_{P}(1), \\
& n^{-2a+b}\hat{G}_{n,2,2}\hat{\Omega}_{n,2,2}\hat{G}^{\top}_{n,2,2} 
= (n^{-4a+2b}\hat{G}_{n,2,2})(n^{6a-3b}\hat{\Omega}_{n, 2,2})(n^{-4a+2b}\hat{G}^{\top}_{n,2,2}) 
= G_{C,2,2}\Omega_{2,2}G^{\top}_{C,2,2}+o_{P}(1).
\end{align*}
Then, 
\[
n^{-2a+b}\big(\hat{G}_{n}\hat{\Omega}_{n}\hat{G}^{\top}_{n}\big)_{2,2} = G_{C,2,2}\Omega_{2,2}G^{\top}_{C,2,2}+o_{P}(1) = (G_{C}\Omega G_{C}^{\top})_{2,2}+o_{P}(1).
\]
Therefore,
\begin{align*}
n \cdot \frac{(\hat{\gamma}_{k}-\gamma_{k0})}{\sqrt{\hat{V}_{1+k,1+k}}}
&= \sqrt{n^{2-2a+b}} \cdot \frac{(\hat{\gamma}_{k}-\gamma_{k0})}{\sqrt{n^{-2a+b}\big(\hat{G}_{n}\hat{\Omega}_{n}\hat{G}^{\top}_{n}\big)_{1+k,1+k} }} \\
&= \sqrt{n^{2-2a+b}} \cdot \frac{(\hat{\rho}-\rho_{0})}{\sqrt{(G_{C}\Omega G_{C}^{\top})_{1+k,1+k}+o_{P}(1)}} \\
&\overset{d}{\longrightarrow} N(0, 1).
\end{align*}
\end{proof}

\begin{proof}[Proof of Proposition \ref{p_one}]
Without covariates, the negative log-likelihood function is 
\[
\mathcal{L}_{n}(\rho_n) = \frac{1}{n^4}\sum_{(i,j,k,l)}\ell_{ijkl}(\rho_n),
\]
where
\begin{align*}
\ell_{ijkl}(\rho_n) 
=~& \mathbb{I}\left(S_{ijkl} = 0 \text{ or } \pm 1\right)\log(1+\exp(2\rho_n)) 
- \mathbb{I}\left(S_{ijkl} = \pm 1\right)\rho_n.
\end{align*}
Then, the first-order condition is
\[
\frac{1}{n^4}\sum_{(i,j,k,l)}\left\{
\frac{2\mathbb{I}\left(S_{ijkl} = 0 \text{ or } \pm 1\right)\exp(2\hat{\rho}_n)}{1+\exp(2\hat{\rho}_n)} 
- 2\mathbb{I}\left(S_{ijkl} = \pm 1\right)
\right\} = 0,
\]
from which we get
\[
\hat{\rho}_n = \frac{1}{2}\log\left(
\frac{\sum_{(i,j,k,l)}\mathbb{I}(S_{ijkl} = 1)}
     {\sum_{(i,j,k,l)}\mathbb{I}(S_{ijkl} = 0)}
\right).
\]
\end{proof}

\begin{Lem}[Le Cam's two-point method (Section 15.2 in \cite{wainwright2019high})]\label{LeCam}
Let $\mathcal{P}$ be a model, let $\theta: \mathcal{P} \rightarrow \Theta$ be a parameter of interest, and let $L: \Theta \times \Theta \rightarrow [0, \infty)$ be a loss function satisfying:
\begin{itemize}
  \item[(i)] $L(\theta_1, \theta_2) = L(\theta_2, \theta_1)$ for all $\theta_1, \theta_2 \in \Theta$;
  \item[(ii)] $L(\theta_1, \theta_2) + L(\theta_2, \theta_3) \geq A L(\theta_1, \theta_3)$ for all $\theta_1, \theta_2, \theta_3 \in \Theta$.
\end{itemize}
Suppose that there exist $P_1, P_2 \in \mathcal{P}$ such that $L(\theta(P_1), \theta(P_2)) \geq \delta > 0$ and $\mathrm{TV}(P_1, P_2) \leq C$. Then, for any estimator $\hat{\theta}$, we must have
\[
\sup_{P \in \mathcal{P}} \mathbb{E}_P L(\hat{\theta}, \theta(P)) \geq \frac{A \delta}{2}(1 - C),
\]
where we write $\mathbb{E}_P$ for the expectation when $X \sim P$.
\end{Lem}

\begin{proof}[Proof of Theorem \ref{Rate}]
For the lower bound of $\hat{\rho}_n$, we consider the following two distributions:

(1) $P_1$: $\mu_{n1} = -a \log n$, $\rho_{n1} = b \log n$, $\alpha_i = \beta_i = 0$ for all $i$ and $\gamma = 0$;

(2) $P_2$: $\mu_{n2} = -a \log n + \epsilon_1$, $\rho_{n2} = b \log n + \epsilon_2$, $\alpha_i = \beta_i = 0$ for all $i$ and $\gamma = 0$, where $\epsilon_1$ and $\epsilon_2$ are to be specified.

The KL divergence between $P_1$ and $P_2$ is
\begin{align*}
\frac{1}{\binom{n}{2}}\text{KL}(P_1, P_2) 
=& \frac{1}{k(\mu_{n1}, \rho_{n1})} \log\left(\frac{k(\mu_{n2}, \rho_{n2})}{k(\mu_{n1}, \rho_{n1})} \right) + \frac{2 \exp(\mu_{n1})}{k(\mu_{n1}, \rho_{n1})} \log\left(\frac{k(\mu_{n2}, \rho_{n2})}{k(\mu_{n1}, \rho_{n1})} \cdot \exp(-\epsilon_1) \right) \\
& + \frac{ \exp(\mu_{n1} + 2\rho_{n1})}{k(\mu_{n1}, \rho_{n1})} \log\left(\frac{k(\mu_{n2}, \rho_{n2})}{k(\mu_{n1}, \rho_{n1})} \cdot \exp(-2\epsilon_1 - \epsilon_2) \right) \\
=& \log\left(\frac{k(\mu_{n2}, \rho_{n2})}{k(\mu_{n1}, \rho_{n1})} \right) - \frac{ \exp(\mu_{n1})}{k(\mu_{n1}, \rho_{n1})} \epsilon_1 - \frac{ \exp(\mu_{n1} + 2\rho_{n1})}{k(\mu_{n1}, \rho_{n1})} (2\epsilon_1 + \epsilon_2),
\end{align*}
where $k(\mu_n, \rho_n) = 1 + 2\exp(\mu_n) + \exp(2\mu_n + \rho_n)$.

Next, we bound the first term:
\begin{align*}
\log\left(\frac{k(\mu_{n2}, \rho_{n2})}{k(\mu_{n1}, \rho_{n1})} \right) 
&= \log\left(1 + \frac{2 \exp(\mu_{n1})(\exp(\epsilon_1) - 1) + \exp(\mu_{n1} + 2\rho_{n1})(\exp(2\epsilon_1 + \epsilon_2) - 1)}{k(\mu_{n1}, \rho_{n1})} \right) \\
&\leq \frac{2 \exp(\mu_{n1})}{k(\mu_{n1}, \rho_{n1})} (\epsilon_1 + \epsilon_1^2) 
+ \frac{ \exp(\mu_{n1} + 2\rho_{n1})}{k(\mu_{n1}, \rho_{n1})} (2\epsilon_1 + \epsilon_2 + (2\epsilon_1 + \epsilon_2)^2),
\end{align*}
since $\log(1 + x) \leq x$ and $\exp(x) \leq 1 + x + x^2$ for $x \leq 1$.

Therefore,
\[
\text{KL}(P_1, P_2) \leq n^{2 - a} \epsilon_1^2 + \frac{1}{2} n^{2 - 2a + b} (2\epsilon_1 + \epsilon_2)^2.
\]

If $a \geq b$, set $\epsilon_1 = 0$ and $\epsilon_2 = \sqrt{n^{-2 + 2a - b}}$. Then $\text{KL}(P_1, P_2) \leq 1/2$. By Pinsker's inequality (Lemma 2.5 in \cite{tsybakov2009nonparametric}), $\text{TV}(P_1, P_2) \leq \sqrt{\text{KL}(P_1, P_2)/2} \leq 1/2$. Letting $L(\theta_{n1}, \theta_{n2}) = (\rho_{n1} - \rho_{n2})^2$, we have $L(\theta_{n1}, \theta_{n2}) \geq n^{-2 + 2a - b}$, so by Lemma~\ref{LeCam},
\[
\sup_{\mathbb{P} \in \mathcal{P}} \mathbb{E}(\hat{\rho}_n - \rho_{n0})^2 \geq C n^{-2 + 2a - b}.
\]

If $a < b$, set $\epsilon_1 = \frac{1}{2} \sqrt{n^{-2 + a}}$ and $\epsilon_2 = -2 \epsilon_1$. Then again $\text{KL}(P_1, P_2) \leq 1/2$, and $L(\theta_{n1}, \theta_{n2}) = (\rho_{n1} - \rho_{n2})^2 \geq \frac{1}{4} n^{-2 + a}$. Hence,
\[
\sup_{\mathbb{P} \in \mathcal{P}} \mathbb{E}(\hat{\rho}_n - \rho_{n0})^2 \geq C n^{-2 + a}.
\]

Thus, in either case,
\[
\sup_{\mathbb{P} \in \mathcal{P}} \mathbb{E}(\hat{\rho}_n - \rho_{n0})^2 \geq C n^{-2 + 2a - \min\{a, b\}}.
\]

\vspace{1em}

For the lower bound of $\hat{\gamma}$, consider:

(1) $P_3$: $\mu_{n3} = -a \log n$, $\rho_{n3} = b \log n$, $\alpha_i = \beta_i = 0$ for all $i$ and $\gamma_3 = 0$;

(2) $P_4$: same as $P_3$ except $\gamma_4 = \epsilon_3 \mathbf{1}_d$.

The KL divergence is
\begin{align*}
\text{KL}(P_3, P_4) 
&= \sum_{i<j} \left\{
\log\left( \frac{k_{ij}(\theta_{n4})}{k_{ij}(\theta_{n3})} \right) 
- \frac{ \exp(\mu_{n3} + 2\rho_{n3})}{k_{ij}(\theta_{n3})} (V_{ij}^{\top} \mathbf{1}_p \epsilon_3)
\right\},
\end{align*}
where $k_{ij}(\theta_n) = 1 + 2\exp(\mu_n) + \exp(2\mu_n + \rho_n + V_{ij}^{\top} \gamma)$.

Using the same bounding argument and assuming $\|V_{ij}\|_2$ is uniformly bounded,
\[
\text{KL}(P_3, P_4) \leq \sum_{i<j} n^{-2a + b} (V_{ij}^{\top} \mathbf{1}_p \epsilon_3)^2 \leq C n^{2 - 2a + b} \epsilon_3^2.
\]

Set $\epsilon_3 = \sqrt{(2C)^{-1} n^{-2 + 2a - b}}$, and let $L(\theta_{n3}, \theta_{n4}) = \| \gamma_3 - \gamma_4 \|_2^2$. Then $L(\theta_{n3}, \theta_{n4}) \geq C n^{-2 + 2a - b}$, and Lemma~\ref{LeCam} implies
\[
\sup_{P \in \mathcal{P}} \mathbb{E}_P \| \hat{\gamma} - \gamma_0 \|_2^2 \gtrsim n^{-2 + 2a - b}.
\]
\end{proof}

\section*{Additional Simulation Results}

In Section~\ref{simulation}, we considered the case where $a > b > 0$. Here, we provide additional simulation results for other sparsity regimes, specifically $a > 0 > b$ and $b > a > 0$, while keeping all other parameters unchanged.

\begin{figure}[htbp]
\centering
\subfloat[Absolute error for $\hat{\rho}_n$]{
\includegraphics[width=.3\textwidth,height=3.5cm]{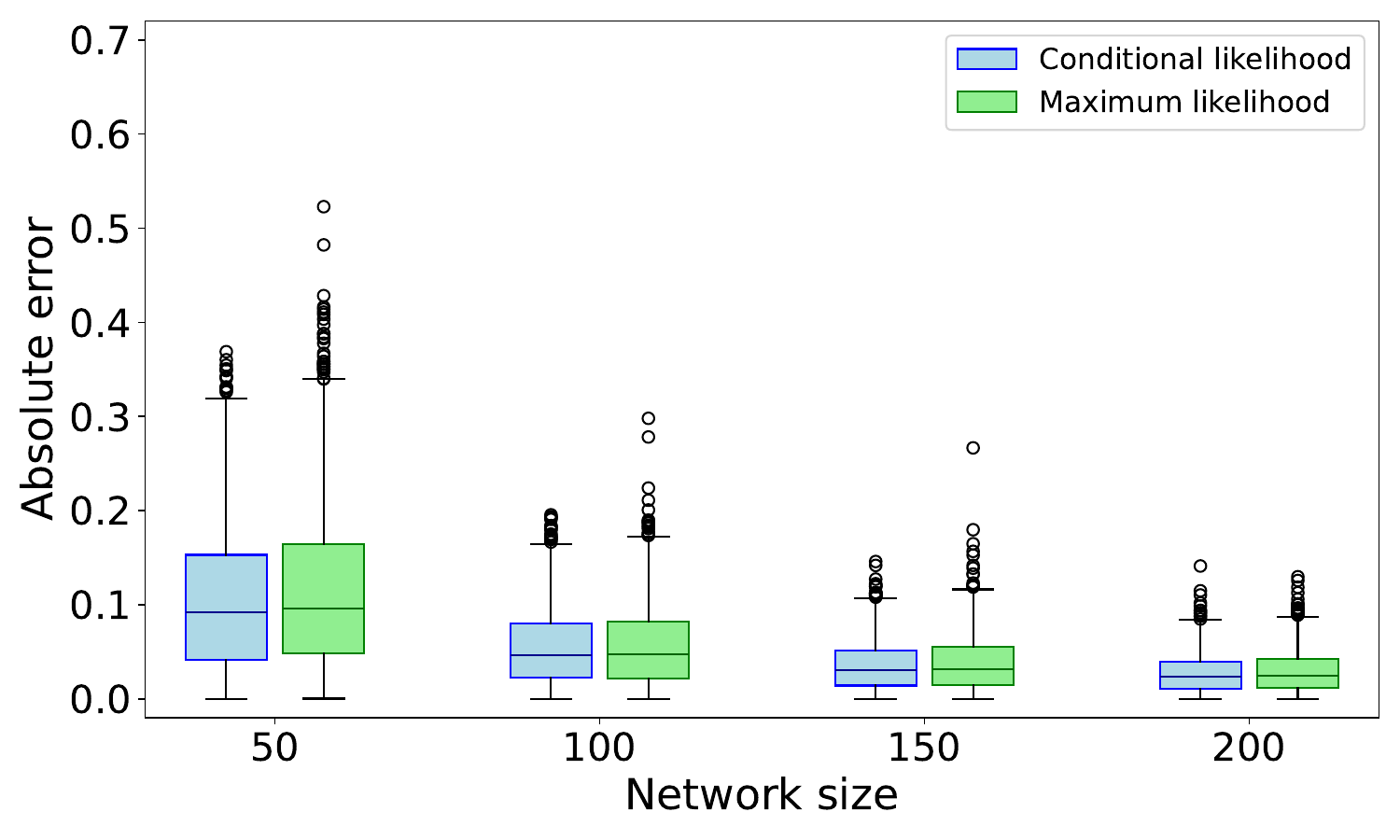}
}
\hfill
\subfloat[Absolute error for $\hat{\gamma}_1$]{
\includegraphics[width=.3\textwidth,height=3.5cm]{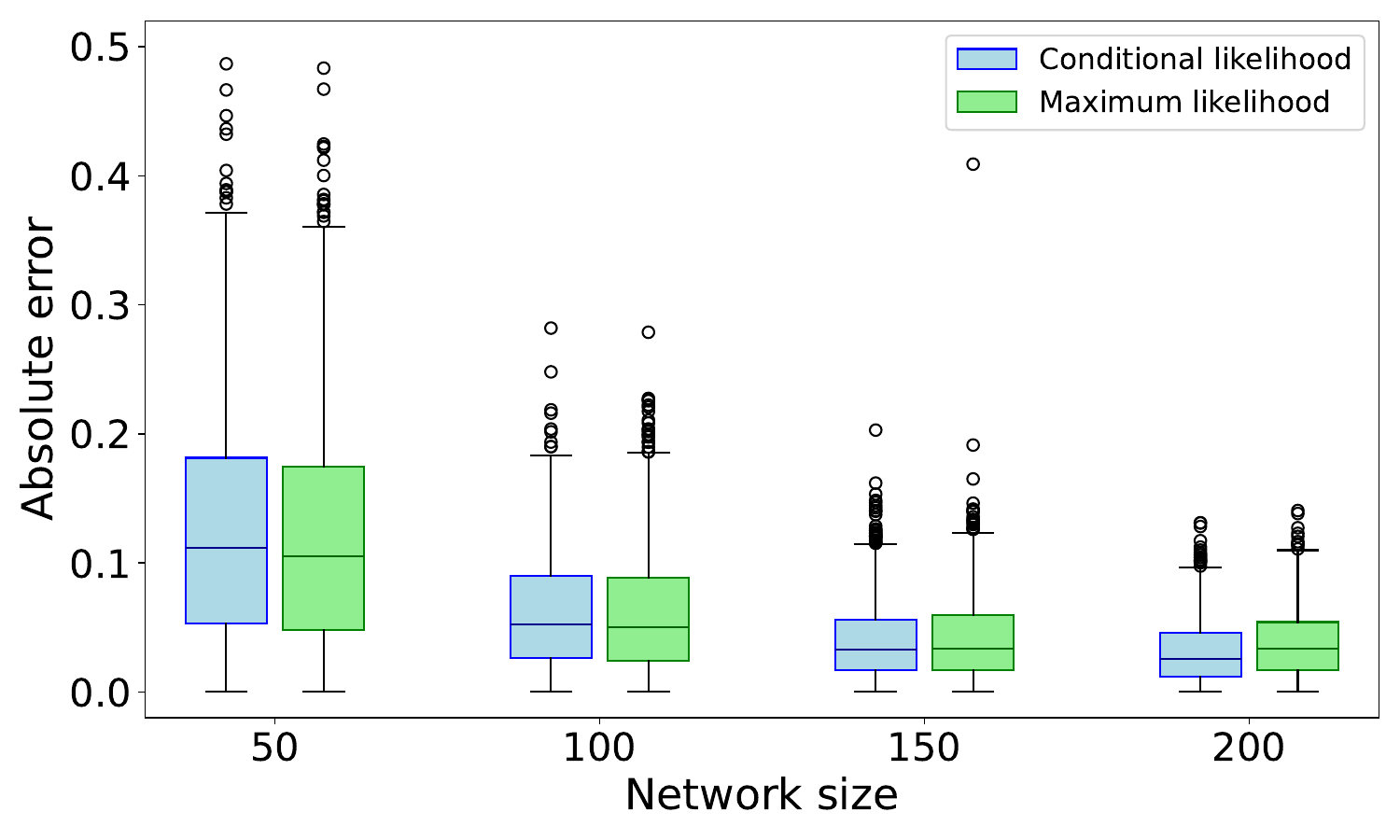}
}
\hfill
\subfloat[Absolute error for $\hat{\gamma}_2$]{
\includegraphics[width=.3\textwidth,height=3.5cm]{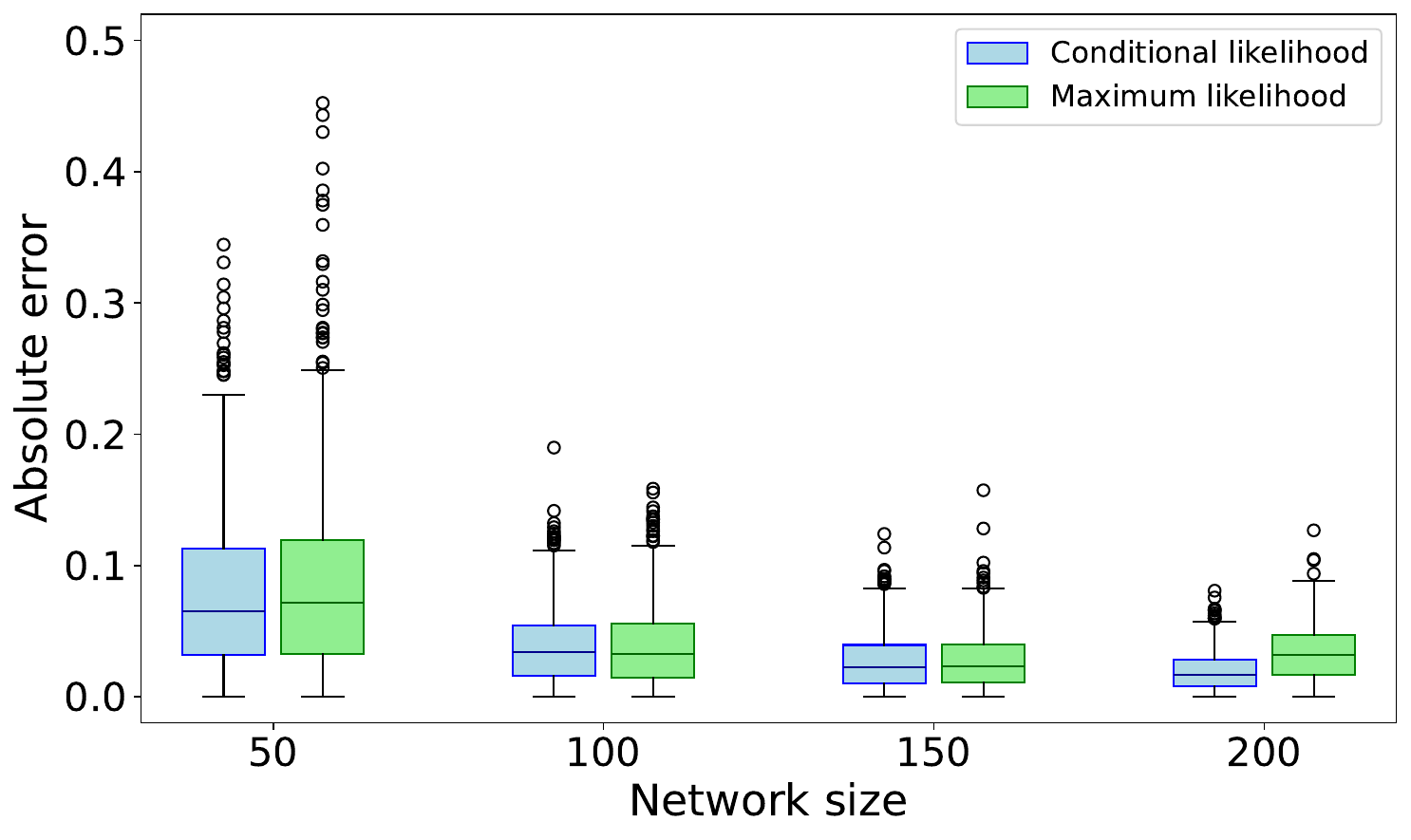}
}
\caption{Absolute estimation error across various network sizes for $a = 0.2$ and $b = -0.1$.}
\label{error_plot_2}
\end{figure}

\begin{figure}[htbp]
\centering
\subfloat[Normal QQ plot for $\hat{\rho}_n$]{
\includegraphics[width=.28\textwidth,height=4.5cm]{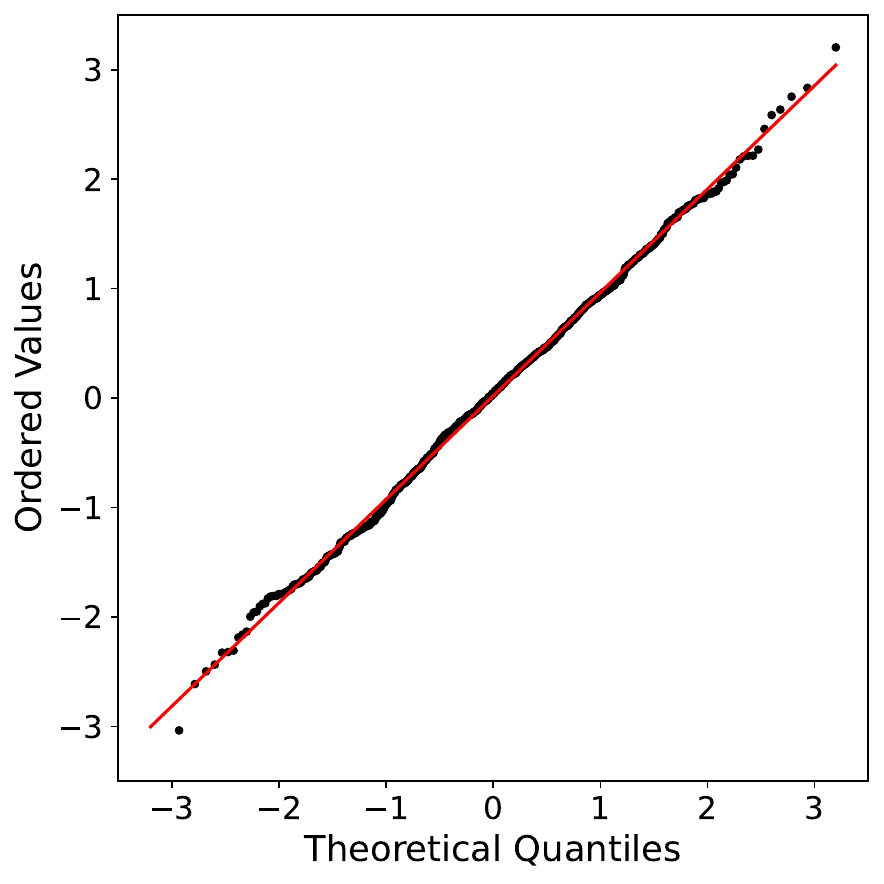}
}
\hfill
\subfloat[Normal QQ plot for $\hat{\gamma}_1$]{
\includegraphics[width=.28\textwidth,height=4.5cm]{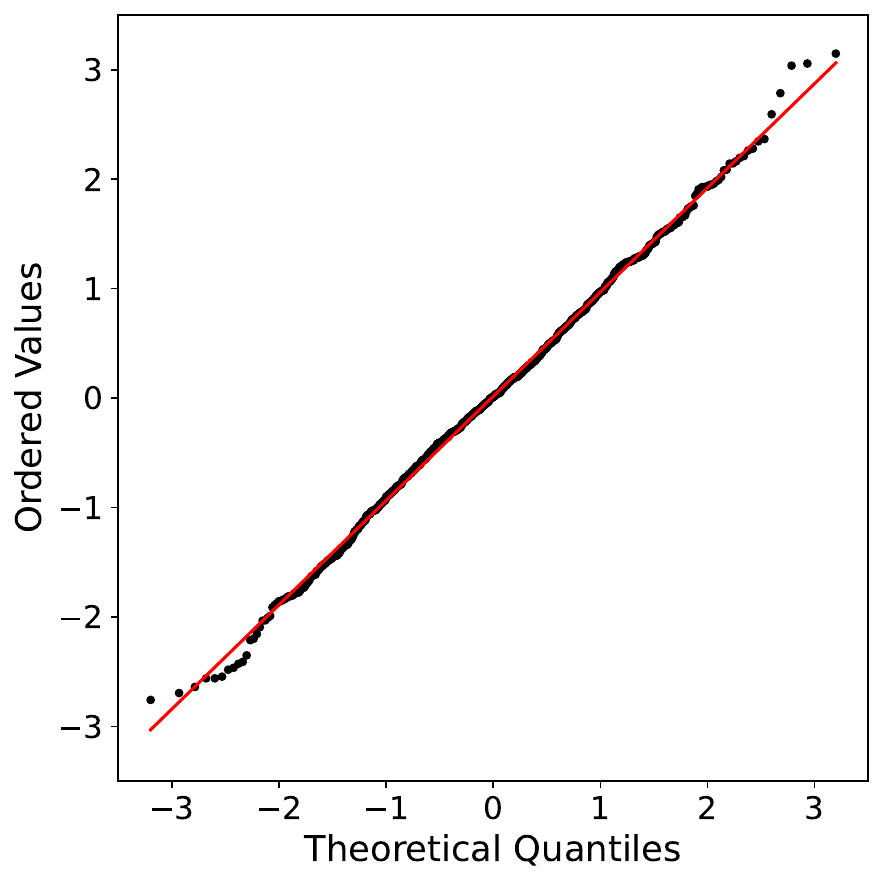}
}
\hfill
\subfloat[Normal QQ plot for $\hat{\gamma}_2$]{
\includegraphics[width=.28\textwidth,height=4.5cm]{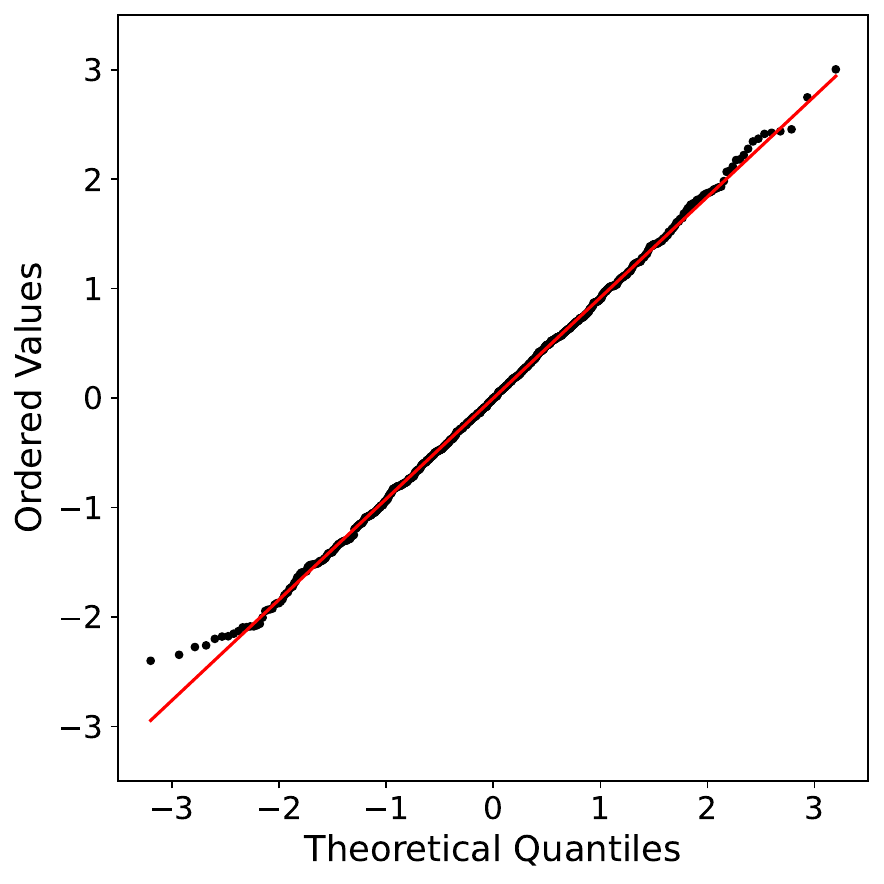}
}
\caption{QQ plots of standardized estimators for $n = 200$, $a = 0.2$, and $b = -0.1$.}
\label{qq2}
\end{figure}

\begin{figure}[htbp]
\centering
\subfloat[Absolute error for $\hat{\rho}_n$]{
\includegraphics[width=.3\textwidth,height=3.5cm]{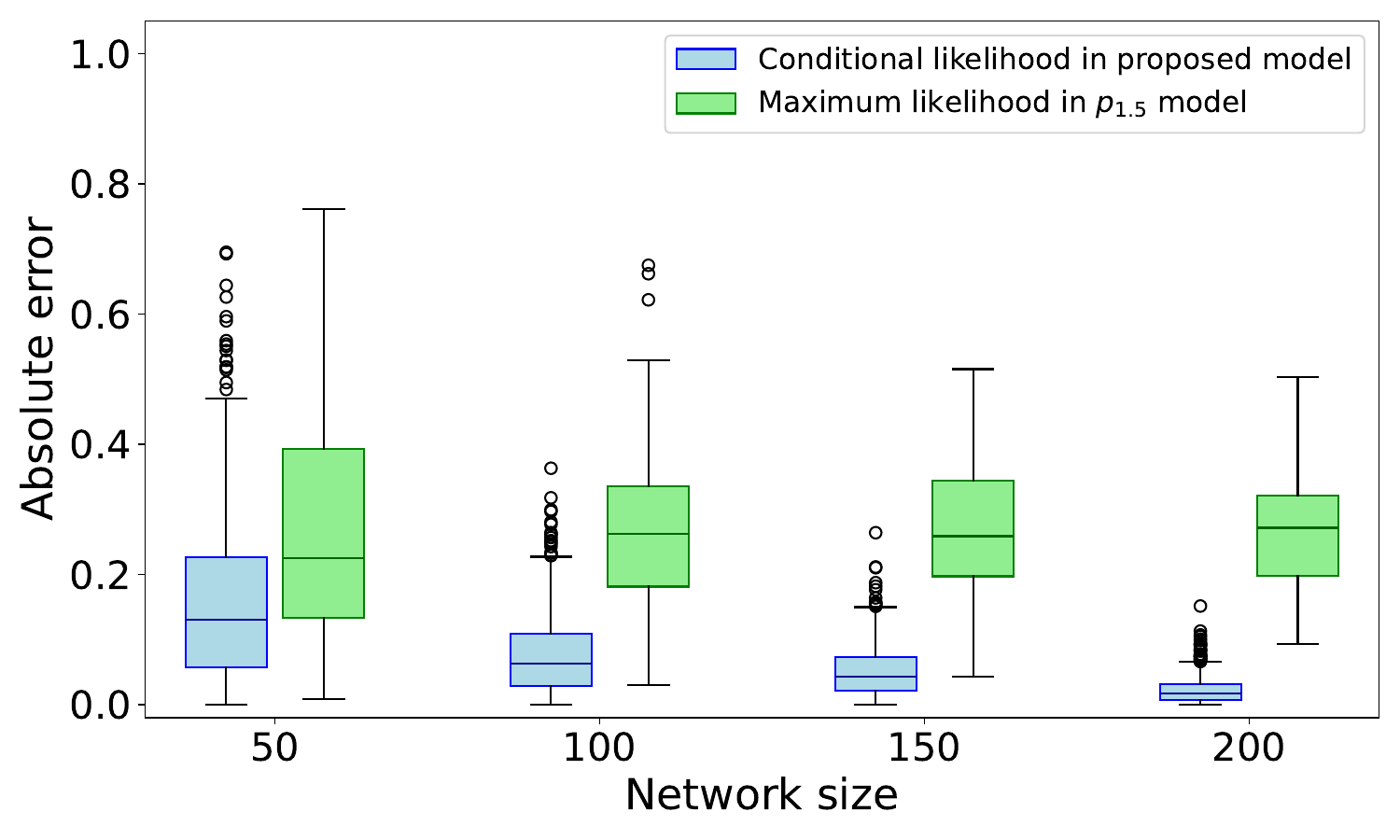}
}
\hfill
\subfloat[Absolute error for  $\hat{\gamma}_1$]{
\includegraphics[width=.3\textwidth,height=3.5cm]{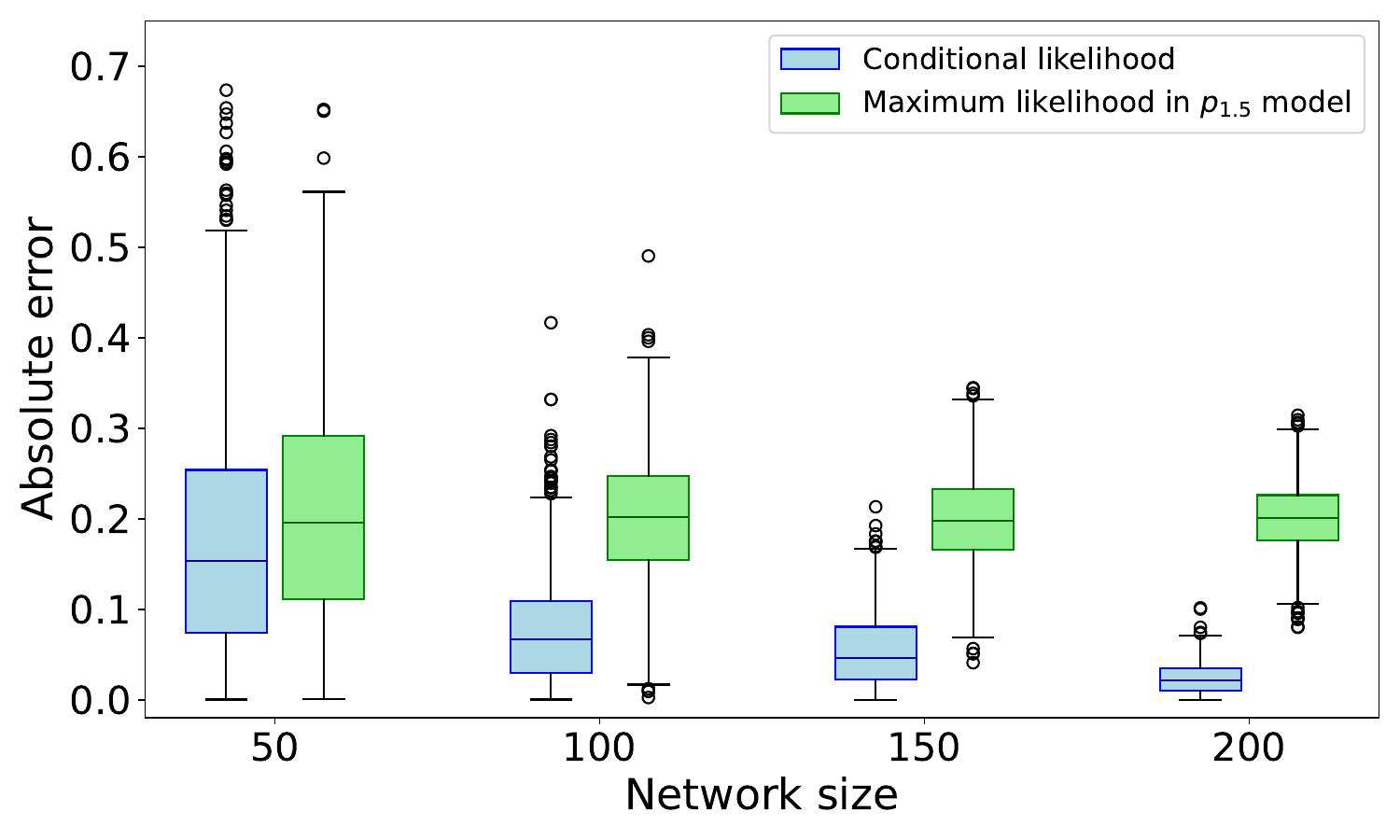}
}
\hfill
\subfloat[Absolute error for  $\hat{\gamma}_2$]{
\includegraphics[width=.3\textwidth,height=3.5cm]{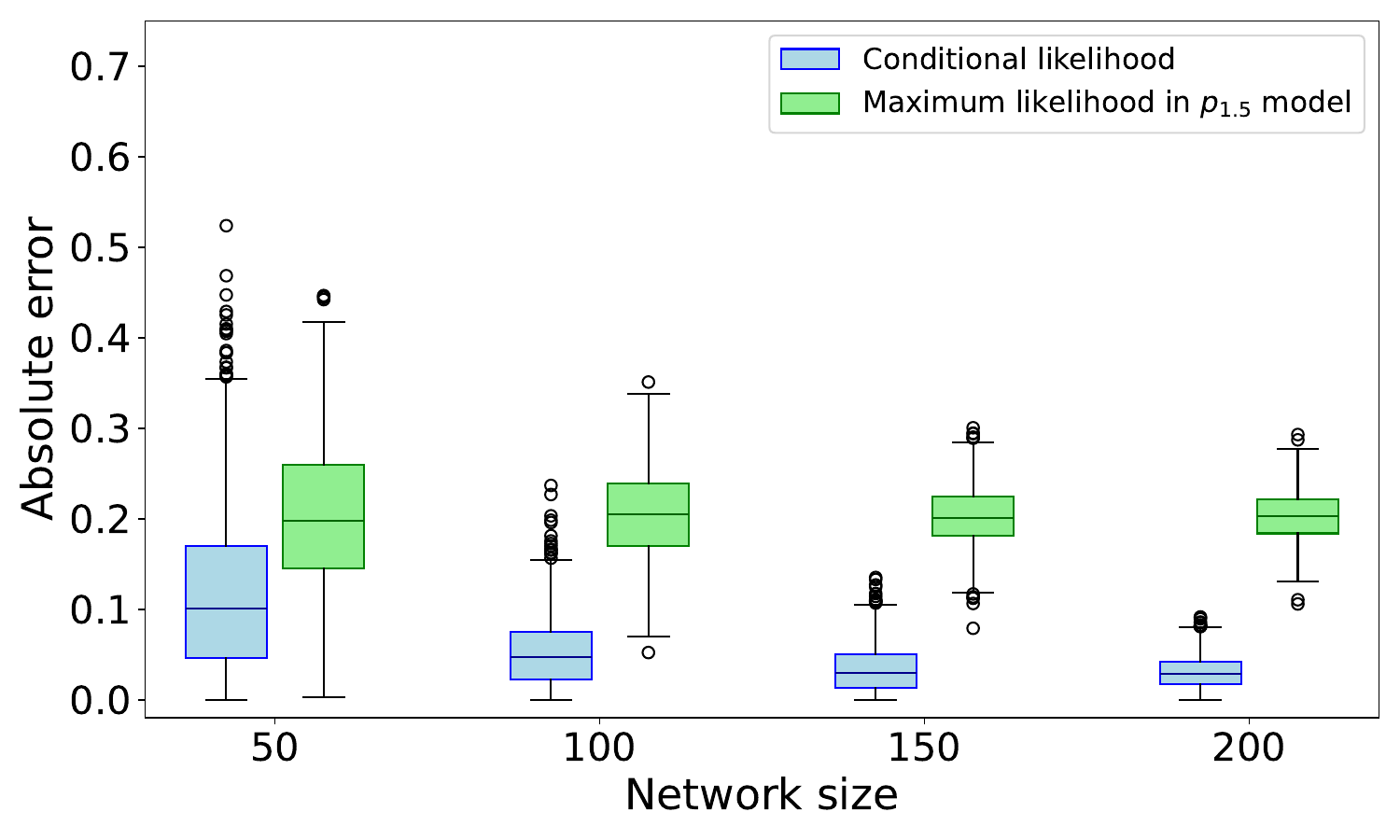}
}
\caption{Comparison with $p_{1.5}$ model across different network sizes when $a = 0.2$, and $b = -0.1$.}
\label{error_p1.5_2}
\end{figure}

\begin{figure}[htbp]
\centering
\subfloat[Absolute error for $\hat{\rho}_n$]{
\includegraphics[width=.3\textwidth,height=3.5cm]{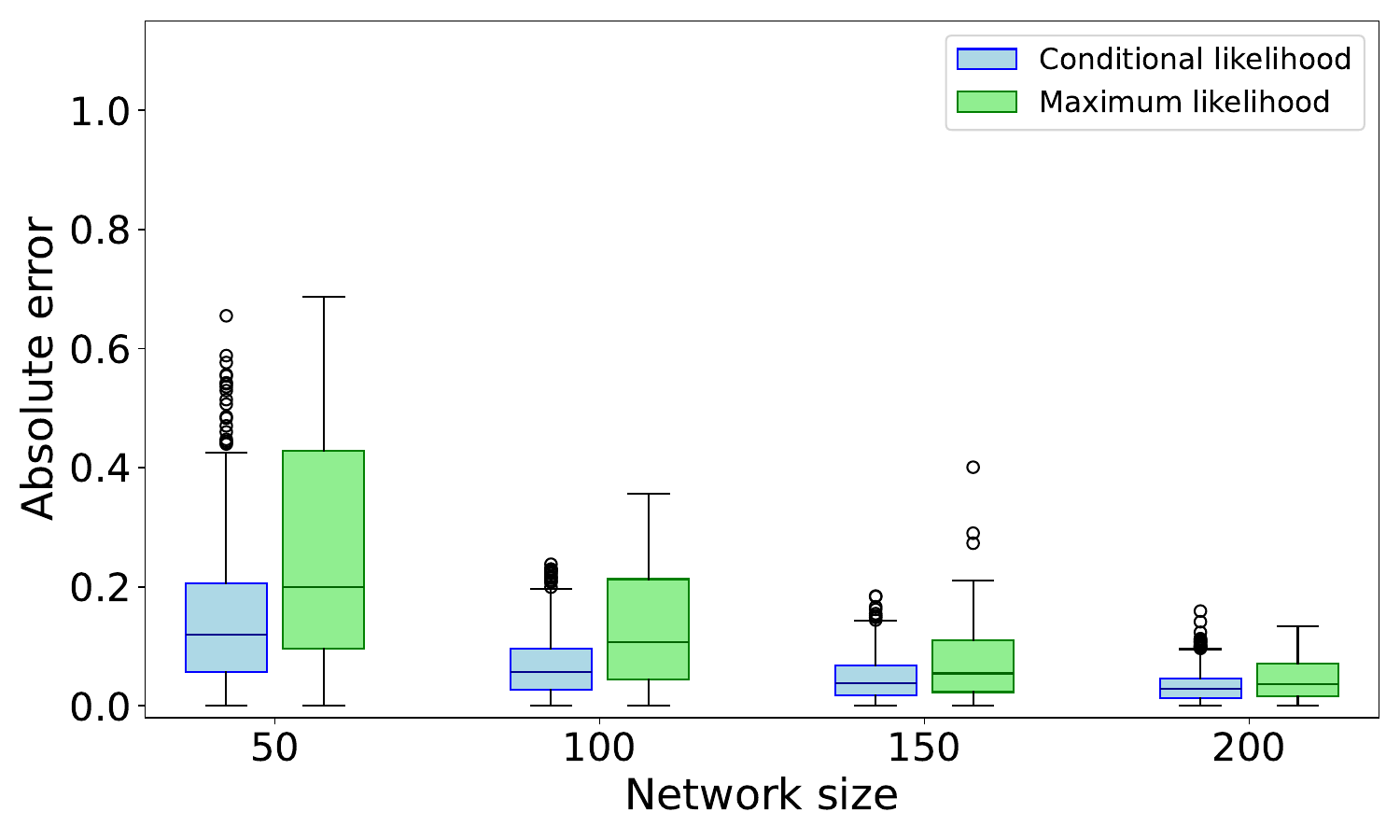}
}
\hfill
\subfloat[Absolute error for $\hat{\gamma}_1$]{
\includegraphics[width=.3\textwidth,height=3.5cm]{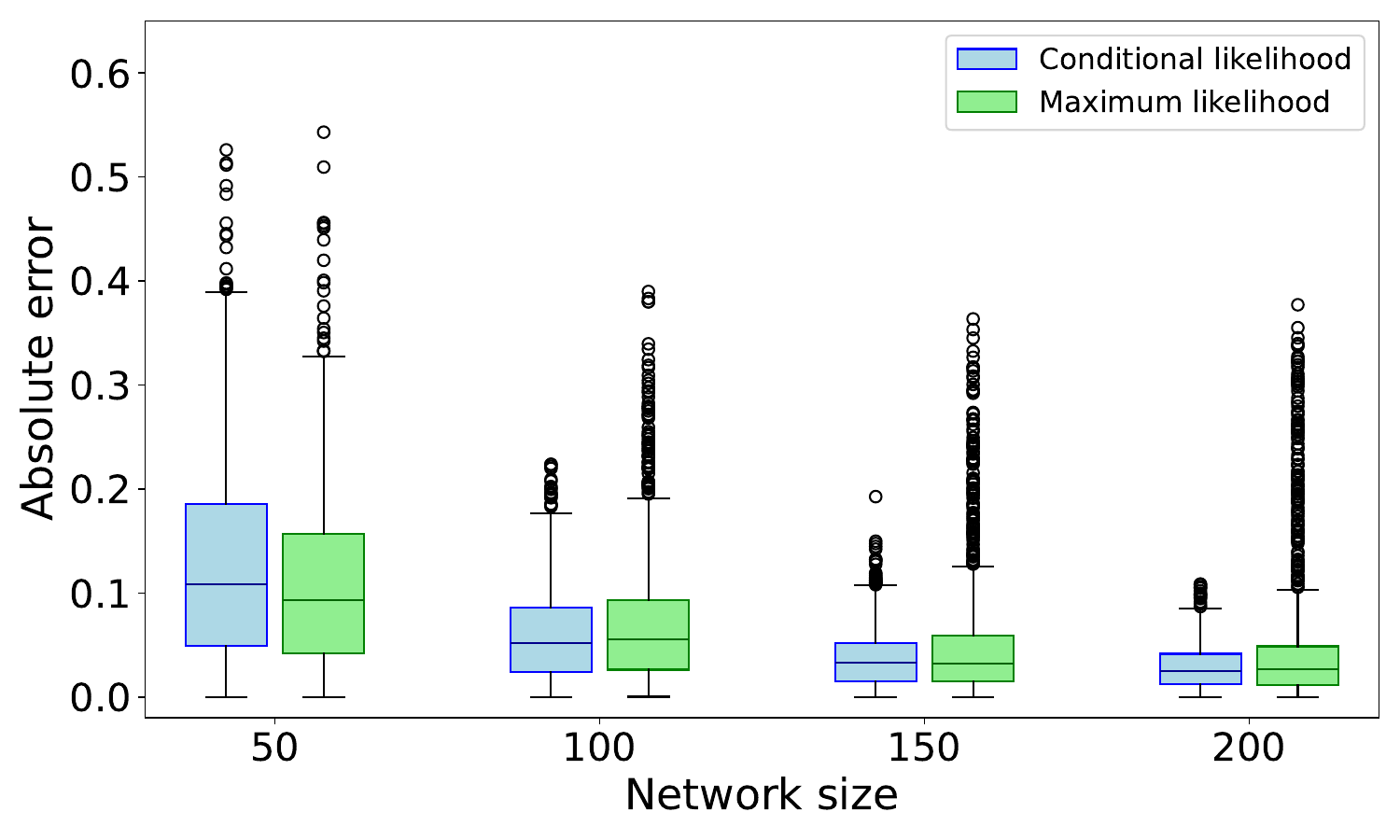}
}
\hfill
\subfloat[Absolute error for $\hat{\gamma}_2$]{
\includegraphics[width=.3\textwidth,height=3.5cm]{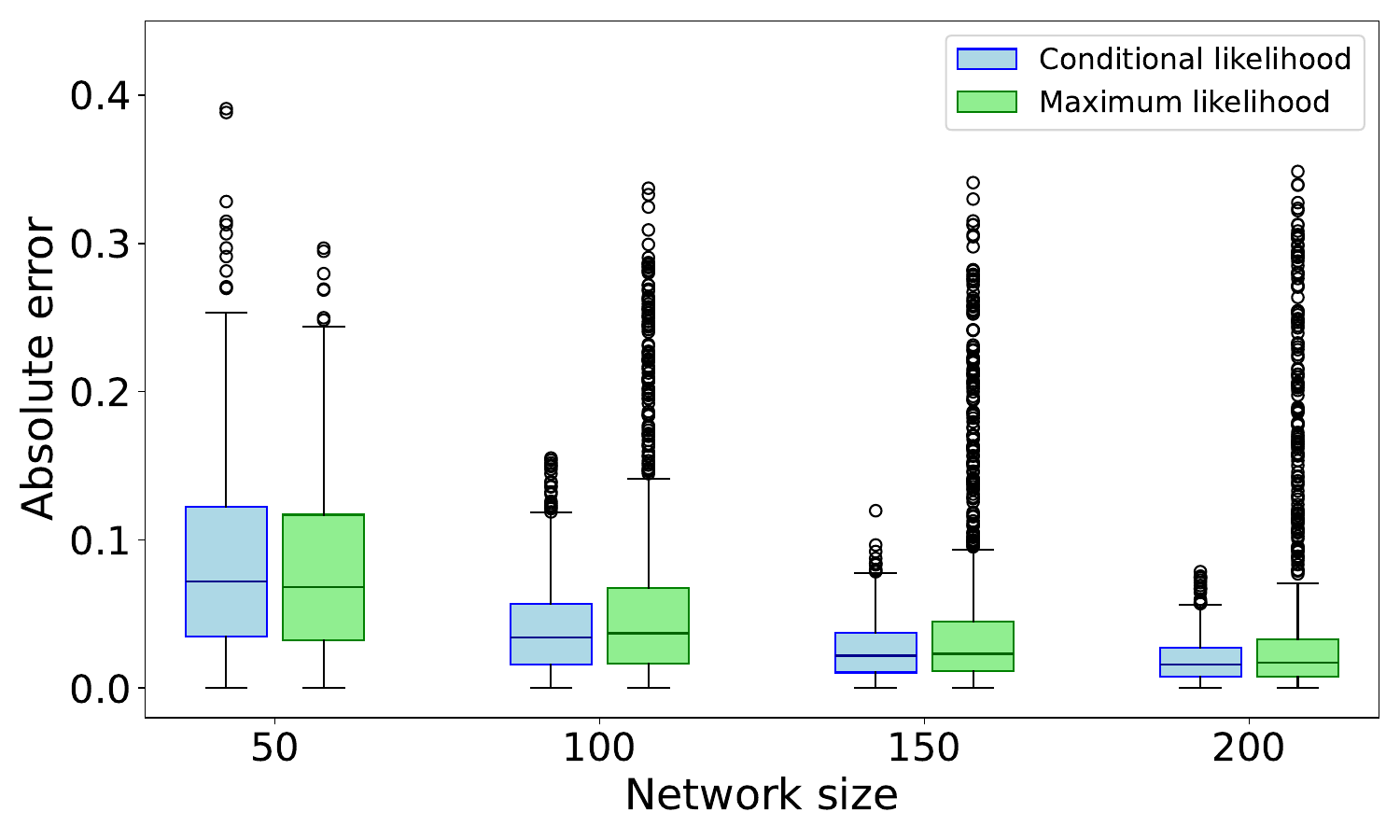}
}
\caption{Absolute estimation error across various network sizes for $a = 0.3$ and $b = 0.5$.}
\label{error_plot_3}
\end{figure}

\begin{figure}[htbp]
\centering
\subfloat[Normal QQ plot for $\hat{\rho}_n$]{
\includegraphics[width=.28\textwidth,height=4.5cm]{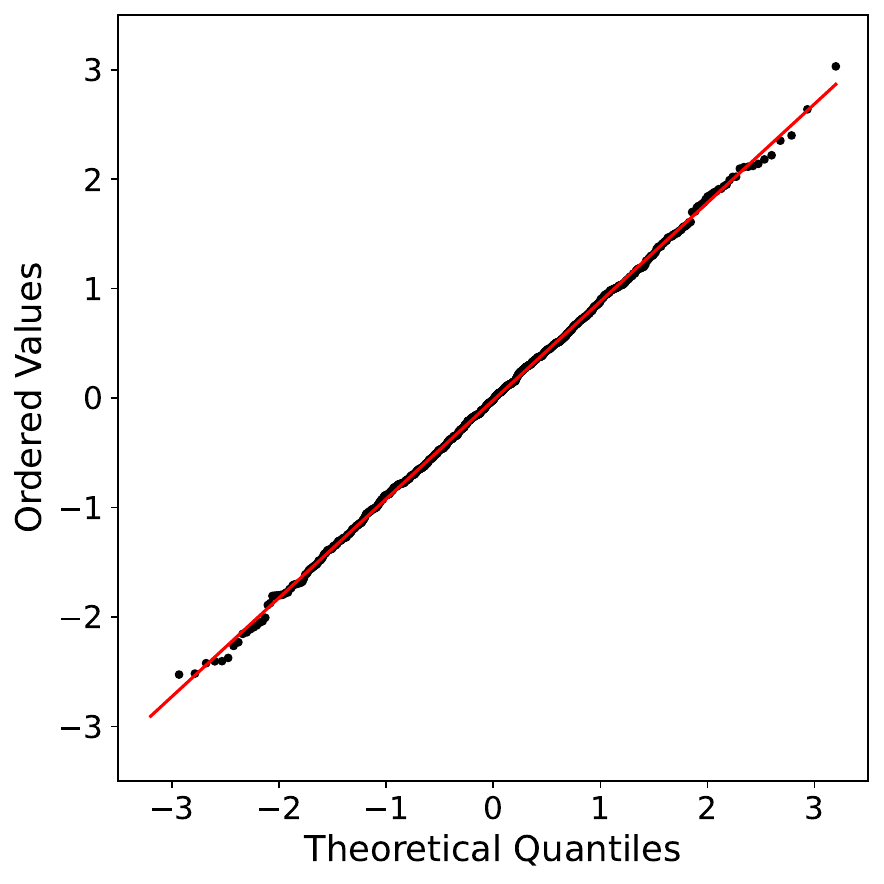}
}
\hfill
\subfloat[Normal QQ plot for $\hat{\gamma}_1$]{
\includegraphics[width=.28\textwidth,height=4.5cm]{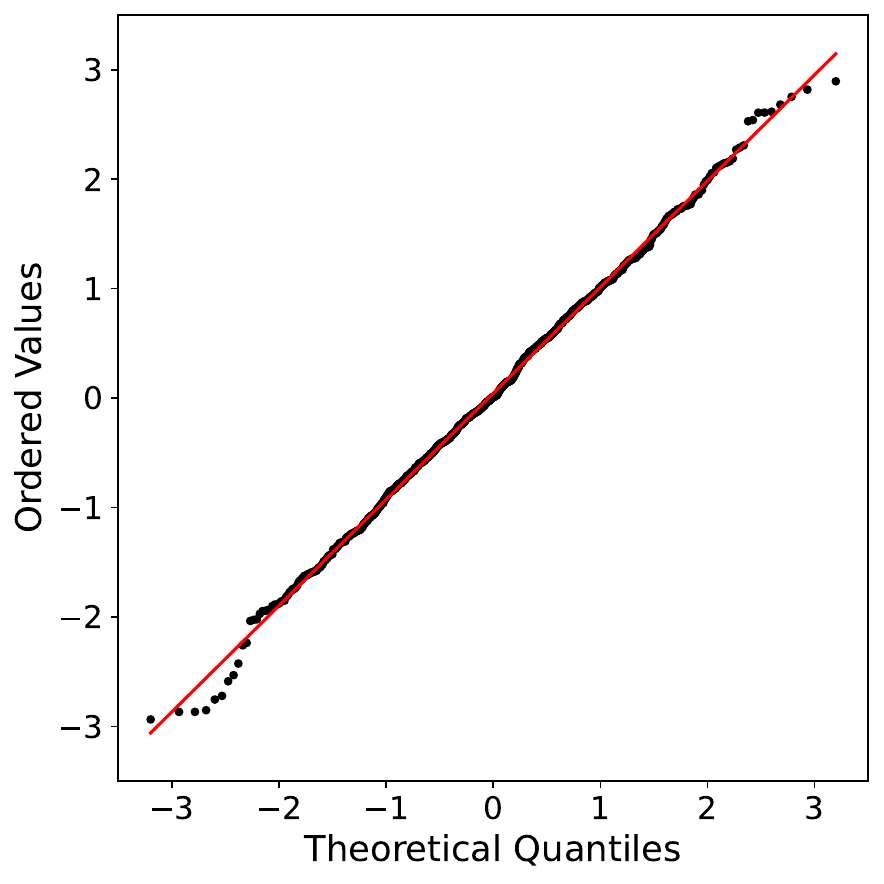}
}
\hfill
\subfloat[Normal QQ plot for $\hat{\gamma}_2$]{
\includegraphics[width=.28\textwidth,height=4.5cm]{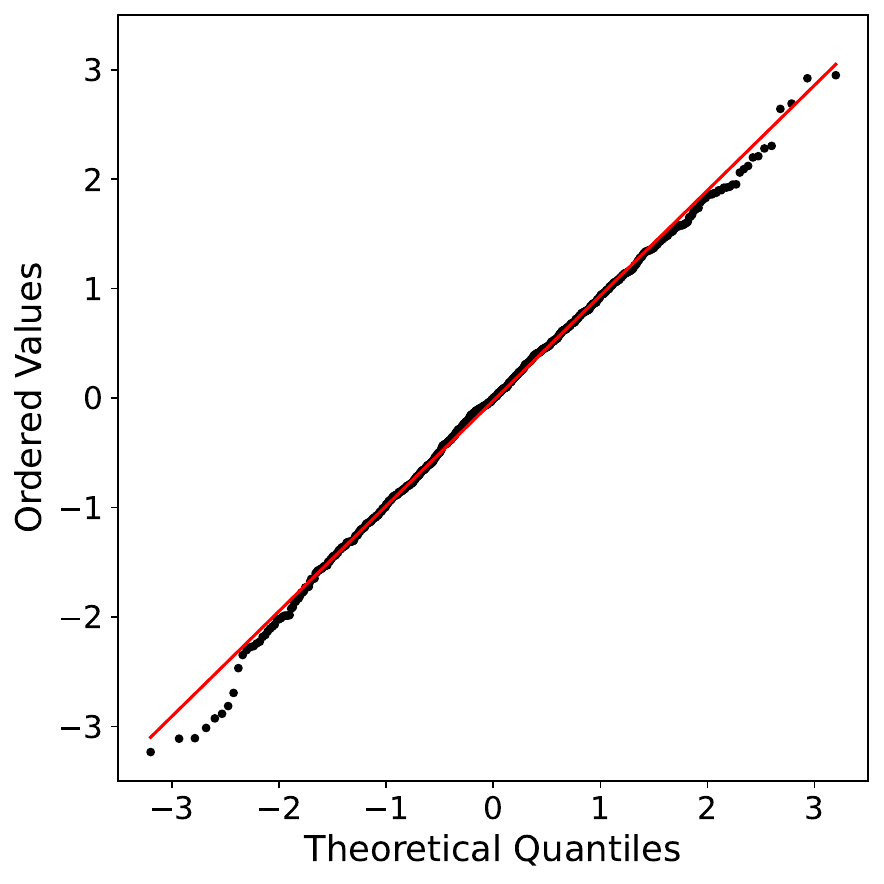}
}
\caption{QQ plots of standardized estimators for $n = 200$, $a = 0.3$, and $b = 0.5$.}
\label{qq3}
\end{figure}

\begin{figure}[H]
\centering
\subfloat[Absolute error for $\hat{\rho}_n$]{
\includegraphics[width=.3\textwidth,height=3.5cm]{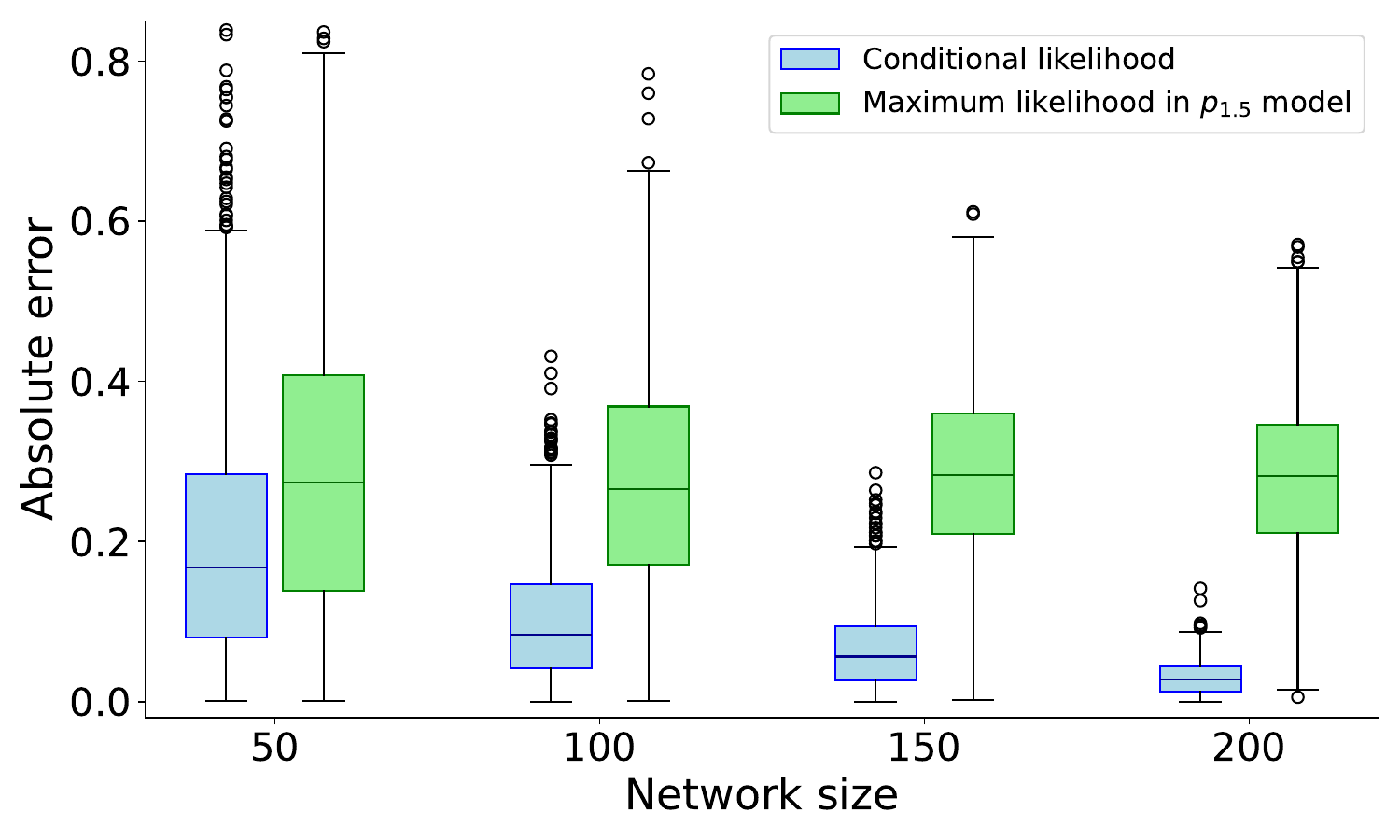}
}
\hfill
\subfloat[Absolute error for  $\hat{\gamma}_1$]{
\includegraphics[width=.3\textwidth,height=3.5cm]{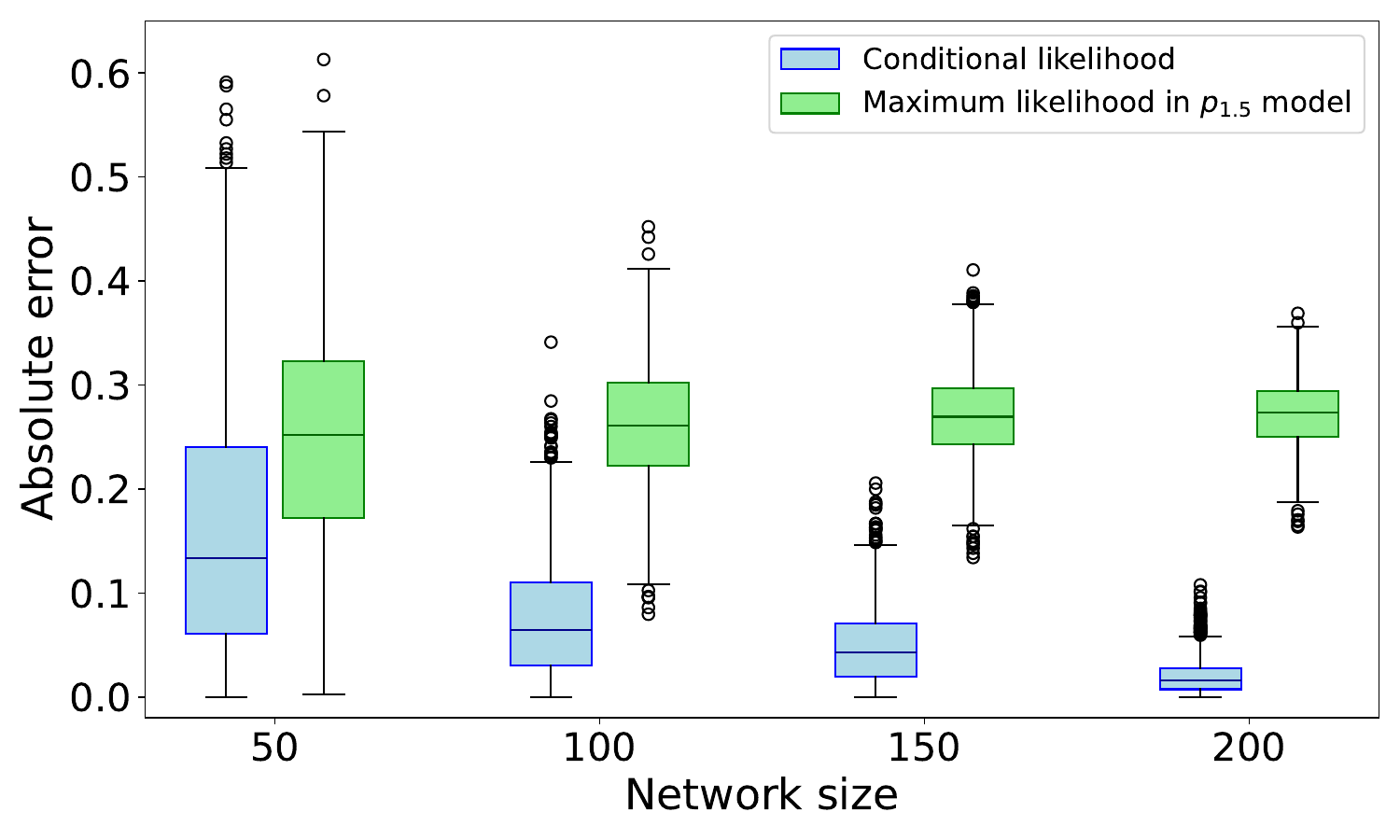}
}
\hfill
\subfloat[Absolute error for  $\hat{\gamma}_2$]{
\includegraphics[width=.3\textwidth,height=3.5cm]{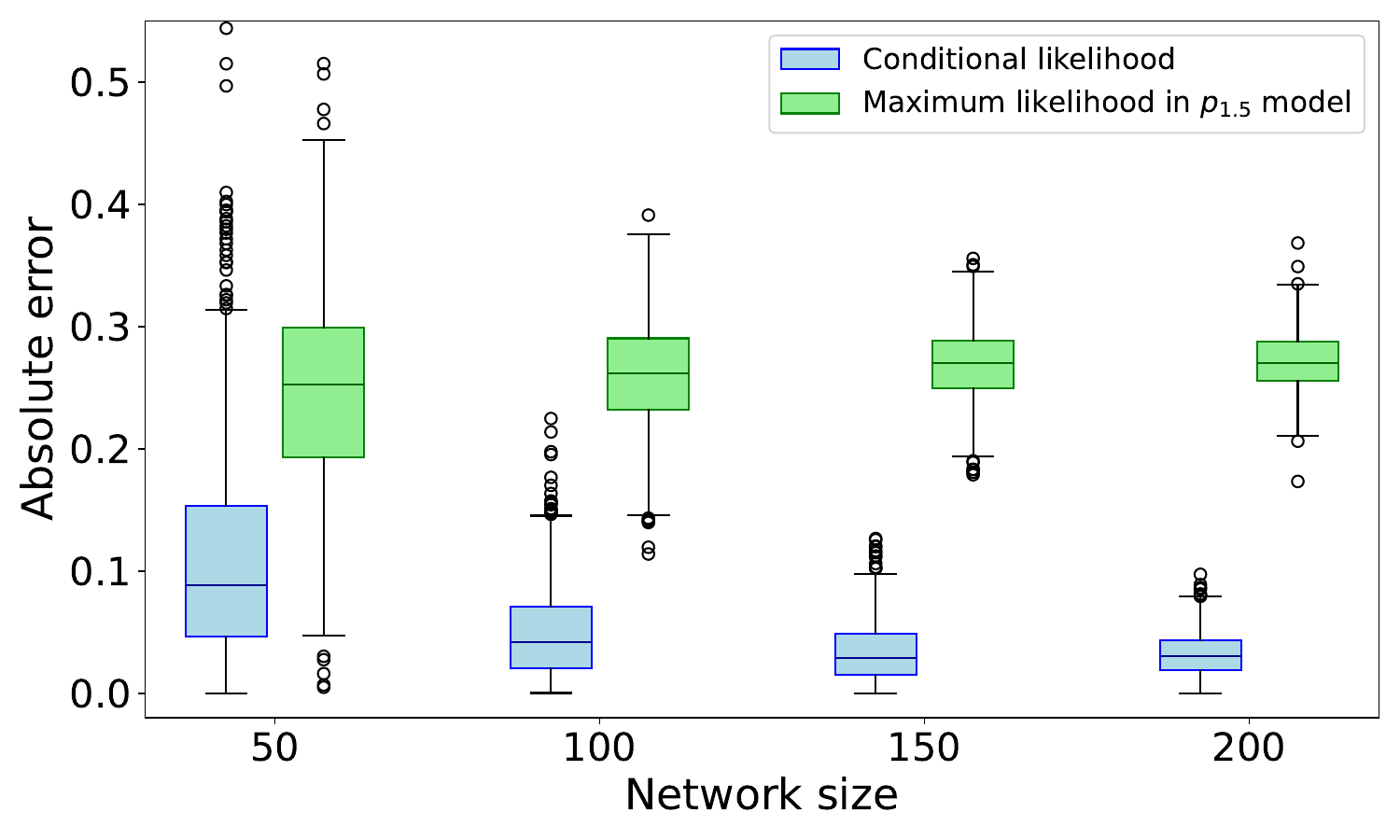}
}
\caption{Comparison with $p_{1.5}$ model across different network sizes when $a = 0.3$, and $b = 0.5$.}
\label{error_p1.5_3}
\end{figure}

Figure~\ref{error_plot_2} shows the absolute estimation errors across different network sizes for the setting $a = 0.2$ and $b = -0.1$, while Figure~\ref{error_plot_3} presents the results for $a = 0.3$ and $b = 0.5$. The corresponding empirical coverage and QQ plots are shown in Table~\ref{Table: coverage2} and Figure~\ref{qq2} for the first setting, and in Table~\ref{Table: coverage3} and Figure~\ref{qq3} for the second. Finally, we compare our model with the $p_{1.5}$ model. The results are detailed in Figure \ref{error_p1.5_2} and \ref{error_p1.5_3}. These findings further support the validity of Theorem~\ref{Inference} and the robustness of our model.

\begin{table}[htbp]
    \centering
    \begin{tabular}{lrrrrrrrrrrrr}
        \toprule
        & \multicolumn{2}{c}{$n = 50$} && \multicolumn{2}{c}{$n = 100$} && \multicolumn{2}{c}{$n = 150$} && \multicolumn{2}{c}{$n = 200$} \\
        \cmidrule{2-3} \cmidrule{5-6} \cmidrule{8-9} \cmidrule{11-12}
        & Coverage & Width && Coverage & Width && Coverage & Width && Coverage & Width \\
        \midrule
        $\hat{\rho}_n$       & 98.2\% & 0.596 && 97.5\% & 0.282 && 95.9\% & 0.187 && 96.8\% & 0.141 \\
        $\hat{\gamma}_1$     & 98.4\% & 0.720 && 97.7\% & 0.327 && 95.9\% & 0.216 && 96.2\% & 0.163 \\
        $\hat{\gamma}_2$     & 98.1\% & 0.468 && 96.2\% & 0.211 && 95.5\% & 0.138 && 97.1\% & 0.104 \\
        \bottomrule
    \end{tabular}
    \caption{Empirical coverage (nominal 95\%) and median lengths of confidence intervals for $\hat{\vartheta}_n$ when $a = 0.2$ and $b = -0.1$.}
    \label{Table: coverage2}
\end{table}

\begin{table}[htbp]
    \centering
    \begin{tabular}{lrrrrrrrrrrrr}
        \toprule
        & \multicolumn{2}{c}{$n = 50$} && \multicolumn{2}{c}{$n = 100$} && \multicolumn{2}{c}{$n = 150$} && \multicolumn{2}{c}{$n = 200$} \\
        \cmidrule{2-3} \cmidrule{5-6} \cmidrule{8-9} \cmidrule{11-12}
        & Coverage & Width && Coverage & Width && Coverage & Width && Coverage & Width \\
        \midrule
        $\hat{\rho}_n$       & 98.0\% & 0.837 && 96.2\% & 0.373 && 96.3\% & 0.241 && 96.0\% & 0.179 \\
        $\hat{\gamma}_1$     & 97.5\% & 0.727 && 95.8\% & 0.317 && 96.6\% & 0.200 && 96.1\% & 0.146 \\
        $\hat{\gamma}_2$     & 98.1\% & 0.485 && 98.1\% & 0.210 && 97.0\% & 0.133 && 96.9\% & 0.097 \\
        \bottomrule
    \end{tabular}
    \caption{Empirical coverage (nominal 95\%) and median lengths of confidence intervals for $\hat{\vartheta}_n$ when $a = 0.3$ and $b = 0.5$.}
    \label{Table: coverage3}
\end{table}

\end{document}